\documentclass[12pt]{extarticle}

\usepackage[sorting=none,style=numeric-comp]{biblatex}
\usepackage{enumitem}
\usepackage{geometry}

\usepackage{fancyhdr}
\usepackage{mathtools}
\usepackage{subfigure}
\usepackage{amssymb}
\usepackage{amsmath}
\usepackage{hyperref}
\usepackage{amsfonts}
\usepackage{amssymb}
\usepackage{caption}
\usepackage{float}
\usepackage{bm}
\usepackage{graphicx}
\usepackage{color}
\usepackage{verbatim}
\usepackage{blindtext, multicol}
\usepackage{multirow}
\usepackage{epigraph}
\usepackage[dvipsnames]{xcolor}
\usepackage{physics}

\newcommand\Tstrut{\rule{0pt}{2.9ex}} 

\usepackage{cancel}
\usepackage{etoolbox}

\newcommand{\CancelTo}[3][]{%
  \ifblank{#1}{}{%
    \renewcommand{\CancelColor}{#1}%
  }
  \cancelto{#2}{#3}%
}

\usepackage[english]{babel}
\usepackage{csquotes}
\DeclareUnicodeCharacter{0301}{\'{i}}
\usepackage[nottoc]{tocbibind}

\usepackage{academicons}
\usepackage{xcolor}
\usepackage{fontawesome5}
\definecolor{orcidlogocol}{rgb}{0.65, 0.807, 0.223}

\newcommand{\orcid}[1]{$\,$\href{https://orcid.org/#1}{\textcolor{orcidlogocol}{\faOrcid}}}
\def \blue {\color{blue}}
\usepackage[T1]{fontenc} 

\def\beq{\begin{equation}}
\def\eeq{\end{equation}}
\def\ber{\begin{eqnarray}}
\def\eer{\end{eqnarray}}
\def\benu{\begin{enumerate}}
\def\eenu{\end{enumerate}}

\def\d{\mathrm{d}}
\def\l{\left}
\def\r{\right}
\def\f{\frac}
\def\mpl{m_{\rm p}}

\def\mpc{\rm Mpc^{-1}}
\def\ns{n_{_{\rm S}}}

\def\Tre{T_{_{\rm re}}}

\def\wre{w_{_{\rm re}}}
\def\Nre{N_{_{\rm re}}}
\def\Rre{\rho_{_{\rm re}}}
\def\Nrd{N_{_{\rm RD}}}
\def\ng{n_{_{\rm GW}}}
\def\Og{\Omega_{_{\rm GW}}}

\def \red {\color{red}}
\def \blue {\color{blue}}
\def \Blue {\color{Blue}}

\def \dgray {\color{darkgray}}
\def \purple {\color{purple}}

\def \Brown {\color{Brown}}
\def \gray {\color{gray}}

\def \lleq {\lower0.9ex\hbox{ $\buildrel < \over \sim$} ~}
\def \ggeq {\lower0.9ex\hbox{ $\buildrel > \over \sim$} ~}

\hypersetup{
    colorlinks = true,
    linkcolor = Blue,
    citecolor = blue,
    urlcolor = Plum
}

\usepackage{authblk}

\begin{document}

\title{\textbf{{\Brown Cosmic Inflation: Background dynamics, \\ Quantum fluctuations and Reheating}}}
\author[]{{\bf {\dgray Swagat S. Mishra}} \orcid{0000-0003-4057-145X} \\ {\small (Email: \href{swagat.mishra@nottingham.ac.uk}{{\blue swagat.mishra@nottingham.ac.uk}})} }
\affil[]{{\small Centre for Astronomy and Particle Theory (CAPT), School of Physics and Astronomy, University of Nottingham, University Park Campus, Nottingham NG7 2RD, United Kingdom.}}

\maketitle

\begin{figure}[H]
\centering
\includegraphics[width=0.9 \textwidth]{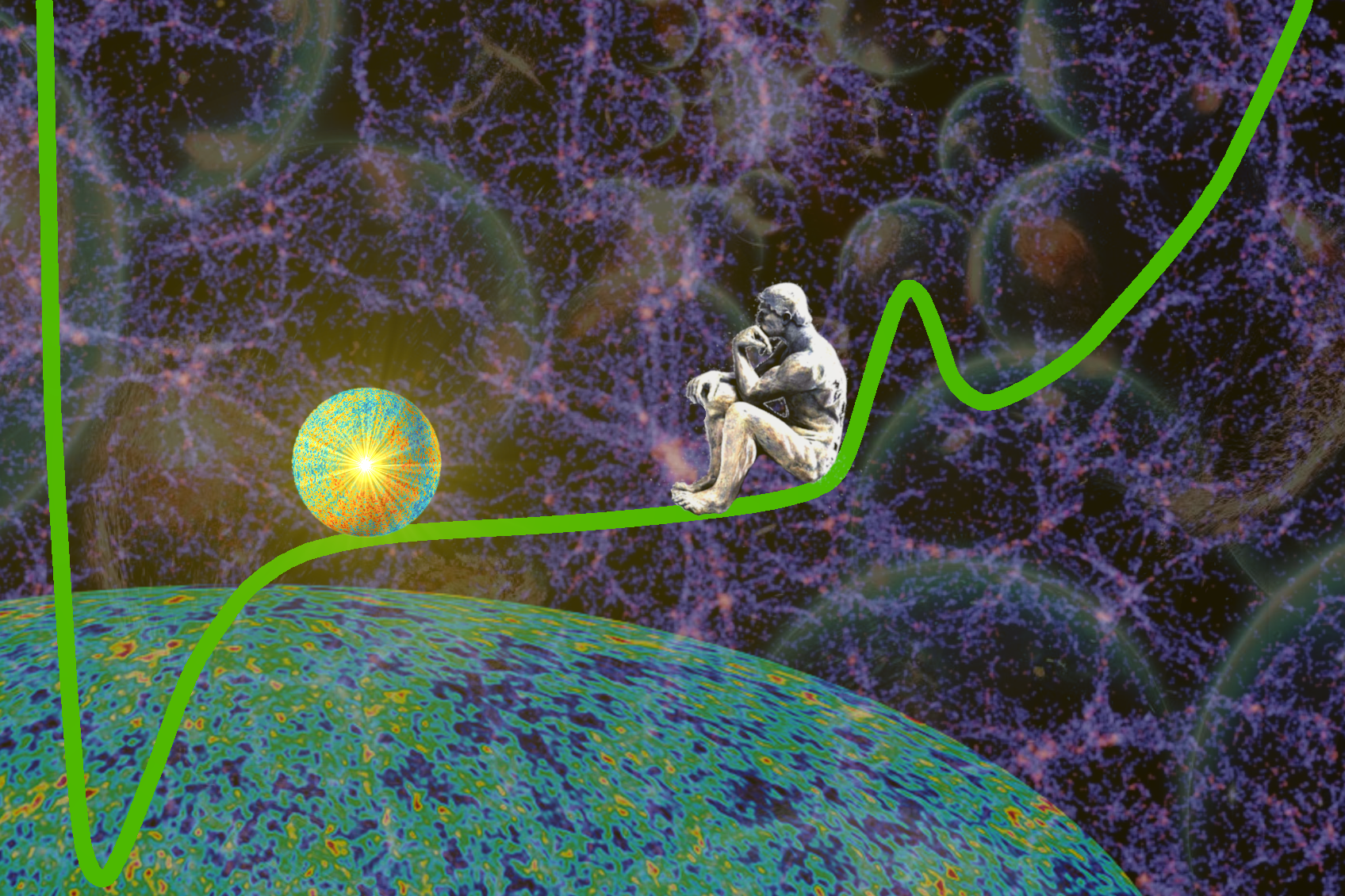} 
\label{fig:cover_CMB_inf}
\captionsetup{labelformat=empty}
\caption{\tiny{ {\gray [NASA/ESA, WMAP, PLANCK teams, Max Planck Institute for Astrophysics, Auguste Rodin.]}}}
\end{figure}

\vspace{0.05in}

\begin{center}
\begin{Large}
{\bf {\Blue Introductory Lecture Notes}}
\vspace{0.1in}

\end{Large}

\end{center}

\bigskip

\begin{figure}[H]
\centering
\includegraphics[width=0.15 \textwidth]{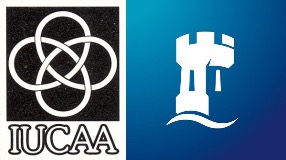} 
\label{fig:logo}
\end{figure}

\newpage

\vspace{0.3in}

\begin{abstract}
Cosmic inflation is a transient period of rapid accelerated expansion of space which has been hypothesized to have taken place prior to the hot Big Bang phase of the universe. It is the leading paradigm of the very early universe that provides  natural initial conditions for the hot Big Bang phase, both at the background and at the perturbation level.  In fact,
quantum vacuum fluctuations during inflation generate both scalar and tensor type primordial perturbations that are correlated over super-Hubble scales. The scalar fluctuations, upon their Hubble-entry, lead to the temperature and density inhomogeneities in the primordial plasma which eventually grow {\em via} gravitational instability to form the large-scale structure of the universe. Furthermore, the inflationary tensor fluctuations  constitute a background of {\em stochastic Gravitational Waves} (GWs) upon their Hubble-entry in the post-inflationary universe.  The correlation functions of   inflationary fluctuations constitute the primary observational tool to probe the high energy physics of the early universe. 

\medskip

Furthermore, in the inflationary paradigm all constituents of the universe at the subsequent radiation-dominated epoch were created during a  process known as {\em reheating}, which is supposed to have taken place  after the end of inflation. In the simplest inflationary scenario, the energy density of the universe was concentrated in a slowly evolving  scalar field, called the {\em inflaton}. After the end of inflation, the (almost) homogeneous inflaton condensate  started to oscillate coherently around the minimum of its effective potential.  The  oscillating inflaton condensate then decayed and transferred all  of its energy to the quanta of  other fields that are coupled to the inflaton. Subsequently,  the universe became thermalized at some high temperature, marking the commencement   of the hot Big Bang phase.  Since reheating is the intermediate stage between inflation and the radiation-dominated hot Big Bang phase,  it is associated with a number of important aspects of the hot Big Bang cosmology. 

\bigskip

 In these lecture notes, we provide a pedagogical introduction to some aspects of the inflationary cosmology including the background scalar field dynamics, generation of primordial  seed perturbations {\em via} quantum fluctuations during inflation, and the process of reheating after inflation in the single-field inflationary paradigm.

\end{abstract}

\vspace{0.1in}

\begin{center}
\begin{large}
{\bf Declaration}
\end{large}
\end{center}

These are introductory lecture notes primarily based on a set of lectures I delivered  as a supplementary course in Cosmology for the PhD coursework at IUCAA (India) in Jan/Feb 2024. Presentation of some of the topics originated from my lectures for the Master students I supervised in between 2017-23. The lectures are primarily intended for PhD students  who have had introductory courses on general relativity, quantum field theory, particle physics and cosmology.  I have made an attempt to compose a range of topics that are somewhat  complementary to many of the existing (excellent) lecture notes in the literature and consequently, a number of key aspects of the inflationary cosmology have not been included in the present version.  The approach here is relatively pedagogical and various concepts  are presented from a personal  perspective. Some topics are covered quite comprehensively, and hence I hope these notes will  also be useful to  researchers in the field.  The present version  may quite possibly contain typos and errors.   {\em I would be grateful to receive comments, suggestions, and typo corrections via email.}

\newpage
\vspace{0.3in}

\begin{center}

        ``Ages on ages, before any eyes could see 
        
        \vspace{0.06in}
        
year after year, thunderously pounding the shore as now.

\vspace{0.06in}

For whom? For what? On a dead planet

\vspace{0.06in}

with no life to entertain.

\bigskip
  
  Never at rest tortured by energy

\vspace{0.06in}
  
wasted prodigiously by the sun, poured into space.

\vspace{0.06in}

A mite makes the sea roar.

\bigskip

Deep in the sea all molecules repeat

\vspace{0.06in}

the patterns of one another till complex new ones are formed.

\vspace{0.06in}

They make others like themselves and a new dance starts.

\bigskip

Growing in size and complexity living things,

\vspace{0.06in}

masses of atoms DNA, protein,

\vspace{0.06in}

dancing a pattern ever more intricate.

\bigskip

Out of the cradle onto dry land,

\vspace{0.06in}

here it is standing: 

\vspace{0.06in}

atoms with consciousness; matter with curiosity.

\vspace{0.06in}

Stands at the sea, wonders at wondering:

\bigskip

 \textit{I, a universe of atoms} --  
 
 \vspace{0.06in}
 
 \textit{An atom in the universe}.''

\end{center}

  \begin{flushright}
        -- Richard P. Feynman
        \end{flushright}  
  
   \begin{small}
   
     \begin{flushright}
      
        \hfill
         \vfill
       

        \begin{flushright}
        {\color{Brown} \textbf{\textit{These lecture notes are dedicated to all the  small town and rural}}}
        \end{flushright}  
        \begin{flushright}
        {\color{Brown} \textbf{\textit{school  children aspiring to accomplish  something of  great}}}
        \end{flushright}  
        \begin{flushright}
        {\color{Brown} \textbf{\textit{value in their lives, despite the circumstances.}}}
        \end{flushright}  
        \medskip
        \bigskip
        \bigskip
        \bigskip
        \bigskip

    \end{flushright}  

\end{small}

\newpage

\section*{Acknowledgements}
 To begin with, I thank  Surhud More for the primary incentive for these lectures. I am  indebted to Yuri Shtanov, Edmund Copeland and Rafid  Mahbub for  spending their valuable time in reading the notes, and for providing insightful comments and suggestions on the preliminary version, which have greatly improved the quality of presentation in the current version.  Special thanks to many of my mentors, colleagues and students, including Varun Sahni, Edmund Copeland, Yuri Shtanov, Alexei Starobinsky, L. Sriramkumar, Sanil Unnikrishnan, Parth Bhargava, Alexey Toporensky, David Stefanyszyn, Anne Green, Paul Saffin, Antonio Padilla, Oliver Gould, Rafid Mahbub, Sanjit Mitra, Kandaswami Subramanian, Tarun Souradeep, Mohammad Sami, Prasant Samantray, Shabbir Shaikh, H. V. Ragavendra and  Bhavana Bhat, for various discussions and exchanges on the topic over the years.

 \medskip

 The cover page image has been designed by Siddharth Bhatt, compiling a number of figures (credit: NASA/ESA, WMAP, PLANCK teams, Max Planck Institute for Astrophysics, Auguste Rodin) using \texttt{GIMP}. All the figures in the main text have been primarily generated by the author, at times with crucial assistance from students and colleagues. In particular, I thank    Sanket Dave, Mohammed Shafi, and Siddharth Bhatt for their help in generating some of the figures.

\medskip

 I am  supported by a STFC Consolidated Grant [No. ST/T000732/1] as a postdoctoral Research Fellow at the University of Nottingham (UK). I am  grateful to IUCAA (India) for their hospitality and to Ranjeev Misra for his kind support. I further thank the accommodating staff members of the  Chef's Way cafe, where a significant portion of these lecture notes were compiled. 

\bigskip

For the purpose of open access, the
author has applied a CC BY public copyright licence to any Author Accepted Manuscript version arising.

\medskip

{\bf Data Availability Statement:} This work is entirely theoretical and has no associated data.
\newpage

 \tableofcontents

\begin{center}
\section*{Units, Notation and Convention}
\end{center}

\fbox{\begin{minipage}{39em}

\bigskip

\begin{itemize}

\item Einstein's Gravity ({\bf GR}) with metric signature $(-,+,+,+)$
\item {\bf Natural Units} $\hbar,c =1$, reduced Planck mass $m_p = \f{1}{\sqrt{8\pi G}}$
\item Hubble parameter $H$, Hubble radius $R_H$, comoving Hubble radius $r_h$, physical size of causal horizon $d_H$, comoving horizon $r_H$ 
\item $3$D spatial vectors are denoted by an overhead arrow mark, \textit{e.g.} the position and momentum  vectors in the comoving coordinates are  denoted as $\vec{x}$, $\vec{k}$, respectively.
\item Inflaton field $\varphi(t,\vec{x})$, homogeneous background field $\phi(t)$ such that $\varphi(t,\vec{x}) = \phi(t) + \delta \varphi(t,\vec{x})$
\item Fourier transformation of a scalar field
     $$ \Psi(t,\vec{x}) = \int  \f{{\rm d}^3 \vec{k}}{(2\pi)^3} ~ \Psi_k(t) ~ e^{i\, \vec{k}.\vec{x}} \quad \Rightarrow \quad \Psi_k(t) = \int\, {\rm d}^3 \vec{x} ~ \Psi(t,\vec{x}) ~ e^{-i\, \vec{k}.\vec{x}} $$
\item Power spectrum of vacuum quantum fluctuations of a field \ 
       $\boxed{ {\cal P}_\Psi(k) \, = \, \f{k^3}{2\pi^2} \, |\Psi_k|^2}$
\item  Variance \ \ \ $\boxed{\sigma_\Psi^2 = \int_{-\infty}^{\infty} \, {\rm d}\Psi ~ \Psi^2 ~ P[\Psi] \equiv \int_{k_{\rm min}}^{k_{\rm max}}  \f{{\rm d}k}{k} ~ {\cal P}_\Psi(k)   } $       
\item Scalar (comoving curvature) fluctuations during inflation $\zeta$,   scalar power spectrum ${\cal P}_{\zeta}(k)$
\item Tensor (transverse and traceless) fluctuations $h_{ij}$,  tensor power spectrum ${\cal P}_{T}(k)$
\item Conformal time ${\rm d} \tau =\f{{\rm d}t}{a(t)}$
\item Derivative \textit{w.r.t} cosmic time $\frac{\partial \Psi}{\partial t} \equiv \dot{\Psi}$ and \textit{w.r.t} conformal time $\frac{\partial \Psi}{\partial \tau} \equiv \Psi^\prime$
\item Derivatives \textit{w.r.t} $\phi$ are denoted as $\f{{\rm d}V(\phi)}{{\rm d}\phi} \equiv V_{,\phi}$, $\f{{\rm d}^2V(\phi)}{{\rm d}\phi^2} \equiv V_{,\phi\phi}$
\end{itemize}
\end{minipage} }

\section{Introduction}
\label{sec:intro}
Remarkable theoretical and observational advancements in cosmology over the past five  decades have led to the emergence of the standard paradigm of the universe, popularly known as the Big Bang Theory, within which the \textit{spatially flat ${\rm \Lambda}$CDM model} is often used as the fiducial/concordance model. This standard paradigm of cosmology describes the evolution of the universe reasonably well, starting from about 1 sec after the Big Bang, all the way until the present epoch.

However, the success of the concordance model relies upon the existence of hitherto unknown matter and energy constituents in the universe, in particular,  a non-baryonic  dark matter component that reinforces the formation of the large scale structure (LSS)  of the universe in the matter-dominated (MD) epoch,  as well as  a negative pressure dark energy component responsible for the observed accelerated expansion of space closer to the present epoch. In the concordance model, the dark matter is  often assumed to be cold and pressureless  on cosmological scales, and dark energy is assumed to be the cosmological constant $\Lambda$. The nature of  dark matter and dark energy are two of the greatest puzzles confronting the standard cosmological paradigm in the $21^{\rm st}$ century.

On the other hand, the success of the standard paradigm also heavily relies upon the  specific  primordial initial conditions at the beginning of the hot Big Bang phase. At the background level, these initial conditions take the form of (i) {\em  extreme spatial flatness} of the universe throughout  its evolution history and (ii) {\em high degree of spatial homogeneity and isotropy} over large  length scales  in the early universe. In particular, the average temperature of the Cosmic Microwave Background (CMB)  seems to be uniform up to a high degree of precision  across all sky. Since the standard Big Bang theory assumes that the universe was born in the state of being  radiation dominated (RD)   with decelerated expansion, both of the aforementioned initial conditions are highly non-trivial (unnatural) and appear to be extremely fine tuned. More importantly, we have also established  the existence of tiny primordial curvature fluctuations in the early universe ({\em via} temperature and polarisation fluctuations in the CMB over a range of angular scales) which seed the formation of the large-scale structure in the late universe {\em via} gravitational instability. The latest CMB observations by the  Planck collaboration \cite{Planck_overview,Planck_inflation} indicate that the initial curvature fluctuations were highly Gaussian, nearly scale-invariant and predominantly adiabatic in nature. Consequently, these specific properties of the primordial  fluctuations, combined with the extreme spatial flatness,  homogeneity and isotropy strongly point towards the existence of a non-standard (neither radiation nor matter dominated) transient epoch in the very early universe prior to the commencement of the  hot Big Bang phase.   In an expanding universe, assuming Einstein's theory of Gravity to hold at early times,  these primordial  initial conditions can be explained only if this transient period exhibited  rapid accelerated expansion of space, known as `{\em cosmic inflation}'.

The  inflationary paradigm \cite{Starobinsky:1980te,Guth:1980zm,Linde:1981mu,Albrecht:1982wi,Linde:1983gd,Linde:1990flp,Baumann_TASI} has emerged as the leading
scenario for describing the very early universe and for setting natural  initial conditions for the hot Big Bang phase.  One of the key predictions of the inflationary scenario is the unavoidable  quantum-mechanical 
production  of  primordial (scalar) curvature perturbations on super-Hubble scales \cite{Mukhanov:1981xt,Hawking:1982cz,Starobinsky:1982ee,Guth:1982ec}. In fact, the inflationary predictions  for the statistical properties of these scalar quantum fluctuations (made in the early 1980s) match extremely well with the latest precision CMB  observations by the Planck collaboration (data released in 2018) over a range of length scales, which consolidates inflation as a feasible    scenario of the pre-hot Big Bang universe \cite{Planck:2018vyg,Planck_overview,Planck_inflation,Tegmark:2004qd,Dodelson:2003ip}. In its simplest realization, inflation is usually assumed to be sourced by a single canonical scalar field $\varphi$ with a shallow self-interaction potential $V(\varphi)$ which is minimally coupled to gravity \cite{Linde:1990flp,Baumann_TASI,Martin:2013tda}.    At early times, when the inflaton is higher up in its potential, the universe inflates nearly exponentially. While, towards the end of inflation, the inflaton begins to roll rapidly down its potential before exhibiting coherent  oscillations around its minimum.

Another crucial  prediction of inflation is the generation of tensor perturbations {\em via} quantum fluctuations of the transverse and traceless part of the metric tensor during the accelerated expansion, which constitute a stochastic background of relic gravitational 
waves (GWs) \cite{Starobinsky:1979ty,Sahni:1990tx,Allen:1987bk}. Scalar and tensor perturbations generated during inflation create  distinctive  imprints on the primordial CMB radiation which can be used to deduce the scalar spectral index $n_{_S}$ and the
tensor-to-scalar ratio $r$ -- two of the most important observables which can be used to rule out competing
inflationary models \cite{Dodelson:1997hr,Martin:2013tda,Planck_overview,Planck_inflation}.

It is well known that the inflationary  GW  power spectrum at large scales provides us with important 
information about the nature of an inflaton field due to its direct relation to the inflaton potential \cite{Sahni:1990tx,Caprini:2018mtu}.
Of greater importance is the fact that their spectrum, $\Og(f)$ (defined as the  fractional energy density of GWs per logarithm interval of their frequency $f$ at the present epoch) and the spectral index 
$\ng = {\rm d}\ln{\Og}/{\rm d}\ln{f}$ at sufficiently small scales can serve
as key probes to the high energy physics of the post-inflationary primordial epoch. The primordial spectrum of relic gravitational radiation at small scales is exceedingly sensitive to the post-inflationary primordial
equation of state (EoS), $w$. In fact, the GW spectrum has distinctly different properties
for stiff/soft equations of state. For a stiff EoS, $w > 1/3$, the GW spectrum shows a blue tilt:
$\ng > 0$, that increases the GW amplitude on small scales. A softer EoS, $w < 1/3$, on
the other hand, leads to a red tilt, whereas the radiation EoS, $w = 1/3$, results in a flat
spectrum with $\ng \simeq 0$.

Another key aspect of inflationary cosmology is  the epoch of {\em reheating} during which the energy stored in the inflaton  is  transferred into the radiative degrees of freedom of the hot Big Bang phase. The post-inflationary history of the universe, prior to the commencement of Big Bang Nucleosynthesis (BBN), remains observationally inaccessible at present, despite a profusion of theoretical progress \cite{Albrecht:1982mp,Kolb:1990vq, Kofman:1994rk,Shtanov:1994ce,Kofman:1996mv,Kofman:1997yn,Greene:1997fu,Amin:2014eta,Lozanov:2019jxc} in this direction. 
It is expected, however,  that the post-inflationary universe in between the end of inflation and  the beginning  of the radiation domination  phase passed through a series of physical epochs \cite{Antusch:2021aiw},
each of which can be characterized by an EoS, $w_i$. Potential relics from this primordial epoch in the form of primordial black holes, oscillons and gravitational waves will provide us with key information about the dynamics of reheating in the upcoming decades.

\bigskip

In the following, we will provide a pedagogical introduction to the inflationary dynamics. After a brief review of the standard cosmological paradigm in Sec.~\ref{sec:standard_BBT}, we will provide a discussion of the initial conditions for the early universe in Sec.~\ref{sec:fine_initial} and demonstrate that, in the standard radiation-dominated universe, the initial conditions need to be highly fine-tuned. Then we will discuss the proposal of the inflationary scenario in the pre-hot Big Bang epoch to address the background initial conditions in Sec.~\ref{sec:inf_resolution}. Sec.~\ref{sec:inf_dyn} is dedicated to  the background scalar field dynamics during inflation. In Sec.~\ref{sec:Inf_correlators}, we will then turn our attention to the most important predictions of the inflationary hypothesis: the generation  of scalar and tensor fluctuations that are correlated over super-Hubble scales.   Sec.~\ref{sec:inf_dyn_obs} is devoted to a discussion on the latest observational constraints on inflation, while the post-inflationary reheating dynamics is introduced in Sec.~\ref{sec:Reheating}. We will provide a discussion of some of the key aspects of inflationary cosmology that have not been considered in detail in these notes in Sec.~\ref{sec:Discussion} and conclude by providing references to  supplementary material on inflation in Sec.~\ref{sec:Supp}. 

Various appendices  provide more technical as well as supplementary information on the subject. Apps.~\ref{app:Hubble_z} and \ref{app:N*_duration} discuss the thermal history and post-inflationary kinematics of the universe, respectively. App.~\ref{app:Hamilton_Jacobi} provides a discussion on the Hamilton-Jacobi formalism, with an application to the power-law inflation. Apps.~\ref{app:MS}, \ref{app:MS_analyt_sol} and \ref{app:massive_dS} are dedicated to the analytical treatment of fluctuations in the inflationary  and  de Sitter spacetimes.

\section{Standard cosmological model: a brief review}
\label{sec:standard_BBT}
As discussed in the introductory section, the  observable universe appears to be spatially flat, homogeneous, and isotropic on large cosmological scales throughout its probed history, as inferred from the LSS  and CMB observations as well as  BBN constraints. The space-time metric of  a nearly homogeneous and isotropic universe exhibits (approximate) spatial translation and rotation isometries and hence, is represented by the Friedmann-Lema\^{\i}tre-Robertson-Walker (FLRW) metric which takes the form
\beq
{\rm d}s^2 = - {\rm d}t^2 + a^2(t)\, \l[ \f{{\rm d}r^2}{1-K \, r^2} + r^2 \, {\rm d}\theta^2 + r^2 \, \sin^2{\theta} \, {\rm d}\phi^2 \r] \, ;
\label{eq:FLRW}
\eeq
where $K=0, \, \pm 1$ characterizes the uniform curvature of constant-time  spatial hypersurface\footnote{The definition of $K$ used here might be different from other lecture notes and books in the literature, \textit{e.g.} some sources use $\kappa = K/a_0^2$ where $a_0$ being the scalar factor at the present epoch.}.  $K=0$ corresponds to a spatially flat $\mathbb{R}^{(3)}$ universe, $K=1$ (positive spatial curvature) corresponds to a spatially closed $\mathbb{S}^{(3)}$ universe  and $K = -1$ (negative spatial curvature) corresponds to a spatially open hyperbolic $\mathbb{H}^{(3)}$ universe, see Refs.~\cite{Kolb:1990vq,Mukhanov:2005sc,Rubakov:2017xzr,Baumann:2022mni}. Specializing to the spatially flat FLRW metric for which $K=0$, we get
\beq
{\rm d}s^2 = - {\rm d}t^2 + a^2(t) \, \delta_{ij}\, {\rm d}x^i {\rm d}x^j \, .
\label{eq:flat_FLRW}
\eeq
The  rate of spatial expansion is described by the {\em Hubble parameter} which is defined as 
\beq
H = \frac{\dot{a}}{a} \, .
\label{eq:Hubble_kinematic}
\eeq
Expansion of the universe results in redshifting of the wavelength (or decreasing of the momentum along the   geodesics) of massless particles such as photons. The redshift is related to the scale factor {\em via}
\beq
1+z = \f{a_0}{a} \, ,
\label{eq:z_a_relation}
\eeq
where $a_0$ is the present day scale factor for which $z=0$. We will not discuss the kinematics of FLRW space-time and refer the interested readers to standard textbooks on the subject, for example Ref.~\cite{Baumann:2022mni}.

Dynamics of space-time, sufficiently below the Planck scale,  is  governed by the Einstein's Field Equations\footnote{We will not discuss modified or extended theories of gravity in these lectures. Interested readers should see Ref.~\cite{Sotiriou:2008rp} and references therein.}
\beq
G_{\mu \nu} \equiv R_{\mu\nu} - \f{R}{2} \, g_{\mu\nu}  = \f{1}{m_p^2} \, T_{\mu \nu} \, .
\label{eq:GR_field}
\eeq
For a homogeneous and isotropic distribution of matter-energy constituents described by the energy-momentum tensor $T^{\mu}_{\;\:\;\nu} \equiv \mathrm{diag}\left(-\rho, \,  p,   \, p, \,  p\right)$, and the metric represented by the  FLRW line element given in Eq.~(\ref{eq:FLRW}), the time-time and space-space components of  Einstein's field equations take the form
\begin{eqnarray}
H^2 &=& \frac{1}{3{m_p^2}}\,\rho - \frac{K}{a^2}\, , \label{eq:Friedmann_1_gen} \\
\frac{\ddot{a}}{a} &\equiv& \dot{H} + H^2  = -\frac{1}{6{m_p^2}}\,(\rho + 3 \, p)  \, .
\label{eq:Friedmann_2_gen}
\end{eqnarray}
 $\rho$ and $p$ are energy density and pressure respectively.  Eqs.~(\ref{eq:Friedmann_1_gen}) and (\ref{eq:Friedmann_2_gen}) are called the Friedmann equations. Eq.~(\ref{eq:Friedmann_2_gen}) determines whether the expansion of the  universe  decelerates or accelerates, given the energy density and pressure of the constituents. Energy-momentum conservation leads to 
 \beq
\dot{\rho} + 3 \, H \, \l(  \rho+p \r) =0 \, ,
\label{eq:Friedmann_3_gen}
 \eeq
 which can also be derived by combining Eqs.~(\ref{eq:Friedmann_1_gen}) and (\ref{eq:Friedmann_2_gen}). In cosmology, we will often work with perfect fluids for which the pressure and density will be related {\em via} an equation of state (EoS) of the form
 \beq
p = w \, \rho \, ,
\label{eq:EoS_gen}
 \eeq
 where the equation of state parameter $w \in \l[ -1, \, 1\r]$. Consequently, we can write Eq.~(\ref{eq:Friedmann_3_gen}) as 
 \beq
\dot{\rho} + 3 \, H \, \l( 1+w \r) \rho =0 \, ,
\label{eq:Friedmann_3_gen_w}
 \eeq 
 whose general solution for  constant $w$  can be written as 
 \beq
\rho(a) = \rho_1 \, \l( \f{a}{a_1} \r)^{-3(1+w)} \, , 
\label{eq:rho_w_gen}
 \eeq
 where $\rho_1$ is the energy density at some reference epoch $a_1$ which is usually taken to be the scale factor of the present epoch.
 Incorporating the above expression into Eq.~(\ref{eq:Friedmann_1_gen}), we obtain
  \beq
a(t) = a_1 \, \l( \f{t}{t_1} \r)^{\f{2}{3(1+w)}} \, ; \quad H(a) = H_1 \, \l( \f{a}{a_1} \r)^{-3(1+w)/2} \, .
\label{eq:a_t_gen}
 \eeq
Note that the EoS parameter for radiation (relativistic constituents), matter (non-relativistic constituents), and 
 a cosmological constant are given by $w_r = \f{1}{3}$, $w_m \simeq 0$, and  $w_{\rm DE} = -1$ respectively, which leads to the following expressions for the energy density
 \beq
\rho(a) \propto \begin{cases} 1/a^4 \, ; &  {\rm RD} \\ 1/a^3 \, ; &  {\rm MD} \, ; \\ {\rm const.} \,  &  {\rm CCD} \, ,\end{cases} 
\label{eq:rho_w}
 \eeq
and the following expressions for scale factor in a spatially flat universe
 \beq
a(t) \propto \begin{cases} t^{1/2} \, ; &  {\rm RD} \\t^{2/3} \, ; &  {\rm MD} \, ; \\ e^{H\, t} \,  &  {\rm CCD} \, . \end{cases} 
\label{eq:a_t}
 \eeq
A pure cosmological constant-dominated (CCD) universe is called the `\textit{de Sitter}' (dS) universe for which the Hubble parameter $H$ is constant and the scale factor expands exponentially. In general, the universe can be filled with a number of different constituents at a given time (which is the case for our own universe). We assume that different constituents with densities $\rho_i$, and EoS parameters $w_i$, do not interact with each other (apart from gravity).  Then the Friedmann Eqs.~(\ref{eq:Friedmann_1_gen}), (\ref{eq:Friedmann_2_gen}) and energy momentum conservation Eq.~(\ref{eq:Friedmann_3_gen}) take the form 
\begin{eqnarray}
H^2 &=& \frac{1}{3{m_p^2}}\, \sum_i \, \rho_i - \frac{K}{a^2}\, , \label{eq:Friedmann_1_gen_i} \\
\frac{\ddot{a}}{a} &\equiv& \dot{H} + H^2  = -\frac{1}{6{m_p^2}}\, \sum_i \, \l( 1+3\, w_i \r) \, \rho_i  \, , \label{eq:Friedmann_2_gen_i} \\
\dot{\rho_i} &=& -3 \, H \, \l( 1+w_i \r) \rho_i  \, ,
\label{eq:Friedmann_3_gen_w_i}
\end{eqnarray}

\begin{figure}[htb]
\begin{center}
\includegraphics[width=0.75\textwidth]{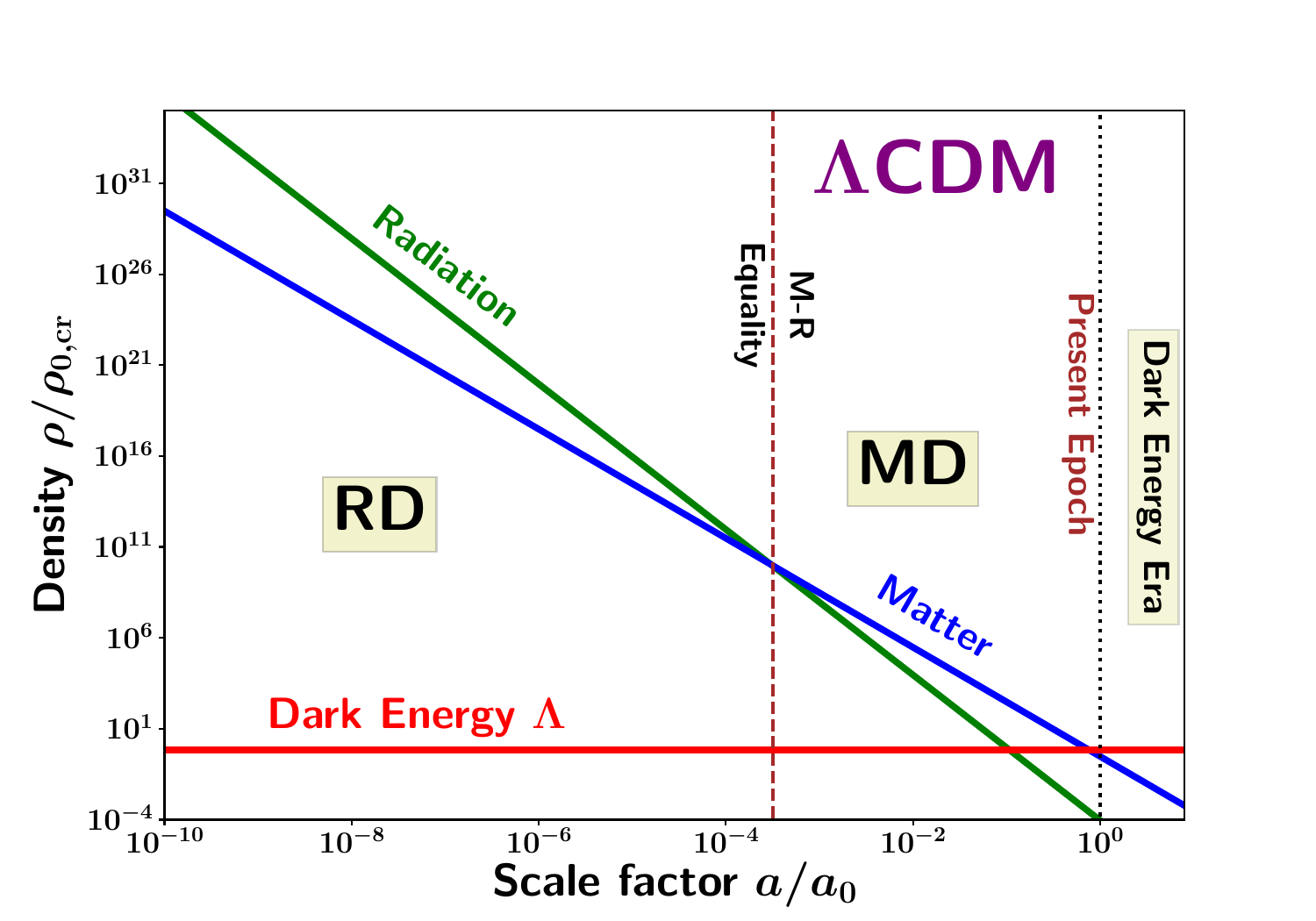}
\caption{The evolution of the density of radiation, matter and dark energy in the context of the flat $\Lambda$CDM concordance cosmology. Note that, while the density of the matter  and radiative degrees of freedom (represented by the solid blue and green curves respectively) evolve with time, the density of the cosmological constant $\Lambda$ (represented by the solid red curve) remains constant.}
\label{fig:cosmic_events}
\end{center}
\end{figure}
 
 A range of cosmological and astrophysical observations relevant to the  standard cosmological paradigm indicate that our universe was radiation dominated at early times during the hot Big Bang plasma phase.  It made a transition to the  matter dominated  epoch around a redshift $z \simeq 3400$, while continuing to exhibit  a decelerated expansion throughout most of its expansion history. However, more recently around the redshift $z\sim 0.7$ the expansion of the  universe started accelerating closer to the present epoch. A cosmological constant $\Lambda$ with $w_{\rm DE} = -1$ explains most cosmological observations very well\footnote{However, note that we are currently confronted  with the {\em Hubble tension} conundrum \cite{Planck:2018vyg,Planck_overview,2021ApJ...919...16F,2022MNRAS.511..662H,2022MNRAS.515L...1W,2022MNRAS.515L...1W,2016ApJ...826...56R,2018ApJ...855..136R,2019ApJ...876...85R,2021ApJ...908L...6R} and a modification to the standard scenario  might have some bearing upon the resolution of this tension.}. All the stable constituents of the Standard Model of particle physics account for  the radiation and baryon fraction of the energy budget of our universe. While pressureless cold dark matter (CDM) and a small positive cosmological constant ($\Lambda$) as the dark energy constitute the dominant energy budget of the universe at the present epoch. This is often referred to as  the standard or  concordance model of  cosmology,  fittingly  known as the {\em flat $\Lambda$CDM model}.

It is convenient to introduce dimensionless density parameters $\Omega_i$ which are defined as
\beq
\Omega_i = \f{\rho_i}{\rho_c}\, ; \quad {\rm with} \quad \rho_c = 3 \, m_p^2 \, H^2 
\eeq 
being the {\em critical energy density}. Hence the Hubble parameter for the flat ${\rm \Lambda}$CDM model can be written as 
 \beq
H = H_0 \, \sqrt{\Omega_{0r}  \l( a_0/a \r)^4 + \Omega_{0m}  \l( a_0/a \r)^3 + \Omega_{0\Lambda}} = H_0 \, \sqrt{\Omega_{0r}  \l( 1+z \r)^4 + \Omega_{0m}  \l( 1+z \r)^3 + \Omega_{0\Lambda}} \, ,
\label{eq:H_a_z_LCDM}
 \eeq
 where $\Omega_{0i}$ is the dimensionless density parameter of a constituent at the present epoch\footnote{In many textbooks and research papers, including Ref.~\cite{Planck_overview}, $\Omega_{0i}$ is often denoted by simply $\Omega_i$. However, in these lecture notes, we make a distinction between $\Omega_{0i}$ and $\Omega_i$, where the former represents the density parameter at the present epoch, while the latter refers to the same at any general epoch.}. Note that, in a spatially flat universe, 
 $$\sum_i \Omega_i (a) =  1$$
 at all cosmological epochs.
 The latest CMB observations reported in Ref.~\cite{Planck_overview} suggest that
 \beq
\Omega_{0r} \simeq 9.1\times 10^{-5} \, ; ~ \Omega_{0m} \simeq 0.315 \, ; ~ \Omega_{0\Lambda} \simeq  0.685 \, .  
\label{eq:Omega_CMB_planck}
 \eeq
The  expression for the Hubble parameter in Eq.~(\ref{eq:H_a_z_LCDM}) is accurate until about $z \simeq 10^9$. At higher redshifts deep inside the radiation dominated epoch the effective number of relativistic degrees of freedom changes, implying there will be corrections to the expression for the Hubble parameter (see App.~\ref{app:Hubble_z}). (This is also true for the Standard Model neutrinos,  some of which are known to be massive, and  hence, represent part of the non-relativistic matter at the present epoch, although they were relativistic (radiation-like) in the early universe.)

\medskip

 In short, the broad outline of our knowledge of modern cosmology  can be succinctly  summarised as follows:

\begin{itemize}
\item The universe is roughly homogeneous and isotropic on large cosmological  length scales ($> 100~{\rm Mpc}$)  and exhibits negligible spatial curvature throughout its (known/probed) expansion history. 

\item The universe has been expanding for the last 13.8 billion years, at least from the time when it was about $1$ second old until the present epoch. The metric   of the universe, at the background level, is well approximated by the  flat  Friedmann-Lema\^{\i}tre-Robertson-Walker (FLRW) line element given in Eq.~(\ref{eq:flat_FLRW}).

\item Observations of the Cosmic Microwave Background (CMB) radiation indicate that prior to the epoch  when our universe was  about  $370,000$ years old, corresponding to a redshift of $z \simeq 1100$, it was in a thermal hot dense plasma state. 

\item The early universe was thermal and  radiation dominated, historically  termed as the {\em hot Big Bang\/} phase,  which  made a transition to the matter dominated epoch when it was around $50,000$ years old, at a redshift $z_{\rm eq} \simeq 3400$. The expansion of the universe was decelerating during both the aforementioned epochs.

\item More recently, the universe has entered into a phase of  accelerated expansion  around the time when it was about $8$ billion years old.

\item In addition to baryons, photons and neutrinos, our universe also contains a gravitationally clustering  non-relativistic substance known as  {\em dark matter} (DM) which is pressureless (and hence `cold') on large extra-galactic scales. Observations suggest that DM is {\em non-baryonic}, in the sense that it  does not interact with the Standard Model constituents at low energy (apart from gravity, of course). DM plays a dominant role  in facilitating the formation of  large-scale structure  in the universe \cite{Bertone:2016nfn,Peebles:2017bzw,Green:2021jrr,Sahni:2004ai}.

\item Additionally, the cosmic recipe also includes a gravitationally non-clustering  negative pressure substance   termed  as  {\em dark energy}  (DE) in order to  fuel the late time accelerated expansion \cite{SupernovaCosmologyProject:1998vns,SupernovaCosmologyProject:1997zqe,SupernovaSearchTeam:1998fmf,Sahni:2004ai,Copeland:2006wr,Sahni:2006pa,}. The present day equation of state of the dark energy is close to that of a cosmological constant  $\Lambda$ \cite{Einstein:1917ce,Zeldovich:1968ehl,Weinberg:1987dv,Weinberg:1988cp,Krauss:1995yb,Sahni:1999gb,Peebles:2002gy,Padmanabhan:2002ji,Carroll:2000fy,Sahni:2004ai,Bousso:2007gp}.  The concordance cosmological model is consequently  termed as the flat  {\rm $\Lambda$}CDM model, or  more colloquially  as the `{\em Big Bang Theory}'. Evolution of the energy density of radiation, matter and dark energy are shown in  Fig.~\ref{fig:cosmic_events}.

\item Starting    with  an initial  spectrum of  almost scale-invariant, adiabatic, Gaussian  density fluctuations \cite{Planck_overview,Planck_inflation} that are correlated on  super-Hubble scales,  the standard cosmological paradigm successfully  describes the growth and formation of  structure in our universe,  at least on large length scales.

\item The nature of these special initial conditions for the concordance cosmology, in conjunction with CMB and LSS observations, strongly point towards a  transient period of  rapid accelerated expansion of space, known as {\em Cosmic Inflation}, which is supposed to have happened  in the very early universe \cite{Starobinsky:1980te,Guth:1980zm,Linde:1981mu,Albrecht:1982wi,Linde:1983gd,Linde:1990flp,Baumann_TASI} prior to the commencement of the radiative hot Big Bang phase.

\end{itemize}

Note that there is a classical singularity in the standard Big Bang theory \cite{Hawking:1970zqf} in the context of classical general relativity  where $a=0$ and the spacetime curvature diverges. We use this  classical singularity to set our initial time $t=0$. However, it is crucial to mention that going backwards in time, the curvature of the spacetime  of the universe will approach Planckian values before hitting the singularity and hence we expect substantial modifications to  Einstein's theory due to quantum gravity effects which are supposed to remove the classical singularity. Hence the existence of the classical singularity must not to be taken literally. For all practical purposes, we will assume that the hot Big Bang phase began at some high temperature below the Planck scale (and much above $1~{\rm MeV}$ in order to ensure successful BBN), namely 
$$1 ~ {\rm MeV} \,  \ll  \, T_i^{{\rm HBB}} \,  \leq \,  10^{19} ~ {\rm GeV} \, .$$

\bigskip

While the standard Big Bang theory has been remarkably successful in describing the evolution of the universe \cite{Planck_overview} both at the background and at the perturbation level, its success  relies  heavily upon the following initial conditions at the beginning of the hot Big Bang phase:
\begin{enumerate}
\item \textit{High degree of spatial flatness:} 
As mentioned before, the background  metric of the universe is consistent with being spatially flat  throughout the history of the universe. In the next section, we will quantify this statement and demonstrate that this leads to a high degree of fine-tuning of the initial expansion rate of the hot Big Bang universe, usually known as the \textit{flatness problem}. 
\item \textit{Extreme homogeneity of the initial hypersurface:} 
The success of BBN as well as the precision observations of the CMB indicate that the early universe was spatially homogeneous and isotropic  over large length scales. For example, the  temperature of the CMB is almost uniform  all across the sky. In the next section, we will demonstrate that the aforementioned fact leads to the so-called \textit{horizon problem}. 
\item \textit{Super-Hubble correlation of seed perturbations:} 
One of the greatest discoveries in modern  cosmology is the fact that the large scale structure (LSS)  of the universe is formed due to the gravitational collapse of tiny initial density fluctuations in the early universe. In fact the statistical properties of these primordial seed fluctuations have been measured very accurately by CMB and LSS observations and they have three key properties \cite{Planck_overview,Planck:2019kim,Planck_inflation}.
\begin{itemize}
\item
Primordial fluctuations are predominantly {\em  adiabatic}, which indicates that initial fluctuations in different constituents in the hot Big Bang phase were almost the same. This  further points to the fact that the seed fluctuations in the early universe emerged out of the perturbations of a single field, namely the comoving curvature fluctuations $\zeta(t,\vec{x})$ which is a gauge invariant quantity (\textit{i.e.} it remains invariant under the  gauge transformation  from one local frame to another, that connects  different ways of slicing and threading spacetime into a homogeneous background part and perturbations around that background.) 
\item
The power spectrum of $\zeta$ is {\em nearly scale-invariant}, namely
$${\cal P}_\zeta  = A_S \left(\frac{k}{k_*} \right)^{n_{_S}-1}  \, ; ~ k_* = 0.05 \, {\rm Mpc}^{-1} \, $$
with 
$$ A_S \simeq 2 \times 10^{-9} \, , ~ n_{_S}-1 \simeq -0.035 ~~ ({\rm small ~ {\red \textbf{red~tilt}}}) \, .$$
\item The primordial seed fluctuations are approximately {\em  Gaussian\/} with probability distribution function given by
$$ P[\zeta]  = {\cal B} \exp \left[\frac{-\zeta^2}{2\sigma_\zeta^2} \l( 1 +  f_{\rm NL} \,  \zeta + ... \r) \right] \, ,$$
where ${\cal B}$ is a normalisation factor. The  variance $\sigma_\zeta^2$ is given by
$$ \sigma_\zeta^2 = \int_{-\infty}^{\infty} \, {\rm d}\zeta ~ \zeta^2 ~ P[\zeta] \equiv \int_{k_{\rm min}}^{k_{\rm max}}  \f{{\rm d}k}{k} ~ {\cal P}_\zeta(k) \simeq 10^{-9} \, . $$
Latest CMB and LSS observations are consistent with $f_{\rm NL} = 0$ which implies that we have not found any deviation from Gaussianity so far. 
\item The temperature power spectrum of the  CMB on large angular scales indicates that the primordial fluctuations exhibit non-vanishing statistical correlations on super-Hubble scales. 
\end{itemize} 
These specific characteristics of the seed fluctuations demand  a physical mechanism to explain their origin. As we will discuss later in these notes, quantum fluctuations during the accelerated expansion in the inflationary scenario provide a natural mechanism to explain these observed properties of the primordial fluctuations.
\end{enumerate}

\section{Fine-tuning of initial conditions for the hot Big Bang}
\label{sec:fine_initial}

\subsection{High degree of  spatial flatness}
\label{sec:flatness_problem}
The fine-tuning of the initial expansion rate of the universe for the hot Big Bang phase can be stated in terms of the  famous  flatness problem. The first Friedmann Eq.~(\ref{eq:Friedmann_1_gen}) for a spatially curved universe can  be written as 
\beq
|\Omega_{\rm tot}(a) - 1 | \equiv |\Omega_K(a)| =  \frac{|K|}{a^2\,H^2} = (aH)^{-2}  \, ,
\label{eq:Om_K_Friedmann}
\eeq
where $\Omega_K$ corresponds to the dimensionless  density parameter associated with the spatial curvature. From  Eq.~(\ref{eq:a_t_gen}), for a single component dominated universe, we have 
\beq
\l( aH \r)^{-1} = \l( a_i H_i \r)^{-1} \, \l( \f{a}{a_i} \r)^{\f{1+3w}{2}} \, .
\label{eq:aH_a}
\eeq
Eq.~(\ref{eq:Om_K_Friedmann}) can then be described more generally for a single component dominated universe as 
\beq
|\Omega_{\rm tot}(a) - 1 | = |\Omega_{\rm tot}(a_i) - 1 | \,  \l( \f{a}{a_i} \r)^{1+3w} ~ \propto ~ a^{1 + 3 w} \, ,
\label{eq:friedmann_flatness_general}
\eeq
which implies
 \beq
 |\Omega_{\rm tot}(a) - 1 | \propto \begin{cases} a^2 \, ; &  {\rm RD}\, , \\ a \, ; &  {\rm MD} \, . \end{cases} 
\label{eq:Om_K_RD_MD}
 \eeq
This shows that  $\Omega_{\rm tot}(a)$ continues to diverge away from $1$ in the radiation and matter-dominated epochs. Given the present day bound on $|\Omega_{\rm tot}(a_0) - 1 |$ (or equivalently, on $|\Omega_K(a_0)|)$ from  the latest CMB observations \cite{Planck:2018vyg},
$$ |\Omega_K(a_0)| \leq 10^{-2} \, ,$$
combined with the fact that $ |\Omega_K(a)| $ has been growing ever since the commencement of the hot Big Bang phase, it seems that the early universe was spatially flat up to an unnaturally high degree of precision. In order to quantify this, let us compute the bound on $ |\Omega_K(a_0)| $ at the beginning of Big Bang Nucleosynthesis, which corresponds to $z_{_{\rm BBN}} \simeq 4 \times 10^{9}$. 
$$ |\Omega_K(a_0)| \leq 10^{-2} \Rightarrow  \f{|\Omega_K(a_0)|}{|\Omega_K(a_{\rm BBN})|}  \, |\Omega_K(a_{\rm BBN})| \leq 10^{-2} \, , $$
which leads to
\beq
|\Omega_K(a_{\rm BBN})| \leq  6 \times 10^{-18} \, .
\label{eq:Om_K_BBN_bound} 
\eeq
Further back in time, at the grand unification (GUT) scale ($z_{_{\rm GUT}} \simeq 10^{29}$), the bound on the curvature parameter becomes
\beq
|\Omega_K(a_{\rm GUT})| \leq  10^{-56} \, .
\label{eq:Om_K_GUT_bound} 
\eeq
Our universe during  most of its expansion history   has been dominated by radiation and matter\footnote{We have ignored the existence of dark energy in the above analysis. Since the universe began to accelerate around $z\simeq 0.7$, the scale factor at that epoch is given by $a_{\rm DE} \simeq a_0/1.7 $. So the universe spends less than an e-fold of expansion being accelerated closer to the present epoch and hence, the effect of dark energy can be be safely ignored while describing the flatness problem. The same applies to the horizon problem as well.}, hence the spatial curvature according to equation (\ref{eq:friedmann_flatness_general}) should be extremely high today, even though it was small at early times. But observations suggest that the spatial curvature at the present epoch is still very small which implies that the spatial curvature in very early times must have been extremely small.  In fact, the simple calculation that we carried out above demonstrates that $|\Omega_{\rm tot}|$ must have been close to $1$ up to several decimal places of accuracy at the beginning of the hot Big Bang phase in the very early universe. Another way to state the fine-tuned spatial  flatness  is that the initial density $\rho_{i}^{\rm tot}$ of all matter-energy components of the hot Big Bang phase must have been close to  the critical density $\rho_c = 3m_p^2 H^2$
up to several decimal places. This also implies that the initial expansion speed $\dot{a}_i$ must be highly fine-tuned in order for $|\Omega_K|$ to remain much smaller than unity at the present epoch.
Such extreme fine-tuning of the initial expansion speed is known as the flatness problem and suggests an early epoch of expansion prior to the hot Big Bang phase which can dynamically drive $\Omega_K$ towards an extremely small value by the time the hot Big Bang phase commences.

Note that if $K=0$ exactly, then there is no flatness problem classically. However, $K=0$ corresponds to a spatially flat universe which is either infinite in extent or has a non-trivial topology. Additionally, if the universe began around the Planck scale, then large quantum gravitational fluctuations might alter the geometry/topology of spatial hypersurfaces. Furthermore, in quantum cosmology,  quantum creation of the universe  usually renders it to be either positively curved \cite{Hartle:1983ai,Hawking:1983hj,Vilenkin:1982de} or negatively curved \cite{Sasaki:1993ha,Sasaki:1993kt}.

\subsection{Horizon problem}
\label{sec:Horizon_problem}
CMB observations indicate that the temperature of the early universe, closer to recombination, is almost uniform across diametrically opposite points in the sky (with very tiny  variations in the  temperature of CMB  of the order $\Delta T/T \simeq 10^{-5}$).  
This suggests that the early universe was incredibly uniform over large distance scales. However,  this observation leads to a potential problem in the standard hot Big Bang theory, which originates  from the fact that light could have travelled only a finite distance from the beginning of the hot Big Bang phase until the emission of  CMB photons around the epoch of recombination. Additionally, the spatial hypersurface at the epoch of BBN is also supposed to be highly uniform and hence the same arguments may be extrapolated  to even earlier epochs.  To illustrate this point explicitly, let us define the physical size of the  causal horizon in cosmology.

\medskip

\begin{figure}[htb]
\centering
\includegraphics[width=0.8\textwidth]{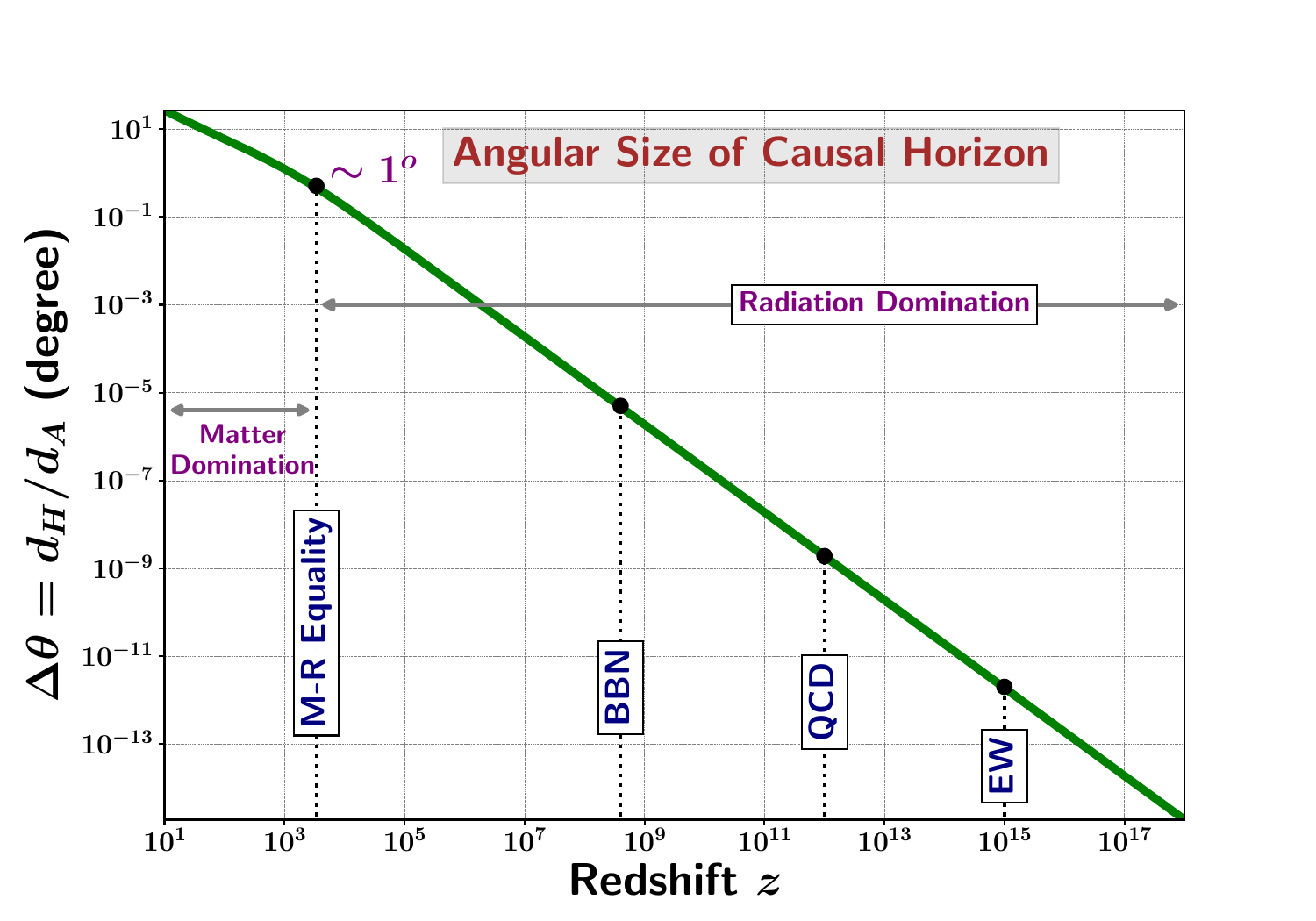} 
\caption{Angular size  of the causal horizon, given by Eq.~(\ref{eq:theta_horizon}), is plotted as a function of redshift for the standard ${\rm \Lambda}$CDM universe.}
\label{fig:horizon_problem_demonstration}
\end{figure}

The  physical size of the {\em causal horizon} (also called the \textit{particle horizon}) at any epoch $a(t)$ is defined as the maximum distance that any signal could have travelled from the beginning of the universe at $t=t_i$ until that epoch, and is given by 
\beq
d_H (t)  = a(t) \, r_H(t) \, .
\label{eq:dH_def}
\eeq
Where the  comoving causal horizon   $r_H(t)$ can be computed by noting that  signals propagate radially in an FLRW universe, so we have ${\rm d}\theta = {\rm d}\phi =0$. Additionally, for the propagation of light (or any other massless particle), ${\rm d}s^2 =0$ in Eq.~(\ref{eq:flat_FLRW}), which leads to 
$${\rm d}r = \f{{\rm d}t}{a(t)} \, ,$$
and hence the expression for the physical size of the causal horizon becomes 
\beq
d_H(t) \equiv a(t) \, \int_{t_i}^{t} \, \f{{\rm d}\tilde{t}}{a(\tilde{t})} = a \, \int_{a_i}^{a} \, \f{{\rm d}\tilde{a}}{\tilde{a}^2 \, H(\tilde{a})} = (1+z) \, \int_{z}^{z_i} \, \f{{\rm d}\tilde{z}}{H(\tilde{z})} \, .
\label{eq:d_H_t_a_z}
\eeq
Note that Eq.~(\ref{eq:d_H_t_a_z}) can also be written as 
\beq
d_H(a)  = a \, \int_{a_i}^{a} \, {\rm d} \, \ln(\tilde{a}) \, \l( \tilde{a}H \r)^{-1}  \, .
\label{eq:d_H_a}
\eeq
For a single component dominated decelerating universe with EoS parameter $w > -1/3$, and assuming $t_i = 0 = a_i$, we can carry out the integration in Eq.~(\ref{eq:d_H_a}) to obtain
\beq
d_H(a) =  d_H^i \, \l( \f{a}{a_i} \r)^{\f{3(1+w)}{2}}  \, ,
\label{eq:d_H_a_w} 
\eeq
where
\beq
 d_H^i = \l( \f{2}{1+3w} \r) \, \f{1}{H_i} \, .
\label{eq:d_H_ai}
\eeq
Since physical length scales grow proportionally to the scale factor in an expanding universe, $L (a) \propto a$, we have
\beq
\f{d_H(a)}{L(a)} \propto a^\f{1+3w}{2}  \, ,
\label{eq:d_H_L_a_w} 
\eeq
which  shows that the physical size of the causal horizon grows faster than length scales in a decelerating universe, for which $w > -1/3$. To be precise, the causal horizon grows as
 \beq
d_H(a) \propto \begin{cases} a^2 \, ; &  {\rm RD}\, , \\ a^{3/2} \, ; &  {\rm MD} \, . \end{cases} 
\label{eq:dH_a_RD_MD}
 \eeq
 Having derived these crucial properties of the causal horizon in a universe with decelerated expansion, let us compute the angular size subtended by the horizon at some epoch $z$ in the sky  with an observer at the present epoch at $z=0$, which is defined as 
 \beq
 \Delta \theta(z) \equiv \f{d_H(z)}{d_A(z)}  \,
\label{eq:theta_horizon}
\eeq
with
\beq
d_H(z) = \frac{1}{a_0\,H_0} \int_{z}^{\infty} \frac{{\rm d}\tilde{z}}{\sqrt{\Omega_{0m}\,(1+\tilde{z})^3 + \Omega_{0r} \, (1+\tilde{z})^4 + \Omega_{0 \Lambda} }} \, ,
\label{eq:d_Horizon}
\eeq
and the angular diameter distance is given by 
\beq
d_A(z) = \frac{1}{a_0\,H_0} \int_{0}^{z}\frac{{\rm d}\tilde{z}}{\sqrt{\Omega_{0m}\,(1+\tilde{z})^3 + \Omega_{0r} \, (1+\tilde{z})^4 + \Omega_{0 \Lambda} }} \, .
\label{eq:d_A}
\eeq
To be concrete, let us compute the angular size subtended by the horizon at recombination with respect to an observer at the present epoch --
 \beq
 \Delta \theta(z_{\rm CMB}) = \f{d_H(z_{\rm CMB})}{d_A(z_{\rm CMB})} = \f{\int_{z_{\rm CMB}}^{\infty} {\rm d}\tilde{z} ~ \l[ \Omega_{0m}\,(1+\tilde{z})^3 + \Omega_{0r} \, (1+\tilde{z})^4 + \Omega_{0 \Lambda} \r]^{-1/2}}{\int_{0}^{z_{\rm CMB}}  {\rm d}\tilde{z} ~ \l[ \Omega_{0m}\,(1+\tilde{z})^3 + \Omega_{0r} \, (1+\tilde{z})^4 + \Omega_{0 \Lambda} \r]^{-1/2}} \simeq 1.16^{\circ}.
\label{eq:theta_horizon_CMB}
\eeq
This implies that the regions in our CMB sky separated by no more than about $1^{\circ}$ angle were in causal contact, hence they could have been in thermal equilibrium  at the time of recombination, thereby allowing them to have the same temperature. However, as stressed before, the entire CMB sky seems to have the same average temperature.  This is known as the horizon problem.  Additionally, since the hot Big Bang phase is assumed to be very uniform also  at epochs earlier than recombination, the horizon problem becomes much more severe as we extrapolate  further back into the past. For example, we can compute the angular size of the horizon at any given  epoch in the early universe  using Eq.~(\ref{eq:theta_horizon}), with the result plotted in  Fig.~{\ref{fig:horizon_problem_demonstration}}.

\medskip

\begin{figure}[htb]
\centering
\includegraphics[width=0.8\textwidth]{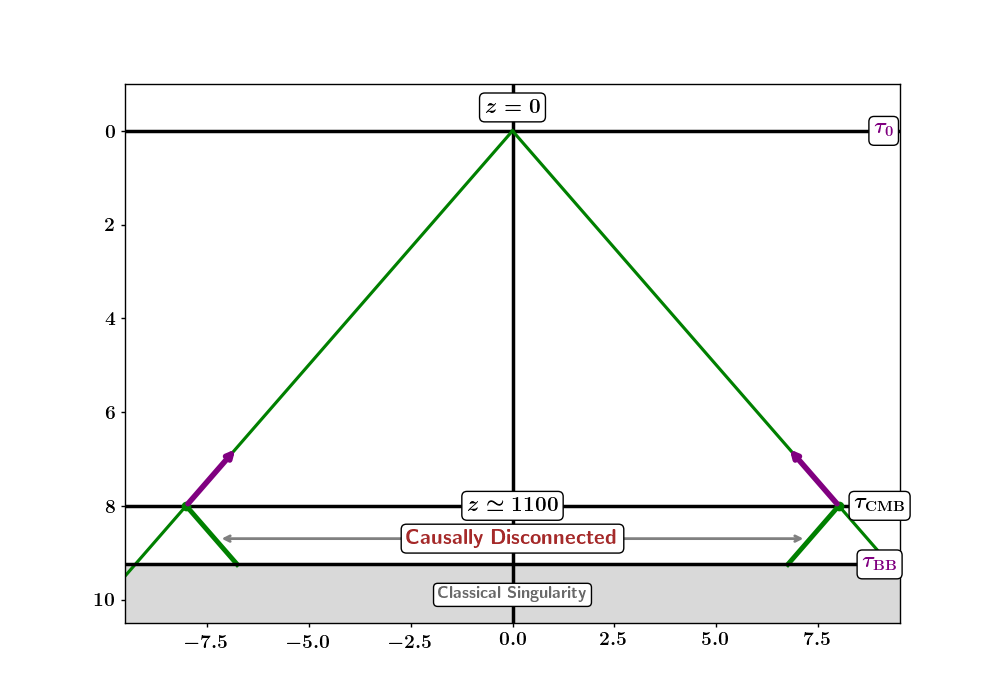} 
\caption{The conformal causal diagram of the universe in the standard cosmology where the universe began  in the hot Big Bang phase being radiation dominated. This figure schematically illustrates the fact that diametrically opposite points in the CMB sky could not have been in causal contact by the time of photon decoupling due to the lack of enough conformal time between the classical Big Bang singularity and the epoch of recombination.}
\label{fig:conformal_horizon_No_inf}
\end{figure}

Another intuitive way to  understand the Horizon problem is  to look at the conformal time elapsed between the beginning of the universe, until a given epoch, defined as,
\beq
\Delta\tau = \int_0^t \, \f{{\rm d}\tilde{t}}{a(\tilde{t})} = \int_0^a \, \f{{\rm d}\tilde{a}}{\tilde{a}^2 \, H(\tilde{a})}  \, \, 
\label{eq:conformal_HP}
\eeq
which is analogous to the comoving causal horizon $r_H = \f{d_H}{a}$, and hence gives us the separation between comoving coordinates that have been in causal contact since the beginning of the universe. The advantage of using conformal time in eliciting  the horizon problem is the following.  The FLRW metric given in Eq.~(\ref{eq:flat_FLRW}) can be written in terms of conformal time as 
\beq
{\rm d}s^2 =a^2(\tau) \l[ - {\rm d}\tau^2 +  \, \delta_{ij}\, {\rm d}x^i {\rm d}x^j \r] \, ,
\label{eq:flat_FLRW_tau}
\eeq
which is conformally equivalent to the metric of the Minkowski space-time. Hence, the causal structure of FLRW space-time  in terms of the conformal time is equivalent to that of the Minkowski space-time. So we can draw the usual space-time diagrams of  Special Relativity with light cones subtending  $45^\circ$ angle with the space and time axes. Hence, for diametrically opposite points in the CMB (which are at a fixed comoving distance from us) to be in causal contact with each other, the conformal time elapsed between the beginning of the universe and recombination should be greater than the conformal time elapsed between recombination and today. However, from Eq.~(\ref{eq:conformal_HP}), we have
 \beq
\Delta\tau(a) \propto \begin{cases} a \, ; &  {\rm RD} \, , \\ a^{1/2} \, ; &  {\rm MD} \, , \end{cases} 
\label{eq:Delta_tau_a_RD_MD}
 \eeq
which indicates that the conformal time grows with scale factor in the standard Big Bang theory and hence its size would be extremely small at the time of recombination, namely,
$$ \Delta\tau\l(a_{\rm BB} \to a_{\rm CMB}\r) \,  \ll \,  \Delta\tau\l( a_{\rm CMB} \to a_0\r) \, .$$
Hence, diametrically opposite points in the CMB sky could not have been in causal contact since the birth of the universe, leading to the Horizon problem. This is illustrated in Fig.~\ref{fig:conformal_horizon_No_inf}.

\bigskip

 So we conclude that, in the standard hot Big Bang theory, the early universe seems to be comprised of an unnaturally large number of causally disconnected regions, each possessing the same average temperature and density. This  again demands for a physical mechanism prior to the standard hot Big Bang phase which can bring the entire  sky into causal contact by the commencement of the hot Big Bang phase.

\subsection{Super-Hubble correlation of primordial  perturbations}
\label{sec:super_Hub_corr}

Towards the end of Sec.~\ref{sec:standard_BBT}, we mentioned that the primordial seed fluctuations of the comoving curvature perturbation $\zeta$ are nearly adiabatic, Gaussian and almost scale-invariant. However, more interestingly, these fluctuations have a non-vanishing two-point  auto-correlation (power spectrum or variance) at angular scales that are larger than the angular size of  the Hubble radius in the CMB sky. This can be easily seen from the low-$l$ (large scale) angular power spectrum of temperature fluctuations in the CMB sky (see Fig.~1 of Ref.~\cite{Planck_inflation}). The physical size of the Hubble radius is defined as
\beq
R_H = \f{1}{H} = \f{1}{H_i} \, \l( \f{a}{a_i}\r)^{\f{3(1+w)}{2}} \, ,
\label{eq:R_H_phys}
\eeq
while the size of the comoving Hubble radius is given by
\beq
r_h = (aH)^{-1} = (a_iH_i)^{-1} \, \l( \f{a}{a_i}\r)^{\f{1+3w}{2}} \, .
\label{eq:r_h_comov}
\eeq
Comparing Eq.~(\ref{eq:d_H_ai}) with Eq.~(\ref{eq:R_H_phys}), we find the relation between the causal horizon and Hubble radius in a decelerating single component dominated universe to be 
\beq
d_H = \l( \f{2}{1+3w} \r) \, R_H .
\label{eq:d_H_R_H_reln}
\eeq
Thus, a super-Hubble correlation  in a decelerating universe also implies a super-Horizon correlation of primordial fluctuations. This naturally demands for a physical explanation.   At this point, it is important to remind the reader of  the crucial distinction between causal horizon and Hubble radius.  Size of the causal horizon is determined by integrating over the expansion history, as can be seen from Eq.~(\ref{eq:d_H_t_a_z}). Consequently,  if the distance between two points in space at a given epoch is larger than the physical size of the causal horizon, then these two points were never in causal contact since the beginning of the universe. While, if two points are separated by a distance greater than the physical size of the Hubble radius at some epoch, then these points are not in causal contact at that epoch. However, this does not necessarily imply they were never in causal contact in the past. 

Hence the resolution of the conundrum of super-Hubble primordial correlations  involves the existence of  an epoch prior to the hot Big Bang phase where the  Hubble radius evolves differently as compared to the causal  horizon. In fact, as we will see, in an accelerating universe the natural spacetime dynamics leads to length scales constantly being stretched outside the Hubble radius without violating causality.

\section{The inflationary hypothesis}
\label{sec:inf_resolution}
 In this section, we will stress that  a sufficiently long epoch of accelerated expansion of space  prior to the hot Big Bang phase naturally addresses all of the above problems and provides  appropriate initial conditions for the hot Big Bang phase.  Hence, {\em the hot Big Bang phase  in the standard Big Bang theory is not the beginning of our universe, but rather the end of an earlier epoch of accelerated expansion}. In fact, this is the key message that these lecture notes are meant to convey.

The resolution of the problems of fine tuning of initial conditions emerges from the key observation that all three initial condition problems of the hot Big Bang stem from one common dynamics of the decelerating universe, that is the fact that the comoving Hubble radius $r_h = (aH)^{-1}$  increases with time, as can be seen from Eqs.~(\ref{eq:Om_K_Friedmann}),~(\ref{eq:d_H_a}) and (\ref{eq:r_h_comov}). Consequently, we can  address all of them by assuming a sufficiently long epoch of expansion history where the comoving Hubble radius decreases with time, namely
\beq
\f{{\rm d}}{{\rm d} t} \l(aH\r)^{-1} < 0 ~ \Rightarrow ~ 1+\f{\dot{H}}{H^2} > 0  ~ \Rightarrow ~ \ddot{a} > 0 ,
\label{eq:eh_dec_acc}
\eeq
which is equivalent to an accelerating or {\em inflating} universe. Let us explicitly demonstrate how inflation addresses the flatness and horizon problems, and generates fluctuations that are correlated over super-Hubble scales.  

\subsection{Addressing the flatness problem}
\label{sec:sol_inf_flatness}
The initial flatness problem of the hot Big Bang phase discussed in  Sec.~\ref{sec:flatness_problem} can be naturally solved  provided if there was a long enough period of inflation prior to the radiation domination, as discussed above. In particular, for a perfect fluid with EoS $w$ to drive inflation, we need 
$$ \f{\ddot{a}}{a} > 0 \quad \Rightarrow \quad 1 \, + \, 3\, w < 0 \quad \Rightarrow  \quad w < -1/3 \, ,$$
leading to 
\beq
|\Omega_{\rm tot}(a) - 1 | = |\Omega_{\rm tot}(a_i) - 1 | \,  \l( \f{a}{a_i} \r)^{-|1+3w|} ~ \propto ~ a^{-|1 + 3 w|} \, .
\label{eq:friedmann_flatness_soln}
\eeq
Hence, during inflation,  $|\Omega_{\rm tot} - 1 |$ or equivalently $|\Omega_K|$ decreases with time which results in $\Omega_{\rm tot}(a)$ being driven towards unity, as long as inflation lasts long enough. In particular, for near exponential inflation sourced by an effective cosmological constant $w\simeq -1$, we have
\beq
|\Omega_{\rm tot}(a_{\rm end}) - 1 | = |\Omega_{\rm tot}(a_i) - 1 | \,  e^{-2\, \Delta N} \, ,
\label{eq:friedmann_flatness_soln_qdS}
\eeq
where $\Delta N = \ln\l( a_{\rm end} / a_i \r)$ is the total number of e-folds of accelerated expansion during inflation. Assuming inflation began at some higher energy scale closer to the GUT scale,  with the curvature density being of the same order as the total density \textit{i.e.} of order unity, 
$$|\Omega_K(a_i)| = |\Omega_{\rm tot}(a_i)-1 |  \approx \mathcal{O}(1)  \, ,$$
Eq.~(\ref{eq:friedmann_flatness_soln_qdS}) leads to
$$ |\Omega_{\rm tot}(a_{\rm end}) - 1 | \approx  \mathcal{O}(1) \times  e^{-2\, \Delta N} \, .$$
We see that $ |\Omega_{\rm tot}(a_{\rm end}) - 1 | \leq 10^{-56}$ for $\Delta N \geq 64.47$. This demonstrates that starting from an order unity curvature term at the GUT scale, about $65$ e-folds of inflation is enough to result in $\Omega_{0K} \leq 10^{-2}$. In other words, if the hot Big Bang is not the beginning of our universe, but rather is the end of a sufficiently  long enough period of inflation, then we expect the spatial curvature to be negligible at the present epoch. 

\subsection{Addressing the horizon  problem}
\label{sec:sol_inf_horizon}
During the rapidly accelerated expansion of space, for which $\f{1+3w}{2} < 0$, the conformal time $\tau$ can be obtained by integrating Eq.~(\ref{eq:conformal_HP}) to be
\beq
\tau = \tau_1 + \f{1}{a_1 H_1} \, \l\vert \f{2}{1+3w}\r\vert \, \l[ 1 - \l(\f{a_1}{a}\r)^{|1+3w|/2}\r] \, , 
\label{eq:conformal_accelerating}
\eeq
where $a_1$ is the scale factor at some reference epoch $\tau_1$. Eq.~(\ref{eq:conformal_accelerating}) shows that the conformal time is negative and diverges at early times, $\tau \to -\infty$ as $a \to 0$. Thus, accelerated expansion yields a substantial amount of conformal time during the early history of the universe prior to $\tau=\tau_1$, which in turn brings together a  large region of comoving space into causal contact going back in the past. 

\begin{figure}[htb]
\centering
\includegraphics[width=0.85\textwidth]{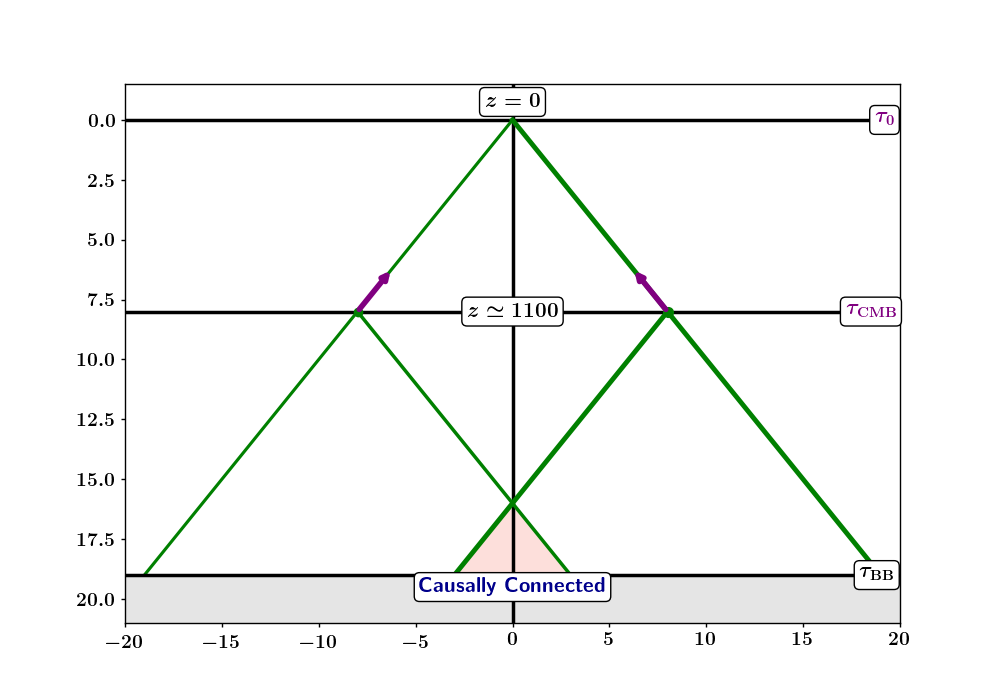} 
\caption{The conformal causal diagram of the universe in the presence of an early stage of exponential inflation before the commencement of the hot Big Bang phase. From this figure, it is clear that diametrically opposite points in the CMB sky could have easily been in causal contact if there was a sufficiently long period of inflation prior to the hot Big Bang phase, which facilitates for a substantial amount of conformal time to have elapsed in the past light cone. }
\label{fig:conformal_horizon_inf}
\end{figure}

In order to see this more explicitly, let us specialise to the case of exponential inflation, for which
$$ \tau = \tau_1 + \f{1}{a_1 H} - \f{1}{aH} \, .$$
Taking $\tau_1 = -1/(a_1 H)$, we derive the expression for the conformal time in  de Sitter spacetime to be
\beq
\boxed{ \tau = -\f{1}{aH} }\, ,
\label{eq:tau_dS}
\eeq
which tells us that the conformal time is large and negative at early times during inflation, while it approaches zero towards the end of inflation, namely, 
$$\tau \, : \,  -\infty \, \longrightarrow \,  0 $$
during inflation. Hence a sufficient amount of exponential expansion in the early universe results in bringing the entire CMB sky in causal contact, as shown schematically  in  Fig.~\ref{fig:conformal_horizon_inf}. 

Alternatively, the physical size of the causal horizon $d_H$ in an accelerated universe becomes much larger than the Hubble radius. In fact at the end of inflation, we have  
$$ d_H \sim e^{\Delta N} \, R_H \, ,$$
where $\Delta N$ is the total number of e-folds of expansion of space during inflation. Hence the horizon size at any epoch in the post inflationary universe is much larger than the Hubble radius, which naturally resolves the horizon problem discussed in Sec.~\ref{sec:Horizon_problem}.

\subsection{Generating super-Hubble fluctuations}
\label{sec:sol_inf_corr}
During the accelerated expansion of space, the comoving Hubble radius, $r_h$, starts shrinking with time, \textit{i.e.}
$$ r_h \propto a^{-\f{|1+3w|}{2}} \, $$
 as can be seen from from Eq.~(\ref{eq:r_h_comov}). Since  comoving length scales  corresponding to the fluctuations remain fixed with time, a shrinking comoving Hubble radius causes length scales, that were initially shorter than the Hubble radius (known as  {\em sub-Hubble modes}), to become larger than the Hubble radius  ({\em super-Hubble}) 
 at late times. Thus the rapid accelerated expansion of space during inflation dynamically stretches the wavelength of  (initial) sub-Hubble  fluctuations to large  super-Hubble scales at late times. A long enough period of inflation  ensures that by the end of inflation, all of the observable length scales in the CMB sky, which started out being sub-Hubble, would have become super-Hubble. On super-Hubble scales the comoving curvature fluctuations $\zeta$ remain frozen/constant, as we will see in Sec.~\ref{sec:inf_dyn_scalar_QF}. \\
 \begin{figure}[htb]
\centering
\includegraphics[width=0.8 \textwidth]{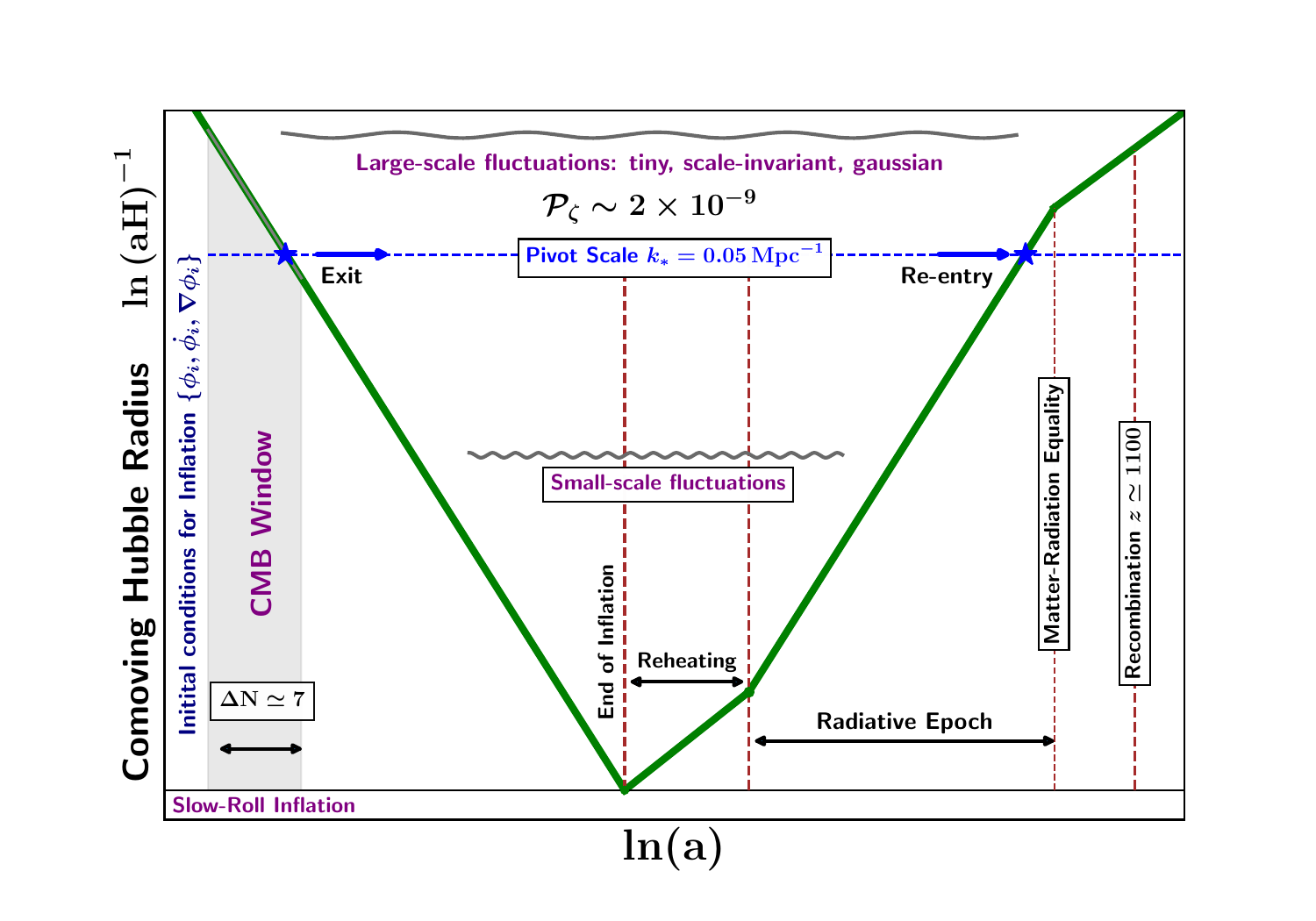} 
\caption{This figure schematically illustrates the evolution of the comoving Hubble radius $(aH)^{-1}$ with scale factor, both plotted on a  logarithmic scale (we have assumed  $H$ to be almost constant during inflation). $(aH)^{-1}$ decreases which causes physical scales to exit the Hubble radius during inflation. As  $(aH)^{-1}$ rises during the post-inflationary epochs,  physical scales begin to re-enter the Hubble radius. The CMB pivot scale, as used by the Planck mission, is set at $k_* = 0.05 \, {\rm Mpc}^{-1}$ and has been depicted by the dashed blue line.  }
\label{fig:comoving_hubble}
\end{figure}

\begin{figure}[htb]
\centering
\includegraphics[width=0.8\textwidth]{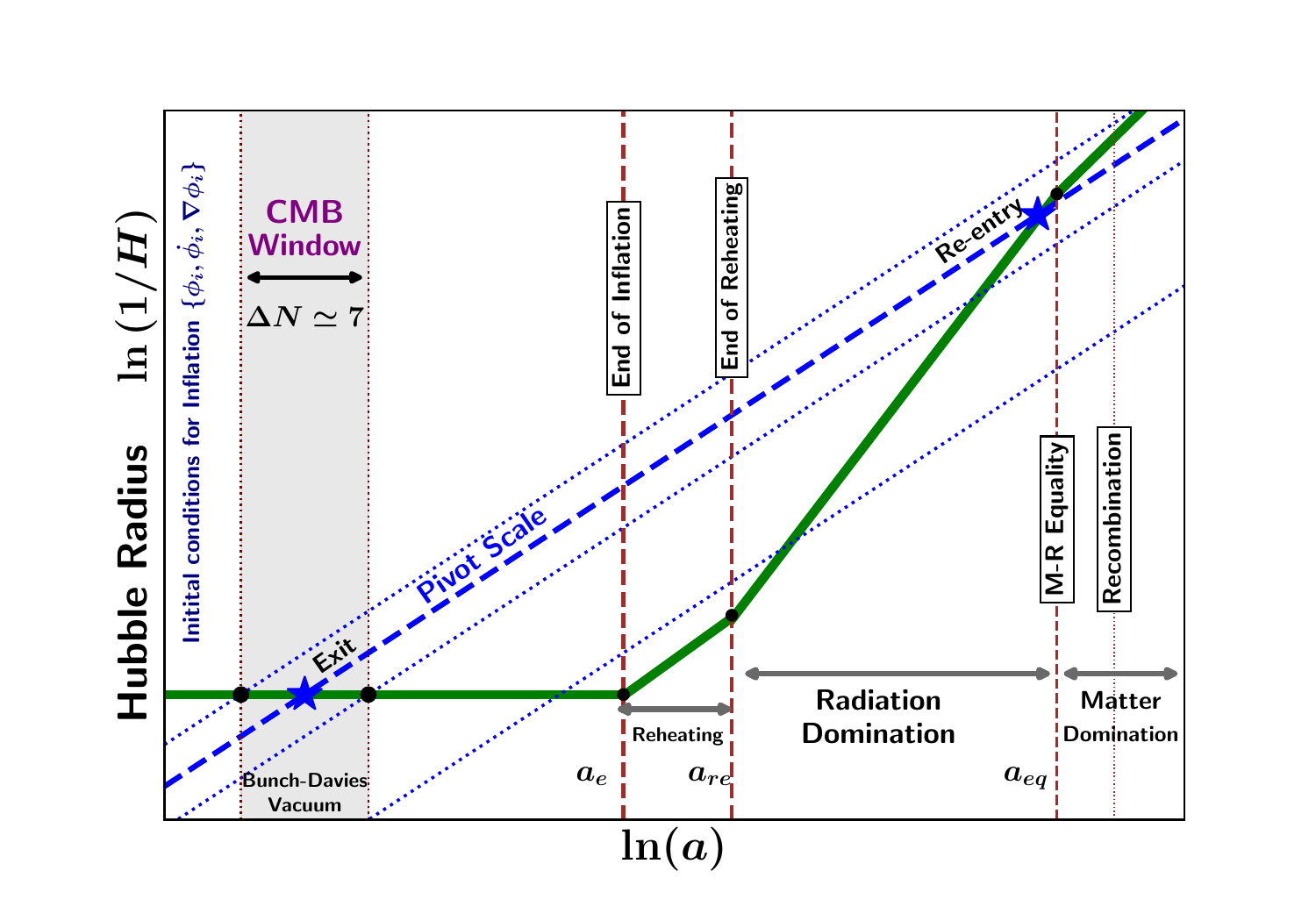} 
\caption{Evolution of the physical Hubble radius $1/H$ is shown  with scale factor (both plotted in logarithmic scale). The physical Hubble radius $1/H$ stays almost constant during inflation which causes physical scales to exit the Hubble radius during inflation. $1/H$ rises in the post-inflationary epoch and  the physical scales begin to re-enter the Hubble radius.  }
\label{fig:physical_hubble}
\end{figure}

 Inflation thus provides a causal mechanism to generate correlated primordial fluctuations on super-Hubble scales. The  final state  of   inflationary fluctuations then acts as the initial seed fluctuations for the hot Big Bang phase. The transition epoch between the end of inflation and the beginning of the hot Big Bang phase is called {\em reheating} (as mentioned before) which can last up to a few e-folds of 
 (decelerated) expansion.     Evolution of the Hubble radius during and after inflation is illustrated schematically in Fig.~\ref{fig:comoving_hubble} using  a comoving scale,  and in Fig.~\ref{fig:physical_hubble} using  the physical scale, assuming exponential expansion of space during inflation. The blue coloured lines indicate fluctuations of different wavelengths, while  the blue stars  indicate the Hubble-exit of the CMB pivot scale during inflation and, later, its Hubble-entry after the end of inflation.  Figs.~\ref{fig:comoving_hubble} and \ref{fig:physical_hubble}  demonstrate that long wavelength fluctuations, which made their Hubble-exit at early times during inflation,  re-entered the Hubble radius at  late times closer to the present epoch. On the other hand, shorter wavelength fluctuations became super-Hubble towards the end of inflation, and subsequently made their Hubble re-entry quite early after  inflation had ended. 

As the comoving Hubble radius began to increase (because of the decelerated expansion of space after inflation) and one by one,  these frozen  fluctuations became sub-Hubble again, they began evolving with time. Eventually, the fluctuations were amplified by gravitational instability to form the LSS of the universe. Note that several sources in the literature extensively refer to Hubble-exit and Hubble-entry as `horizon exit' and `horizon entry'. However, in these lecture notes, we have made an attempt to avoid such phrases and  stick to  the more accurate  phrases: Hubble-exit and Hubble-entry. 

 Hubble-exit and entry can be understood intuitively as follows:  during inflation, accelerated expansion rapidly stretched the wavelengths of fluctuations. Eventually the frequencies of oscillations of these modes fell below the rate of expansion of space. Or equivalently,  we say the modes became super-Hubble. Similarly after the end of inflation, the rate of expansion of space fell rapidly. Hence fluctuations which were super-Hubble can eventually began to oscillate once the rate of expansion fell below their frequencies of oscillations. Or equivalently, we  say the fluctuation modes became sub-Hubble 
 after their Hubble re-entry. 

 In general, a variety of functional forms of the scale factor $a(t)$ which supports accelerated expansion of space, \textit{i.e.} $\ddot{a} > 0$, can address the initial conditions problem of the hot Big Bang theory, as long as inflation lasts long enough. However, from the latest CMB observations we know that if inflation happened, then the accelerated expansion  of space must have been nearly exponential, $a(t) \propto e^{Ht}$, where $H$ is almost constant during inflation. Hence we will primarily focus on the  {\em quasi-de Sitter} expansion of space during inflation. Typically a period of quasi-de Sitter inflation lasting for at least 60--70 e-folds of expansion is enough to address the initial conditions for the hot Big Bang phase, as discussed in App.~\ref{app:N*_duration}.  The comoving Hubble radius plotted in Fig.~\ref{fig:comoving_hubble} is particularly convenient to refer to because of the symmetry between its  fall during inflation and rise in the radiation dominated epoch. In fact, Fig.~\ref{fig:comoving_hubble} demonstrates that  in order to ensure that  the largest observable length scales are sub-Hubble during inflation, the universe must  spend approximately same number of e-folds of inflationary evolution  as it does in the post-inflationary epoch.

\bigskip

In Einstein's gravity, an accelerated expansion of space  can be sourced by a negative pressure substance satisfying $w < -1/3$, as can be seen from Eq.~(\ref{eq:Friedmann_2_gen_i}). A natural candidate to drive inflation is a singlet scalar field $\varphi$ with self-interaction potential $V(\varphi)$ possessing   a low enough effective mass $m_{\rm eff}^2  \equiv \f{{\rm d}^2V(\varphi)}{{\rm d}\varphi^2} \lesssim H^2$. Since by now the curious  reader might have become bored with preceding long and detailed  discussions  on inflationary  kinematics,  let us move forward to delve into the rich physics of single field  inflationary dynamics without further delay.

\section{Scalar field dynamics of inflation}
\label{sec:inf_dyn}
In these lecture notes, we focus on the simplest inflationary  scenario sourced by  a single  scalar field  $\varphi$, known as the {\em inflaton field}, with a self-interaction potential $V(\varphi)$ which is minimally coupled to gravity. The full system of   {\bf Gravity\,+\,Scalar field}    is described by the action 
\beq
S[\varphi]=\int {\rm d}^{4}x\, \sqrt{-g}\; {\cal L}(X,\varphi) \, ,
\label{eq:Action_full}
\eeq
where
 the Lagrangian density ${\cal L}(X, \varphi)$
is a function of the field $\varphi$ and the
kinetic term
\beq
X \equiv \frac{1}{2} \, \partial_{\mu}\varphi\; \partial^{\mu}\varphi = \f{1}{2} \, \partial_{\mu}\varphi \; \partial_{\nu}\varphi \, g^{\mu\nu}  \, .\label{eq:X-phi}
\eeq
Varying Eq.~(\ref{eq:Action_full}) with respect to $\varphi$ results in the
Euler-Lagrange field equation for the evolution of the scalar field, given by
\beq
\frac{\partial {\cal L}}{\partial \varphi} - \left(\frac{1}{\sqrt{-g}}\right)\partial_{\mu}\left(\sqrt{-g}\frac{\partial {\cal L}}{\partial \left(\partial_{\mu}\varphi\right)}\right) = 0 \, .
\label{eq:EOM1}
\eeq
The energy-momentum tensor of the scalar field associated with the space-time translational invariance of the action in Eq.~(\ref{eq:Action_full}) is given by
\beq
T^{\mu\nu}
=\left(\f{\partial{\cal L}}{\partial X}\right)\, \left(\partial^{\mu}\varphi\; \partial^{\nu}\varphi\right)
- g^{\mu\nu}\, {\cal L}~.
\label{eq:SET}
\eeq
In Sec.~\ref{sec:inf_dyn_QF}, we will study linear perturbation theory during inflation. For that purpose, we will split\footnote{The splitting is made explicit in Sec.~\ref{sec:inf_dyn_QF}.} the system into time-dependent uniform background and  space (and time) dependent perturbations. We start with the discussion of the background scalar field $\phi(t)$ and space-time dynamics, described by the FLRW metric defined below.  

\subsection{Background dynamics during inflation}
\label{inf_dyn_background}
Specializing to a spatially flat FLRW universe
and a homogeneous part $\phi(t)$ of the  scalar field $\varphi(t,\vec{x})$, one finds
\beq 
{\rm d}s^2  =  -{\rm d}t^2 + a^{2}(t) \;  \left[{\rm d}x^2 + {\rm d}y^2 + {\rm d}z^2\right] \, ,
\label{eq:FRW}
\eeq
\beq
T^{\mu}_{\;\:\;\nu} \equiv \mathrm{diag}\left(-\rho_{_{\phi}}, \,  p_{_{\phi}} \,    \, p_{_{\phi}}, \,  p_{_{\phi}}\right) \, ,
\eeq
where the energy density $\rho_{_{\phi}}$, and pressure $p_{_{\phi}}$, of the homogeneous inflaton field  are given by
\begin{eqnarray}
\rho_{_{\phi}} &=& \left(\f{\partial {\cal L}}{\partial X}\right)\, (2\, X)- {\cal L}\label{eq:rho-phi} \, , \\
p_{_{\phi}} &=& {\cal L}\label{eq:p-phi},
\end{eqnarray}
with  $X = -\f{1}{2}\dot{\phi}^2$.
Evolution of the scale factor $a(t)$ is governed by the Friedmann equations
\begin{eqnarray}
 \left(\frac{\dot{a}}{a}\right)^{2}  \equiv H^2 = \f{1}{3m_p^2} \, \rho_{_{\phi}},\label{eq:Friedmann eqn1}\\
\frac{\ddot{a}}{a} \equiv \dot{H} + H^2 = -\f{1}{6m_p^2}\, \left(\rho_{_{\phi}} + 3\,p_{_{\phi}}\right),\label{eq:Friedmann eqn2}
\end{eqnarray}
where $H \equiv \dot{a}/a$ is the Hubble parameter and  $\rho_{_{\phi}}$ satisfies the conservation equation
\beq
{\dot \rho_{_{\phi}}} = -3\, H \left(\rho_{_{\phi}} + p_{_{\phi}}\right) ~.
\label{eq:conservation eqn}
\eeq

\medskip

\begin{figure}[hbt]
\begin{center}
\includegraphics[width=0.75\textwidth]{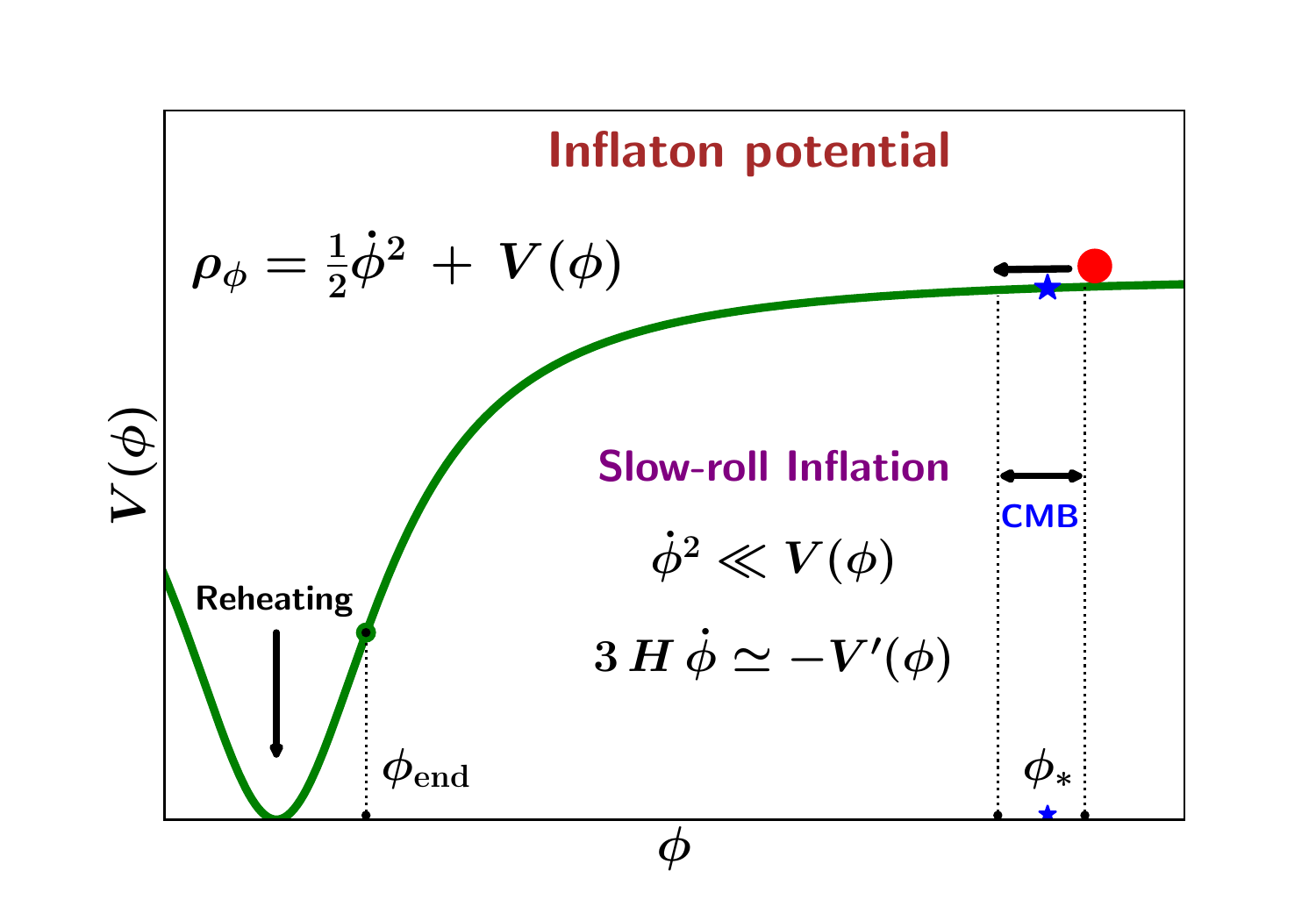}
\caption{This figure schematically depicts a prototypical inflaton potential (solid green curve).  The `CMB window' represents field values corresponding to the Hubble-exit epochs of  scales  $k\in \left[0.0005,0.5\right]~{\rm Mpc}^{-1}$ that are observationally accessible with the latest CMB missions.}
\label{fig:inf_pot_toy_vanilla}
\end{center}
\end{figure}

In the standard inflationary paradigm, inflation is sourced
 by a  scalar field $\phi$ with a  potential $V(\phi)$ (see Fig.~\ref{fig:inf_pot_toy_vanilla}) which is minimally coupled to gravity. For  a canonical scalar field, the Lagrangian density takes the form 
\beq
{\cal L}(X,\phi) = - X - V(\phi) \, .
\label{eq:Lagrangian}
\eeq
Substituting Eq.~(\ref{eq:Lagrangian}) into
Eq.~(\ref{eq:rho-phi}) and Eq.~(\ref{eq:p-phi}), the expressions for energy density and pressure become
\begin{eqnarray}
\rho_{_{\phi}} &=& \frac{1}{2}{\dot\phi}^2 +\;  V(\phi) \, ,\nonumber\\
p_{_{\phi}} &=& \frac{1}{2}{\dot\phi}^2 -\; V(\phi) \, , ~~
\label{eq:p-model}
\end{eqnarray}
consequently
the two Friedmann Eqs.~(\ref{eq:Friedmann eqn1}), (\ref{eq:Friedmann eqn2})  and the equation of motion Eq.~(\ref{eq:conservation eqn}) become
\begin{align}
H^2 = \frac{1}{3m_p^2} \, \rho_{\phi} = \frac{1}{3m_p^2} \left[\frac{1}{2}{\dot\phi}^2 +V(\phi)\right],
\label{eq:friedmann1}\\
\dot{H} \equiv \frac{\ddot{a}}{a}-H^2 = -\frac{1}{2m_p^2}\, \dot{\phi}^2 \, ,
\label{eq:friedmann2}\\
{\ddot \phi}+ 3\, H {\dot \phi} + V_{,\phi} = 0 \, .
\label{eq:phi_EOM}
\end{align}
The evolution of various physical quantities  during inflation is usually described with respect to the number of e-folds of expansion which is given by $N = \ln(a/a_i)$, where $a_i$ is some arbitrary  epoch at very early times during inflation.  It is informative to write and solve the Friedmann equations in terms of the number of e-folds  as follows:
\ber
H^2 &=& \frac{V(\phi)}{3\,m_p^2 -  \frac{1}{2}\l(\frac{{\rm d}\phi}{{\rm d}N}\r)^2} \, ;\label{eq:Friedmann_1_N} \\
\frac{{\rm d}H}{{\rm d}N} &=& -H \,\frac{1}{2m_p^2}\l(\frac{{\rm d}\phi}{{\rm d}N}\r)^2 \label{eq:Friedmann_2_N} \, ; \\
\frac{{\rm d}^2\phi}{{\rm d}N^2} &=& -\frac{{\rm d}\phi}{{\rm d}N} \l(  3 + \f{1}{H}\, \f{{\rm d}H}{{\rm d }N} \r) - \frac{1}{H^2} \, V_{,\phi}\label{eq:Friedmann_3_N} \, .
\eer
A better physical  quantity to depict  any epoch with scale factor $a$ during the inflationary expansion is the    {\em  number of e-folds  before the end of inflation} which is defined as 
\beq
N_e(a)  = \ln \l( \frac{a_e}{a} \r) =\int_{t}^{t_e} H(t) \, {\rm d}t,
\label{eq:efolds}
\eeq
where $H(t)$ is the Hubble parameter during inflation, and $a_e$ denotes the scale factor at the  
end of inflation, hence
$N_e  = 0$ corresponds to the end of inflation. 
 Typically a period of quasi-de Sitter (exponential) inflation\footnote{For a detailed account on the geometry of de Sitter spacetime including  its cosmological significance, see Refs.~\cite{Spradlin:2001pw,Bousso:2002fq,Anninos:2012qw}.}  lasting for at least 60--70 e-folds is required in order to address the problems of the standard hot Big Bang model discussed in Sec.~\ref{sec:fine_initial}. We denote $N_*$ as the number of e-folds (before the end of inflation) when the CMB pivot scale 
 \beq
 k_*=(aH)_*=0.05~\mpc
 \label{eq:k_pivot_CMB}
 \eeq
left the comoving Hubble 
radius during inflation. Typically  $N_* \in [50,\,60]$ depending upon the details of reheating after inflation (see Ref.~\cite{Liddle:2003as} and App.~\ref{app:N*_duration}).
The quasi-de Sitter like phase corresponds to the inflaton field rolling slowly down the 
potential $V(\phi)$. For a variety of  functional forms of the inflaton potential $V(\phi)$,  there exists a {\em slow-roll regime} of inflation,
ensured by the presence of the Hubble friction term \cite{Belinsky:1985zd,Mukhanov:2005sc,Baumann_TASI,Remmen:2013eja,Brandenberger:2016uzh, Mishra:2018dtg}  in Eq.~(\ref{eq:phi_EOM}).
 This slow-roll regime can be conveniently characterised by the first  two kinematic Hubble slow-roll parameters $\epsilon_H$, $\eta_H$, defined by ~\cite{Liddle:1994dx,Baumann_TASI}  
\ber
\epsilon_H &=& -\frac{\dot{H}}{H^2}= -\frac{{\rm d} \, \ln{H}}{{\rm d}N} = \frac{1}{2m_p^2} \, \frac{\dot{\phi}^2}{H^2}  \, ;
\label{eq:epsilon_H}\\
\eta_H &=& -\frac{\ddot{\phi}}{H\dot{\phi}}=\epsilon_H  - \frac{1}{2\epsilon_H} \, \frac{{\rm d}\epsilon_H}{{\rm d}N}  \, ,
\label{eq:eta_H}
\eer
where the slow-roll regime of inflation corresponds to 
\beq
\epsilon_H,~|\eta_H| \ll 1 \, .
\label{eq:slow-roll_condition}
\eeq
 Note that $\epsilon_H \geq 0$, while $\eta_H$ may be negative or positive depending on whether $\epsilon_H$ is increasing or decreasing with time. Using the definition of the Hubble parameter, $H=\dot{a}/a$, we have $\ddot{a}/a=\dot{H}+H^2=H^2(1 + \dot{H}/H^2)$. From the expression for $\epsilon_{H}$ in Eq.~(\ref{eq:epsilon_H}), it is easy to see that 
\beq
\f{\ddot{a}}{a} =  H^2  \, \big( 1 - \epsilon_H \big) \, ,
\label{eq:inf_acc_cond}
\eeq
which implies that the universe accelerates, ${\ddot a} > 0$, when $\epsilon_{H} < 1$.
Using equation (\ref{eq:friedmann1}), the expression for  $\epsilon_{H}$ in Eq.~(\ref{eq:epsilon_H}) reduces to $\epsilon_{H} \simeq \f{3}{2}\f{\dot{\phi}^2}{V}$  when
${\dot\phi}^2 \ll V$.  In fact, under the slow-roll conditions in Eq.~(\ref{eq:slow-roll_condition}), the Friedmann equations given by Eqs. (\ref{eq:friedmann1}) and (\ref{eq:phi_EOM}) take the form
\ber
H^2 &\simeq& \f{V(\phi)}{3 \, m_p^2}  \, ;  \label{eq:friedmann_SR1} \\
\dot{\phi} &\simeq& - \f{V_{,\phi}}{3 \, H} \, .
\label{eq:friedmann_SR2}
\eer
The last expression is a consequence of $|\eta_H| \ll 1$, which ensures that the inflaton speed does not change too rapidly, resulting in a long enough period of accelerated expansion before the end of inflation. The classical slow-roll dynamics corresponding to Eq.~(\ref{eq:friedmann_SR2}) during inflation is similar to the slow terminal motion of an object where the  attractive force and the restraining  drag/friction force are almost balanced with each other. The slow-roll conditions in Eqs.~(\ref{eq:friedmann_SR1}) and (\ref{eq:friedmann_SR2}) can be combined to obtain the {\em slow-roll trajectory} described by 
\beq
\boxed{ \dot{\phi} \simeq  - \f{m_p}{\sqrt{3}} \, \f{V_{,\phi}}{\sqrt{V(\phi)}}} \, .
\label{eq:SR_trajectory}
\eeq
It is well known that the slow-roll trajectory is actually a local attractor for  a number of  different models of inflation, see  Refs.~\cite{Belinsky:1985zd,Mukhanov:2005sc,Baumann_TASI,Remmen:2013eja,Brandenberger:2016uzh, Mishra:2018dtg}. The slow-roll regime of inflation is  described  more systematically   in terms of the {\em Hubble flow parameters} $\epsilon_n$ defined by
\beq
\epsilon_{n+1} = \f{{\rm d}\ln \, \epsilon_n}{{\rm d}N} \, ; \quad {\rm with} \quad \epsilon_1 = \epsilon_H \, .
\label{eq:Hubble_flow}
\eeq
Accordingly, from Eq.~(\ref{eq:eta_H}), the second Hubble flow parameter $\epsilon_2$ is related\footnote{Note  that many references use the symbol $\eta$ to represent $\epsilon_2$.} to $\epsilon_H$ and $\eta_H$ \textit{via}  
\beq
\epsilon_2 = 2 \l( \epsilon_H - \eta_H \r) \, .
\label{eq:epH_ep2_rel}
\eeq
Apart from the aforementioned kinematic slow-roll parameters, the slow-roll regime  is also often characterised  by the dynamical potential slow-roll parameters  \cite{Baumann_TASI}, defined by
\ber
\epsilon_{_V}  &=& \frac{m_{p}^2}{2}\left (\frac{V_{,\phi}}{V}\right )^2 \, ; \label{eq:SR_epV} \\
\eta_{_V} &=& m_{p}^2 \, \left( \frac{V_{,\phi\phi}}{V} \right) \, .
\label{eq:SR_etaV}
\eer
For small values of these parameters  $\epsilon_{H} \ll 1, \, \eta_{H} \ll 1$, from Eqs.~(\ref{eq:epsilon_H}) and  (\ref{eq:friedmann_SR2}) one finds $\epsilon_{H}  \simeq V_{,\phi}^2/(18 \, m_p^2 \, H^4)$, which, using Eq.~(\ref{eq:friedmann_SR1}) becomes
 $$\epsilon_{H} \simeq \frac{m_{p}^2}{2}\left (\frac{V_{,\phi}}{V}\right )^2 =  \epsilon_{_V} \,  .$$
 Similarly, taking a time derivative of Eq.~(\ref{eq:friedmann_SR2}) and incorporating it into Eq.~(\ref{eq:friedmann_SR1}), we get
 $$\eta_H = -\frac{\ddot{\phi}}{H\dot{\phi}} \simeq \eta_V - \epsilon_V \, .$$
The end of inflation is marked by
$$ \epsilon_H (\phi_e) = 1 \simeq \epsilon_{_V}(\phi_e) \, . $$
Furthermore, under the slow-roll approximations, since $\epsilon_H \simeq \epsilon_{_V}$, the Hubble-exit of the CMB pivot scale $k_*$ is denoted by the number of e-folds before the end of inflation $N_*$, which can be obtained from Eq.~(\ref{eq:efolds}) to be
\beq
\boxed{ N_* = \int_{t_*}^{t_e} {\rm d}t \, H(t)  ~ = ~ \frac{1}{m_p}\,\int_{\phi_e}^{\phi_*}\,\frac{d\phi}{\sqrt{2\,\epsilon_H}} ~ \simeq ~ \frac{1}{m_p}\,\int_{\phi_e}^{\phi_*}\,\frac{d\phi}{\sqrt{2\,\epsilon_{_V}(\phi)}} } \, .
\label{eq:N_*}
\eeq
The above formula is quite useful in computing the value of $\phi_*$ for a given potential that supports slow roll, as we will see in Sec.~\ref{sec:inf_models}.

\medskip

 Before proceeding further, we  remind the reader of the
distinction between the \textit{quasi-de Sitter} (qdS) and  \textit{slow-roll} (SR)  approximations.   

\begin{itemize}
\item {\bf Quasi-de Sitter} inflation corresponds to the condition $\epsilon_H \ll 1 ~ \Rightarrow ~ \f{1}{2} \, \dot{\phi}^2 \ll V(\phi) \, .$
\item {\bf Slow-roll} inflation corresponds to both $\epsilon_H,\, |\eta_H| \ll 1 \, .$ 
\end{itemize}
Note that  one can deviate from the slow-roll regime by having $|\eta_H| \geq 1$ while still maintaining the qdS expansion by keeping $\epsilon_H \ll 1$, which is exactly what happens during the so-called {\em ultra slow-roll} (USR) inflation \cite{Kinney:2005vj,Motohashi:2014ppa,Dimopoulos:2017ged,Pattison:2018bct,Mishra:2023lhe}. This distinction will not  be important for the rest of these lecture notes in the present version. Under either of the aforementioned assumptions, the conformal time, $\tau$, is given by Eq.~(\ref{eq:tau_dS}) to be $\tau \simeq -1/(aH)$.
\begin{figure}[hbt]
\begin{center}
\includegraphics[width=0.75\textwidth]{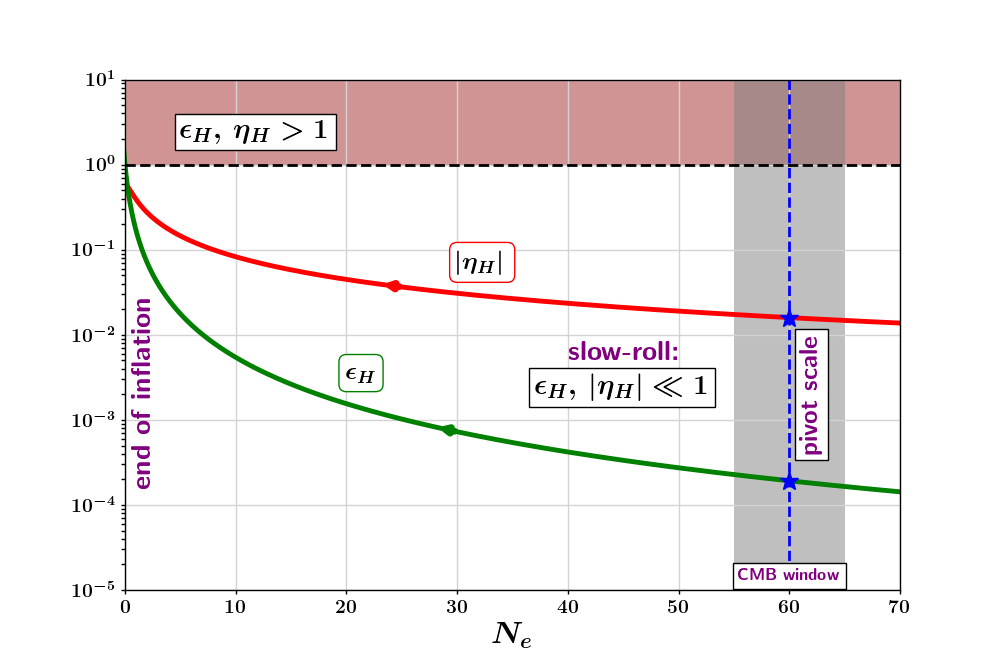}
\caption{The evolution of the  slow-roll parameters $\epsilon_H$ and $\eta_H$ is shown  as a function of the number of e-folds before the end of inflation $N_e$ for Starobinsky potential (\ref{eq:pot_Star}). From this plot, it is easy to notice that at early times when $N_e \gg 1$, the slow-roll conditions are satisfied {\it i.e} $\epsilon_H,\, |\eta_H| \ll 1$. However, the  slow-roll conditions are violated  towards the end of inflation (marked by $N_e = 0$ and $\epsilon_H = 1$).}
\label{fig:inf_Strb_SRparams}
\end{center}
\end{figure}

 \medskip
 
 During inflation, since the inflaton evolves monotonically, one can describe the evolution of the background dynamics  by re-writing Eqs.~(\ref{eq:friedmann1})--(\ref{eq:phi_EOM}) such that  $\phi$  behaves as a time variable. Inflationary dynamics in terms of a $\phi$-clock is known as the {\em Hamilton-Jacobi} formalism~\cite{Liddle:1994dx} of inflation, discussed in App.~\ref{app:Hamilton_Jacobi}.  
 
\bigskip
We conclude that a scalar field with a sufficiently shallow potential $V(\phi)$ with $\epsilon_{_V} \ll 1$ leads to nearly exponential (qdS) expansion of the background space-time during inflation, which successfully addresses the fine tuning of initial conditions for the background evolution, such as the flatness and the horizon problems, discussed in Secs.~\ref{sec:flatness_problem} and \ref{sec:Horizon_problem}. As inflation progresses, since the friction coefficient $H$ falls with time (although slowly) the slow-roll parameter $\epsilon_H$ (also $\epsilon_{_V}$) increases and eventually becomes greater than unity, thus terminating the accelerated expansion of space. Towards the end of inflation, slow-roll conditions are violated (see Fig.~\ref{fig:inf_Strb_SRparams}). After the end of inflation, the inflaton field oscillates around the minimum of the potential. Subsequently, the oscillating inflaton condensate decays into matter and radiation, {\em via} a process called {\em reheating}, leading to the commencement of the hot Big Bang phase. We will return to the topic of reheating in Sec.~\ref{sec:Reheating}.

\bigskip

Next, we move on to discuss how quantum fluctuations during inflation provides a natural mechanism for generating the initial seeds of structure formation in the universe.

\subsection{Quantum fluctuations during Inflation}
\label{sec:inf_dyn_QF}
Our system is a canonical scalar field minimally coupled to gravity whose action is given by 
\beq
S[g_{\mu\nu},\varphi] = \int \,  {\rm d}^4x \,  \sqrt{-g} \,  \l( \, \f{m_p^2}{2} \, R - \f{1}{2}  \, \partial_{\mu}\varphi \,  \partial_{\nu}\varphi  \, g^{\mu\nu}-V(\varphi)\r) .
\label{eq:Action_phi_g_munu}
\eeq
In perturbation theory, we split the metric and inflaton field into their corresponding homogeneous  background pieces  and fluctuations, namely
$$g_{\mu\nu}(t,\vec{x}) = \bar{g}_{\mu\nu}(t) + \delta g_{\mu\nu}(t,\vec{x}) \, ; \quad \varphi(t,\vec{x}) = \phi(t) + \delta\varphi(t,\vec{x}) \, .$$
Note that the perturbed metric $\delta g_{\mu\nu}$ has 10 degrees of freedom, out of which only two are independent, while the rest are fixed by gauge freedom, and the Hamiltonian and momentum constraints. The  perturbed  line element in the {\em Arnowitt-Deser-Misner (ADM) formalism} \cite{Arnowitt:1962hi,Maldacena:2002vr} can be written as   
\beq
 {\rm d}s^2 = -  \alpha^2 \, {\rm d}t^2 + \gamma_{ij} \l( {\rm d}x^i + \beta^i \, {\rm d}t \r) \l( {\rm d}x^j + \beta^j \, {\rm d}t \r) \, ,
\label{eq:metric_ADM}
\eeq
where $\alpha = 1 + \delta\alpha$ and $\beta^i$ are the lapse and shift functions, while $\gamma_{ij}$ are the dynamical metric perturbations.

In these lecture notes, we will skip the detailed discussion of relativistic cosmological perturbation theory (which is an extremely important topic) and direct the interested readers to Refs.~\cite{Mukhanov:1990me,Malik:2008im,Baumann_TASI,Lyth:2009zz,Gorbunov:2011zzc,Sriramkumar:2009kg,Baumann:2022mni}. For the purpose of  pedagogy and simplicity, we work in the  comoving gauge\footnote{See  App.~\ref{app:MS} for a brief discussion on the perturbed metric in both comoving and spatially flat gauge choices.}  defined by
\beq 
\delta\varphi(t,\vec{x}) = 0 \, ; \quad \gamma_{ij}(t,\vec{x}) = a^2 \bigl[ \bigl(  1 + 2 \, \zeta(t,\vec{x}) \bigr) \, \delta_{ij} \, + \, h_{ij}(t,\vec{x}) \bigr]  \, .
\label{eq:metric_momov_gauge}
\eeq
 In particular, two gauge-invariant light fields are guaranteed to exist in the single field inflationary scenario, which are
\begin{enumerate}
\item Scalar-type {\em comoving curvature perturbations}~~ $\zeta(t,\vec{x}) \, ;$
\item Tensor-type {\em transverse and traceless perturbations}~~  $ h_{ij}(t,\vec{x})\, . $
\end{enumerate}
The scalar fluctuations would eventually induce density and temperature perturbations in the hot Big Bang phase, and subsequently the large scale structure of the universe, while the tensor fluctuations propagate as  gravitational waves (GWs) at late times and constitute a stochastic cosmological  background of GWs. 

\medskip
The perturbed action in the comoving gauge can be expressed as 
\beq
S[g_{\mu\nu},\varphi] = S_{B}[\bar{g}_{\mu\nu},\phi] + S^{(2)}[\zeta] + S^{(2)}[h_{ij}] + S_{\rm int}[\zeta, h_{ij}] \, ,
\label{eq:Action_perturbed_scalar-tensor}
\eeq
where $S_{B}[\bar{g}_{\mu\nu},\phi]$ is the background action, while $S^{(2)}[\zeta], \,  S^{(2)}[h_{ij}]$ are the  quadratic actions for the scalar and tensor fluctuations (which results in linear field equations for the fluctuations). The term $ S_{\rm int}[\zeta, h_{ij}]$ is the action for fluctuations beyond the linear order which we do not discuss in these lecture notes and direct the interested readers to Refs.~\cite{Baumann_TASI,Maldacena:2002vr,Baumann:2018muz}.  

The background action $S_{B}[\bar{g}_{\mu\nu},\phi]$ leads to the familiar Friedmann Equations 
$$H^2=\left(\frac{\dot{a}}{a}\right)^{2} = \left(\frac{1}{3m_p^2}\right)\rho_{_{\phi}},$$
$$\dot{H} = -\f{1}{2 m_p^2}\dot{\phi}^2 ~,$$ 
which result in an accelerated expansion of space for a suitable potential $V(\phi)$ at the background level. Under slow-roll condition $\epsilon_H \ll 1$, we get  quasi-dS expansion $a(t) \sim e^{Ht}$.

The quadratic action  for  the (linear) scalar-type fluctuations   in the comoving gauge is
$$
S^{(2)}[\zeta] = \f{1}{2} \int {\rm d}\tau \, {\rm d}^3\vec{x} ~  \l( a m_p \sqrt{2\epsilon_H} \r)^2  \,  \l[ (\zeta')^2 - (\partial_i \zeta)^2 \r]  \, ,
$$
and for the  (linear) tensor-type perturbations is
\[S^{(2)}[h_{ij}] = \f{1}{2} \int {\rm d}\tau \, {\rm d}^3\vec{x} ~  \,  \l(\f{a \, m_p}{2}\r)^2  \l[   (h'_{ij})^2 - (\partial_l h_{ij})^2  ~.\r]\] 

\bigskip

 At linear order in perturbation theory, the scalars, vectors and tensors are decoupled (the so-called {\em SVT decomposition theorem}),  and hence their evolution during inflation can be studied separately, as discussed in Ref.~\cite{Baumann_TASI}. We will begin with scalar fluctuations and then move on to study tensor fluctuations\footnote{It is important to note that vector perturbations are not created during inflation and any pre-existing vector perturbations decay rapidly due to the exponential expansion during inflation \cite{Baumann_TASI}. Hence, we do not discuss  vector perturbations in these notes.}. In the following, we describe the evolution of perturbations during inflation generated by quantum fluctuations at linear order in perturbation theory. Note that the fluctuations are categorized into scalars, vectors and  tensors depending on their transformation properties  under the rotation group  ${\rm SO}(3)$ on the three dimensional spatial hypersurfaces.

 In the following section, we will delve into the study of scalar and tensor fluctuations during inflation after providing  a brief introduction to the  statistics of quantum vacuum fluctuations.

\section{Correlators of inflationary fluctuations}
\label{sec:Inf_correlators}
The primary observables in cosmology are the late time {\rm cosmological correlators} \cite{Baumann:2018muz} which are in general the $N$-point correlation functions of different  physical quantities, such as  fluctuations in  temperature and density contrast.  For primordial fluctuations, which are the seeds of the late-time correlators, we will primarily be interested in the comoving curvature fluctuations $\zeta(t,\vec{x})$ and the transverse and traceless tensor fluctuations $h_{ij}(t,\vec{x})$. 

 As discussed in Ref.~\cite{Baumann:2018muz}, an accurate measurement  of the late time correlators  of observables ${\cal O}_i(t,\vec{x})$ (in the form of  CMB temperature and polarisation anisotropies as well as large scale clustering of matter in the  universe) along with the evolution equations and matter-energy constituents, would eventually enable  us to determine the $N$-point correlators of the initial fluctuations, denoted by $\Psi_i(t_{\rm ini},\vec{x})$ at the beginning of the hot Big Bang phase, and consequently at the end of inflation hypersurface. Schematically, this is represented as
$$\boxed{ \langle \, \hat{\bm{\Psi}}_1(t,\vec{x}) \, \hat{\bm{\Psi}}_2(t,\vec{x}) \, \hat{\bm{\Psi}}_3(t,\vec{x}) \, ...\,  \rangle_{_{\rm ini}} ~ \bm{\longleftarrow} ~ \langle   \, \bm{{\cal O}}_1(t,\vec{x}) \, \bm{{\cal O}_2}(t,\vec{x}) \, \bm{{\cal O}_3}(t,\vec{x})  ... \rangle_{_{\rm late}}}$$

\subsection{Statistics of vacuum quantum  fluctuations}
\label{sec:Inf_stats}

In this subsection, we provide a succinct discussion on  the statistics of cosmological fluctuations at the end of inflation with a particular focus on the 2-point auto-correlator and the power spectrum. For a detailed account on the subject, we direct the readers to Refs.~\cite{1986ApJ...304...15B,Baumann:2018muz} and the  \href{https://github.com/ddbaumann/cosmo-correlators/blob/main/LectureNotes-July2023.pdf}{\purple {\bf lecture notes}}. 

We begin with the standard Fourier transform convention used throughout these notes, namely
\ber
\Psi(t,\vec{x}) &=& \int  \f{{\rm d}^3 \vec{k}}{(2\pi)^3} ~ \Psi_{\vec{k}}(t) ~ e^{i\, \vec{k}.\vec{x}} \quad \, ; \label{eq:Fourier_trans} \\
\quad \Psi_{\vec{k}}(t) &=& \int\, {\rm d}^3 \vec{x} ~ \Psi(t,\vec{x}) ~ e^{-i\, \vec{k}.\vec{x}}  \, , \label{eq:Fourier_inv_trans}
\eer
where $\Psi_k$ is the Fourier transformation of $\Psi$. If $\Psi(t,\vec{x})$ fluctuations are drawn from a statistical ensemble with a probability distribution  function (PDF) $P[\Psi]$, then the variance of the fluctuations is defined as  
\beq
\sigma_\Psi^2 (t) \equiv \int \, {\rm d}\Psi \, P[\Psi] \, \Psi^2  = \langle \, \Psi(t,\vec{x}) \, \Psi(t,\vec{x}) \,  \rangle \, ,
\label{eq:Variance_Psi}
\eeq
where $\langle ... \rangle$ denotes statistical ensemble average with respect to the PDF $P[\Psi]$. The power spectrum can be written using the Fourier decomposition given in Eq.~(\ref{eq:Fourier_trans}) as  
\beq
\sigma_\Psi^2 (t) = \int  \f{{\rm d}^3 \vec{k}_1}{(2\pi)^3} \, \int  \f{{\rm d}^3 \vec{k}_2}{(2\pi)^3} \,  \langle \, \Psi_{\vec{k}_1} (t) \, \Psi_{\vec{k}_2}(t)  \,  \rangle \,  e^{i\, \l( \vec{k}_1 + \vec{k}_2 \r).\vec{x}} .
\label{eq:Variance_Psi_Fourier_1}
\eeq
The 2-point correlator in the momentum space is defined as
\beq
  \langle \, \Psi_{\vec{k}_1} (t) \, \Psi_{\vec{k}_2}(t)  \,  \rangle \equiv  \f{2\pi^2}{k^3} \, {\cal P}_\Psi(t,\vec{k}_1) \,  \, (2\pi)^3 \, \delta_D^{(3)} \l(\vec{k}_1 + \vec{k}_1 \r)   \, ,
\label{eq:Def_power_spectrum_gen}
\eeq
where $ {\cal P}_\Psi(t,\vec{k}_1)$ is  called the power spectrum of fluctuations $\Psi$. Assuming the background spacetime to be {\em homogeneous and isotropic}, so that the power spectrum only depends on $k = |\vec{k}_1|$, we get 
\ber
 \langle \, \Psi_{\vec{k}_1} (t) \, \Psi_{\vec{k}_2}(t)  \,  \rangle & \equiv &  \f{2\pi^2}{k^3} \, {\cal P}_\Psi(t,k)   \, (2\pi)^3 \, \delta_D^{(3)} \l(\vec{k}_1 + \vec{k}_1 \r) \, ; \label{eq:Def_power_spectrum_FLRW} \\
 \sigma_\Psi^2 (t)   &=&  \int  {\rm d} \ln k \, {\cal P}_\Psi(t,k)  \, ,
\label{eq:Variance_power_spectrum_FLRW} 
\eer
which shows that the power spectrum ${\cal P}_\Psi(t,k)$ is the change in the variance  per logarithm interval of $k$. Similarly, in order to study the higher-point statistics, we need to compute the skewness  and the kurtosis; or equivalently, in Fourier space,  the {\em bispectrum} $ \langle \, \Psi_{\vec{k}_1}\Psi_{\vec{k}_2}\Psi_{\vec{k}_3} \,  \rangle$ and  the {\em trispectrum} $ \langle \, \Psi_{\vec{k}_1}\Psi_{\vec{k}_2}\Psi_{\vec{k}_3} \, \Psi_{\vec{k}_4} \,  \rangle$ respectively.  In the present version of these lecture notes, we  primarily focus on the 2-point correlators or the power spectra of scalar and tensor fluctuations. However, we provide a brief discussion on higher-point correlators in Sec.~\ref{sec:Non_Gaussianity}.

In order to characterize linear quantum fluctuations during inflation, we begin with the quantum field (in the  Heisenberg picture)
\beq
\hat{\Psi}(t,\vec{x}) = \int \, \f{{\rm d}^3 \vec{k}}{(2\pi)^3} \, \hat{\Psi}_{\vec{k}}(t) \, e^{i\vec{k}.\vec{x}} = \int \, \f{{\rm d}^3 \vec{k}}{(2\pi)^3} \, \l[ f_k(t) \, e^{i\vec{k}.\vec{x}} \, \hat{a}_{\vec{k}} + f^{*}_k(t) \, e^{-i\vec{k}.\vec{x}} \, \hat{a}_{\vec{k}}^{\dagger}\r]  \, ,
\label{eq:Psi_quantum}
\eeq
with 
\beq
\hat{\Psi}_{\vec{k}}(t) =  f_k(t)  \, \hat{a}_{\vec{k}} + f^{*}_k(t) \, \hat{a}_{-\vec{k}}^{\dagger} \, ,
\label{eq:Psi_k_quantum}
\eeq 
where  $\hat{a}_{\vec{k}}, \, \hat{a}_{\vec{k}}^{\dagger}$ are  the creation and annihilation operators respectively,  satisfying the usual commutation relation
\beq
\l[ \hat{a}_{\vec{k}_1}, \, \hat{a}_{\vec{k}_2}^{\dagger} \r] = (2\pi)^3 \, \delta_{\rm D}^3 \l( \vec{k}_1 - \vec{k}_2 \r)  \, ,
\label{eq:a_k_quant}
\eeq 
and  the vacuum state $| 0 \rangle$ is defined by 
\beq
\hat{a}_{\vec{k}} \, | 0 \rangle = 0  \, .
\label{eq:vacuum_annihilation}
\eeq
Similarly, the mode functions $f_k(t)$ satisfy the field equations for $\hat{\Psi}(t,\vec{x})$ and hence, they only depend upon $k=|\vec{k}|$. These mode functions are canonically normalized to
\beq
f_k \, \dot{f}_k^{\ast}  - \dot{f}_k \, f_k^{\ast} = i \, \hbar \, ,
\label{eq:mode_functions_normalisation_QFT}
\eeq
 so that the quantum field $\hat{\Psi}$ and its conjugate momentum $\hat{\Pi}_\Psi  = \dot{\hat{\Psi}}$ satisfy the usual canonical commutation relation 
\beq
\l[ \hat{\Psi}  (t,\vec{x}_1), \, \hat{\Pi}_\Psi (t,\vec{x}_2)\r] \, = \, i \, \hbar ~ \delta_D^{(3)}\l( \vec{x}_1 - \vec{x}_2\r) \, .
\label{eq:CCR_Psi}
\eeq
It is important to stress that the splitting of $\hat{\Psi}_{\vec{k}}(t)$ in terms of $\hat{a}_{\vec{k}}$ and $ \hat{a}_{\vec{k}_2}^{\dagger}$  in Eq.~(\ref{eq:Psi_k_quantum}) is not unique and  there exists a family of mode functions $f_k$ satisfying Eqs.~(\ref{eq:vacuum_annihilation}) and (\ref{eq:mode_functions_normalisation_QFT}), as discussed in Ref.~\cite{Allen:1985ux}. Hence, in contrast to the Minkowski spacetime  fluctuations, there is no unique vacuum state $|0\rangle$ for the inflationary fluctuations at this stage.

  For vacuum quantum fluctuations, using Eqs.~(\ref{eq:Psi_k_quantum}) and (\ref{eq:vacuum_annihilation}), we get
  $$\langle \,  0  \, | \, \hat{\Psi}_{\vec{k}_1} (t) \, \hat{\Psi}_{\vec{k}_2}(t)  \, | \,  0 \, \rangle \, 
 = \langle \,  0  \, | \, f_{\vec{k}_1}  f_{\vec{k}_2}^{\ast} \,   \hat{a}_{\vec{k}_1}  \hat{a}_{-\vec{k}_2}^{\dagger} \, | \,  0 \, \rangle \, ,$$
 which, using Eq.~(\ref{eq:a_k_quant}), yields 
 \beq
\boxed{  \langle \, \Psi_{\vec{k}_1} (t) \, \Psi_{\vec{k}_2}(t)  \,  \rangle \equiv  \langle \,  0  \, | \, \hat{\Psi}_{\vec{k}_1} (t) \, \hat{\Psi}_{\vec{k}_2}(t)  \, | \,  0 \, \rangle = |f_k(t)|^2 \,  (2\pi)^3 \, \delta_D^{(3)} \l(\vec{k}_1 + \vec{k}_1 \r) } \,  .
\label{eq:PS_quan_vacuum}
 \eeq
 Using Eq.~(\ref{eq:Def_power_spectrum_FLRW}), we obtain the expression for the  power spectrum of vacuum quantum fluctuations to be 
 \beq
\boxed{  {\cal P}_{\Psi} (t, k) =    \frac{k^3}{2\pi^2} \, |f_k(t)|^2 } \, .
\label{eq:PS_Psi}
\eeq
We stress that Eq.~(\ref{eq:Def_power_spectrum_FLRW}) is the general definition of the power spectrum, while the definition provided by Eq.~(\ref{eq:PS_Psi}) is only valid for vacuum quantum fluctuations.  The appearance of the term  $\delta_D^{(3)} (\vec{k}_1 + \vec{k}_1 )$ in Eq.~(\ref{eq:PS_quan_vacuum}) is a consequence of translational invariance (homogeneity) of the background space, since the linear correlators respect the background isometries. 

Sometimes the cosmological  power spectrum is defined as 
\beq
 \tilde{{\cal P}}_{\Psi} (t, k) =  \f{2\pi^2}{k^3} \,  {\cal P}_{\Psi} (t, k)  \, ,
 \label{eq:PS_def_alternate}
\eeq
where $\tilde{{\cal P}}_{\Psi} (t, k)$ is referred to as the {\em dimensionful power spectrum}, while ${\cal P}_{\Psi} (t, k)$ is referred to as  the {\em dimensionless\/}  power spectrum. In these notes, we will use Eq.~(\ref{eq:Def_power_spectrum_FLRW}) as the definition of power spectrum, rather than Eq.~(\ref{eq:PS_def_alternate}), in order to avoid confusion. However, it is important to keep this distinction in mind\footnote{The dimensionful power spectrum $\tilde{{\cal P}}_{\Psi} (t, k)$ is usually denoted as ${\cal P}_{\Psi} (t, k)$, while the dimensionless power spectrum ${\cal P}_{\Psi} (t, k)$ is denoted as $\Delta_{\Psi}^2 (t, k)$. However, we will stick to the notations we defined above throughout these lecture notes.}.   

\subsection{Scalar quantum  fluctuations during inflation}
\label{sec:inf_dyn_scalar_QF}

\begin{figure}[htb]
\begin{center}
\includegraphics[width=0.75\textwidth]{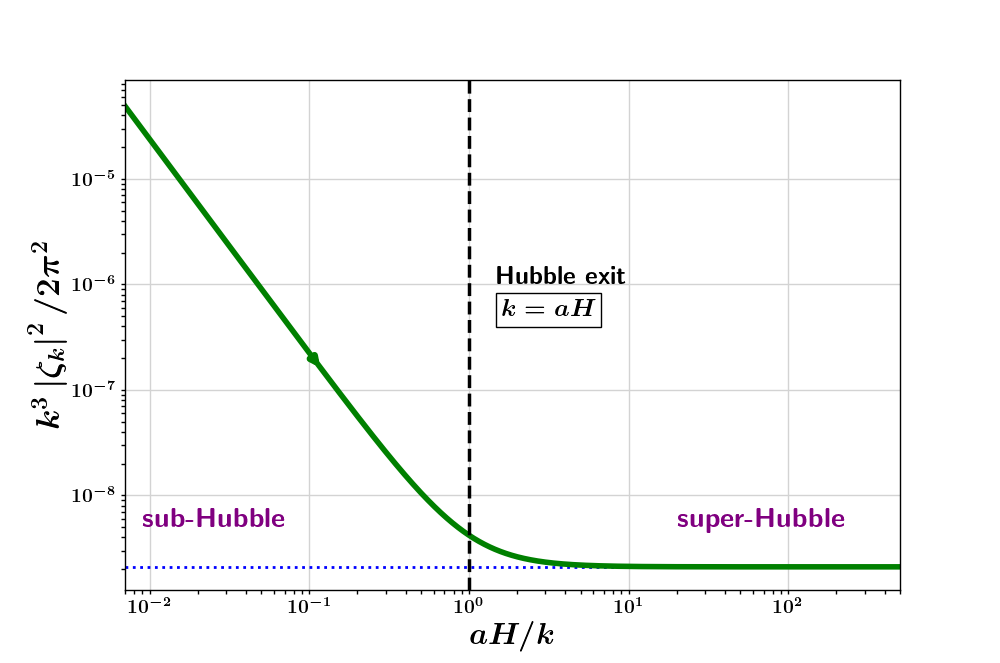}
\caption{Evolution of the scalar power spectrum $\f{k^3}{2\pi^2} |\zeta_k|^2$  is plotted by numerically solving Eq.~(\ref{eq:fourier_zeta}) for a mode exiting the Hubble radius at about 60 e-folds before the end of inflation (for the Starobinsky potential given in Eq.~{\ref{eq:pot_Star}}).  At early times when the mode is sub-Hubble, {\it i.e.} $k \gg aH$, the power decreases as ${\cal P}_{\zeta} \sim (aH)^{-2}$ as expected.  After the Hubble-exit,  the power  freezes to a constant in the super-Hubble regime when $k \ll aH$. We note down  its value  after the mode-freezing as the super-Hubble scale power corresponding to that mode. Repeating the procedure for a range of scales $k$ yields us the power spectrum of scalar fluctuations. The same numerical analysis can be carried out for tensor fluctuations.}
\label{fig:inf_Strb_horizonexit}
\end{center}
\end{figure}

The effective action for the inflationary scalar fluctuations $\zeta$ in the comoving gauge $\delta\varphi = 0$ at linear order in perturbation theory (hence, the quadratic action) is given by \cite{Maldacena:2002vr,Baumann:2018muz} (see App.~\ref{app:MS})
\beq
\boxed{ \, S^{(2)}[\zeta(\tau, \vec{x})] = \f{1}{2} \int {\rm d}\tau \, {\rm d}^3\vec{x} ~~  z^2 \l[  \,   (\zeta')^2 - (\partial_i \zeta)^2 \,  \r] \, } \, ,
\label{eq:Action_quad_zeta}
\eeq
where the `\textit{pump term}' $z$, and its derivatives are given by
\ber
    z &=& am_p \sqrt{2\epsilon_H}  \, ; \label{eq:z_pump} \\
\f{z'}{z} &=& aH \, \l(1+\epsilon_H-\eta_H \r) \, ; \label{eq:z'_pump_der} \\
\f{z''}{z} &=&  \l( aH \r)^2 \l[ 2 + 2 \, \epsilon_H - 3 \, \eta_H + 2\, \epsilon_H^2 + \eta_H^2 -3 \, \epsilon_H \, \eta_H - \f{1}{aH} \, \eta_H' \r] \, . \label{eq:z''_pump_dder}
\eer
We stress that the above expressions are exact at linear order in perturbation theory and {\em have not been truncated\/} at quadratic order in slow-roll parameters $\epsilon_H, \, \eta_H$.
Upon change of variable 
\beq
v \equiv z  \, {\zeta} \, ,
\label{eq:MS_variable}
\eeq
the scalar action~(\ref{eq:Action_quad_zeta}) takes the form 
\beq
\boxed{ \, S^{(2)} [v] = \frac{1}{2} \, \int  \, {\rm d}\tau \,  {\rm d}^3 \vec{x} \, \left[\left(v'\right)^2 - \left(\partial_i v\right)^2+\frac{z''}{z}v^2\right] \, } \, , 
\label{eq:Action_quad_v}
\eeq
 The variable $v$, which itself is a scalar  field like $\zeta$, is called the {\em Mukhanov-Sasaki variable\/}\footnote{Note that the original Mukhanov-Sasaki variable, as defined in Refs.~\cite{Sasaki:1986hm,Mukhanov:1988jd}, is 
  ${\cal Q}(t,\vec{x}) = \f{1}{a(t)} \, v(t,\vec{x}) \, .$} in the  literature \cite{Sasaki:1986hm,Mukhanov:1988jd}. Eq.~(\ref{eq:Action_quad_v}) represents the action of a canonical massive scalar field in Minkowski spacetime. The field equation for $v$ is given by
 \beq
v'' \, + \, \left(-\nabla^2 \, + \,  {\cal M}_{\rm eff}^2\right) v = 0 \, ,
\label{eq:MS_field}
\eeq
where the time-dependent {\em effective (tachyonic) mass term} is given by \cite{Mishra2023}
\ber
\boxed{ {\cal M}_{\rm eff}^2(\tau) \equiv - \f{z''}{z} = - \l( aH \r)^2 \l[ 2 + 2 \, \epsilon_H - 3 \, \eta_H + 2\, \epsilon_H^2 + \eta_H^2 -3 \, \epsilon_H \, \eta_H - \f{1}{aH} \, \eta_H' \r] }\, . 
\label{eq:M_eff_MS}
\eer
The Fourier modes $v_k$ satisfy the  {\em Mukhanov-Sasaki equation}, given by  
\beq
\boxed{ v''_k \, + \, \left(k^2+ {\cal M}_{\rm eff}^2\right) v_k=0} \, .
\label{eq:MS_modes}
\eeq
Specializing to  slow-roll inflation, $\epsilon_H,\, |\eta_H|, \, |\eta_H'| \ll 1$, the above equation in the quasi-de Sitter limit from Eq.~(\ref{eq:tau_dS})  reduces to
\beq
v''_k \, + \, \l( \, k^2 - 2 a^2 H^2 \,  \r) v_k=0 ~ \Rightarrow ~ \boxed{ v''_k \, + \, \l( \, k^2 - \f{2}{\tau^2} \,  \r) v_k=0 } \, ,
\label{eq:MS_modes_dS}
\eeq
whose general solution is given by\footnote{In Eq.~(\ref{eq:MS_modes_dS}), we have dropped the slow-roll parameters in our analysis. If we keep the slow-roll parameters, then the general solution can be written in terms of Bessel or Hankel functions of order $\nu$ which is given in terms of the slow-roll parameters as shown in App.~\ref{app:MS_analyt_sol} and Ref.~\cite{Mishra2023}. For qdS approximation, where we drop all the slow-roll parameters, $\nu=3/2$ and the solution gets reduced to the simple form given in Eq.~(\ref{eq:MS_modes_sol_dS}).} 
\beq
v_k(\tau) = C_1 \f{1}{\sqrt{2k}} \l(1-\f{i}{k\tau}\r) e^{-ik\tau} + C_2 \f{1}{\sqrt{2k}} \l(1+\f{i}{k\tau}\r) e^{+ik\tau}  \, .
\label{eq:MS_modes_sol_dS}
\eeq
The above expression will be crucial in our computation of  the power spectrum of inflationary fluctuations below. Note that our primary goal is to compute the power spectrum of vacuum quantum fluctuations of $\zeta$ at late times  (when all the observable modes $\zeta_k$ are super-Hubble), which is defined by   
\beq
 {\cal P}_\zeta \equiv   \frac{k^3}{2\pi^2} \, |{\zeta}_k|^2 \Big|_{k\ll aH}  = \frac{k^3}{2\pi^2} \, \frac{|v_k|^2}{z^2} \Big|_{k\ll aH}  \, .
\label{eq:MS_Ps}
\eeq
Hence, before quantitatively computing the power spectrum, let us understand the behaviour of each mode $\zeta_k$ in the super-Hubble regime. For this purpose, we write the following  Euler-Lagrange field equation for $\zeta$ corresponding to the quadratic action~(\ref{eq:Action_quad_zeta}):
$$\zeta'' + 2\left(\frac{z'}{z}\right){\zeta}' - \nabla^2\zeta = 0 \, ,$$
which leads to the equation for the mode functions  $\zeta_k$ to be 
\beq
\boxed{ {\zeta}''_k + 2\left(\frac{z'}{z}\right){\zeta}'_k +k^2 {\zeta}_k =0  } \, .
\label{eq:fourier_zeta}
\eeq
 At sufficiently late times when a mode is super-Hubble, namely $k\ll aH$ or equivalently, $-k\tau \ll 1$, we have
 $${\zeta}''_k + 2\left(\frac{z'}{z}\right){\zeta}'_k  \simeq 0 \, .$$
From Eq.~(\ref{eq:z'_pump_der}), we know that $z'/z = aH (1+\epsilon_H -\eta_H)$. Specializing to slow-roll  approximations and dropping $\epsilon_H, \, \eta_H$, we have $z'/z \simeq aH = -1/\tau$, leading to
 $${\zeta}''_k \simeq \f{2}{\tau} \, {\zeta}'_k \, ,$$
 whose solution is given by
 \beq
 \zeta_k \simeq A_k + B_k \, \tau^2 = A_k + \f{B_k}{(aH)^2}\, ,
 \label{eq:zeta_k_frozen}
 \eeq
with $A_k,B_k$ being two integration constants. This shows that on super-Hubble scales, $\zeta_k$ comprises of a constant mode and a decaying mode. Note that at late times, since $\tau \to 0$, the decaying mode can be ignored, which leads to the conclusion that $\zeta_k$ remains constant or  {\em frozen} outside the Hubble radius. This is demonstrated in Fig.~\ref{fig:inf_Strb_horizonexit} for the Starobinsky potential given in Eq.~(\ref{eq:pot_Star}).

The conservation of $\zeta_k$ for adiabatic perturbations on super-Hubble scales is one of the most important consequences of the rapidly accelerated expansion during inflation. This allows us to translate/extrapolate the  power spectra of inflationary fluctuations (at large cosmological scales) at the end of inflation through the unknown reheating  history of the universe (discussed in Sec.~\ref{sec:Reheating}), all the way until these fluctuations enter the Hubble radius closer to matter-radiation equality and recombination epochs. Thus, it enables inflation to make firm predictions about the late time structure in the universe. It is important to stress that the conservation of $\zeta_k$ for adiabatic perturbations on super-Hubble scales is also valid  beyond Einstein's gravity and potentially beyond linear perturbation theory, as demonstrated in Refs.~\cite{Wands:2000dp,Lyth:2004gb}.

 \bigskip
 The aforementioned classical dynamics of $\zeta_k$ provides us with intuitive insights into the qualitative properties  of super-Hubble scalar fluctuations during inflation. In order to quantitatively compute the power spectrum, we need to consider vacuum quantum fluctuations during inflation, for which we will treat the fields $\hat{\zeta}$ and $\hat{v}$  as  quantum fields. Since the action for $\hat{v}$ in Eq.~(\ref{eq:Action_quad_v}) is in  the canonically normalized form,  we will apply the method of canonical quantization discussed in Sec.~\ref{sec:Inf_stats} to quantize it. Working in the Heisenberg picture, we expand $\hat{v}$ in terms of creation and annihilation operators as follows:
\beq
\hat{v}(\tau,\vec{x}) = \int \, \f{{\rm d}^3 \vec{k}}{(2\pi)^3} \, \hat{v}_k(\tau) \, e^{i\vec{k}.\vec{x}} = \int \, \f{{\rm d}^3 \vec{k}}{(2\pi)^3} \, \l[ v_k(\tau) \, e^{i\vec{k}.\vec{x}} \, \hat{a}_{\vec{k}} + v^{*}_k(\tau) \, e^{-i\vec{k}.\vec{x}} \, \hat{a}_{\vec{k}}^{\dagger}\r] \, , 
\label{eq:zeta_quantum}
\eeq
where the Fourier modes $\hat{v}_k$  are those of the one-dimensional time dependent quantum harmonic oscillators, that can be written as 
\beq
\hat{v}_k(\tau) =  v_k(\tau)  \, \hat{a}_{\vec{k}} + v^{*}_k(\tau) \, \hat{a}_{-\vec{k}}^{\dagger} \, ,
\label{eq:zeta_k_quantum}
\eeq 
with $\hat{a}_{\vec{k}}, \, \hat{a}_{\vec{k}}^{\dagger}$ being the  annihilation and creation operators respectively,  satisfying the usual commutation relations
\beq
\l[ \hat{a}_{\vec{k}_1}, \, \hat{a}_{\vec{k}_2}^{\dagger} \r] = (2\pi)^3 \, \delta_{\rm D}^3 \l( \vec{k}_1 - \vec{k}_2 \r)  \, .
\label{eq:a_k_quantum}
\eeq 
The mode functions $v_k(\tau)$ satisfy  
the Mukhanov-Sasaki Eq.~(\ref{eq:MS_modes}), namely 
\beq
v''_k \, + \, \left(k^2-  \f{z''}{z}\right) v_k=0 \, ,
\label{eq:MS_modefunctions}
\eeq
and hence they only depend on the magnitude $k=|\vec{k}|$ of the comoving frequency.

Given a mode $k$, at sufficiently early times when it is deep in the  sub-Hubble  regime i.e $k\gg aH$,  we can assume $v_k$ to be the fluctuations of a massless field in Minkowski spacetime. Consequently,  we impose  the Bunch-Davies vacuum  initial condition  \cite{Bunch:1978yq}
\beq
\boxed{ v_k\big\vert_{-k\tau \gg 1}= \frac{1}{\sqrt{2k}} \, e^{-ik\tau} } \, ,
\label{eq:Bunch_Davies}
\eeq
on each cosmologically observable mode at sufficiently early times during inflation. At this stage, it is important to point out that we will be using the  results from quantum field theory(QFT) at two important places, namely: (i) in the definition of power spectrum given in Eq.~(\ref{eq:MS_Ps}), as discussed in Sec.~\ref{sec:Inf_stats}, and (ii)  in choosing the initial quantum vacuum which is characterised  by the Bunch-Davies  mode functions given in  Eq.~(\ref{eq:Bunch_Davies}).

\bigskip

\begin{figure}[htb]
\begin{center}
\includegraphics[width=0.75\textwidth]{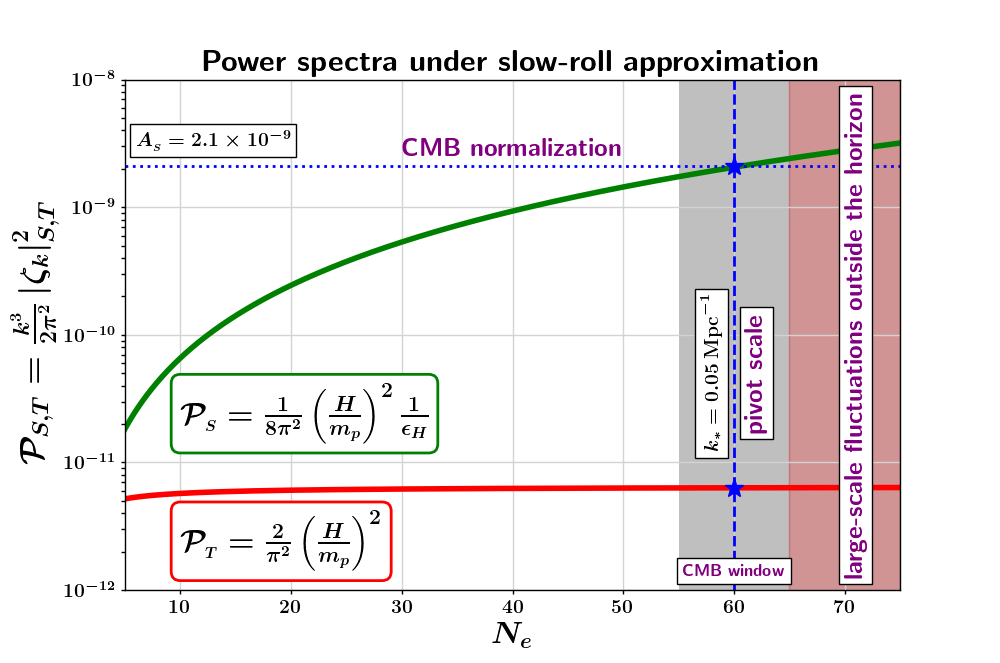}
\caption{The  power spectra of scalar and tensor  quantum fluctuations  (computed using  the slow-roll formulae given by Eqs.~(\ref{eq:P_S_SR}) and  (\ref{eq:P_T_dS}) respectively) are shown for comoving modes exiting the Hubble radius at different number of e-folds $N_e$ before the end of inflation  for the Starobinsky potential  in Eq.~(\ref{eq:pot_Star}). The CMB window (shown as the grey shaded region) corresponds to comoving  modes in the range $k_{\rm CMB} \in [0.0005, 0.5]~{\rm Mpc}^{-1}$ that are being probed by the current CMB missions.   Fluctuations over larger scales (shown as the red shaded region) are outside the observable universe at present.}
\label{fig:inf_Strb_SRpower}
\end{center}
\end{figure}

As discussed before, during inflation  the comoving Hubble radius falls, causing modes to become super-Hubble i.e $k\ll aH$ and Eq.~(\ref{eq:MS_modes}) dictates that $|v_k| \propto z$ and hence $\zeta_k$ approaches a constant value.  By solving the Mukhanov-Sasaki equation we can estimate the dimensionless  primordial power spectrum of $\zeta$  using Eq.~(\ref{eq:MS_Ps}). 
Imposing standard normalisation from QFT given in Eq.~(\ref{eq:mode_functions_normalisation_QFT}) and the  Bunch-Davies initial conditions (\ref{eq:Bunch_Davies}) on the general solution given in Eq.~(\ref{eq:MS_modes_sol_dS}),  we get 
\begin{enumerate}
\item {\em Normalisation} $\Rightarrow ~~  |C_1|^2 - |C_2|^2  = 1$ (where $\hbar =1$)\,;
\item {\em Bunch-Davies  condition}   $ \Rightarrow ~~ v_k(\tau)  \longrightarrow \f{1}{\sqrt{2k}} e^{-ik\tau}$ for $k|\tau| \gg 1  \, ,$
\end{enumerate}
leading to
$$ C_1 = 1,~C_2 = 0 \, ,$$
which then fixes the vacuum state $| 0 \rangle$, and  yields the expression for the quasi-dS  mode functions of the Mukhanov-Sasaki variable to be 
\beq
\boxed{v_k(\tau) =  \f{1}{\sqrt{2k}} \l(1-\f{i}{k\tau}\r) e^{-i \, k\tau} } \, ,
\label{eq:MS_vk_SR}
\eeq
with
$$ \Rightarrow \l \vert v_k \r \vert^2     =   \f{1}{2k} \, \f{ \l(1+k^2 \tau^2\r)}{k^2\tau^2} \, . $$
Incorporating the Bunch-Davies initial condition imposed mode functions from Eq.~(\ref{eq:MS_vk_SR}), we 
obtain the scalar power spectrum, defined in Eq.~(~\ref{eq:MS_Ps}),  to be 
\beq
{\cal P}_\zeta(k) =  \frac{1}{8\pi^2}\left(\frac{H}{m_p}\right)^2\frac{1}{\epsilon_H} \l[ 1 + \l(\f{k}{aH}\r)^2 \r] \, ,
\label{eq:P_S_SR_full}
\eeq
which on super-Hubble scales, $k \ll aH$, is given by the (famous) expression
\beq
\boxed{ {\cal P}_\zeta(k)\Big \vert_{k\ll aH} = \frac{1}{8\pi^2}\left(\frac{H}{m_p}\right)^2\frac{1}{\epsilon_H} }~,
\label{eq:P_S_SR}
\eeq
which shows that there is finite power on super-Hubble scales which depends on the Hubble parameter $H$ and the first slow-roll parameter $\epsilon_H$ during inflation. The scalar and tensor power spectra determined using  the  slow-roll approximations have been plotted in Fig.~\ref{fig:inf_Strb_SRpower} for the Starobinsky potential given in Eq.~(\ref{eq:pot_Star}). From Eqs.~(\ref{eq:P_S_SR_full}) and (\ref{eq:P_S_SR}), it is easy to notice that  the frozen value of the scalar power spectrum at late times  $k \ll aH$ is half of its value at the Hubble-crossing $k=aH$,  as can be seen in Fig.~\ref{fig:inf_Strb_horizonexit}. However, since $H^2/\epsilon_H$ is  almost constant during slow-roll, this implies,
$$ \f{H^2}{\epsilon_H} \bigg\vert_{k = aH}  \, \simeq \,  \f{H^2}{\epsilon_H} \bigg\vert_{k \ll aH} \, ,$$
we can use the corresponding values of $H$ and $\epsilon_H$ at the Hubble crossing of a particular mode to compute the power spectrum, as long as we are using Eq.~(\ref{eq:P_S_SR}), instead of Eq.~(\ref{eq:P_S_SR_full}). This is known as the {\em Hubble-crossing formalism} \cite{kinney2009tasi,Kinney:2005vj}. 

 In order to compare the inflationary predictions with CMB observations, it is instructive to write  the scalar power spectrum given in Eq.~(\ref{eq:P_S_SR}) in the form of a power-law as 
\beq
 {\cal P}_{\zeta}(k) = A_S \l( \frac{k}{k_*} \r)^{n_{_S} - 1} \, ,
  \label{eq:PS_scalar_power-law}
\eeq
where, $ A_S $ is the amplitude of scalar fluctuations at the pivot scale $k=k_*$. The \textit{scalar spectral index}  $ n_{_S} - 1 $ is defined as
\beq
n_{_S} -  1 = \f{{\rm d}\ln {\cal P}_\zeta}{{\rm d} \ln k} = \f{{\rm d}\ln{\cal P}_\zeta}{{\rm d}N} \, \f{{\rm d}N}{{\rm d} \ln k} \, ,
\label{eq:nS_def}
\eeq 
which is to be computed at the Hubble-crossing $k = aH$ of a given mode. Hence we get
$$ \f{{\rm d} \ln k}{{\rm d}N} = 1 + \f{{\rm d} \ln H}{{\rm d}N} = 1 - \epsilon_H \, ,$$
leading to
\beq
\Rightarrow \f{{\rm d}N}{{\rm d} \ln k} = \f{1}{1-\epsilon_H} \simeq 1+\epsilon_H + {\cal O}\l( \epsilon_H^2\r) \, .
\label{eq:N_k_reln}
\eeq
Similarly, from Eq.~(\ref{eq:P_S_SR}), we get
$$\f{{\rm d}\ln{\cal P}_\zeta}{{\rm d}N} = 2 \, \f{{\rm d} \ln H}{{\rm d}N} - \f{{\rm d} \ln \epsilon_H}{{\rm d}N} \, , $$
which, using Eq.~(\ref{eq:eta_H}) becomes
\beq
\Rightarrow \f{{\rm d}\ln{\cal P}_\zeta}{{\rm d}N} = 2 \, \eta_H - 4 \, \epsilon_H \, .
\label{eq:P_zeta_der_reln}
\eeq
Incorporating Eqs.~(\ref{eq:N_k_reln}) and (\ref{eq:P_zeta_der_reln}) into Eq.~(\ref{eq:nS_def}) we obtain the expression for the scalar spectral index $n_{_S}-1$ to be
\beq
\boxed{ n_{_S} - 1  = 2  \, \eta_H - 4 \, \epsilon_H } \, .
\label{eq:nS_SR}
\eeq
Since $\epsilon_H,\, |\eta_H| \ll 1$ during slow-roll inflation, Eq.~(\ref{eq:nS_SR}) shows that $n_{_S} - 1 \simeq 0$, implying the power spectrum is almost scale-invariant with a tiny spectral tilt.  It is possible that in  some specific case,   the inflationary dynamics yields $\eta_H =  2 \epsilon_H$, leading to an {\em exact scale-invariant/Harrison-Zeldovich} power spectrum with $n_{_S} = 1$. However, this will require a particular fined tuned functional form of the potential  $V(\phi)$,  as shown in Ref.~\cite{Starobinsky:2005ab}.

This is in sharp contrast to the quantum fluctuations of a massless free scalar field in Minkowski space for which the mode functions are given by
 \beq
\boxed{v_k^M(t) =  \f{1}{\sqrt{2k}}  e^{-ikt} } \, .
\label{eq:MS_vk_Minkowski}
\eeq
Hence the power spectrum is given by
\beq
\boxed{ {\cal P}_M(k) = \left(\frac{k}{2\pi}\right)^2}~,
\label{eq:Ps_Minkowski}
\eeq
which has a prominent {\Blue {\bf blue tilt}} ${\cal P}_M(k) \propto \left(\frac{k}{2\pi}\right)^2$ on all scales,  because of which the power is strongly suppressed on large scales and has negligible macroscopic  consequence. 

\begin{figure}[htb]
\begin{center}
\includegraphics[width=0.8\textwidth]{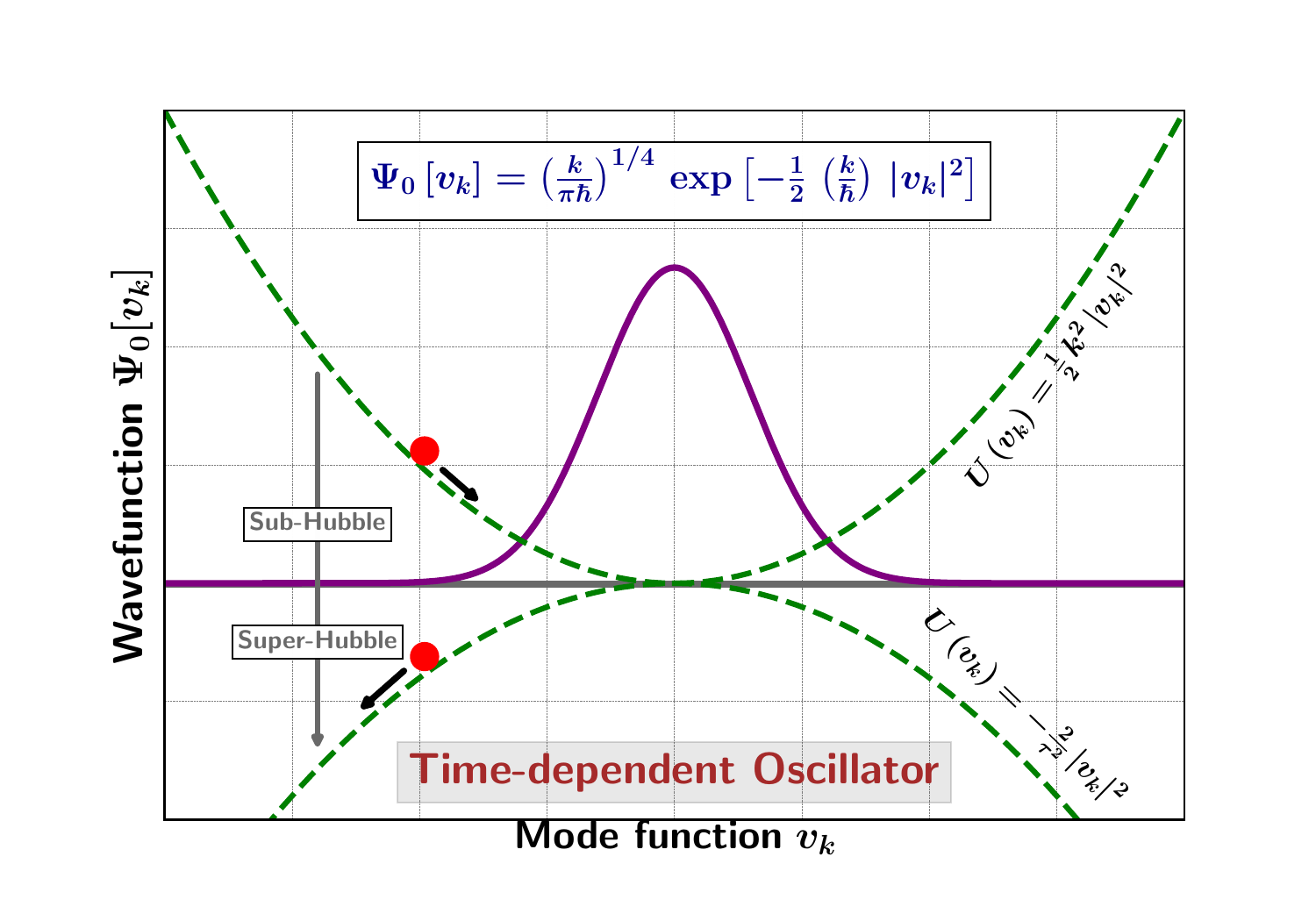}
\caption{The time-dependent effective potential and the ground state wave function are plotted in green and purple  respectively for  a  harmonic oscillator  with a  time dependent frequency as a representative of the Mukhanov-Sasaki  mode functions  $v_k$, as  described by Eq.~(\ref{eq:MS_modefunctions}).}
\label{fig:SHO_excitation}
\end{center}
\end{figure}

The key difference between inflationary fluctuations and fluctuations in Minkowski spacetime is the presence of the tachyonic mass term in Eq.~(\ref{eq:MS_modes}), which forces the mode functions $v_k$ to grow linearly with scale factor, thus making sure that $\zeta_k$ is conserved on super-Hubble scales. In fact the growth/instability of $v_k$ can be understood from the fact that Eq.~(\ref{eq:MS_modes}) describes an {\em inverted oscillator\/}  at late times, when $k^2\tau^2 \ll 1$, featuring a maximum at the centre \cite{Guth:1985ya} as shown in  Fig.~\ref{fig:SHO_excitation}. Accordingly, $\zeta_k$ behaves like an over-damped  oscillator whose damping term increases quickly/non-adiabatically as the mode becomes super-Hubble.
  To see this qualitatively, let us consider the effective frequency/mass term of the Mukhanov-Sasaki variable, namely
 \ber
\Omega_v^2(k,\tau) &=& k^2 + {\cal M}_{\rm eff}^2(\tau) = k^2 - \f{2}{\tau^2}\, ; \label{eq:MS_Omega_k}\\
\Omega_v^{ '}(k,\tau) &=& \f{2}{\tau^3 \, \Omega_v} \, , \label{eq:MS_Omega_k_der}
 \eer
for which, one obtains 
 \beq
\bigg\vert\f{\Omega'}{\Omega^2}\bigg\vert = \bigg\vert\f{2}{\l( k^2\tau^2 -2  \r)^{3/2}}\bigg\vert \,  .
\label{eq:MS_adiabaticity_violation}
 \eeq
 The above expression demonstrates that $\big\vert\f{\Omega'}{\Omega^2}\big\vert \geq 1$ when $-k\tau\leq 2$, indicating {\em violation of adiabaticity}, as the mode approaches the Hubble radius, generating large excitations of the $\hat{v}$ quanta. The effective mass term of the $\hat{v}$ field becomes tachyonic after Hubble crossing.   Hence,  the initial Bunch-Davies type vacuum fluctuations appear to be in a highly excited state from the point of view of the late time Hamiltonian when the modes are super-Hubble. In the following, we will explicitly quantify the above qualitative arguments. Since this excited state corresponds to large occupation number of particles, the super-Hubble  quantum fluctuations can be considered as classical in this regard \cite{Riotto:2002yw,Guth:1985ya}.  The issue  of  quantum-to-classical transition is rather complex and is a topic of intense investigation (see Refs.~\cite{Kiefer:1998qe,Kiefer:1998jk,Kiefer:1998pb,Burgess:2006jn,Kiefer:2007zza,Kiefer:2008ku,Martin:2012ua,Lyth:2006qz,Lochan:2014dca,Burgess:2014eoa,Martin:2015qta,Burgess:2017ytm,Martin:2018zbe,Colas:2023wxa,Bhattacharyya:2024duw}). 

The quadratic action for $\hat{\zeta}$ in Eq.~(\ref{eq:Action_quad_zeta}) can be written in terms of the Fourier modes by proceeding  as follows. Incorporating the Fourier decomposition  of $\hat{\zeta}$
$$ \hat{\zeta}(\tau,\vec{x}) = \int  \f{{\rm d}^3 \vec{k}}{(2\pi)^3} ~ \hat{\zeta}_{\vec{k}}(\tau) ~ e^{i\, \vec{k}.\vec{x}} $$
into the action in Eq.~(\ref{eq:Action_quad_zeta}), we get
$$ S^{(2)}[\hat{\zeta}(\tau, \vec{x})] = \f{1}{2} \int {\rm d}\tau \,  z^2   \int  \f{{\rm d}^3 \vec{k}_1}{(2\pi)^3} \int  \f{{\rm d}^3 \vec{k}_2}{(2\pi)^3} \int {\rm d}^3\vec{x} \, e^{i \l( \vec{k}_1 + \vec{k}_2 \r).\vec{x}}  \l[ \hat{\zeta}'_{\vec{k}_1} \hat{\zeta}'_{\vec{k}_2} - \l( i\, \vec{k}_1\r). \l(i \, \vec{k}_2 \r)\hat{\zeta}_{\vec{k}_1}  \hat{\zeta}_{\vec{k}_2} \r]  \, . $$
Using the definition of Dirac delta function
\beq
\delta_D^{(3)}(\vec{k}_1 + \vec{k}_2) = \f{1}{(2\pi)^3} \,  \int {\rm d}^3\vec{x} \, e^{i \l( \vec{k}_1 + \vec{k}_2 \r).\vec{x}}  \, ,
\label{eq:Dirac_delta_def}
\eeq
 we obtain the expression for the quadratic  action of $\hat{\zeta}$ to be 
\beq
S^{(2)}[\hat{\zeta}] = \f{1}{2} \int {\rm d}\tau \int  \f{{\rm d}^3 \vec{k}}{(2\pi)^3} \, z^2 \l[ \l( \hat{\zeta}_k'\r)^2 - k^2 \, \hat{\zeta}_k^2 \r]   \, .
\label{eq:Action_quad_zeta_Fourier}
\eeq
Similarly the quadratic action for the Mukhanov-Sasaki variable $\hat{v}$ from Eq.~(\ref{eq:Action_quad_v}) becomes 
\beq
S^{(2)}[\hat{v}] = \f{1}{2} \int {\rm d}\tau \int  \f{{\rm d}^3 \vec{k}}{(2\pi)^3}  \l[ \l( \hat{v}_k'\r)^2 - \l( k^2 + {\cal M}_{\rm eff}^2 \r) \hat{v}_k^2 \r]   \, .
\label{eq:Action_quad_MS_Fourier}
\eeq
From the above actions written in terms of Fourier modes, we can compute the vacuum expectation values of any relevant physical quantity associated with the quantum field $\hat{v}$, since we have fixed the vacuum $|0 \rangle$ \textit{w.r.t} to the Bunch-Davies condition imposed mode functions given in Eq.~(\ref{eq:MS_vk_SR}), re-written as 
\ber
v_k(\tau) &=&  \f{1}{\sqrt{2k}} \l(1-\f{i}{k\tau}\r) e^{-i \, k\tau} \, ; \label{eq:v_k_MS_sol_final} \\
\zeta_k(\tau) &=&  \f{-\tau}{\sqrt{2k}} \l(\f{H}{m_p} \r) \, \f{1}{\sqrt{2\epsilon_H}} \l(1-\f{i}{k\tau}\r) e^{-i \, k\tau} \, .
\eer
The conjugate momentum of the Mukhanov-Sasaki field is $\hat{\Pi}_{v} = \hat{v}'$, while its Lagrangian density is given by
$${\cal L}_v^{(2)} = \f{1}{2} \l[  \left(\hat{v}'\right)^2 - \left(\partial_i \hat{v}\right)^2 - {\cal M}_{\rm eff}^2 \hat{v}^2 \r] \, .$$
Hence the Hamiltonian corresponding to  the Mukhanov-Sasaki field is 
\beq
\hat{H}_v^{(2)} \equiv \int {\rm d}^3\vec{x} \, \l( \hat{v} \, \hat{\Pi}_{v} - {\cal L}^{(2)} \r) = \f{1}{2} \int {\rm d}^3\vec{x} \, \l[  \hat{\Pi}_{v}^2 + \left(\partial_i \hat{v}\right)^2 +{\cal M}_{\rm eff}^2 \hat{v}^2 \r] \, ,
\label{eq:MS_Hamiltonian}
\eeq
which takes the following form in terms of Fourier modes $\hat{v}_k$:
\beq
 \hat{H}_v^{(2)}= \f{1}{2} \int  \f{{\rm d}^3 \vec{k}}{(2\pi)^3}  \l[ \l( \hat{v}_k'\r)^2 + \l( k^2 + {\cal M}_{\rm eff}^2 \r) \hat{v}_k^2 \r]   \, .
\label{eq:MS_Hamiltonian_Fourier}
\eeq
Expanding the quantum field $\hat{v}$ and its momentum $\hat{\Pi}_{v}$ in terms of creation and annihilation operators, as given in Eq.~(\ref{eq:zeta_quantum}), we obtain
\ber
\hat{v}_k(\tau) &=&  v_k(\tau)  \, \hat{a}_{\vec{k}} + v_k^{\ast}(\tau) \, \hat{a}_{-\vec{k}}^{\dagger} \, ; \label{eq:v_k_quantum} \\
\hat{\Pi}_{v}(\tau) &=&  v'_k(\tau)  \, \hat{a}_{\vec{k}} + v_k'^{\ast}(\tau) \, \hat{a}_{-\vec{k}}^{\dagger} \, . \label{eq:v_der_quantum}
\eer 
Using the above Eqs.~(\ref{eq:v_k_quantum}), (\ref{eq:v_der_quantum})  in Eq.~(\ref{eq:MS_Hamiltonian_Fourier}), we obtain
\begin{align*}
\hat{H}_v^{(2)} = \f{1}{2} \int  \f{{\rm d}^3 \vec{k}}{(2\pi)^3}  \left[ \l( \l(v_k'\r)^2 + \Omega_k^2 v_k^2 \r)\hat{a}_{\vec{k}}\hat{a}_{\vec{k}} +   \l( \l(v_k'^{\ast}\r)^2 + \Omega_k^2 v_k^{\ast \, 2} \r)\hat{a}_{-\vec{k}}^{\dagger}\hat{a}_{-\vec{k}}^{\dagger} \right. \\ \left. + \l( |v_k'|^2 + \Omega_k^2 |v_k|^2 \r) \l( \hat{a}_{\vec{k}}\hat{a}_{-\vec{k}}^{\dagger} + \hat{a}_{-\vec{k}}^{\dagger}  \hat{a}_{\vec{k}}  \r)  \right] ,
\end{align*}
where the time dependent frequency is given by 
$$ \Omega_k^2 = k^2 + {\cal M}_{\rm eff}^2(\tau) = k^2 - \f{2}{\tau^2}\, .$$
Using the commutation relation from Eq.~(\ref{eq:a_k_quantum}), we get
$$ \hat{a}_{\vec{k}} \hat{a}_{-\vec{k}}^{\dagger} = (2\pi)^3 \, \delta_D^{(3)}(0) +  \hat{a}_{-\vec{k}}^{\dagger}\hat{a}_{\vec{k}} \, ,  $$
which leads to
\begin{align}
 \hat{H}_v^{(2)} = \f{1}{2} \int  \f{{\rm d}^3 \vec{k}}{(2\pi)^3} \left[ \l( \l(v_k'\r)^2 + \Omega_k^2 v_k^2 \r)\hat{a}_{\vec{k}}\hat{a}_{\vec{k}} \, + \,   \l( \l(v_k'^{\ast}\r)^2 + \Omega_k^2 v_k^{\ast \, 2} \r)\hat{a}_{-\vec{k}}^{\dagger}\hat{a}_{-\vec{k}}^{\dagger} \right. \nonumber \\
 \left. + \l( |v_k'|^2 + \Omega_k^2 |v_k|^2 \r) \l( 2 \hat{a}_{-\vec{k}}^{\dagger} \hat{a}_{\vec{k}} + (2\pi)^3 \delta_D^{(3)}(0)   \r) \right]  \, .
\label{eq:MS_Hamiltonian_Fourier1}
\end{align}
The energy of the quantum field is given by the vacuum expectation value of the Hamiltonian, namely,
\beq
E_v \equiv \langle \, 0 \, | \, \hat{H}_v^{(2)} \, | \, 0 \, \rangle =  \f{1}{2} \, (2\pi)^3 \, \delta_D^{(3)}(0) \, \int  \f{{\rm d}^3\vec{k}}{(2\pi)^3}   \,   \l[ \, |v_k'|^2 + \Omega_k^2 |v_k|^2 \,  \r] \,  .
\label{eq:MS_Energy}
\eeq
The appearance of the  Dirac delta function indicates that formally the total vacuum energy of the quantum field is divergent, as expected. A better physical quantity to compute  is the vacuum energy density, $\rho_v$. Noting that 
$$(2\pi)^3\delta^{(3)}(0)=\lim_{|\vec{k}|\to 0} \, \int  \, {\rm d}^3\vec{x}  \, e^{i \, \vec{k}.\vec{x}} =  \int \,  {\rm d}^3\vec{x} \, ,$$ 
 the vacuum energy density  can be obtained from Eq.~(\ref{eq:MS_Energy}) to be
\beq
\rho_v =  \f{1}{2}  \, \int  \f{{\rm d}^3\vec{k}}{(2\pi)^3}   \,   \l[ \, |v_k'|^2 + \Omega_k^2 \,  |v_k|^2 \,  \r] \,  . \label{eq:MS_Energy_density} 
\eeq
At this stage, it is important to obtain simplified  expressions for $v_k$ and $v_k'$, starting from Eq.~(\ref{eq:v_k_MS_sol_final}), as  
\begin{align}
v_k(\tau) &=  \f{1}{\sqrt{2k}} \l(1-\f{i}{k\tau}\r) e^{-i \, k\tau}  ~ \Rightarrow ~ |v_k(\tau)|^2 = \f{1}{2k} \l(1+\f{1}{k^2\tau^2} \r) \, ; \label{eq:MS_magnitude} \\
v_k'(\tau) &=  \f{1}{\sqrt{2k}} \, \f{1}{\tau} \, \l[ -1 + i \l(\f{1}{k\tau} - k\tau\r)\r] e^{-i \, k\tau} ~ \Rightarrow ~  |v_k'(\tau)|^2 = \f{k}{2} \l(1 - \f{1}{k^2\tau^2} + \f{1}{k^4\tau^4} \r) \, , \label{eq:MS_der_magnitude}
\end{align}
which can be used to compute the integrand of Eq.~(\ref{eq:MS_Energy_density}) to be  
\beq
|v_k'|^2 + \Omega_k^2 |v_k|^2  \simeq |v_k'|^2 + \l( k^2 - \f{2}{\tau^2} \r) |v_k|^2 = \f{k}{2} \l( 2 - \f{2}{k^2\tau^2} - \f{1}{k^4\tau^4} \r) \, .
\label{eq:MS_der_comb}
\eeq
It is instructive to obtain the following  sub-Hubble and super-Hubble limits of the above expression:
\ber
|v_k'|^2 + \Omega_k^2 |v_k|^2\bigg\vert_{-k\tau \to \infty} &=&  k \, ; \label{eq:MS_der_comb_subH} \\
|v_k'|^2 + \Omega_k^2 |v_k|^2\bigg\vert_{-k\tau \to 0} &=& -\f{k}{2} \l( \f{1}{k^4\tau^4} \r) \, . \label{eq:MS_der_comb_supH}
\eer
Incorporating the expression in Eq.~(\ref{eq:MS_der_comb_supH}) into Eq.~(\ref{eq:MS_Energy_density}), we obtain the energy density of the super-Hubble (IR) part of the  $\hat{v}$ field to be 
\begin{align}
\rho_v\big\vert_{\rm IR} (\tau) = - \f{1}{2} \int_{k_{\rm IR}}^{-1/\tau} \f{{\rm d}^3\vec{k}}{(2\pi)^3}   ~  \f{k}{2} \, \l(  \f{1}{k^4 \tau^4} \r) = - \f{1}{4\pi^2}\f{1}{\tau^4} \int_{k_{\rm IR}}^{-1/\tau}  {\rm d} \, \l(\ln  k \r) =  - \f{1}{4\pi^2} \f{1}{\tau^4} \ln\l(\f{-1}{k_{\rm IR}\tau}\r)\,  , 
\label{eq:MS_Energy_density_supH}
\end{align}
where $k_{\rm IR}$ is a cut-off introduced to regulate the IR divergence (since inflation may have a beginning \cite{Borde:2001nh}, $k_{\rm IR}$ corresponds to the comoving mode making its Hubble-exit at the beginning of inflation). Eq.~(\ref{eq:MS_Energy_density_supH}) demonstrates that  the energy density of the super-Hubble part of $\hat{v}$ diverges at late times, $-\tau \to 0$, which can be attributed to the fact that the effective mass term is tachyonic, as described previously in Fig.~\ref{fig:SHO_excitation} with the analogy of an inverted oscillator. (Note that the energy density corresponding to the sub-Hubble (UV) part of the $\hat{v}$ field in Eq.~(\ref{eq:MS_der_comb_subH})  leads to the usual UV   divergence associated with quantum fluctuations in Minkowski spacetime.) 

Similarly we compute the expectation value of the occupation number density of particles of a given mode $k$ to be 
\beq
\langle \, 0 \, | \, \hat{n}_v(k) \, | \, 0 \, \rangle \equiv  \f{1}{2} \,  \Big\vert \f{|v_k'|^2 + \Omega_k^2 |v_k|^2}{k}\Big\vert_{-k\tau \to 0} -\f{1}{2} \simeq \f{1}{4} \, \l(\f{-1}{k\tau}\r)^4  \to \infty ,
\label{eq:MS_number-density_k}
\eeq
 which diverges, indicating abundant excitations of $\hat{v}$ quanta, as discussed before. The corresponding energy density and particle number density for the $\hat{\zeta}$ field can be obtained by $\hat{\zeta} = \hat{v}/(am_p\sqrt{2\epsilon_H})$. Accordingly, the finite and frozen power spectrum of $\hat{\zeta}$ on super-Hubble scales can be justified in this Heisenberg picture. Note that similar arguments are also applicable for the tensor-type fluctuations discussed in Sec.~\ref{sec:inf_dyn_tensor_QF}.  See App.~\ref{app:massive_dS} for a discussion on the quantum fluctuations of a massive scalar field in de Sitter spacetime and their relation to the inflationary scalar fluctuations.

\bigskip

Before concluding the treatment of scalar fluctuations, let us stress that the computation of  the  inflationary scalar power spectrum given in Eq.~(\ref{eq:P_S_SR}) is one of the key achievements of modern theoretical cosmology. This is certainly one of the  most important applications of quantum field theory to curved spacetime in physics, generating the seeds for the large-scale cosmological structure in the universe\footnote{In April 2014 (following the announcement of B-mode polarization signal by BICEP2 in Ref.~\cite{BICEP2:2014owc}), Juan Maldacena gave a set of 4 lectures on the computation of inflationary quantum fluctuations at the International Centre for Theoretical Physics (ICTP), Trieste in Italy. At the beginning of the first lecture, he stressed the importance of learning these computations for a theoretical physicist in the form of a funny allegory (here is  the \href{https://indico.ictp.it/event/a13193/session/3/contribution/17/material/video/}{\purple 
 {\bf link to the  first lecture}}). I was fortunate to be present in the classroom as a young student who was inspired by those  lectures and later  taught myself the calculations over the period of a few years (and I still continue to learn new ways of computing and understanding them).}. These results were first obtained in the early 1980s around the time of the famous Nuffield conference organised by Stephen Hawking and colleagues (see Ref.~\cite{Gibbons:1984hx}) and were published in Refs.~\cite{Mukhanov:1981xt,Hawking:1982cz,Starobinsky:1982ee,Guth:1982ec}. For a historical account of the subject, see the popular Ref.~\cite{Guth:1997wk}.

\subsection{Tensor quantum fluctuations during inflation}
\label{sec:inf_dyn_tensor_QF}
The gauge-invariant  tensor fluctuations $h_{ij}(\tau,\vec{x})$ during inflation are described by the quadratic action \cite{Maldacena:2002vr,Baumann:2018muz}
\beq
    S^{(2)}[h_{ij}] = \frac{1}{2} \int {\rm d}\tau~{\rm d}^3\vec{x}~\left(\frac{a m_p}{2}\right)^2 \left[ h_{ij}'^2 - (\partial_l h_{ij})^2\right] \, ,
    \label{eq:Action_quad_tensor}
\eeq
where the  transverse and traceless tensor field  $h_{ij}(\tau,\vec{x})$  satisfies 
\beq
\partial^i \,  h_{ij} = 0 ~~\text{(transverse)}\, ; \quad  h_i^i  = 0 ~~ \text{(traceless)} \, , 
\label{eq:tensor_transverse_traceless}
\eeq
and  can be decomposed into its two orthogonal  polarization components,
\begin{align}
    h_{ij}(\tau,\vec{x}) &= \frac{1}{\sqrt{2}} \begin{bmatrix} h_+ & h_{\times} & 0 \\ h_{\times} & -h_+ & 0 \\ 0 & 0 & 0 \end{bmatrix}  = \frac{1}{\sqrt{2}} \begin{bmatrix} 1 & 0 & 0 \\ 0 & -1 & 0 \\ 0 & 0 & 0 \end{bmatrix} h_+(\tau,\vec{x}) + \frac{1}{\sqrt{2}} \begin{bmatrix} 0 & 1 & 0 \\ 1 & 0 & 0 \\ 0 & 0 & 0 \end{bmatrix} h_{\times}(\tau,\vec{x}) \label{eq:h_k_tensor} \\
    &= \frac{1}{\sqrt{2}}~\epsilon^+_{ij} \,  h_+(\tau,\vec{x}) + \frac{1}{\sqrt{2}}~\epsilon^{\times}_{ij} \,  h_{\times}(\tau,\vec{x}) \, , \label{eq:h_k_tensor_polarisation}
\end{align}
with $\epsilon^{\lambda}_{ij} \epsilon^{ij \, \lambda'} = 2 \delta^{\lambda \lambda'};~\lambda=\{+, \times \}$. The action can be rewritten as 
\beq
    S^{(2)}[h_+,\, h_\times] = \frac{1}{2} \int {\rm d}\tau~{\rm d}^3\vec{x}~\left(\frac{a m_p}{2}\right)^2 \, \sum_{\lambda=+,\times} \, \left[ ({h_\lambda}')^2 - (\partial_l h_\lambda)^2\right] \, .
    \label{eq:Action_tensor_canonical}
\eeq
  We define  the two Mukhanov-Sasaki  variables for the tensor perturbations to be
\beq
v_\lambda = \left(\f{a m_p}{2} \right) h_\lambda \, ,
\label{eq:MS_tensor_v}
\eeq
in terms of which the action becomes
\beq
   \boxed{ \,  S^{(2)}[v_+, \, v_\times] = \frac{1}{2} \int {\rm d}\tau~{\rm d}^3\vec{x}~ \, \sum_{\lambda=+,\times} \, \left[ ({v_\lambda}')^2 - (\partial_i v_\lambda)^2 + \frac{a''}{a} \, {v_\lambda}^2\right] \, } \, ,
\eeq
which is equivalent to the action of two massless scalar fields in de Sitter spacetime, as can be inferred by comparing this with Eq.~(\ref{eq:Action_quad_v}). The full power spectrum of tensor fluctuations is defined  by 
\begin{align}
    {\cal P}_T(k) &= \frac{k^3}{2 \pi^2} \left(\abs{h_+}^2 + \abs{h_{\times}}^2\right) = \frac{k^3}{2 \pi^2} \left(\frac{2}{a m_p}\right)^2 \left(\abs{v_+}^2 + \abs{v_{\times}}^2\right) \, .
\end{align}
Borrowing the result from Eq.~(\ref{eq:MS_vk_SR}) in Sec.~(\ref{sec:inf_dyn_scalar_QF}), we obtain
\beq
{\cal P}_T(k) = \frac{2}{\pi^2} \left(\frac{H}{m_p}\right)^2 \left(1 + k^2 \tau^2\right) \, ,
\label{eq:P_T_dS_full}
\eeq 
which on super-Hubble scales $|k\tau| \to 0$ takes the form
\beq
\boxed{ {\cal P}_T(k)\Big \vert_{k\ll aH} = \frac{2}{\pi^2} \left(\frac{H}{m_p}\right)^2 } \, .
\label{eq:P_T_dS}
\eeq 
The tensor power spectrum only depends upon the Hubble parameter during inflation, and hence encodes direct information about the energy scale of inflation. In contrast, the scalar power spectrum in Eq.~(\ref{eq:P_S_SR}) depends  both on the  Hubble parameter $H$ and the first slow-roll parameter $\epsilon_H$, as stressed before. The tensor power spectrum\footnote{Some papers in the literature use a different normalisation for the tensor field,  instead of using the canonically normalised fields $\lbrace v_+, \, v_\times \rbrace$. While the field equations are the same for both choices of normalisation, the action will differ by additional multiplicative factors, therefore the power spectra will also differ. For this reason, we recommend using the canonically normalised field variables, in terms of which the tensor action in Eq.~(\ref{eq:Action_tensor_canonical}) takes the form of two massless fields in de Sitter spacetime.} in de Sitter spacetime was originally computed by Starobinsky in his pioneering work~\cite{Starobinsky:1979ty} in 1979, a year before the inception of the inflationary hypothesis. 

Similar to the scalar power spectrum in Eq.~(\ref{eq:PS_scalar_power-law}), we can also write the tensor power spectrum as a power law, that is, 
\beq
{\cal P}_T(k) = A_T \l( \frac{k}{k_*} \r)^{n_{_T}}
\label{eq:PS_tensor_power-law}
\eeq
where $ A_T $ is the amplitude of the tensor modes at the pivot scale $ k_* $  which is chosen to be $ k_* = 0.05 \, \rm Mpc^{-1} $ and $ n_{_T} $ is the \textit{tensor spectral index}.
An important quantity is the tensor-to-scalar ratio which is defined as,
\beq
r = \frac{A_T}{A_S} = \frac{{\cal P}_T(k_*)}{{\cal P}_{\zeta}(k_*)} .
\label{eq:tensor_scalar_ratio}
\eeq
Using Eqs.~(\ref{eq:P_S_SR}) and (\ref{eq:P_T_dS}), we have
\beq
\label{eq:r_H_SR}
r = 16 \, \epsilon_H 
\eeq
\beq
\label{eq:tensor_spectral_index_Hubble_slow-roll}
n_{_T} = -2 \,  \epsilon_H 
\eeq
Thus, a  {\em consistency relation} can be established between the tensor spectral index and the tensor-to-scalar ratio as,
\beq
r= -8 \, n_{_T} \, ,
\eeq
which can be used as a smoking gun test of the single field slow-roll inflationary paradigm. 

By this stage, a careful reader must have noticed that the scalar spectral index  is denoted as $n_{_S}-1$, while the tensor spectral index  is denoted as $n_{_T}$. Such  a difference in  the notation is an unfortunate historical artifact. Since scale-invariance is an approximate symmetry during inflation, we should have denoted the scalar spectral index by $n_{_S}$,  where $|n_{_S}| \ll 1$ is a small number. In fact, we had originally intended to use this improved notation throughout these lecture notes. However, this might have created potential confusion for the reader since most other sources use $n_{_S}-1$. For this reason\footnote{And for the reason that it may be notationally convenient for observers  to present their measurements as $n_{_S} \simeq 0.965$ (which is an order 1 number), rather than as $n_{_S}\simeq -0.035$ (which is a small number), as pointed out to the author  by Daniel Baumann.}, we have (reluctantly) reverted to use the standard  $n_{_S}-1$  notation for  the scalar spectral index. Furthermore, notice that in Fig.~\ref{fig:inf_ns_r_latest}, we labelled the $X$-axis as scalar spectral index, although we plot $n_{_S}$, rather than $n_{_S}-1$. Again, we have reluctantly  given in to such a convention in order to be consistent with Fig.~8 of Ref.~\cite{Planck_inflation}. 

\bigskip 

Before moving on to study the dynamics of different popular inflationary models, it is important to remind the readers that we are working in natural units, where $\hbar, \, c  = 1$. Hence they do not appear in the expressions for the power spectra of quantum fluctuations in Eqs.~(\ref{eq:P_S_SR}) and (\ref{eq:P_T_dS}). Putting  $\hbar$ and $c$  back into the expressions for power spectra, we obtain
\ber
 {\cal P}_\zeta(k)\Big \vert_{k\ll aH} &=& \f{1}{8\pi^2} \, \l( \frac{\hbar}{c^5} \r) \, \left(\frac{H}{m_p}\right)^2 \, \frac{1}{\epsilon_H}  \, , \label{eq:P_S_SR_hbar} \\
 {\cal P}_T(k)\Big \vert_{k\ll aH} &=& \frac{2}{\pi^2} \, \l( \frac{\hbar}{c^5} \r) \,  \left(\frac{H}{m_p}\right)^2 \, . \label{eq:P_T_dS_hbar}
\eer

\subsection{Beyond linear perturbation theory: primordial non-Gaussianity}
\label{sec:Non_Gaussianity}
In Secs.~\ref{sec:inf_dyn_scalar_QF} and \ref{sec:inf_dyn_tensor_QF}, we computed the power spectra corresponding to the two-point auto-correlator of the scalar and tensor fluctuations at  linear order in perturbation theory, using the quadratic actions given by Eqs.~(\ref{eq:Action_quad_zeta}) and (\ref{eq:Action_quad_tensor}). At linear order, the scalar and tensor fluctuations behave like free (non-interacting) quantum fields and hence their wave functions are Gaussian~\cite{Celoria:2021vjw}. Therefore, the computation of their skewness  coming from the three-point (and higher odd-point) correlators using the quadratic action yields vanishing non-Gaussianity.

Even though the inflationary fluctuations are highly Gaussian, we expect a minimal amount of non-Gaussianity to be present because -- (i) Einstein's field equations are inherently  non-linear, and (ii) the inflaton's self-interaction potential $V(\phi)$ is usually not quadratic.    In order to compute them carefully, we need to move beyond the free (quadratic) actions  and work with the interaction action~(\ref{eq:Action_perturbed_scalar-tensor}). In the following, we briefly mention some of the key results on primordial non-Gaussianity in the context of single field slow-roll inflation.   
 
A systematic computation of primordial non-Gaussianity using the standard techniques of interacting QFT, known as the {\em In-In formalism}, was  carried out in a pioneering work in  Ref.~\cite{Maldacena:2002vr}. For example, using the  cubic order  action, we can compute  the (interacting) vacuum expectation value of the {\em bispectrum} ${\cal B}_\zeta\l( k_1, k_2, k_3\r)$,  defined as 
\beq
\langle \, \zeta_{\vec{k}_1} \, \zeta_{\vec{k}_2} \, \zeta_{\vec{k}_3} \, \rangle = (2\pi)^3 \, \l(\f{2\pi^2}{k_1 \, k_2 \, k_3}\r)^2 \, {\cal B}_\zeta\l( k_1, \, k_2, \, k_3\r) \, \delta_D^{3}\l( \vec{k}_1 + \vec{k}_2 + \vec{k}_3\r) \, ,
\label{eq:Bispectrum_def}
\eeq
where the appearance of the Dirac delta function is a manifestation of the translational invariance (homogeneity) of   space during inflation, namely $\vec{k}_1 + \vec{k}_2 + \vec{k}_3 = 0$ which form a closed triangle of vectors. This also indicates that the magnitude of  the bispectrum depends on the shape of this triangle. In particular, in the {\em squeezed limit} where one of the modes ($k_1$) is long and becomes super-Hubble  very early compared to the other two, \textit{i.e.} $k_1 \, \ll \, k_2, \, k_3$, the bispectrum takes the form~\cite{Maldacena:2002vr,Creminelli:2003iq}
\beq
\boxed{ {\cal B}_\zeta\l( k_1, \, k_2, \, k_3\r) \big\vert_{k_1 \, \ll \, k_2,k_3} ~ = ~ \l(1-n_{_S}\r)  {\cal P}_\zeta(k_2) {\cal P}_\zeta(k_3) } \, ,
\label{eq:Maldacena_consistency}
\eeq
which is known as the (Maldacena) {\em single field consistency relation} for the bispectrum.  Since $1-n_{_S} \ll 1$, the bispectrum is predicted to be very small (although necessarily non-zero) in the squeezed limit. 

The relative strength of the bispectrum is usually defined in the {\em equilateral configuration} 
 with $k_1 = k_2 = k_3 = k$ as~\cite{Baumann:2018muz}
 \beq
f_{\rm NL} (k) = \f{5}{18} \, \f{{\cal B}_\zeta\l( k, \, k, \, k\r)}{{\cal P}_\zeta^2(k)} \, ,
\label{eq:def_f_NL}
 \eeq
which is a measure of the fractional non-linearity  present in the scalar fluctuations.  Note that the factor $5/18$ is again an artifact of historical notation. The latest CMB  observations~\cite{Planck:2019kim,} are consistent with $f_{\rm NL} =0$, establishing  the highly Gaussian nature of primordial fluctuations and providing further support for the single field inflationary paradigm.  In parallel with the bispectrum of the scalar fluctuations $\zeta$, one can also study the bispectrum of tensor fluctuations $h_{ij}$, as well as the 3-point cross-correlation between the two. Additionally, one can also compute the {\em trispectrum}, corresponding to the four-point correlator, of the scalar and tensor fluctuations. 

\medskip

The  dynamics of primordial  interactions during inflation are encoded in the properties of primordial non-Gaussianity, which  can be inferred from the future CMB and LSS missions. Since inflation was supposed to have occurred at ultra-high energy scales, the branch of early universe cosmology that studies the structure of higher point correlators has been termed as `{\em cosmological collider physics}' \cite{Arkani-Hamed:2015bza}.  A significant improvement in the sensitivity of observational cosmology is anticipated in the coming few decades, especially in the context of galaxy clustering and $21 ~{\rm cm}$ intensity fluctuations, which will provide crucial constraints on primordial non-Gaussianity. Since our treatment of inflationary non-Gaussianity was very minimal, we refer the readers to Refs.~\cite{Creminelli:2003iq,Chen:2010xka,Komatsu:2010hc,Wang:2013zva,Lee:2016vti,Meerburg:2019qqi}  for a detailed analysis on the subject. 

\subsection{Post-inflationary evolution of primordial fluctuations}
\label{sec:evolution_fluct_post}
In the preceding subsections, we computed the primordial scalar and tensor power spectra on super-Hubble scales from inflation which serve as initial seed fluctuations for the post-inflationary universe.   In the following, we briefly discuss  the evolution of primordial scalar and tensor fluctuations upon their Hubble-entry in the post-inflationary epochs.  We begin with scalar fluctuations, which induce temperature and density fluctuations in the CMB, and  later seed the large scale structure of the universe. We then move on to discuss  the evolution of tensor fluctuations which propagate as primordial GWs and constitute a stochastic GW background at the present epoch.
\subsubsection{Evolution of scalar fluctuations after inflation}
\label{sec:scalar_evolution}
After the end of inflation, the super-Hubble comoving curvature perturbations $\zeta_k$ begin to re-enter the Hubble radius when their comoving wavenumber becomes $k = aH$. Upon their Hubble-entry they begin to oscillate and evolve with time. They subsequently induce the observed temperature anisotropies of the CMB, which ultimately leads to the large scale clustering of galaxies in the universe and the formation of the cosmic web.

In the context of the CMB, the primary observable is the two-point auto-correlator of temperature fluctuations $\delta T/\bar{T}$ along two different  directions $\hat{n}_1, \, \hat{n}_2$ (unit vectors) in the sky, separated by an angle $\cos{\theta} = \hat{n}_1 . \hat{n}_2 $. The angular power spectrum ${\cal C}_l^{\rm TT}$ can be conveniently defined by expanding the temperature auto-correlator  in terms of   spherical harmonics as 
\beq
\left\langle \, \f{\delta T (\hat{n}_1)}{\bar{T}} \, \f{\delta T (\hat{n}_2)}{\bar{T}} \, \right\rangle \Bigg\vert_{\tau_{\rm CMB}}= \sum_l \,  \l( \f{2l+1}{4\pi} \r) \, {\cal C}_l^{\rm TT}(\tau_{\rm CMB}) \, \mathbb{P}_l(\cos{\theta}) \,  ,
\label{eq:T_Pl_angular}
\eeq
where $\mathbb{P}_l(\cos{\theta})$ are the  Legendre polynomials \cite{NIST:DLMF}. 
The angular power spectrum ${\cal C}_l^{\rm TT}$ of the  CMB temperature fluctuations can  be related to the primordial power spectrum of comoving curvature fluctuations $\zeta$ \textit{via} \cite{Baumann:2018muz}
\beq
\boxed{ \, C_l^{\rm TT}(\tau_{\rm CMB}) = \f{4\pi}{(2l+1)^2} \, \int {\rm d}\ln{k} ~ {\cal T}_l^2(k,\tau_{\rm CMB}) \, {\cal P}_\zeta(k) \, } \, ,
\label{eq:CMB_Cl_TT}
\eeq
where the {\em transfer function} ${\cal T}_l(k,\tau_{\rm CMB})$ encapsulates the evolution of $\zeta_k$ starting from the epoch of  its Hubble-entry until recombination, followed by a projection onto the CMB sky. By correctly computing the transfer function using linear perturbation theory in the post-inflationary universe, one can determine ${\cal C}_l^{\rm TT}$. Note that the angular power spectrum is sensitive to both the initial primordial fluctuations and the subsequent evolution of the universe until the present epoch, when the CMB observations are made. Thus, it contains crucial information about both the early and late universe.  Apart from temperature fluctuations, one can also study the fluctuations in the polarisation of  the CMB, which captures information about both scalar and tensor fluctuations in the early universe. The latest CMB observations by the Planck mission and BICEP/Keck collaboration have  imposed stringent constraints on the primordial scalar and tensor fluctuations, which we discuss in Sec.~\ref{sec:Observation_implications}.

\medskip

The physics  of  CMB fluctuations as well as the  formation of the LSS of the universe are two of the most important fields of research in cosmology. We will not elaborate further on this topic  in these lecture notes since it requires a larger dedicated space, but rather we direct the interested readers to Refs.~\cite{Dodelson:2003ft,Mukhanov:2005sc,Baumann:2022mni,Baumann:2018muz}. 
\subsubsection{Evolution of tensor fluctuations after inflation}
\label{sec:tensor_evolution}
The  primordial tensor modes discussed in Sec.~\ref{sec:inf_dyn_tensor_QF}, which become super-Hubble  during inflation, make their Hubble-entry at late times when $k=aH$, manifesting themselves as  stochastic GWs \cite{Starobinsky:1979ty,Caprini:2018mtu,Figueroa:2019paj}.  The physical frequency of these stochastic GWs at the present epoch is given by
\beq
f = \f{1}{2\pi}\l(\f{k}{a_0}\r)=\f{1}{2\pi}\l(\f{a}{a_0}\r)H~,
\label{eq:GW_f}
\eeq 
where $a$, $H$ correspond to the scale factor and Hubble parameter of the universe during the epoch when the corresponding    tensor mode made its Hubble-entry. The characteristic frequencies of the relic GWs that become sub-Hubble prior to 
matter-radiation equality are large enough to enable them to be detected by GW observatories in the near future, such as  the  Laser Interferometer Gravitational-Wave Observatory (LIGO)~\cite{LIGOScientific:2003jxj}, Laser Interferometer Space Antenna (LISA)~\cite{LISACosmologyWorkingGroup:2022jok}, Big Bang Observer (BBO)~\cite{Harry:2006fi} and Pulsar Timing Array (PTA)~\cite{NANOGrav:2023hvm,EPTA:2023xxk,2023,Antoniadis_2022}, to name a few. While longer wavelength GWs can be detected \textit{via} their signature on  the power spectrum of  B-mode polarisation of the CMB (see Refs.~\cite{Kamionkowski:1996zd,Seljak:1996gy,Kamionkowski:1996ks,Planck_inflation,BICEP:2021xfz}).  Using the expression for the Hubble parameter  $H$ in terms of the  temperature $T$ (corresponding to the Hubble-entry of GWs) from App.~\ref{app:Hubble_z}, we obtain 
\beq
\f{H}{\mpl} = \l(\f{\rho}{3\mpl^4}\r)^{\f{1}{2}} = \l(\f{\f{\pi^2}{30}\,g_T\,T^4}{3\mpl^4}\r)^{\f{1}{2}} = \pi \, \l(\f{g_T}{90}\r)^{\f{1}{2}} \, \l(\f{T}{\mpl}\r)^2 \, ,
\label{eq:GW_H_T}
\eeq
where $g_T$ is the effective number of relativistic degrees of freedom in the energy density  at the  temperature $T$. Using entropy conservation (see Refs.~\cite{Kolb:1990vq,Mishra:2021wkm} and App.~\ref{app:Hubble_z}), one obtains
\beq
\f{a}{a_0} = \l(\f{a_{\rm eq}}{a_0}\r)\l(\f{g_{\rm eq}^s}{g_T^s}\r)^{1/3}\l(\f{T_{\rm eq}}{T}\r) \, ,
\label{eq:GW_a_T}
\eeq
where $g_T^s$ and $g_{\rm eq}^s$  are the effective number of relativistic degrees of freedom in the entropy density at some temperature $T$ and at the matter-radiation equality, respectively. Substituting $H$ from Eq.~(\ref{eq:GW_H_T}) and  $a/a_0$ from Eq.~(\ref{eq:GW_a_T}) in Eq.~(\ref{eq:GW_f}), we obtain the following important  expression for the present day frequency of GWs in terms of their Hubble-entry  
temperature
\beq
\boxed{ \, f = 7.36\times 10^{-8}\, {\rm Hz}\,\l(\f{g_0^s}{g_T^s}\r)^{\f{1}{3}}\,\l(\f{g_T}{90}\r)^{\f{1}{2}}\,\l(\f{T}{{\rm GeV}}\r) \, } ~.
\label{eq:GW_f_master}
\eeq
Values of $f$ corresponding to relic  GWs that became sub-Hubble at a number  of  important cosmic epochs are 
shown in table \ref{table:3}.

\begin{table}[htb]
\begin{center}
 \begin{tabular}{||c|c|c|c|c||} 
 \hline\Tstrut
 \bf{Epoch} & \bf{Temperature} $T$ & \bf{GW Present day  {\large f} (in Hz)}\\ [1ex] 
 \hline\hline\Tstrut
 
 Matter-radiation equality & $\sim 0.8$~eV  & $1.14\times 10^{-17}$ \\ [1.2ex] 
 \hline\Tstrut
 CMB pivot scale entry & $\sim 5$~eV  & $8.5\times 10^{-17}$ \\ [1.2ex]
 \hline\Tstrut
 Big Bang Nucleosynthesis & $\sim 1$~MeV  & $1.8\times 10^{-11}$ \\ [1.2ex]
  \hline\Tstrut
 QCD phase transition & $\sim 150$~MeV  & $2.95\times 10^{-9}$ \\ [1.2ex]
 \hline\Tstrut
 Electroweak symmetry breaking & $\sim 100$~GeV  & $2.7\times 10^{-6}$ \\ [1.2ex]
 \hline
\end{tabular}
\captionsetup{
	justification=raggedright,
	singlelinecheck=false
}
\caption{Present day frequencies of relic GWs have been tabulated  for five different temperature  scales, associated with the Hubble-entry of the respective primordial tensor modes. In order to probe
 the epoch of reheating using relic GWs, the physical frequency corresponding to tensor modes which 
become sub-Hubble during reheating must satisfy  $f>f_{_{\rm BBN}} \simeq 10^{-11}$ Hz 
so that reheating terminates before the commencement of BBN. }
\label{table:3}
\end{center}
\end{table}

The evolution equation for the Fourier modes of the tensor fluctuations, defined in Eq.~(\ref{eq:h_k_tensor}), can be obtained from the action in Eq.~(\ref{eq:Action_tensor_canonical}) to be
\beq
\boxed{ \l({h}_k^\lambda\r)'' + 2 \, \left(\frac{a'}{a}\right) \l( {h}_k^\lambda\r)' + k^2 \, {h}_k^\lambda =0  } \, ,
\label{eq:fourier_h_k}
\eeq
with $\lambda=\{+, \times \}$. Since  $a\propto t^{2/\l( 1+3w \r)}$, integrating $\int{\rm d} \tau = \int \f{{\rm d}t}{a(t)}$,  it is easy to obtain the expression for the scale factor $a(\tau)$ to be 
\beq
a(\tau) =  a_i \l[1+ \f{1+3w}{2} \, a_i H_i \,  (\tau-\tau_i)\r]^{2/(1+3w)} \, ,
\label{eq:a_tau_post_inf}
\eeq
where $w$ is the EoS of the post-inflationary epoch; with $w = 0, \, 1/3, \, w_{_{\rm re}}$ corresponding to the EoS parameters during matter domination, radiation domination and reheating epochs respectively. The general solution to Eq.~(\ref{eq:fourier_h_k})  can be  expressed in terms of Bessel/Hankel functions (see Refs.~\cite{Sahni:1990tx,Caprini:2018mtu,Figueroa:2019paj}) and  contains two integration constants, which can be fixed by  imposing the initial conditions for $\lbrace h_k^\lambda,\, (h_k^\lambda)' \rbrace$ on super-Hubble scales to match those of the  inflationary super-Hubble tensor fluctuations.  By evolving the modes across the history of the universe, one can obtain the final expression for $h_k^\lambda$ at the present epoch. For a complete derivation, we direct the readers to Refs.~\cite{Sahni:1990tx,Caprini:2018mtu}.

The present day {\em spectral density of stochastic GWs}, defined in terms of the  critical density, at the present epoch $\rho_{_{0c}}$  is~\cite{Starobinsky:1979ty,Caprini:2018mtu,Figueroa:2019paj}
\beq
\boxed{ \, \Og (f) \equiv \f{1}{\rho_{_{0c}}}\f{\d\rho_{\rm GW}^0(f)}{\d\ln{f}} = \f{1}{\rho_{_{0c}}}\f{\d\rho_{\rm GW}^0(k)}{\d\ln{k}} \, } \, ,
\label{eq:GW_Omega_def}
\eeq 
where the Fourier space  energy density of GWs is   defined as 
\begin{equation}\label{eq; expectation of rgo_GW}
    \l<0\l|\hat{\rho}_{_{\rm GW}}(\tau,\vec{x})\r|0\r>= 2\times \frac{\mpl^2}{8a^2 (\tau)} \int \d \ln k\; \frac{ k^3}{2\pi^2}\l[\l|(h_k ^{\lambda})' (\tau) \r|^2 + k^2 \l|h_k ^{\lambda} (\tau) \r|^2 \r] = \int \d \ln k \; \rho_{_{\rm GW}}(\tau, k) \, .
\end{equation}
Deep inside the sub-Hubble regime where $k \tau \gg 1$, the tensor modes $h_k ^{\lambda}(\tau)$ oscillate rapidly with frequency $k$. During such rapid oscillations, it can be shown~\cite{Boyle_2008} that  $\l|(h_k ^{\lambda})' (\tau) \r|^2 = k^2 \l| h_k ^{\lambda}(\tau) \r|^2$.
Hence the Fourier space  energy density of GWs becomes
\begin{equation}\label{eq:final_form_of_rho-GW}
    \rho_{_{\rm GW}}(\tau, k) = \frac{ \mpl^2}{4 \, a^2 (\tau)} \, \frac{ k^5}{\pi^2}  \l|h_k ^{\lambda} (\tau) \r|^2 \, .
\end{equation}
Accordingly, the present day  spectral density of stochastic GWs, defined in Eq.~(\ref{eq:GW_Omega_def}), is  given  by the following expressions \cite{Sahni:1990tx,Sahni:2001qp,Figueroa:2019paj,Ema:2020ggo,Mishra:2021wkm}:
\begin{alignat}{2}
\Og^{(\rm RD)}(f) &= \frac{1}{24} \, {\cal P}_T(f)\,\Omega_{0r}\, ,
&&f_{\rm eq} < f \leq f_{\rm re} \, , \label{eq:GW_spectrum_1b} \\
\Og^{(\rm re)} (f) &= \Og^{(\rm RD)}
\left (\frac{f}{f_{\rm re}}\right )^{2\left (\frac{w-1/3}{w+1/3}\right )}, \quad
&&f_{\rm re} < f \leq f_{\rm e} \, .
\label{eq:GW_spectrum_1c}
\end{alignat}
Here $f_{\rm eq}$, $f_{\rm re}$, $f_e$ refer  to the present day frequency of relic GWs corresponding to tensor modes that became sub-Hubble during the following stages:  
the epoch of matter-radiation equality ($f_{\rm eq}$), at the end of reheating 
(commencement  of the radiation dominated epoch, $f_{\rm re}$) and at the end of inflation ($f_e$). 
The superscripts `RD' and  `re' in $\Og$ refer to the radiative epoch and the epoch of reheating respectively.  
Note that $f_{\rm re}>f_{\rm BBN} \simeq 10^{-11}~{\rm Hz}$ in order for the universe to be in a thermal radiation dominated phase before the commencement of the BBN. Regarding the EoS $w = \wre$ appearing in Eq.~(\ref{eq:GW_spectrum_1c}) during the epoch of reheating, it is important to keep in mind the following points:
\begin{itemize}
\item
In the case of perturbative reheating  for $V(\phi)\propto \phi^{2n}$, the value of $\wre \equiv \l<w_{_{\phi}}\r>$
is given by $\l<w_{_{\phi}}\r> = \frac{n-1}{n+1}$, as discussed in Sec.~\ref{sec:reheating_perturbative}.
\item In the case of non-perturbative reheating, the physics of the reheating
epoch can be quite complex, as discussed in Sec.~\ref{sec:reheating_nonperturbative}. In this case $\wre$ is sometimes assumed 
to be a constant, for the sake of simplicity \cite{Cook:2015vqa,Dai:2014jja,Munoz:2014eqa}.
\end{itemize}

From  Eqs.~(\ref{eq:GW_spectrum_1b})~and~(\ref{eq:GW_spectrum_1c}), 
  it follows that the 
spectral density of stochastic GWs corresponding to modes that became sub-Hubble prior to 
matter-radiation equality is
\ber
\mbox{{\bf Radiative epoch:}} ~~   \Og^{(\rm RD)}(f)  &=& \l(\frac{1}{24}\r)\,r\,A_{_S}\,\l(\f{f}{f_*}\r)^{n_{_T}}\,\Omega_{0r} ~,  ~~ f_{\rm eq} < f \leq f_{\rm re}\, ,
\label{eq:GW_spectrum_2a}\\
\mbox{{\bf During reheating:}} ~~ \Og^{(\rm re)} (f)  &=& \Og^{(\rm RD)}(f)\, \left (\frac{f}{f_{\rm re}}\right )^{2\left (\frac{w-1/3}{w+1/3}\right )},~~ f_{\rm re} < f \leq f_{\rm e}\, \, ,
\label{eq:GW_spectrum_2b}
\eer
where we have used ${\cal P}_T(f) = {\cal P}_T(f_*)\left (\frac{f}{f_*}\right )^{n_{_T}}$, with $A_T \equiv {\cal P}_T(f_*) =  r \, A_{_S}$ from Eq.~(\ref{eq:tensor_scalar_ratio}). Recall that 
$f_*$ is the physical frequency (of GW) corresponding to the CMB pivot scale comoving wave number $k_*$, as can be seen from the definition in Eq.~(\ref{eq:GW_f}). 

Eqs.~(\ref{eq:PS_tensor_power-law}),~(\ref{eq:GW_spectrum_1b})~and~(\ref{eq:GW_spectrum_1c}) allow us to define a local post-inflationary
gravitational wave  spectral index as follows:
\beq
\ng = \frac{\d\ln{\Og(k)}}{\d\ln{k}} = \frac{\d\ln{\Og(f)}}{\d\ln{f}} ,
\eeq
where
\beq
\ng =n_{_T} + 2\,\left (\frac{w-1/3}{w+1/3}\right ) ,
\label{eq:index1}
\eeq
which implies $\ng > n_{_T}$ for $w > 1/3$, $\ng = n_{_T}$ for $w = 1/3$, and
$\ng < n_{_T}$ for $w < 1/3$,
where $w$ is the background EoS and is given by $w=0$ during matter domination,
$w=1/3$ during radiation domination and by $w = w_{\rm re}$ during reheating.
Since $n_{_T} \simeq -2\epsilon_H$, the latest CMB constraints  on  the tensor-to-scalar ratio $r = 16\,\epsilon_H \leq 0.036$, as discussed in Sec.~\ref{sec:Observation_implications}, imply 
$\vert n_{_T} \vert \leq 0.0045$ (also, see Ref.~\cite{Mishra:2022ijb}). Hence $n_{_T}$  is a very small quantity that does not generate an appreciable change in $\Og(k)$ over a 30 order of magnitude variation in $k$ (and hence in $f$). Therefore Eq.~(\ref{eq:index1}) effectively reduces to
\beq
\ng \simeq 2\,\left (\frac{w-1/3}{w+1/3}\right )~.
\label{eq:spectral_index}
\eeq
Thus the post-inflationary EoS has a direct bearing on the spectral index of relic gravitational
radiation with 
\ber
\ng \geq 0 ~~ {\rm for} ~~ w > 1/3\nonumber\\
 \ng \simeq 0 ~~ {\rm for} ~~ w = 1/3 \nonumber\\
\ng \lesssim 0 ~~ {\rm for} ~~ w< 1/3
\label{eq:GW_ng_cases}
\eer
 which illustrates the extreme sensitivity of the 
GW spectral index to the background EoS of the universe.

Setting $n_{_T} = 0$ for simplicity, one obtains
\begin{alignat}{3}
&{\bullet} ~~\text{Matter domination} \ (w=0) &&\Rightarrow ~~ \ng(k)\bigg\vert_{\rm MD} \ \  &&\simeq ~ - 2 \, , \\
&{\bullet} ~~ \text{Radiation domination} \ (w=1/3) \ \  &&\Rightarrow ~~ \ng(k)\bigg\vert_{\rm RD} \ \  &&\simeq ~~0 \,.
\end{alignat}
\noindent
$\bullet$~~During the pre-hot Big Bang epoch the GW spectrum depends upon the EoS during reheating.
In the context of perturbative reheating, during coherent oscillations of the inflaton around the minimum of the potential $V(\phi)\propto \phi^{2n}$, (and using the standard result $w_{\rm re} = \langle w_\phi \rangle = \f{n-1}{n+1}$ from Eq.~(\ref{eq:EOS_avg_final}) in Eq.~(\ref{eq:spectral_index})) one finds
\beq
\ng(k)\bigg\vert_{\rm OSC} = 2\l(\frac{n-2}{2n-1}\r) \, .
\eeq
However, in marked contrast to perturbative reheating, in models with non-perturbative reheating the 
reheating/preheating epoch can be a complex affair with explosive (resonant) particle production, backreaction
and non-equilibrium field theory all playing a significant role until thermalization is finally reached, as discussed in Sec.~\ref{sec:reheating_nonperturbative}. For
the sake of  simplicity this epoch is usually characterised (see \cite{Dai:2014jja,Creminelli:2014fca,Munoz:2014eqa,Cook:2015vqa}) by a constant effective EoS parameter, $\wre$, so that the general
formulae in Eqs.~(\ref{eq:GW_spectrum_2a})--(\ref{eq:index1})    
also have bearing on this scenario.
\begin{figure}[htb]
\begin{center}
\includegraphics[width=0.49\textwidth]{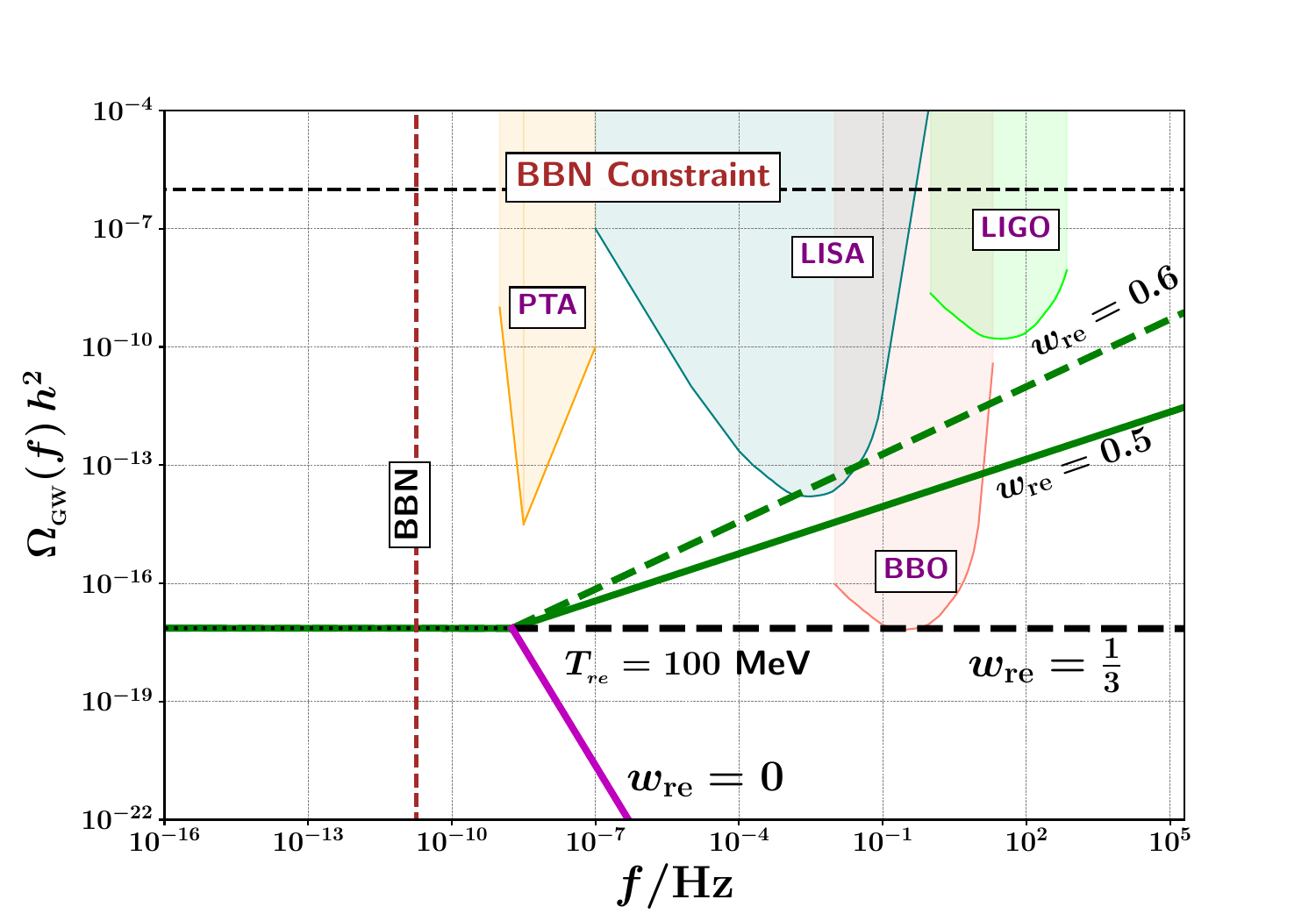}
\includegraphics[width=0.49\textwidth]{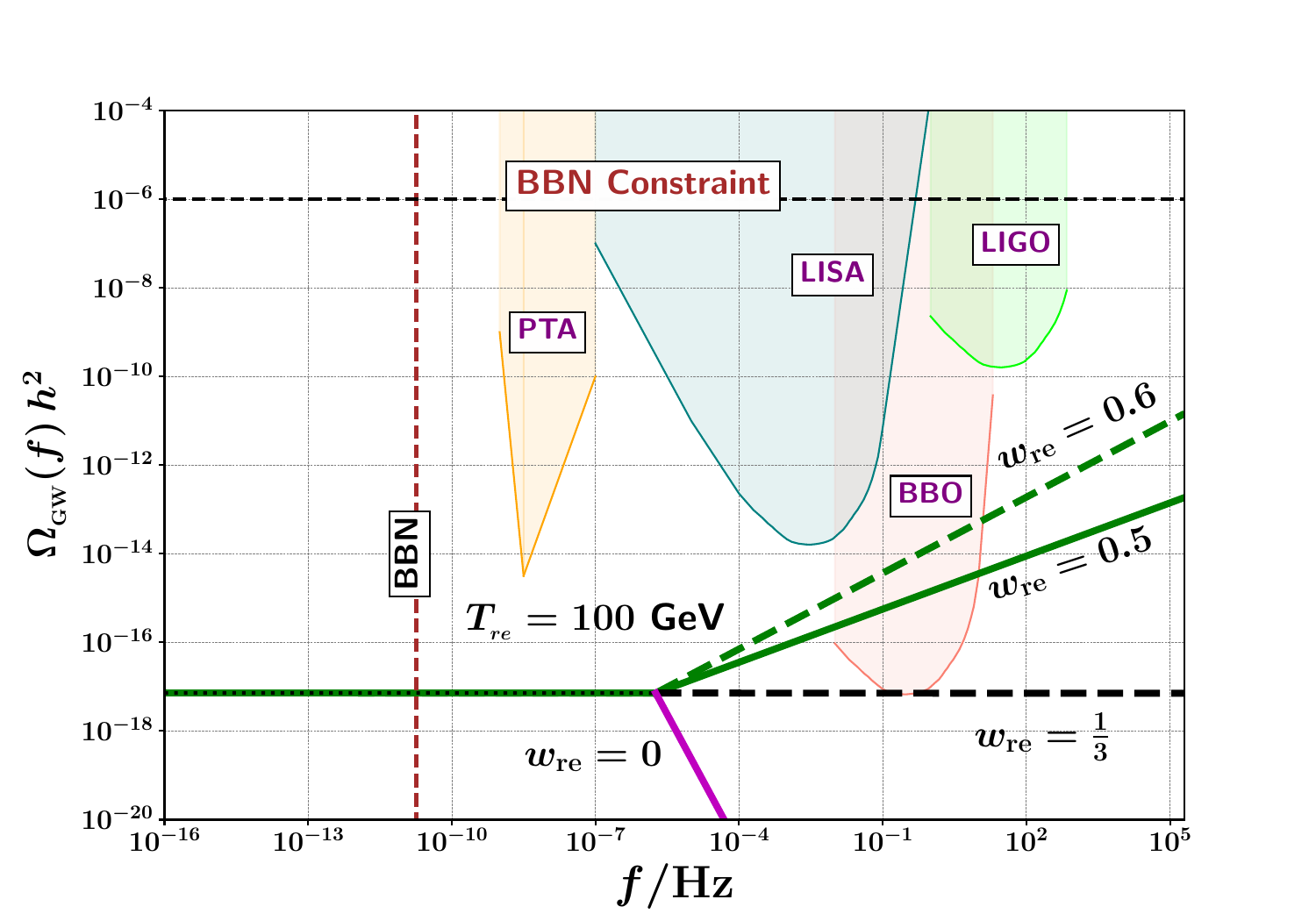}
\caption{The spectral energy density of relic gravitational waves is shown   from  the perspective of 
 ongoing and near future GW observatories such as the advanced LIGO, LISA, BBO and PTA. The {\bf left panel} depicts the spectrum of  relic GWs corresponding  to reheating EoS $\wre = 0, ~1/3, ~1/2, ~3/5$, plotted in solid purple, dashed black, solid green  and dashed  green curves respectively, for  a fixed reheating temperature $\Tre=100$ MeV and tensor-to-scalar ratio $r\simeq 0.003$. 
The {\bf right panel} shows the same but with the higher reheating temperature  $\Tre=100$ GeV. The figure demonstrates that primordial GWs with a blue-tilted spectral energy density, $\ng>0$, corresponding to relatively  stiff  EoS during reheating, will be easier to detect in the near future.}
\label{fig:GW_plateau}
\end{center}
\end{figure} 

Primordial GWs from inflation have not been observed so far, and we currently have an  upper bound on the tensor-to-scalar ratio from CMB observations as $r \leq 0.036$. However, Eqs.~(\ref{eq:GW_spectrum_2a})~and~(\ref{eq:GW_spectrum_2b}) indicate that the amplitude of the GW spectral  density depends on the tensor-to-scalar ratio $r$, while their tilt at high frequency scales depends upon the reheating EoS $w_{_{\rm re}}$. The predicted spectral energy density of  GWs   has been illustrated  in Fig.~\ref{fig:GW_plateau} in the light of both ongoing and  near future GW observatories, such as the  aLIGO~\cite{LIGOScientific:2003jxj},  LISA~\cite{LISACosmologyWorkingGroup:2022jok}, BBO~\cite{Harry:2006fi} and PTA~\cite{NANOGrav:2023hvm,EPTA:2023xxk,2023,Antoniadis_2022}. The left panel shows the spectrum of  relic GWs corresponding  to reheating EoS $w_{_{\rm re}} = 0, \, 1/3, \, 1/2, \, 3/5$, plotted in solid purple, dashed  black, solid green and dashed green curves respectively, for  a fixed (and quite low) reheating temperature $\Tre=100$ MeV and tensor-to-scalar ratio $r\simeq 0.003$. The right panel depicts the same but with a higher reheating temperature  $\Tre=100$ GeV. From Fig.~\ref{fig:GW_plateau}, it is clear that primordial GWs with a blue-tilted spectral energy density, $\ng>0$, corresponding to relatively  stiff  EoS during reheating, will be easier to detect in the near future \cite{Figueroa:2019paj}.

\section{Observational constraints on single field inflation}
\label{sec:inf_dyn_obs}
\subsection{Implications of observational constraints}
\label{sec:Observation_implications}
The latest CMB observations by the Planck mission and  BICEP/Keck collaboration in Refs.~\cite{BICEP:2021xfz,Planck_inflation} have imposed stringent  constraints on the scalar spectral index $n_{_S}$ and  tensor-to-scalar-ratio $r$ at large angular scales, given by
\beq
\boxed{~ n_{_S} \, \in \, \l[ \, 0.957, \, 0.976 \, \r] \, , \quad r(k_*) \, \leq \, 0.036 ~} ~~ \mbox{at} ~~ 95\% ~ \mbox{C.L} \, .
\label{eq:CMB_ns_r_constraint}
\eeq
Additionally, the  CMB observations are consistent with $n_{_S}$ being a constant, without exhibiting any scale dependence or running. The amplitude of scalar fluctuations at the CMB pivot scale has been measured to be \cite{Planck_inflation}
\beq
{\cal P}_\zeta(k_*) \equiv A_S = 2.1\times 10^{-9} \, ,
\label{eq:A_S_constraint}
\eeq
The constraint $r \leq 0.036$ gets translated to a constraint on  the amplitude of the tensor modes 
\beq
{\cal P}_T(k_*) \equiv A_T \leq 7.56 \times 10^{-11} \, ,
\label{eq:A_T_constraint}
\eeq
and hence, using Eq.~(\ref{eq:P_T_dS}), the Hubble parameter during (slow-roll) inflation is constrained  to be
\beq
\boxed{~ H_{\rm inf} \leq 1.93 \times 10 ^{-5}  \, m_p = 4.69 \times 10^{13} \ \rm{GeV} ~} \, .
\label{eq:H_inf_constraint}                                                                                                                               \eeq
We can find the energy scale during inflation using the above parameters and the Friedmann equation $H^2_{\rm inf} \simeq \rho_{\rm inf}/{3 m_p^2}$,
\beq
    \label{eq:inf_Energy_scale_inflation}
    E_{\rm inf} = (\rho_{\rm inf})^{1/4} \implies E_{\rm inf} \leq 1.4 \times 10^{16} \ \rm{GeV} \, ,
\eeq
implying that we are working with scales below the Planck scale ($\sim 10^{18} {\rm GeV}$) which is consistent with our description of inflation using General Relativity. During slow-roll inflation, using Eq.~\eqref{eq:r_H_SR}, constraints can be placed on $\epsilon_H$, $n_{_T}$ and $\eta_H$, namely $\epsilon_H \leq 0.00225$, $|n_{_T}| \leq 0.0045$, and $|\eta_H| \simeq 0.02$, respectively. These constraints further imply $w_{\phi} \leq -0.9985$, 
 which is consistent with a state of exponential expansion.  For further details on the implications of the latest observations for the inflationary paradigm, see  Ref.~\cite{Mishra:2022ijb}.

\begin{figure}[htb]
\begin{center}
\includegraphics[width=0.75\textwidth]{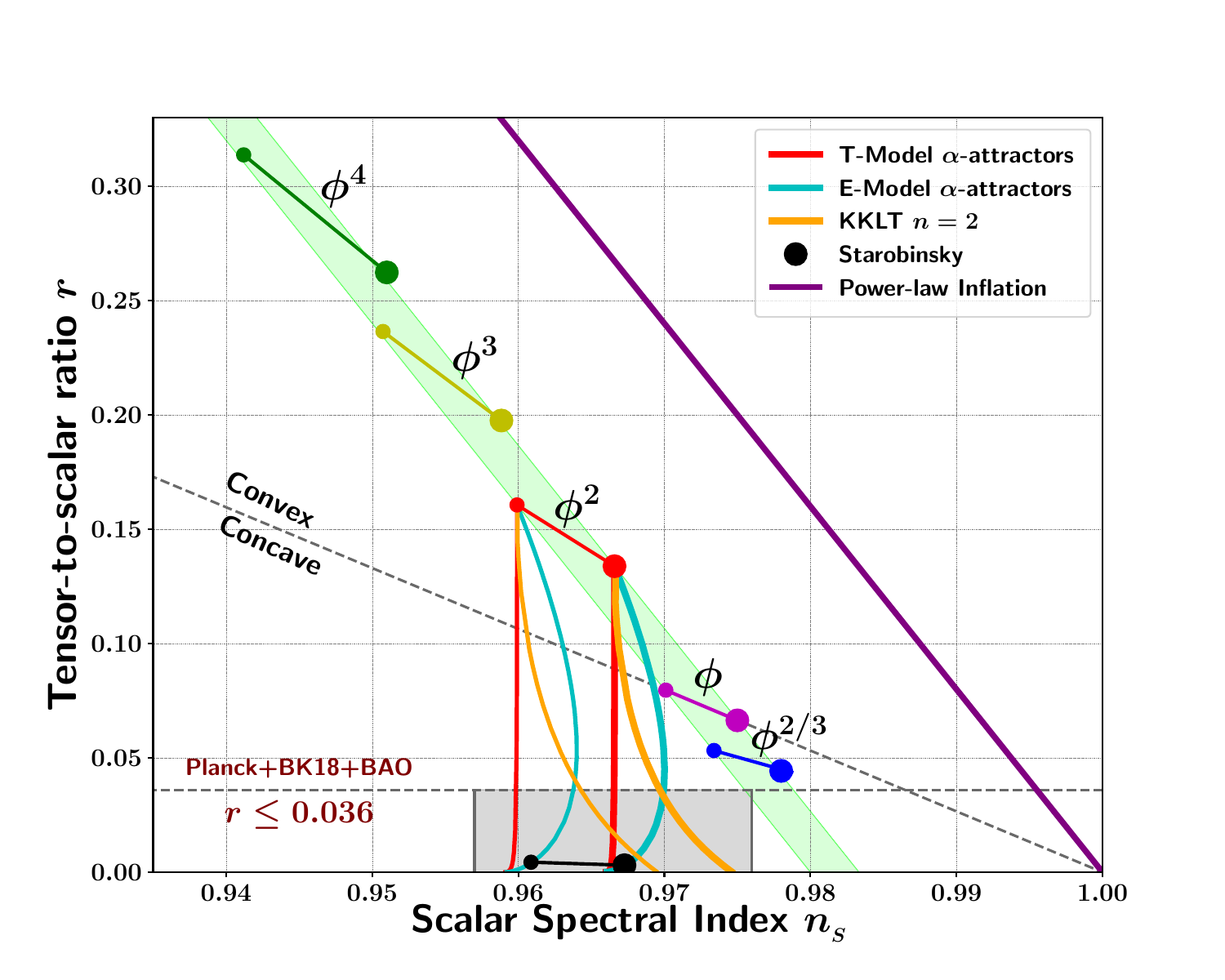}
\caption{This figure plots the  tensor-to-scalar ratio $r$ versus  the  scalar spectral index $n_{_S}$  for a number of 
inflationary  potentials (the thinner and thicker dots and  curves correspond to $N_* = 50,~60$ respectively). 
These  include  predictions of  plateau potentials such as the Starobinsky model, the Standard Model Higgs inflation, the  T-model and E-model $\alpha$-attractors as well as KKLT inflation. The CMB 2$\sigma$ bound 
$0.957 \leq n_{_S} \leq 0.976$
 and
 the upper bound on the tensor-to-scalar ratio  $r\leq 0.036$ are
 indicated by the shaded grey coloured region. Given the upper bound on $r$, it is easy to see that observations appear to favour concave potentials    over the convex ones.}
\label{fig:inf_ns_r_latest}
\end{center}
\end{figure}

The number of e-folds elapsed between the time when the pivot scale left the Hubble radius until the end of inflation is given by,
\beq
    \label{eq:inf_efolds_elapsed}
    N_* = \frac{1}{m_p} \int^{\phi_*}_{\phi_e} \frac{d \phi}{\sqrt{2\epsilon_H}},
\eeq
where $\phi_*$ and $\phi_e$ are the field values when the pivot scale exits the horizon and at the end of inflation respectively. Since $\epsilon_H$ is almost constant during inflation, the above Eq.~\eqref{eq:inf_efolds_elapsed} can be approximated as,
\beq
    \label{eq:inf_efolds_elapsed_difference}
    N_* \simeq  \frac{\Delta \phi}{m_p} \times \frac{1}{\sqrt{2\epsilon_H}}~,
\eeq
with $\Delta \phi = \phi_* - \phi_e $.
 Using the expression for the tensor-to-scalar ratio from Eq.~(\ref{eq:r_H_SR}), we can write the field displacement as 
\beq
\boxed{~ \f{\Delta\phi}{m_p} \, \simeq \, 2.12 \, \l(\f{N_*}{60}\r) \l(\f{r}{0.01}\r)^{1/2} ~} \, ,
\label{eq:Lyth_bound}
\eeq
which is known as the  \textit{Lyth bound}. The above expression implies that in order to generate large, potentially observable, tensor fluctuations, the  field displacement during single field inflation must  be super-Planckian, namely  $ \Delta \phi = \phi_* - \phi_e  > m_p $, which has important implications for inflationary model building \cite{Baumann_TASI,Baumann:2014nda}. Models of inflation in which the field displacement is super-Planckian are called {\em large-field models}. From an effective field theory (EFT) perspective, such a super-Planckian field displacement in the large-field models imply that an infinite
number of terms of the form $a_n \, \l(\phi/\Lambda_c\r)^n \, V(\phi)$ contribute to the inflationary action~(\ref{eq:Action_phi_g_munu})  with each term contributing almost equally, namely $a_n \sim \mathcal{O}(1)$ for all $n$ (the cut-off scale $\Lambda_c \leq m_p$). Hence, the large-field models, although interesting from the observational perspective, are relatively difficult to construct in the EFT, and require additional  symmetry constraints from the fundamental UV physics~\cite{Baumann:2014nda}.

Note that the BICEP/Keck collaboration have actually constrained the tensor-to-scalar ratio $r$ at the scale $k=0.002 ~{\rm Mpc}^{-1}$.  The tensor-to-scalar ratio $r$ at any $k \in [0.0005, \, 0.5]$ can be approximated as,
\beq
    r(k) = \frac{{\cal P}_T(k)}{{\cal P}_S(k)} = \frac{{\cal P}_T(k_*)}{{\cal P}_S(k_*)} \left[ \frac{\left( k/k_* \right)^{n_{_T}}}{\left( k/k_* \right)^{n_{_S} - 1}} \right] = \frac{A_T}{A_S} \l(\frac{k}{k_*} \r)^{1-n_{_S}+n_{_T}}  \, , 
\eeq
 showcasing the near scale-invariance of the ratio. Knowing the fact that the spectral index $|n_{_T}| \ll 1$, we obtain
 $$ r\big\vert_{k=0.002~{\rm Mpc}^{-1}} \simeq r\big\vert_{k=0.05~{\rm Mpc}^{-1}} \,  \l(\frac{0.002}{0.05} \r)^{0.035} \simeq 0.89 \times r\big\vert_{k=0.05~{\rm Mpc}^{-1}} \, .$$
 Hence, the constraint on $r$ at $k=0.002~{\rm Mpc}^{-1}$ by the BICEP/Keck collaboration can be directly translated into the constraint at the Planck  pivot scale $k_*=0.05~{\rm Mpc}^{-1}$ within $10\%$ accuracy.

As discussed in Sec.~\ref{sec:inf_dyn}, the aforementioned  CMB observations~\cite{Planck_inflation} by the Planck collaboration span a range of comoving frequency scales $k \in [0.0005, \, 0.5] ~{\rm Mpc}^{-1}$. Assuming the scalar power spectrum to be strictly of the power-law form given in Eq.~(\ref{eq:PS_scalar_power-law}) with  spectral index $n_{_S}-1$, we can express the variance of $\zeta$ using Eq.~(\ref{eq:Variance_power_spectrum_FLRW}) as
\beq
\sigma_\zeta^2 \, \big\vert_{\rm CMB} =A_S \,  \int_{y_{\rm min}}^{y_{\rm max}} \, \f{{\rm d}\,y}{y} \,  y^{n_{S} - 1} \, ,
\label{eq:sigma_CMB_intermediate}
\eeq
where $y = k/k_*$ and $k_* = 0.05~{\rm Mpc}^{-1}$. Since $k_{\rm min} = 0.0005~{\rm Mpc}^{-1}$ and $k_{\rm max} = 0.5~{\rm Mpc}^{-1}$, we find $y_{\rm min} = 0.1$ and $y_{\rm max} = 10$. Assuming the spectral index $n_{_S}$ to be almost constant, as suggested by the CMB observations, and using the value of $A_S$ from Eq.~(\ref{eq:A_S_constraint}), we obtain the variance of CMB fluctuations to be 
\beq
\sigma_\zeta^2 \, \big\vert_{\rm CMB} \simeq 9.7\times 10^{-9}  \, ,
\label{eq:sigma_CMB_final}
\eeq
and the corresponding standard deviation  becomes $\sigma_\zeta \simeq 9.9\times 10^{-5}$. Hence, the typical size of curvature fluctuations in the CMB sky is often quoted as $\Delta\zeta \simeq 10^{-4}$. 
\subsection{Dynamics of popular inflationary models}
\label{sec:inf_models}
In this section, we discuss how to compute the inflationary observables $\lbrace n_{_S}, \, r \rbrace$ for a given model  which supports slow-roll inflation. Traditionally, convex potentials which are steeper than $V(\phi) \propto \phi$ received  more  attention because they exhibit large field displacement during inflation and generate  large (and easier to observe) tensor-to-scalar ratios.  Amongst these large field models, the most popular ones were  the quadratic potential $V(\phi) \propto \phi^2$ and the quartic potential $V(\phi) \propto \phi^4$  proposed in Ref.~\cite{Linde:1983gd} in the context of {\em chaotic inflation}. Therefore most lecture notes  on inflation  provide a detailed discussion on these two chaotic\footnote{The reason for referring to  these large field models as {\em chaotic inflation} has to do with the type of initial conditions for inflation in these models, see Refs.~\cite{Linde:1983gd,Belinsky:1985zd,Brandenberger:2016uzh,Linde:2017pwt,Mishra:2018dtg}.} inflationary potentials.  
 
 However, the BICEP/Keck bound on the tensor-to-scalar ratio strongly favours concave potentials,  including   asymptotically flat potentials, which are shallower than $V(\phi)\propto \phi$ (see Fig.~\ref{fig:inf_ns_r_latest}). Hence we will primarily focus on computing the inflationary observables of a number of asymptotically flat potentials.  An asymptotically flat potential  has the general functional  form
 \beq
 V(\phi) = V_0 \, {\cal F}\l(\lambda \f{\phi}{m_p}\r)~,
 \label{eq:pot_plateau_toy}
 \eeq 
 such that ${\cal F}(\phi) \longrightarrow 1$ at large field values $\lambda\phi \gg m_p$,
where $\lambda$ is a free parameter.  Such a potential, schematically represented in Fig.~\ref{fig:inf_pot_toy_vanilla}, usually features one or two plateau-like wings  for large field values away from the minimum of the potential, depending on whether the potential is asymmetric or symmetric.  Additionally, an asymptotically flat  potential might approach the plateau either exponentially or algebraically.  Below we briefly  discuss  a number of important plateau potentials in light of the latest CMB observations, keeping  one example of plateau models belonging to one of the following categories.

\begin{enumerate}
\item Symmetric or asymmetric plateau potentials.

\item Single or double parameter plateau potentials.

\item Potentials with an exponential or algebraic approach to the plateau. 

\end{enumerate}

Before diving into asymptotically flat potentials, we first discuss monomial potentials which have the simplest functional form. We also use the monomial potential to explicitly demonstrate how to carry out the computation of inflationary observables for a given potential.
\subsubsection{Monomial potentials}
\label{sec:monomial_potentials}
Monomial potentials are described by the functional form
\beq
V(\phi) = V_0 \l(\frac{\phi}{m_p}\r)^p \, , \quad {\rm with} \quad p>0 \, ,
\label{eq:pot_monomial}
\eeq
for which the potential slow-roll parameters,  given in Eqs.~(\ref{eq:SR_epV}) and (\ref{eq:SR_etaV}), take the form
\beq
\epsilon_{_V}(\phi) = \f{p^2}{2} \, \l(\f{m_p}{\phi}\r)^2 \, ; \quad \eta_{_V}(\phi) = p(p-1) \, \l(\f{m_p}{\phi}\r)^2  \, .
\label{eq:epsilon_eta_monomial}
\eeq
Under the slow-roll approximations, the value of the inflaton field at the end of inflation, $\phi_e$, is given by 
$$ \epsilon_{_V}(\phi_e) = 1 \, ,$$
which, using Eq.~(\ref{eq:epsilon_eta_monomial}) leads to
\beq
\f{\phi_e}{m_p} = \frac{p}{\sqrt{2}} \, .
\label{eq:phi_end}
\eeq
We would next like to compute the value of the inflaton field  $\phi_*$ during the Hubble-exit of the CMB pivot scale $k_*$ for a given potential. Incorporating  Eq.~(\ref{eq:epsilon_eta_monomial}) into Eq.~(\ref{eq:N_*}), and using Eq.~(\ref{eq:phi_end}), we obtain 
$$
N_* = \frac{1}{p} \, \int_{\phi_e}^{\phi_*}\,   \frac{\phi \, {\rm d}\phi}{m_p^2}  =  \frac{1}{2p}\l[\l(\f{\phi_*}{m_p}\r)^2 - \f{p^2}{2}\r] \,
$$
leading to
\beq
\f{\phi_*}{m_p} = \l(2\,p\,N_* + \frac{p^2}{2}\r)^{1/2} \, ,
\label{eq:phi_*_power_law}
\eeq
which yields
\beq
\epsilon_{_V}(\phi_*) =\frac{p}{4 \, N_* + p } \, ; \quad \eta_{_V}(\phi_*) = \frac{2\,(p-1)}{4 \, N_* + p} \, .
\label{eq:eV_etaV_phi}
\eeq
In the slow-roll limit, $\epsilon_H \simeq \epsilon_{_V}$ and $\eta_H \simeq \eta_{_V} - \epsilon_{_V}$. Hence the expressions for the scalar spectral index from Eq.~(\ref{eq:nS_SR}) and tensor-to-scalar ratio from Eq.~(\ref{eq:r_H_SR}) become
\beq
n_{_S} - 1 =  2\,\eta_{_V}(\phi_*) - 6 \, \epsilon_{_V}(\phi_*) \, ; \quad  r = 16 \, \epsilon_{_V}(\phi_*) \, ,
\label{eq:nS_r_V}
\eeq
which for the monomial potential, using Eq.~(\ref{eq:eV_etaV_phi}), become
\ber
n_{_S} &=& \frac{4\,N_* - 3 \, p}{4\,N_* + p} \, ; \label{eq:nS_monomial} \\
r &=& \frac{16 \, p}{\l(4\,N_* + p\r)} \, .
\label{eq:r_monomial}
\eer
 The convex-concave divide  can be determined by looking at the $\lbrace n_{_S}, \, r \rbrace$ predictions of the linear potential $V(\phi) \propto \phi$. For $p=1$ in Eq.~(\ref{eq:pot_monomial}),  using  Eqs.~(\ref{eq:nS_monomial}) and  (\ref{eq:r_monomial}) we obtain
 $$n_{_S} = \f{4 \, N_* -  3}{4 \, N_*+1} \, ; \quad r = \f{16}{4 \, N_*+1} \, ,$$
 yielding
 \beq
r = 4 \l(1- n_{_S} \r) \, ,
\label{eq:ns_r_linear}
 \eeq
 which is the dashed grey line, marked as a divider between the predictions of convex and concave potentials in Fig.~\ref{fig:inf_ns_r_latest}.
 In order to  determine the value of $V_0$ (for different $p$), we compute the scalar power spectrum from Eq.~(\ref{eq:P_S_SR}) under the slow-roll approximations in Eqs.~(\ref{eq:friedmann_SR1})~and~(\ref{eq:friedmann_SR2}). The slow-roll power spectrum at the pivot scale takes the form
\beq
{\cal P}_\zeta (\phi_*) =  \f{1}{24\pi^2} \, \l[ \f{V(\phi_*)}{m_p^4} \r] \, \f{1}{\epsilon_{_V}(\phi_*)} \, .
\eeq
Using Eqs.~(\ref{eq:phi_*_power_law}) and (\ref{eq:eV_etaV_phi}) and imposing the CMB normalization, we obtain
$$ \f{1}{24\pi^2} \, \f{V(\phi_*)}{m_p^4} \, \f{1}{\epsilon_{_V}(\phi_*)} =  2.1 \times 10^{-9} \, ,$$
which yields
\beq
\frac{V_0}{m_p^4} \simeq  2.5 \times 10^{-7} \, \l[ \frac{p^2}{\l(2\,p\,N_* + p^2/2\r)^{\frac{p}{2}+1}}  \r] \, ,
\label{eq:V0_monimial}
\eeq
where we typically expect $N_* \in [50, \, 60]$ depending upon the reheating history, as discussed in App.~\ref{app:N*_duration}. For a given value of $p$, the monomial potential in Eq.~(\ref{eq:pot_monomial}) contains only a single parameter $V_0$, which is completely fixed by the CMB normalisation to be  Eq.~(\ref{eq:V0_monimial}). The predictions for the CMB observables $\lbrace n_{_S}, \, r \rbrace$ of the monomial potential have been shown by the lime coloured shaded region, corresponding to different values of $p$, in Fig.~\ref{fig:inf_ns_r_latest}.  Unfortunately, the latest CMB observations~\cite{Planck_inflation,BICEP:2021xfz}  strongly disfavour the entire family of monomial potentials. Hence, we move on to discuss a number of asymptotically flat potentials next that are favoured by the latest observations. 
\subsubsection{Asymptotically flat potentials}
\label{sec:asymp_flat__potentials}
For most functional forms  ${\cal F}(\phi)$ of the potential in Eq.~(\ref{eq:pot_plateau_toy}), it is difficult to analytically integrate  Eq.~(\ref{eq:N_*}) and invert it to determine $\phi_*$. For potentials that approach the plateau exponentially, the inversion of Eq.~(\ref{eq:N_*}) often involves the Lambert function~\cite{NIST:DLMF,Martin:2013tda}. Therefore one either makes reasonable  analytical approximations  or computes $\phi_*$ numerically \cite{Bhatt:2022mmn}. In the following, we will not  expand upon the techniques used to compute $\lbrace n_{_S}, \, r \rbrace$, rather we will simply  state the known results. The predictions for  $\lbrace n_{_S}, \, r \rbrace$ for different  asymptotically flat potentials, as plotted in Fig.~\ref{fig:inf_ns_r_latest}, were obtained numerically in order to be more accurate.
\subsubsection*{\underline{Starobinsky potential}}
\begin{figure}[htb]
\begin{center}
\includegraphics[width=0.68\textwidth]{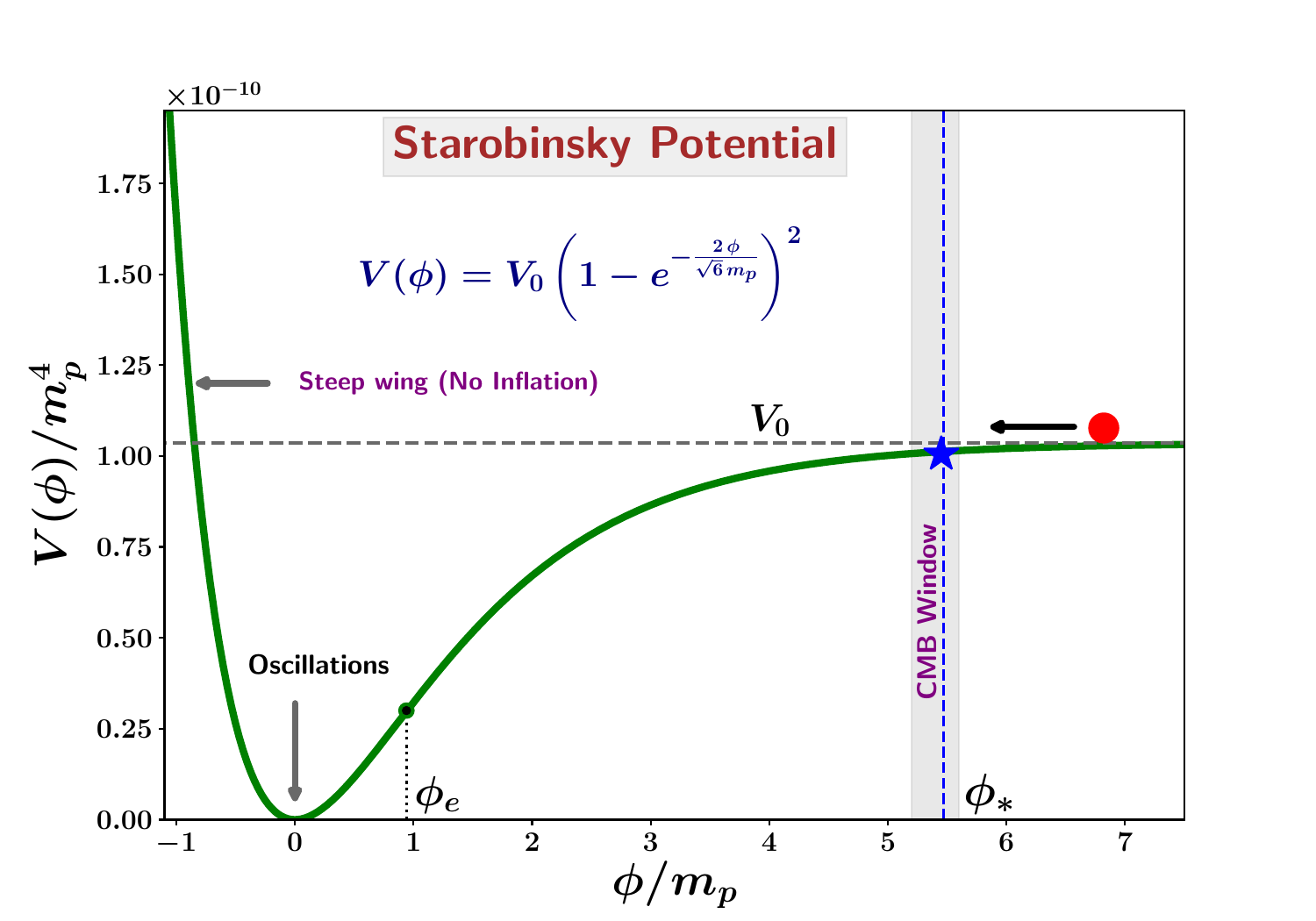}
\caption{This figure shows the Starobinsky potential given by Eq.~(\ref{eq:pot_Star}) with the CMB pivot scale  $k_* = 0.05~{\rm Mpc}^{-1}$ labelled using a blue coloured star, and  the grey colour CMB window $k_{\rm CMB} \in \l[0.0005, \, 0.5\r]~{\rm Mpc}^{-1}$  in the field space. It is clear that the CMB window constitutes only a tiny portion of the available field space  between $\phi_{\rm CMB}$ and the end of inflation $\phi_e$.}
\label{fig:inf_pot_star}
\end{center}
\end{figure}

The potential for Starobinsky inflation \cite{Starobinsky:1980te,Whitt:1984pd}  takes the form\footnote{ Note that Starobinsky inflation was originally formulated as a modified gravity theory  with  the Jordan frame Lagrangian $f(R) = R + R^2/6m^2$ which contains an additional scalar degree of freedom compared to general relativity. Upon a conformal transformation of the Jordan frame metric, one can arrive at the Einstein frame Lagrangian where the extra scalar degree of freedom takes the form of  a canonical scalar field, known as the `scalaron', see Refs.~\cite{Whitt:1984pd,Mishra:2018dtg,Shtanov:2022pdx}.} 
\beq
V(\phi) =  V_0 \, \l( 1 - e^{-\f{2}{\sqrt{6}}\f{\phi}{m_p}} \r)^2 \, ,
\label{eq:pot_Star}
\eeq
which is shown in Fig.~\ref{fig:inf_pot_star}. 

The left wing of the potential is too steep to support inflation (because on the left wing, $\epsilon_{_V}(\phi) > 1$), while the inflationary right wing is asymptotically flat. This potential features a single parameter $V_0$, which is related to the scalaron mass  $m$ by 
$V_0 = \f{3}{4} \, m^2 m_p^2$ \cite{Mishra:2018dtg}, and whose value is completely fixed by the CMB normalization (\ref{eq:A_S_constraint}). The predictions of $\lbrace n_{_S},r \rbrace$ for the Starobinsky potential  are given by 
\ber
n_{_S} \simeq 1 - \f{2}{N_*} \, , ~~ r \simeq  \f{12}{N_*^2} \, , ~{\rm for}~ N_* \gg 1 \, ;
\label{eq:ns_r_star}
\eer
which are shown by black coloured dots in Fig.~\ref{fig:inf_ns_r_latest}. As per standard convention, the smaller and larger  black dots  represent    $\lbrace n_{_S},r \rbrace$ predictions  corresponding to $N_* = 50, \, 60$ respectively. It is important to note that  the predictions of Starobinsky inflation lie at the centre of the observationally  allowed region of $\lbrace n_{_S},r \rbrace$ (shown in grey in Fig.~\ref{fig:inf_ns_r_latest}), making it  one of the most popular inflationary models at present. 
\subsubsection*{\underline{$\alpha$-attractor potentials}}
The $\alpha$-attractors belong to a broad class of superconformal inflationary models featuring  a parameter $\alpha$ in the inflaton potential, which is  inversely proportional to the curvature of the  Kahler manifold in supergravity~\cite{Kallosh:2013yoa,Kallosh:2019jnl}. Two of the most popular models amongst them are the  T- and the E-model  $\alpha$-attractors, because their predictions for  $\lbrace n_{_S}, \, r \rbrace$  are consistent with the latest CMB observations. We begin with  the inflationary predictions of the T-model, before moving on to the E-model. 
\begin{enumerate}
\item \textbf{T-model $\alpha$-attractor}:
The T-model $\alpha$-attractor potential \cite{Kallosh:2013yoa,Kallosh:2019jnl}  is a symmetric plateau potential of  the functional form 
\beq
V(\phi)  = V_0 \, \tanh^{p}{\l(\lambda \f{\phi}{m_p}\r)} \, , 
\label{eq:pot_Tmodel}
\eeq
 where $p$ is a positive real number and  $\lambda$ is related to  $\alpha$ by  $\lambda^2 = \f{1}{6\alpha}$, see Ref.~\cite{Mishra:2022ijb}. For a given value of $p$, the T-model potential has two parameters, namely $V_0$ and $\lambda$. As usual, the value of $V_0$ is fixed by the CMB normalization while $\lambda$ determines the predicted value of  $\lbrace \ns,r \rbrace$ for this potential. 
 
\begin{figure}[htb]
\begin{center}
\includegraphics[width=0.68\textwidth]{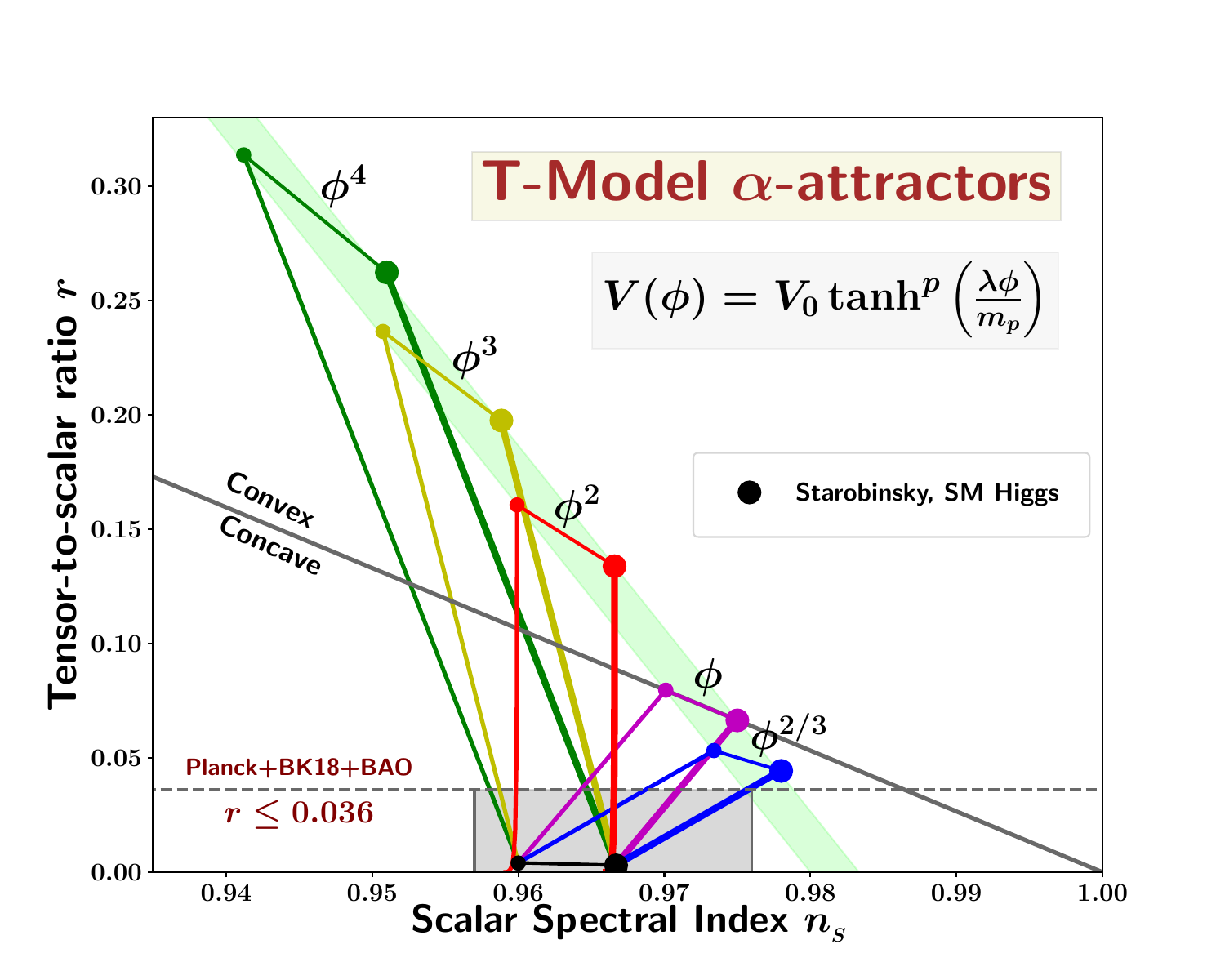}
\caption{This figure is a plot of  the  tensor-to-scalar ratio $r$, versus  the  scalar spectral index $\ns$, for the  T-model  $\alpha$-attractor in Eq.~(\ref{eq:pot_Tmodel})  (the thinner and thicker   curves correspond to $N_* = 50,\, 60$ respectively). The green, olive, red, magenta and blue curves correspond to $p=4,3,2,1,2/3$, in Eq.~(\ref{eq:pot_Tmodel}), respectively. The latest CMB 2$\sigma$ bound 
$0.957 \leq \ns \leq 0.976$
 and
 the upper bound on the tensor-to-scalar ratio  $r\leq 0.036$ are
 indicated by the grey coloured shaded region. It is important to note that upon increasing the value of $\lambda$, the predictions of the T-model potential for different values of $p$  converge towards the cosmological attractor at the centre, described by Eq.~(\ref{eq:ns_r_Tmodel_att}).  For $\lambda \gtrsim 0.1$, the predictions of the T-model become compatible with the latest CMB $2\sigma$ bound.}
\label{fig:inf_ns_r_Tmodel}
\end{center}
\end{figure}
 Predictions of the  simplest T-model potential, which  corresponds to $p=2$ in Eq.~(\ref{eq:pot_Tmodel}), are shown by the red coloured curves in Fig.~\ref{fig:inf_ns_r_latest}. As we vary  value of $\lambda$ from $\lambda: 0\longrightarrow \infty$, the $r$ versus $\ns$ values trace out  continuous  curves   in Fig.~\ref{fig:inf_ns_r_latest} (the thinner and thicker red curves correspond to $N_* = 50,\, 60$ respectively). We notice  that  in one limit, namely $\lambda \ll 1$, the CMB predictions for   $\lbrace \ns,r \rbrace$ of the T-model match with that of the quadratic  potential   $V(\phi) \propto \phi^2$, owing to the fact that the potential in Eq.~(\ref{eq:pot_Tmodel}) for $p=2$ behaves like $V(\phi) \propto \phi^2$ for $\lambda \phi \ll m_p$. In fact this is true in general  for the T-model potential with any value of $p$ in  (\ref{eq:pot_Tmodel}),  leading to the behaviour $V(\phi) \propto \phi^p$  for $\lambda\phi \ll m_p$, which is demonstrated in Fig.~\ref{fig:inf_ns_r_Tmodel}. However in the opposite limit, namely $\lambda \geq 1$ (which results in $\exp\l(\lambda\phi_*/m_p\r)  \gg 1$), the predictions of the  T-model become  \cite{Kallosh:2019jnl} 
\beq
\ns \simeq 1 - \f{2}{N_*} \, ; \quad r \simeq  \f{2}{\lambda^2 N_*^2} \, , ~{\rm for}~ N_* \gg 1\, ,
\label{eq:ns_r_Tmodel_att}
\eeq  
which is independent of $p$ (see Ref.~\cite{Mishra:2021wkm}). 
Due to this property these models are called `\textit{cosmological attractors}' \cite{Kallosh:2019jnl}. 

\item \textbf{E-model $\alpha$-attractor}:

The E-model $\alpha$-attractor potential \cite{Kallosh:2013yoa,Kallosh:2019jnl}  is an asymmetric plateau potential of  the functional form 
\beq
V(\phi) =  V_0 \, \l( 1 - e^{-\lambda\f{\phi}{m_p}} \r)^{p} \, , 
\label{eq:pot_Emodel}
\eeq
 where $p$ is a positive real number. For $p=2$ and $\lambda = 2/\sqrt{6}$, the E-model potential coincides with the Starobinsky potential\footnote{In the E-model potential,  $\lambda$ is related to the $\alpha$ parameter of $\alpha$-attractors \cite{Kallosh:2019jnl} by $\lambda^2 = \f{2}{3\alpha}$.}. In the limit $\lambda \geq 1$ (which results in $\exp\l(\lambda\phi_*/m_p\r)  \gg 1$), the predictions of the E-model become  \cite{Kallosh:2019jnl} 
\beq
\ns \simeq 1 - \f{2}{N_*} \, , ~~ r \simeq  \f{8}{\lambda^2 N_*^2} \, , ~{\rm for}~ N_* \gg 1 \, ,
\label{eq:ns_r_Emodel_att}
\eeq  
which  again is  independent of $p$ (see \cite{Mishra:2021wkm}). The CMB predictions $\lbrace \ns,r \rbrace$   of the E-model potential   are shown by the cyan colour curves in Fig.~\ref{fig:inf_ns_r_latest} for the case $p=2$. (the thinner and thicker cyan curves correspond to $N_* = 50,\, 60$ respectively).
\end{enumerate}

\medskip 

In all three of the aforementioned  models, the potentials approach the  plateau, $V(\phi) \to V_0$,  exponentially as $\phi \to \infty$. In the following we briefly discuss the strongly motivated D-brane KKLT potential which instead  approaches the plateau  algebraically.

\subsubsection*{\underline{D-brane KKLT potential}}
The D-brane KKLT inflation \cite{Kachru:2003aw,Kachru:2003sx,Kallosh:2019jnl} potential (which is analogous to the polynomial $\alpha$-attractor potential \cite{Kallosh:2022feu}) has the following general form
\beq
V(\phi) = V_0 \, \l[\f{\phi^n}{\phi^n+M^n} \r]\, , 
\label{eq:pot_KKLT}
\eeq
where  $n$ is a positive integer and  $M$ is a fundamental  scale of the theory. This is a symmetric plateau potential which approaches the  plateau behaviour algebraically, in contrast to  the exponential approach to plateau behaviour exhibited  by the T-model  potential.

\begin{figure}[htb]
\begin{center}
\includegraphics[width=0.7\textwidth]{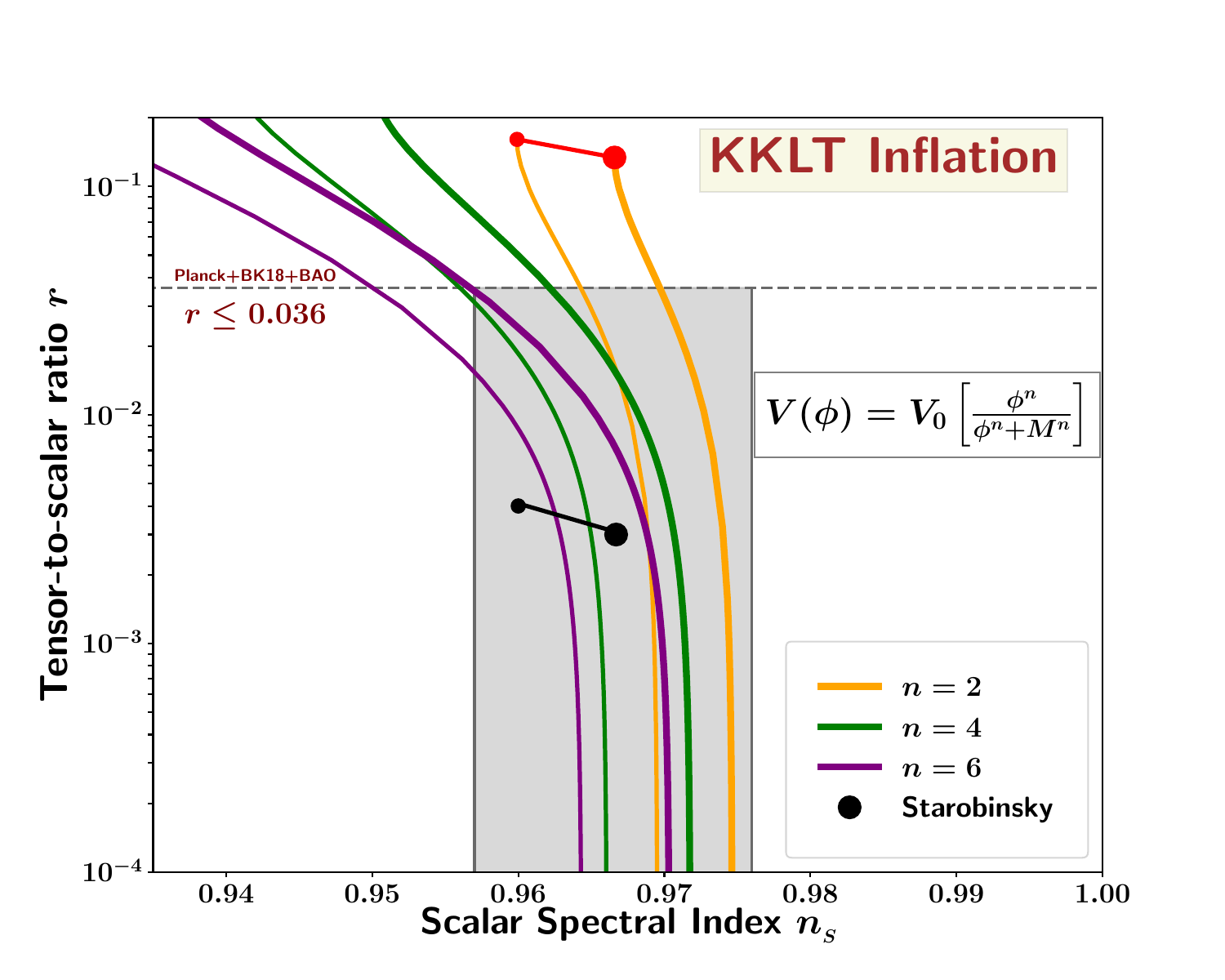}
\caption{This figure is a plot of  the  tensor-to-scalar ratio $r$ versus  the  scalar spectral index $n_{_S}$    for  KKLT inflation (the thinner and thicker curves correspond to $N_* = 50,\, 60$ respectively).    The latest CMB 2$\sigma$ bounds on the scalar spectral index $0.957 \leq n_{_S} \leq 0.976$
 and
 the upper bound on the tensor-to-scalar ratio  $r\leq 0.036$ are
 indicated by the grey coloured shaded region. Upon decreasing the value of parameter $M$ in the KKLT potential, the tensor-to-scalar ratio $r$ decreases and hence the model satisfies the CMB bound for smaller values of $M$.}
\label{fig:inf_ns_r_KKLT}
\end{center}
\end{figure}
In the limit $M\gg m_p$, the CMB window (around the field value $\phi=\phi_*$ corresponding to the Hubble-exit of the pivot scale $k_*$) belongs to the segment of the potential where it has  a monomial  power law form $V(\phi) \propto \phi^n$, and hence its predictions are strongly disfavoured by observations. In the opposite limit, namely $M\ll m_p$, the CMB window belongs to the plateau-like segment, $V(\phi) \simeq V_0$, whose predictions satisfy the CMB data very well. Figure \ref{fig:inf_ns_r_KKLT} shows the $r$ versus $n_{_S}$ plot for the KKLT potential in Eq.~(\ref{eq:pot_KKLT}) for three different values of $n$ (the thinner and thicker  curves correspond to $N_* = 50,\, 60$ respectively).  From this figure, it is easy to infer that  the  predictions of the KKLT potential (\ref{eq:pot_KKLT}) for $\lbrace n_{_S}, r \rbrace$  do not  reach a common attractor regime for different values of $n$ (in contrast to those of the T-model and E-model), rather they  cover a large horizontal  portion of the observationally allowed region of $n_{_S}$, as highlighted in Ref.~\cite{Kallosh:2019jnl}.  The relevance  of plotting $r$  on a logarithmic scale was highlighted in Ref.~\cite{Iacconi:2023mnw}.

A large class  of asymptotically flat potentials can be  categorized  into different {\em universality classes}, depending on their predictions of  $\lbrace n_{_S}, \, r \rbrace$. In this scheme,  the T- and E-model $\alpha$-attractors belong to the same universality class, while the D-brane KKLT inflation belongs to a different class. We will not elaborate further  on this point and direct the interested readers to Refs.~\cite{Roest:2013fha,Mukhanov:2013tua,Mukhanov:2014uwa,Gobbetti:2015cya,Creminelli:2014nqa,Martin:2016iqo}.

\bigskip

The number of single field inflationary models proposed  in the literature  is rather large at present, out of which  we have chosen only a few to illustrate their inflationary dynamics in these lecture notes. For a more extensive list of inflationary models, see Ref.~\cite{Martin:2013tda}. Furthermore, we have discussed  the aforementioned models from a phenomenological point of view, without much reference to their microscopic/fundamental origin. Inflationary model building from fundamental physics, such as String Theory, is an important and active area of research, see  Ref.~\cite{Baumann_TASI,Baumann:2014nda}.

\section{Post-inflationary inflaton dynamics and reheating}
\label{sec:Reheating}
The universe makes a transition from the  accelerated expansion during inflation to a decelerated expansion at the end of inflation when $\epsilon_H = 1 \, \Rightarrow  \,  \dot{\phi}^2 = V(\phi)$. After the end of inflation, the inflaton field oscillates around the minimum of the potential. The time-averaged Hubble parameter of this decelerating universe keeps  falling as $\langle H \rangle \propto 1/t$ on time scales longer than the  period oscillation, and hence at sufficiently late times, we can ignore the expansion of space while dealing with the inflaton oscillations \cite{Turner:1983he}.  Such oscillations are called {\em coherent oscillations}. We will begin with a discussion on the coherently oscillating inflaton field after inflation, in the absence of any external coupling. Then we will move on to  study the decay of the inflaton in the presence of an  external coupling. 
\subsection{Coherently oscillating scalar field}
\label{sec:phi_coherent}
During coherent oscillations of a scalar field, since the expansion of the universe can be ignored on shorter time scales, the density of the field
\beq
\rho_\phi \equiv \f{1}{2}\dot{\phi}^2 + V(\phi) \simeq V(\phi_0) \, ,
\label{eq:rho_tot_coherent}
\eeq
is almost a constant over  oscillation time scales. Here $\phi_0(t)$ is the amplitude of the oscillations which  
decreases slowly  due to the expansion of the universe  on time scales that are much longer than the time  period of oscillations of the inflaton field. Such an approximation is usually known  as the  \textit{adiabatic approximation}. Using Eq.~(\ref{eq:rho_tot_coherent})  the inflaton speed can be expressed as 
\beq
\dot{\phi} =\sqrt{ 2\, \big( V(\phi_0)-V(\phi)\big)} \, .
\label{eq:phi_dot_coherent}
\eeq
Let us calculate the time averaged equation of state $\langle w_\phi \rangle$ and time period of 
 coherent oscillations  of the inflaton around a  monomial potential
\beq
V(\phi) = V_0 \l( \f{\phi}{m_p}\r)^{2n} \, ; \quad {\rm with} \quad n> 0\, .
\label{eq:pot_monomial_n}
\eeq
\begin{figure}[hbt]
\begin{center}
\subfigure[][]{
\includegraphics[width=0.487\textwidth]{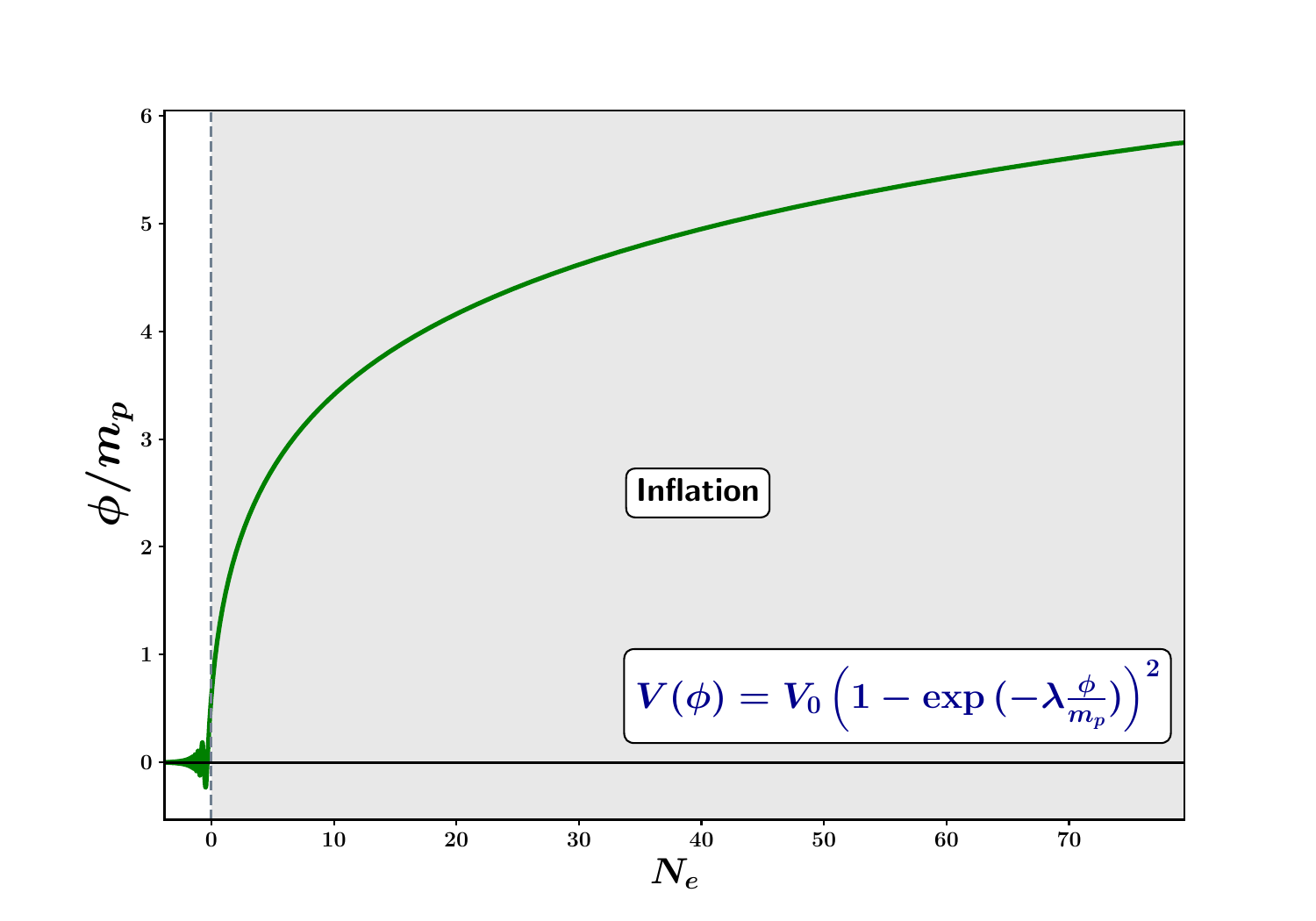}}
\subfigure[][]{
\includegraphics[width=0.487\textwidth]{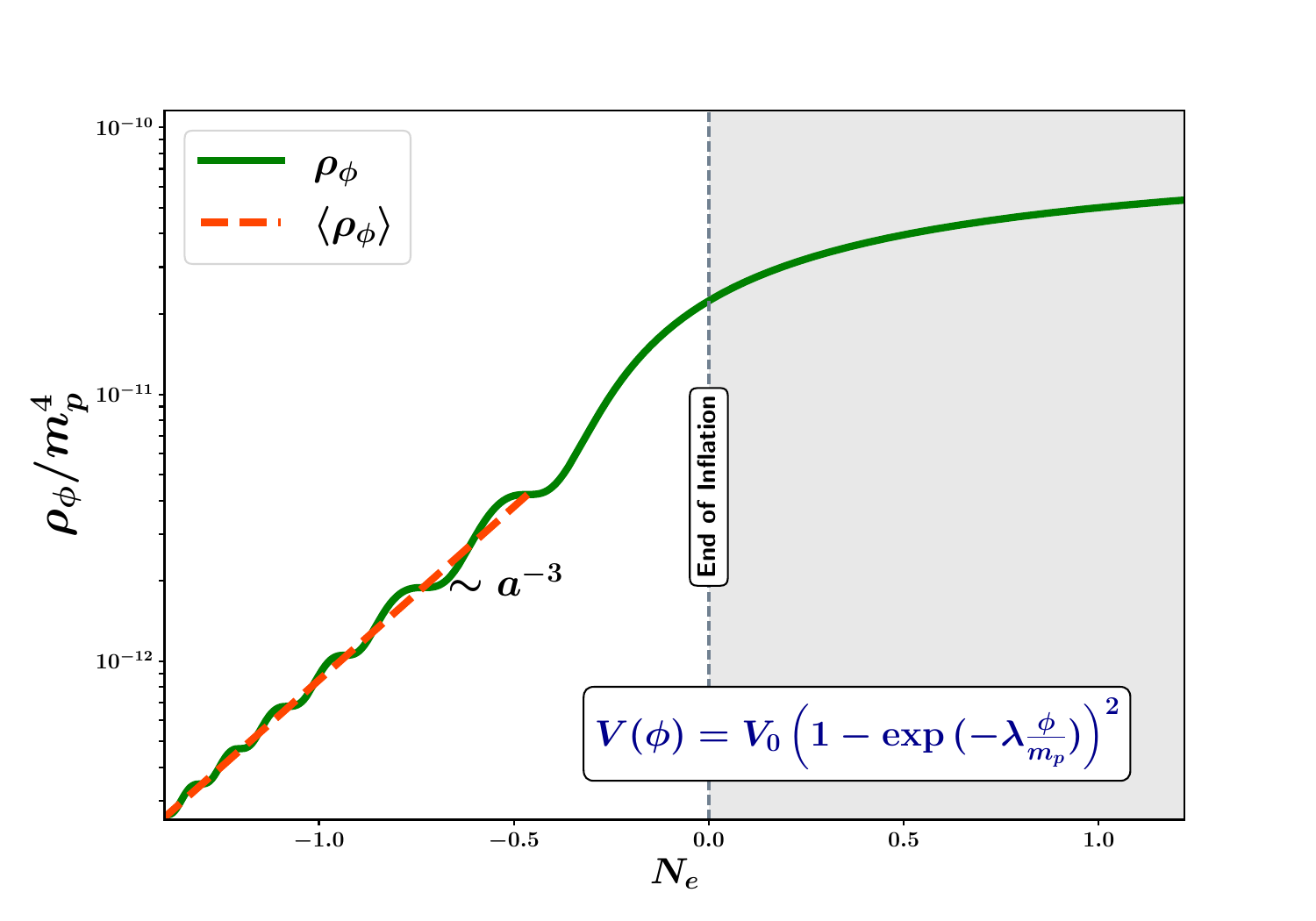}}
    \caption{Evolution of the inflaton field dynamics ({\bf left panel}) and  energy density ({\bf right panel}) is shown as a function of the number of e-folds before the end of inflation, $N_e$.  After the end of inflation, the inflaton $\phi$ begins to oscillate around the minimum of the potential. At late times, the oscillations becomes coherent and the time-averaged equation of state  becomes $\langle w_{\phi} \rangle = 0$, leading to  a pressureless matter-like behaviour. Consequently, the time averaged  energy density  of the inflaton falls as $\rho_\phi \propto 1/a^3$.}
    \label{fig:phi_rho_star}
\end{center}
\end{figure}
\subsubsection{Time-averaged equation of state}
\label{sec:w_avg_osc}
The equation of state of a scalar field is defined as
\begin{equation}
 w_\phi = \frac{p_\phi}{\rho_\phi} = \frac{\f{1}{2} \, \dot{\phi}^2 - V(\phi)}{\f{1}{2} \, \dot{\phi}^2 + V(\phi)} \, ,
 \label{eq:EOS_phi}
\end{equation}
from which we get
$$\Rightarrow  1 + w_{\phi} = \frac{\dot{\phi}^2}{\rho_{\phi}} = 2 \, \frac{V(\phi_0) - V(\phi)}{V(\phi_0)}\, ,$$
leading to
\beq
 1 + w_{\phi} =  2 \, \l[ 1 - \f{ V(\phi)}{V(\phi_0)} \r]  \, .
\label{eq:EOS_w+1}
\eeq
For the monomial potential in Eq.~(\ref{eq:pot_monomial_n}), the above expression becomes  
\beq
1 + w_{\phi} = 2 \, \l[ 1 - \l( \f{ \phi}{\phi_0} \r)^{2n} \r] \, .
\label{eq:EOS_monomial}
\eeq
Averaging the above quantity over a period of an oscillation ($T_{\rm osc}$) gives us 
$$ \langle  1 + w_\phi \rangle \equiv 1 + \langle  w_\phi \rangle  =  \f{\int_0^{T_{\rm osc}} \, {\rm d}t \, \l( 1 + w_\phi \r) }{\int_0^{T_{\rm osc}} \, {\rm d}t} = \f{2 \, \int_{-\phi_0}^{\phi_0} \, {\rm d}\phi \, \l(  \f{ 1 + w_\phi }{\dot{\phi}} \r)  }{2 \int_{-\phi_0}^{\phi_0} \, {\rm d}\phi \, \l( \f{1}{\dot{\phi}} \r)} \, . $$ 
Incorporating 
$$\dot{\phi} = \sqrt{2\, V(\phi_0) \l( 1 - \f{V(\phi)}{V(\phi_0)}  \r)} = \sqrt{2\, V(\phi_0) \l( 1 -  \l( \f{ \phi}{\phi_0} \r)^{2n} \r)} \, , $$
 and using Eq.~(\ref{eq:EOS_monomial}), we get
$$1 + \langle  w_\phi \rangle = \f{2 \, \int_{-\phi_0}^{\phi_0} \, {\rm d}\phi \, \f{ 1 - \l( \phi/\phi_0 \r)^{2n}}{\sqrt{2\, V(\phi_0) \l(1 - \l( \phi/\phi_0 \r)^{2n} \r)} }  }{ \int_{-\phi_0}^{\phi_0} \, {\rm d}\phi \, \f{ 1}{\sqrt{2\, V(\phi_0) \l(1 - \l( \phi/\phi_0 \r)^{2n} \r)} } } \, .$$
Defining $y = \phi/\phi_0$ so that  $\phi \to \phi_0 \, \implies \, y \to 1$, we get
$$ 1 + \langle  w_\phi \rangle = \f{2 \, \int_{-1}^{1} \, {\rm d}y \, \f{ 1 - y^{2n}}{\sqrt{1 - y^{2n}} }  }{ \int_{-1}^{1} \, {\rm d}y \, \f{ 1}{\sqrt{1 - y^{2n}} } }$$
which can be written as 
\beq
 1 + \langle  w_\phi \rangle = \f{2 \, \int_{0}^{1} \, {\rm d}y \, \l(1 - y^{2n} \r)^{1/2}}{ \int_{0}^{1} \, {\rm d}y \,  \l(1 - y^{2n} \r)^{-1/2}}  \, .
\label{eq:EOS_avg_y}
\eeq
Using the integral relations
\beq
 \int_{0}^{1} \, {\rm d}y \, \l(1 - y^{2n} \r)^{1/2} = \f{\Gamma\l(\f{1}{2}\r)\Gamma\l(1+\f{1}{2n}\r)}{2 \, \Gamma\l(\f{3}{2} + \f{1}{2n}\r)}  \, ;  \quad  \int_{0}^{1} \, {\rm d}y \,  \l(1 - y^{2n} \r)^{-1/2} =\f{\Gamma\l(\f{1}{2}\r)\Gamma\l(1+\f{1}{2n}\r)}{\Gamma\l(\f{1}{2} + \f{1}{2n}\r)} \, ,
\label{eq:int_Gamma_monomial}
\eeq
we get
$$1 + \langle  w_\phi \rangle = \f{\Gamma\l(\f{1}{2} + \f{1}{2n}\r)}{\Gamma\l(\f{3}{2} + \f{1}{2n}\r)} = \f{2n}{n+1} \, ,$$
leading to the well-known final expression for the time-averaged equation of state, as first obtained in Ref.~\cite{Turner:1983he}, to be
\beq
\boxed{ ~ \langle  w_\phi \rangle = \f{n-1}{n+1} ~  } \, . 
\label{eq:EOS_avg_final}
\eeq
Hence, for oscillations around   a quadratic potential ($n = 1$), $\langle  w_\phi \rangle = 0$ (matter-like) and around  a quartic potential ($n = 2$), $\langle  w_\phi \rangle = 1/3$ (radiation-like), while $\langle  w_\phi \rangle \leq -1/3$ for $n \leq 1/2$, which leads to an accelerated expansion. 

\subsubsection{Period of an anharmonic  oscillator}
\label{sec:f_anharmonic_osc}
The time period of oscillations is defined as
\beq 
T_{\rm osc} = \int_0^{T_{\rm osc}} {\rm d}t = 2 \, \int^{\phi_0}_{-\phi_0} \frac{{\rm d}\phi}{\dot{\phi}} ,
\label{eq:Time_period_gen}
\eeq
which can be written for the monomial potential~(\ref{eq:pot_monomial_n}), using Eq.~(\ref{eq:phi_dot_coherent}),  as 
$$T_{\rm osc} = \f{2}{\sqrt{2V(\phi_0)}}  \int_{-\phi_0}^{\phi_0} \, {\rm d}\phi \f{1}{\l[1 - \l(\phi/\phi_0 \r)^{2n} \r]^{1/2}} \, .$$
Using the variable redefinition $y = \phi/\phi_0$, we obtain
$$T_{\rm osc} =  \f{2\phi_0}{\sqrt{2V(\phi_0)}}  \int_{-1}^{1} \, {\rm d}y \f{1}{\l[1 - y^{2n} \r]^{1/2}} =  \f{4\phi_0}{\sqrt{2V(\phi_0)}}  \int_{0}^{1} \, {\rm d}y \l[1 - y^{2n} \r]^{-1/2}  $$
Now using the integral identity in Eq.~(\ref{eq:int_Gamma_monomial}), we obtain
$$ T_{\rm osc} = 2\sqrt{2\pi} \, \f{\phi_0}{\sqrt{V(\phi_0)}} \f{\Gamma\l(1 + \f{1}{2n}\r)}{\Gamma\l(\f{1}{2} + \f{1}{2n}\r) }  \, , $$
which, using Eq.~(\ref{eq:pot_monomial_n}), yields the final expression
\beq
\boxed{~ T_{\rm osc} =  2\sqrt{2\pi} \, \f{m_p}{\sqrt{V_0}} \,  \f{\Gamma\l(1 + \f{1}{2n}\r)}{\Gamma\l(\f{1}{2} + \f{1}{2n}\r)~ } \, \l(\f{\phi_0}{m_p}\r)^{1-n} } \, .
\label{eq:Time_monomial_period_final}
\eeq
Eq.~(\ref{eq:Time_monomial_period_final}) clearly demonstrates that for $n=1$, the period of oscillation is independent of the amplitude $\phi_0$. Note that  in  deriving Eqs.~(\ref{eq:EOS_avg_final}) and (\ref{eq:Time_monomial_period_final}), we never used the assumption that $n$ is an integer  in Eq.~(\ref{eq:pot_monomial}). Hence our primary results given in Eqs.~(\ref{eq:EOS_avg_final}) and (\ref{eq:Time_monomial_period_final}) are valid for all $n \in \mathbb{R}^+$.
Also notice that for $n>1$, the period of oscillation becomes shorter as we increase  the amplitude of the oscillations $\phi_0$, while for $n< 1$, the period becomes longer with higher values of  $\phi_0$.  It means that the  quadratic potential is a special case (simple harmonic oscillator) for which the frequency of oscillations is completely independent of the amplitude, as we learned a long way back in our introductory physics course~\cite{Kleppner_Kolenkow_2013}, thanks to the great Galileo Galilei who first made this crucial observation.

\bigskip

\begin{figure}[hbt]
\begin{center}
\includegraphics[width=0.75\textwidth]{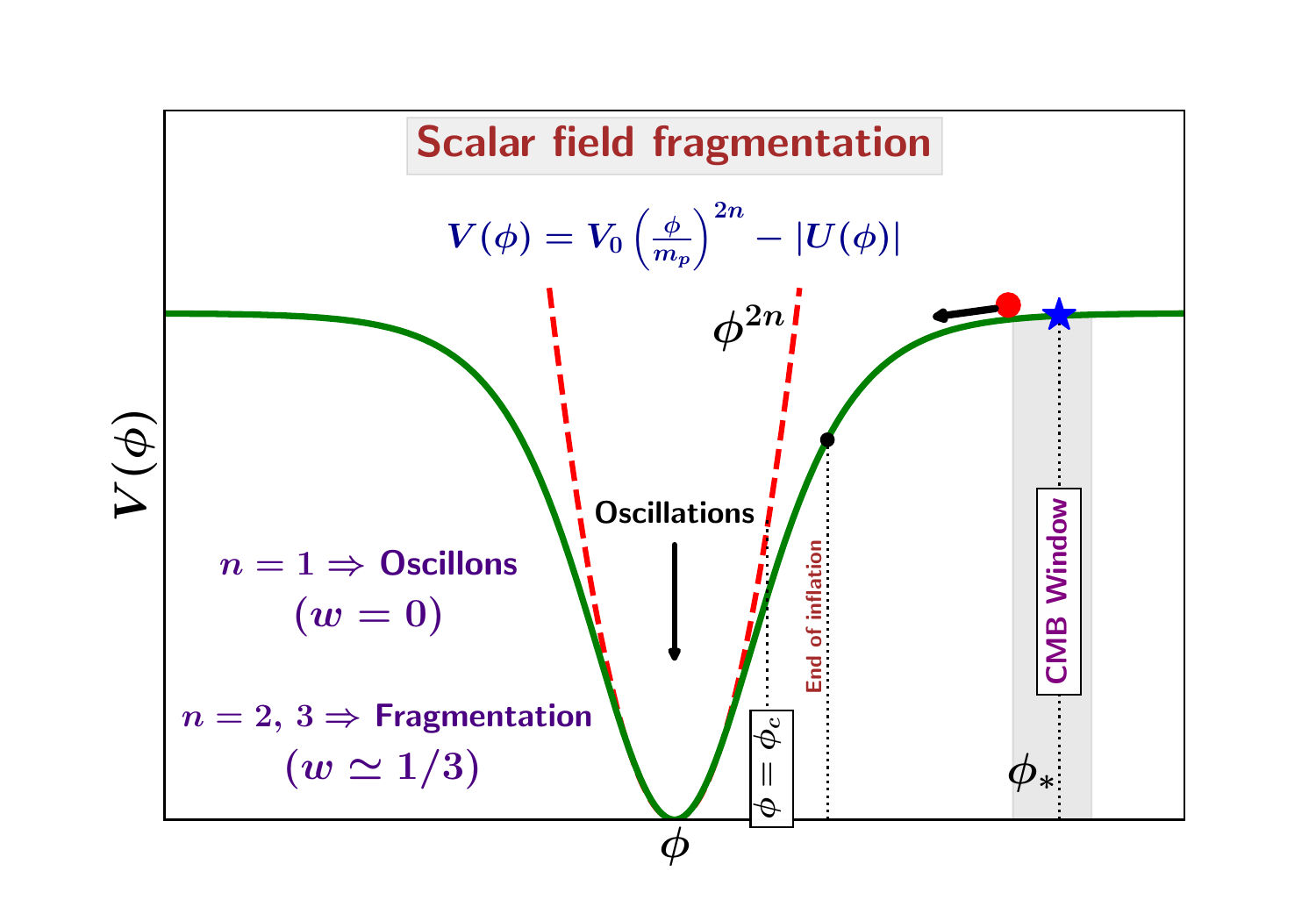}
\caption{This figure schematically illustrates a typical asymptotically flat inflaton potential of the form~(\ref{eq:pot_ing_flat_frag}). Here $\phi_c$ is the critical field value around which the potential flattens away from its monomial form $V(\phi) \sim \phi^{2n}$. The attractive self-interaction $U(\phi)$ results in the fragmentation of the oscillating inflaton condensate due to non-linear effects. For $n=1$, these fragments behave as non-relativistic oscillons, while for $n>1$, the fragments behave as transients which quickly decay into scalar radiation.  }
\label{fig:pot_ing_flat_frag}
\end{center}
\end{figure}

Before moving on to discuss the decay of the inflaton, let us stress that  Eqs.~(\ref{eq:EOS_avg_final}) and (\ref{eq:Time_monomial_period_final}) are also (more or less) applicable to asymptotically flat potentials because the coherent oscillations after the end of inflation are only sensitive to the functional form of the minimum of the potential, rather than to the inflationary  plateau wing\footnote{If the amplitude of oscillations is large enough, then corrections to Eqs.~(\ref{eq:EOS_avg_final}) and (\ref{eq:Time_monomial_period_final}) are required.}.  However, if the plateau is flat enough, then during the first few oscillations, the inflaton field explores some parts of the plateau, as a result of which the inflaton field behaves like a self-interacting oscillating condensate. To be precise, consider oscillations around the minimum of a plateau potential  of the functional form
\beq
V(\phi) = V_0 \l( \f{\phi}{m_p} \r)^{2n} - |U(\phi)| \, , 
\label{eq:pot_ing_flat_frag}
\eeq
where the attractive self-interaction term, $U(\phi)$, is  due to the flattening of the potential at large field values. The potential is schematically plotted in Fig.~\ref{fig:pot_ing_flat_frag}.

The  inflaton condensate oscillating around the potential~(\ref{eq:pot_ing_flat_frag})  fragments to form dense scalar field lumps due to the presence of the non-linear attractive self-interaction $U(\phi)$. For $n=1$, \textit{i.e.}, oscillations around a predominantly quadratic potential (which flattens for large field values), these scalar field lumps behave like quasi-solitonic objects whose localised energy density remains fixed while the field value oscillates with time around a fixed radial profile, as demonstrated in Refs.~\cite{Amin:2011hj,Lozanov:2017hjm}. Such  quasi-breather scalar field lumps/configurations are known as {\em pulsons} or {\em oscillons}. Oscillons are quite long lived and have interesting cosmological implications. However, in these pedagogical lecture notes, we do not elaborate further on oscillons and direct the interested readers to Refs.~\cite{Amin:2011hj,Zhou:2013tsa,Lozanov:2014zfa,Lozanov:2017hjm,Lozanov:2019ylm,Mahbub:2023faw,Kim:2017duj,Kim:2021ipz,Amin:2018xfe,Zhang:2020bec}. For $n>1$, the fragments behave as transients which quickly decay into scalar radiation with $w \simeq 1/3$ as discussed in Ref.~\cite{Lozanov:2017hjm}.

\bigskip

In the above analysis, we ignored the coupling of the inflaton field to other (scalar, spinor and tensor) degrees of freedom. In the presence of external coupling, the inflaton decays into the particles of the coupled fields which eventually leads to reheating of the universe to the hot Big Bang phase, as we now begin to explore. 

\subsection{External coupling and  decay of the inflaton condensate}
\label{sec:inf_decay}
Due to its external coupling, the inflaton condensate $\phi$ decays into one or more offspring fields ($\chi, \, \psi, \, A_\mu$ \textit{etc}), which then interact and  decay into further particles before finally thermalizing  leading to the commencement of Big Bang Nucleosysthesis (BBN). Reheating is supposed to be (directly or indirectly) the origin of (almost) all matter in the hot Big Bang universe \cite{Kofman:1996mv}.  During the post-inflationary oscillations, the inflaton decay can  occur either  perturbatively,
 due to the decay of individual massive inflaton particles, or non-perturbatively {\em via} the mechanism of   parametric resonance due to the coherent nature of the inflaton field oscillations. Which of these two ways is predominant depends upon the amplitude of inflaton oscillations as well as the  nature of the coupling between the inflaton and other bosonic/fermionic degrees of freedom.

\subsubsection{Perturbative reheating}
\label{sec:reheating_perturbative}

In the  perturbative reheating framework, the inflaton field  (after the end of  inflation) is assumed to be comprised   of a large number of massive inflaton particles (of low momenta), each of which can decay independently into the quanta of the coupled offspring fields, and the corresponding decay rate $\Gamma$ can be computed using the standard Feynman rules~\cite{Kofman:1996mv}.  The perturbative theory of reheating was originally developed in the context of the new inflationary
scenario in \cite{Albrecht:1982mp}.
Phenomenologically, it amounts to adding a friction term $\Gamma {\dot\phi}$ to the classical
equation of  motion of the inflaton field oscillating around the minimum of its potential~ \cite{Albrecht:1982mp,Kolb:1990vq}, \textit{i.e.}
\beq
{\ddot\phi} + \l( 3H + \Gamma \r) {\dot\phi} + V_{,\phi}(\phi) = 0 \, .
\label{eq:EOM_damp}
\eeq
 The perturbative theory of reheating works well if either (a) the inflaton decays only into fermions
$\psi$ through
a  Yukawa-type interaction $h\psi{\bar\psi}\phi$,  with a weak coupling constant $h$ satisfying  $h^2 \ll m_\phi/m_p$, or (b)
the coupling of the inflaton to  a scalar boson $\chi$, described by the interaction $\frac{1}{2}g^2\phi^2\chi^2$,  is weak with
$g \ll 3\times 10^{-4}$, making particle production {\em via} non-perturbative effects (parametric resonance) ineffective \cite{Kofman:1996mv}.

In the perturbative regime, since inflaton decay is a slow process, the reheating EoS is given by the time averaged EoS of the inflaton field during its oscillations around the minimum of the potential, \textit{i.e} $w_{\rm re} = \langle w_{\phi} \rangle$. For monomial potentials, $V(\phi) \propto \phi^{2n}$, the equation of state during reheating is given by Eq.~(\ref{eq:EOS_avg_final}) to be
$$
w_{\rm re} = \langle w_{\phi} \rangle = \f{n-1}{n+1} \, .
$$ 
At the beginning of reheating, $H \gg \Gamma$, otherwise significant particle production  during inflation will modify the inflaton potential and might spoil the CMB predictions\footnote{However, successful inflation can be achieved in the presence of dissipative effects, for example in the {\em warm inflationary scenario}, see Ref.~\cite{Berera:2023liv}.}. Hence reheating in the perturbative scenario gets completed only when the (decreasing) expansion rate
becomes equal to the decay rate at late times so that $H \simeq \Gamma$. Following thermalization, the reheating temperature can be computed by considering 
$$\Gamma \simeq H  = \sqrt{\f{1}{3m_p^2} \, \rho(T_{\rm re})} \Rightarrow \Gamma =  \sqrt{\f{1}{3m_p^2} \, \f{\pi^2}{30} \, g_*(T_{\rm re}) \, T_{\rm re}^4}$$
leading to
\beq
\boxed{~ T_{\rm re} \simeq  \l(\f{\pi^2 \, g_*}{90}\r)^{-1/4} \, \Big(\Gamma \, \mpl \Big)^{1/2} =  2.4 \times 10^{18} \times   \l(\f{\pi^2 \, g_*}{90}\r)^{-1/4} \, \Big(\f{\Gamma}{m_p} \Big)^{1/2} ~{\rm GeV}  ~ } \, ,
\label{eq:T_reheating_perturbative}
\eeq
which, intriguingly, is independent of both the duration of inflation and the properties of the inflaton potential 
 $V(\phi)$. As stressed above, since the coupling between the inflaton and the external fields  can alter, {\em via} radiative corrections, the shape of $V(\phi)$ during inflation, this places strong constraints on the total decay rate~\cite{Kofman:1996mv,Kofman:1997yn,Allahverdi:2010xz}. This in turn implies that the reheating temperature  in perturbative models will be relatively low, typically  $T_{\rm re} < 10^9$ GeV (see the introductory \href{ https://www.desy.de/~westphal/workshop_seminar_fall_2010/reheating.pdf}{\purple {\bf lecture notes}}), and hence the duration of reheating $N_{\rm re}$ is quite long and reheating is inefficient in the perturbative regime \cite{Kofman:1996mv}.  

\subsubsection{Non-perturbative reheating}
\label{sec:reheating_nonperturbative}

For inflationary models in which the main source of reheating is through the decay of the inflaton
into bosons, non-perturbative effects might become important to the coherent oscillations of the inflaton field. The scenario is analogous to particle production in the presence of time dependent  external fields.  To be concrete, we will focus on the scenario of the inflaton coupled to a single massless scalar offspring field $\chi$ through an interaction ${\cal I}(\varphi,\chi) $, with the model being described by the action
\begin{equation}
    \label{eq:reh_Action}
    S[g_{\mu\nu}, \, \varphi, \, \chi] = \int {\rm d}^4x~\sqrt{-g} \, \left[ \frac{m_p^2}{2} \, R - \frac{1}{2} \,  \partial_{\mu} \varphi \, \partial_{\nu}  \varphi \, g^{\mu\nu} - \frac{1}{2} \,  \partial_{\mu} \chi \, \partial_{\nu}  \chi \, g^{\mu\nu} - V(\varphi) -{\cal I}(\varphi,\chi)  \right] \, .
\end{equation}
The corresponding field equations are given by
\begin{align}
    \label{eq:reh_EOM_phi}
    \ddot{\varphi} -\frac{\nabla^2}{a^2} \varphi+ 3H\dot{\varphi} + V_{, \varphi} + {\cal I}_{, \varphi}= 0~, \\ \label{eq:reh_EOM_chi}
    \ddot{\chi} -\frac{\nabla^2}{a^2} \chi+ 3H\dot{\chi}  + {\cal I}_{, \chi} = 0 \, ,
\end{align}
with 
\beq
H^2 \, = \, \f{1}{3\, m_p^2} \, \left[  \f{1}{2} \, \dot{\varphi}^2  +   \f{1}{2} \, \f{\vec{\nabla}\varphi}{a} . \f{\vec{\nabla}\varphi}{a} +  V(\varphi)  +   \f{1}{2} \, \dot{\chi}^2  +  \f{1}{2} \, \f{\vec{\nabla}\chi}{a} . \f{\vec{\nabla}\chi}{a}  +   {\cal I} (\varphi, \, \chi)  \right] \, .
\label{eq:Hubble_fields}
\eeq  
Eqs.~(\ref{eq:reh_EOM_phi})--(\ref{eq:Hubble_fields}) indicate that the 
 dynamics of reheating is a complicated, non-linear and   non-thermal process. Hence, various approximations corresponding to neglecting different terms in the above set of equations are usually carried out in the literature, some of which have been indicated by the  arrow marks in the following (to be made more explicit later). 
$$\ddot{\varphi} -\CancelTo[\color{gray}]{0}{\frac{\nabla^2}{a^2}} \varphi + \CancelTo[\color{gray}]{0}{3H\dot{\varphi}} + V_{, \varphi}(\varphi) + \CancelTo[\color{gray}]{0}{{\cal I}_{, \varphi}}= 0 \, $$ 
$$\ddot{\chi} -\frac{\nabla^2}{a^2} \chi+ \CancelTo[\color{gray}]{0}{3H\dot{\chi}}  + {\cal I}_{, \chi} = 0 \, ,$$
$$ H^2 \, = \, \f{1}{3\, m_p^2} \, \left[  \f{1}{2} \, \dot{\varphi}^2  +   \f{1}{2} \, \CancelTo[\color{gray}]{0}{\f{\vec{\nabla}\varphi}{a} . \f{\vec{\nabla}\varphi}{a}} +  V(\varphi)  +   \f{1}{2} \, \CancelTo[\color{gray}]{0}{\dot{\chi}^2}  +  \f{1}{2} \, \CancelTo[\color{gray}]{0}{\f{\vec{\nabla}\chi}{a} . \f{\vec{\nabla}\chi}{a}}  +   \CancelTo[\color{gray}]{0}{{\cal I} (\varphi, \, \chi)}  \right] \, . $$
One also makes further assumptions by considering a particular form of the potential, for example, previously we considered a monomial potential given in Eq.~(\ref{eq:pot_monomial_n}).   In these notes, we will focus on a quadratic-quadratic inflaton coupling to the offspring field  $\chi$ of the form
\begin{align}
    {\cal I} (\varphi, \, \chi)  = \f{1}{2} \, g^2 \, \varphi^2 \, \chi^2  \, .
    \label{eq:Interact_term}
\end{align}
In the post inflationary epoch,  the universe goes through inflaton decay,  thermalization and reheating {\em via} a sequence of successive stages  as discussed  below (see Refs.~\cite{Kofman:1994rk,Shtanov:1994ce,Kofman:1996mv,Kofman:1997yn,Greene:1997fu,Amin:2014eta,Lozanov:2019jxc}).
\subsubsection*{(1) \underline{Preheating}:}
The early stage of inflaton decay, known as {\em preheating}, involves the phenomenon of  parametric resonance brought about by the coherent
oscillations of the inflaton $\phi$ around the minimum of its potential \cite{Kofman:1997yn}. In order to understand this regime, one usually  makes  a number of (physically reasonable) approximations. Since the inflaton behaves as a homogeneous condensate at the end of inflation, its gradient term in Eq.~(\ref{eq:reh_EOM_phi}) may be dropped. Furthermore, since we are working in the {\em cold inflationary regime} for which particle production can be ignored during inflation, we assume that the $\chi$ field is in its vacuum at the beginning of reheating. The energy budget of the universe in this regime  is dominated by the inflaton condensate, hence $\chi$ can be treated as a test field. 

 A convenient way to characterise the dynamics during the early stages of preheating is to linearize  the field fluctuations  in Eqs.~(\ref{eq:reh_EOM_phi})-(\ref{eq:reh_EOM_chi})  and  carry out  the standard  Floquet analysis. To begin with, we   split  each of the field contents in the Eqs.~(\ref{eq:reh_EOM_phi})-(\ref{eq:Hubble_fields}) into a  homogeneous background  and small fluctuations around that background, namely
\ber
\varphi(t,\vec{x}) &=& \phi(t) + \delta\varphi(t,\vec{x}) \, ; \label{eq:phi_split} \\
\chi(t,\vec{x}) &=& \bar{\chi}(t) + \delta\chi(t,\vec{x}) \, . \label{eq:chi_split}
\eer
As discussed before, since the $\chi$ field is expected to be in its vacuum state at the end of inflation, we have $\bar{\chi}(t) \simeq 0$, leading to $\delta\chi(t,\vec{x}) = \chi(t,\vec{x})$. From hereon, we will simply denote the  fluctuations $\delta\chi$ as $\chi$. We use the following approximations to retain only terms that are linear in the fluctuations  $\delta\varphi$ and $\chi$.
\begin{enumerate}
\item The energy  density of the system is predominantly contained in the homogeneous  inflaton condensate $\phi(t)$, \textit{i.e.}
$$|\phi(t)| \, \gg \, |\delta\varphi(t,\vec{x})| , \, |\chi(t,\vec{x})|  \, ;$$
and
$$\rho_\phi \, \gg \, \rho_\chi, \, \rho_{\delta\varphi} \, .$$
\item Since $\bar{\chi}(t)=0$, the interaction term ${\cal I}(\varphi,\chi)$   becomes
$${\cal I}(\varphi,\chi) = {\cal I}(\phi+\delta\varphi,\chi) \simeq \f{1}{2} \, g^2 \, \phi^2(t) \, \chi^2 \, , $$
which can be dropped from the Friedmann Eq.~(\ref{eq:Hubble_fields}) at linear order. Furthermore, we find 
$${\cal I}_{,\varphi} \equiv g^2 \, \varphi \, \chi^2 = g^2 \, \l( \phi(t)+\delta\varphi \r) \, \chi^2 \simeq g^2 \, \phi(t) \, \chi^2  \, , $$ which can be dropped from the evolution Eq.~(\ref{eq:reh_EOM_phi}) at  linear order. Similarly, the last term of the left hand side of Eq.~(\ref{eq:reh_EOM_phi}) becomes
$${\cal I}_{,\chi} \equiv g^2 \, \varphi^2 \, \chi = g^2 \, \l( \phi(t)+ \delta\varphi \r)^2 \, \chi \simeq g^2 \, \phi^2(t) \, \chi $$
at  linear order.
\item Under the linear approximation, the inflaton potential becomes 
$$ V(\varphi) = V(\phi+\delta\varphi) \simeq V(\phi) + V_{,\phi}(\phi) \, \delta\varphi$$
and its derivative becomes
$$V_{,\varphi}(\varphi) = V_{,\varphi}(\phi+\delta\varphi) \simeq V_{,\phi}(\phi)  +  V_{,\phi\phi}(\phi)  \, \delta\varphi \, .$$
\end{enumerate}
 We again stress that these approximations are valid during the early stages of preheating, before backreactions from $\delta\varphi$ and $\chi$ become significant, \textit{i.e.}, before $\rho_{\delta \varphi, \chi} \sim \rho_\phi$. Under the aforementioned approximations, Eqs.~(\ref{eq:reh_EOM_phi})-(\ref{eq:reh_EOM_chi}) become
\ber 
    \ddot{\phi} + 3H\dot{\phi} + V_{,\phi} &=& 0  \, ; \label{eq:phi_no_grad} \\
    {\delta\ddot\varphi}  + 3H{\delta\dot\varphi} +  \l[ - \frac{\nabla^2}{a^2} + V_{,\varphi\varphi}(\phi) \r] \delta\varphi &=& 0 \, ; \label{eq:delphi_linear} \\
    \ddot{\chi}  + 3H\dot{\chi} + \l[ - \frac{\nabla^2}{a^2} +  g^2 \, \phi^2 
    \r]  \, \chi &=& 0 \, , \label{eq:chi_linear}
\eer
and  the  Friedmann Eq.~(\ref{eq:Hubble_fields}) becomes  
\begin{gather}
    \label{eq:Hubble_no_grad}
    H^2 = \frac{1}{3\, m_p^2} \,  \left[  \frac{1}{2} \, \dot{\phi}^2 +  V(\phi) \right] \, .
\end{gather}
Note that in Eqs.~(\ref{eq:phi_no_grad}) and (\ref{eq:Hubble_no_grad}) we have dropped the   $\delta\varphi$ terms in order to describe the background dynamics in terms of the purely homogeneous condensate at the end of inflation. Hence, they represent the background equations, \textit{w.r.t} which fluctuations are defined. The corresponding equations for the evolution of the Fourier modes  $\delta\varphi_k$ and $\chi_k$
take the following form
\ber
    \ddot{\delta\varphi_k}+ 3H\dot{\delta\varphi_k} +  \left[ \frac{k^2}{a^2} + V_{,\phi\phi}(\phi)  \right] \delta\varphi_k &=& 0 \, ;  \label{eq:delphi_k_linear} \\ 
    \ddot \chi_k + 3H\dot \chi_k  +   \left[ \frac{k^2}{a^2} + g^2 \, \phi^2 \right]\chi_k &=& 0 \, . \label{eq:chi_k_linear}
\eer
Eqs. \eqref{eq:delphi_k_linear} and \eqref{eq:chi_k_linear} describe  two independent parametric oscillators with time-dependent damping terms  $3H\dot{\delta\varphi_k}$ and $3H\dot{\chi_k}$
and parametric frequencies of the form
$$\Omega^2_{\delta\varphi}(k,t) = \frac{k^2}{a^2} + V_{,\phi\phi}(\phi) \, ; \quad \Omega^2_\chi(k,t) = \frac{k^2}{a^2} + g^2 \, \phi^2(t) \, .$$  
For simplicity, we ignore the expansion of the background space ($H \simeq 0, \, a =1$),  which is a reasonable assumption on (the inflaton oscillation) time scales that are much shorter compared to the time scale of expansion of space,  $H^{-1}$.  Since we will not discuss  inflaton  fragmentation, we  further  ignore the inflaton fluctuations $\delta\varphi$. Consequently, the system is reduced to
\ber
\ddot{\chi}_k + \left[ k^2 + g^2 \,  \phi^2(t)\right] \chi_k &=& 0 \, ; \label{eq:Preheating_chi_k} \\
\ddot{\phi} \,  + \,  V_{,\phi}(\phi) &=& 0 \,  .  \label{eq:Preheating_phi_SHO} 
\eer
Eq.~(\ref{eq:Preheating_chi_k}) describes the motion of an oscillator with a time dependent frequency $\Omega_k^2(t) = k^2 + g^2 \phi^2(t)$. Since $\phi(t)$ is oscillating,  $\Omega_k^2(t)$ also oscillates with time. Thus,   Eq.~(\ref{eq:Preheating_chi_k}) describes the motion of a parametric oscillator and  can lead to  exponential growth of $\chi_k$ for  certain ranges  of $k$ due to  a phenomenon called \textit{parametric resonance}. 
For   inflaton oscillations around a quadratic potential 
$$V(\phi) = \f{1}{2} \, m^2 \, \phi^2 \, ,$$
the solution to Eq.~(\ref{eq:Preheating_phi_SHO}) is given by
\beq
\phi(t) = \phi_0 \, \cos{\l(mt\r)} \, ,
\label{eq:soln_preheating_phi_SHO}
\eeq
for which  Eq.~(\ref{eq:Preheating_chi_k}) simplifies to
$$
    \ddot \chi_k + \left[ \l( k^2 + \frac{g^2\phi^2_0}{2} \r) + \frac{g^2\phi^2_0}{2} \, \cos{(2mt)} \right] \chi_k = 0 \, ,
$$
which, under the  change in variable 
$$T = mt + \f{\pi}{2} \, $$ 
takes the form of the  {\em Mathieu Equation}
\beq
    \boxed{~ \frac{{\rm d}^2 \chi_k}{{\rm d}T^2} + \left[A_k - 2q \, \cos(2T) \, \right]\chi_k = 0~ }
     \label{eq:Mathieu}
\eeq
where  the {\em resonance parameters} $q, \, A_k$ are given by
\beq
q = \f{g^2}{4}\l(\f{\phi_0}{m}\r)^2 \, ;  \quad A_k = \l( \f{k}{m} \r)^2 + 2 \, q \, .
\label{eq:mathieu_q_Ak}
\eeq
In  Floquet theory \cite{mclachlan1947theory, magnus2004hill,NIST:DLMF}, the Mathieu equation admits solutions of the form
\beq 
    \chi_k(T) = \mathcal{M}_k^{(+)}(T)~e^{\mu_k T}+ \mathcal{M}_k^{(-)}(T)~e^{-\mu_k T}
    \label{eq:sol_Mathieu}
\eeq
where $\mathcal{M}_k^{(+)}$ and $\mathcal{M}_k^{(-)}$ are periodic functions and $\mu_k$ is called the Floquet/Mathieu exponent. For certain ranges of values of $k$ if  ${\Re (\mu_k)} \neq 0$, the solution to the Mathieu equation, namely Eq.~(\ref{eq:sol_Mathieu}),  grows exponentially with time leading to  \textit{parametric resonance}. 
\begin{figure}[H]
\begin{center}
\includegraphics[width=0.49\textwidth]{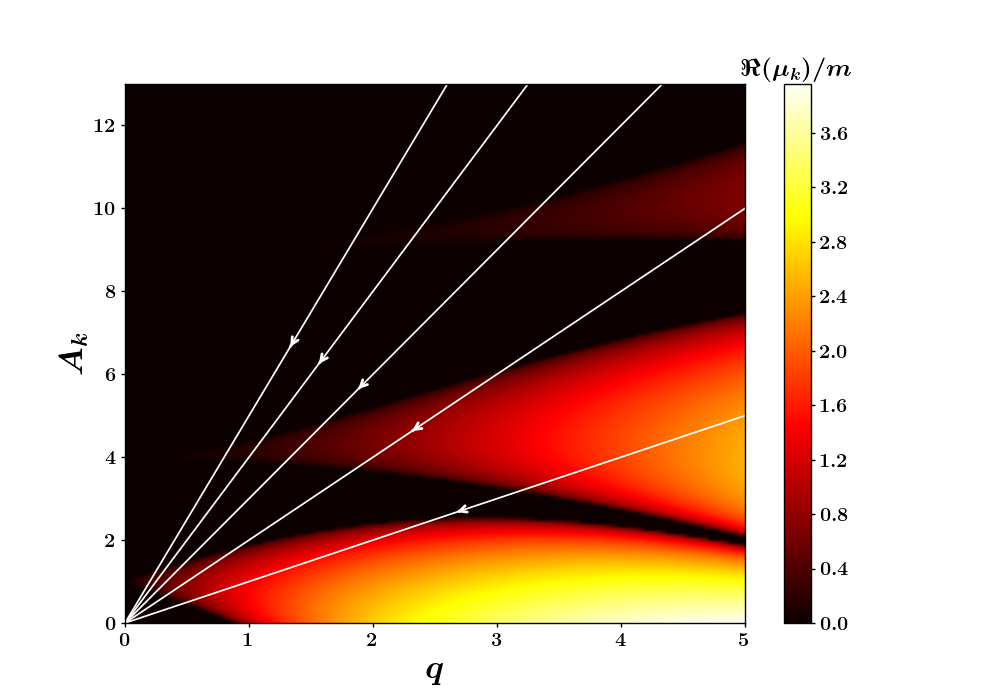}
\includegraphics[width=0.49\textwidth]{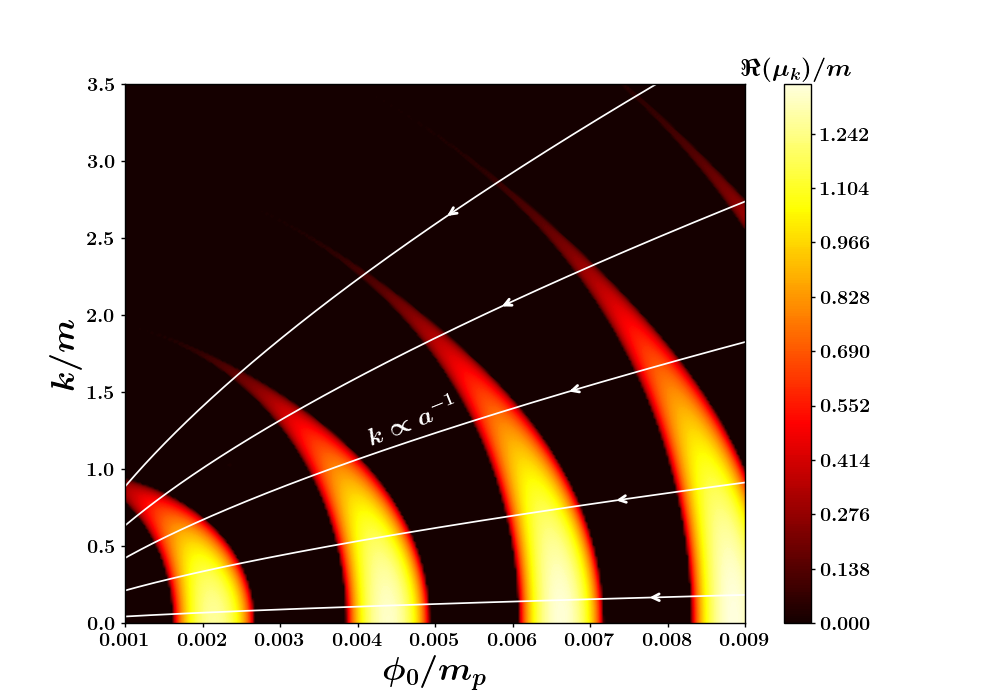}
\caption{The instability charts of the Mathieu  Eq.~(\ref{eq:Mathieu}) are shown here. The magnitude of the Floquet exponent $\mu_k$ depends upon $A_k = (k/m)^2 + 2q$, where $q = g^2\phi^2_0/(4m^2)$ being the resonance parameter. The parameter space $\lbrace q, \, A_k \rbrace$  can be divided into two parts- unstable (\textbf{yellow}) and stable (\textbf{maroon}). Resonant production of $\chi$-particles from the inflaton happens when  $A_k$ and $q$ lie in the unstable region,  corresponding to the Floquet exponent $\Re(\mu_k) \neq 0$. The white lines indicate the redshifting of physical momenta $k_p \propto k/a$, and the inflaton amplitude  $\phi_0(t)$ at longer time scales due to the expansion of the universe. The {\bf left panel} depicts the instability chart in terms of the Mathieu parameters $\lbrace q, \, A_k \rbrace$, while the {\bf right panel} shows the same in terms of the physical parameters $\lbrace \phi_0, \, k \rbrace$.}
    \label{fig:Resonance_bands}
\end{center}
\end{figure} 

The \textit{instability chart} of the Mathieu equation is shown in  Fig.~\ref{fig:Resonance_bands}, where the magnitude of ${\Re (\mu_k)}$ is plotted as a function of the parameters $\phi_0$ and $k$ (both being positive). The regions in the chart are broadly classified into `stable' where ${\Re (\mu_k)} = 0$ and `unstable' where ${\Re (\mu_k)} \neq 0$. The solutions to Eq.~(\ref{eq:Mathieu}) in the unstable regions contain an exponential instability $\chi_k \propto \exp(\mu_k T)$, which corresponds to an exponential growth in the  occupation number density $n_\chi(k)$ of the produced particles~\cite{Kofman:1997yn}, \textit{i.e.}
$$n_\chi(k) \equiv \f{1}{2  \Omega_\chi} \l[ \bigg\vert\f{{\rm d}{\chi_k}}{{\rm d T}}\bigg \vert^2 + \Omega_\chi^2 \, |\chi_k|^2 \r] - \f{1}{2}  \, \propto \,  e^{2\mu_k  T} \, ,$$ 
which can be interpreted as explosive particle production. 

In accordance with the band structure of  these stable and unstable regions, parametric resonance can be further classified into  \textit{narrow} and \textit{broad} regimes.
\begin{figure}[H]
\begin{center}
\subfigure[][]{
\includegraphics[width=0.46\textwidth]{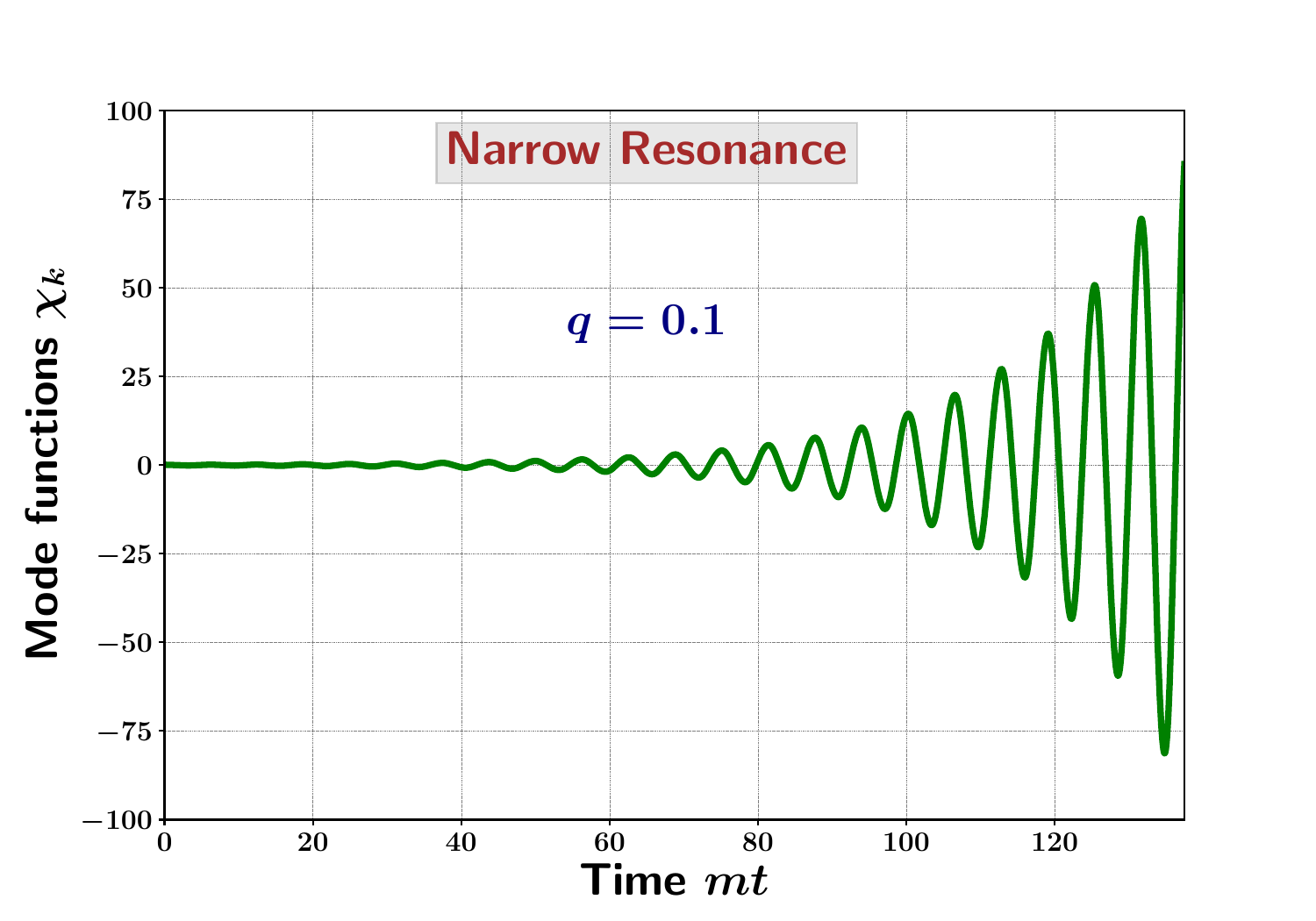}}
\subfigure[][]{
\includegraphics[width=0.46\textwidth]{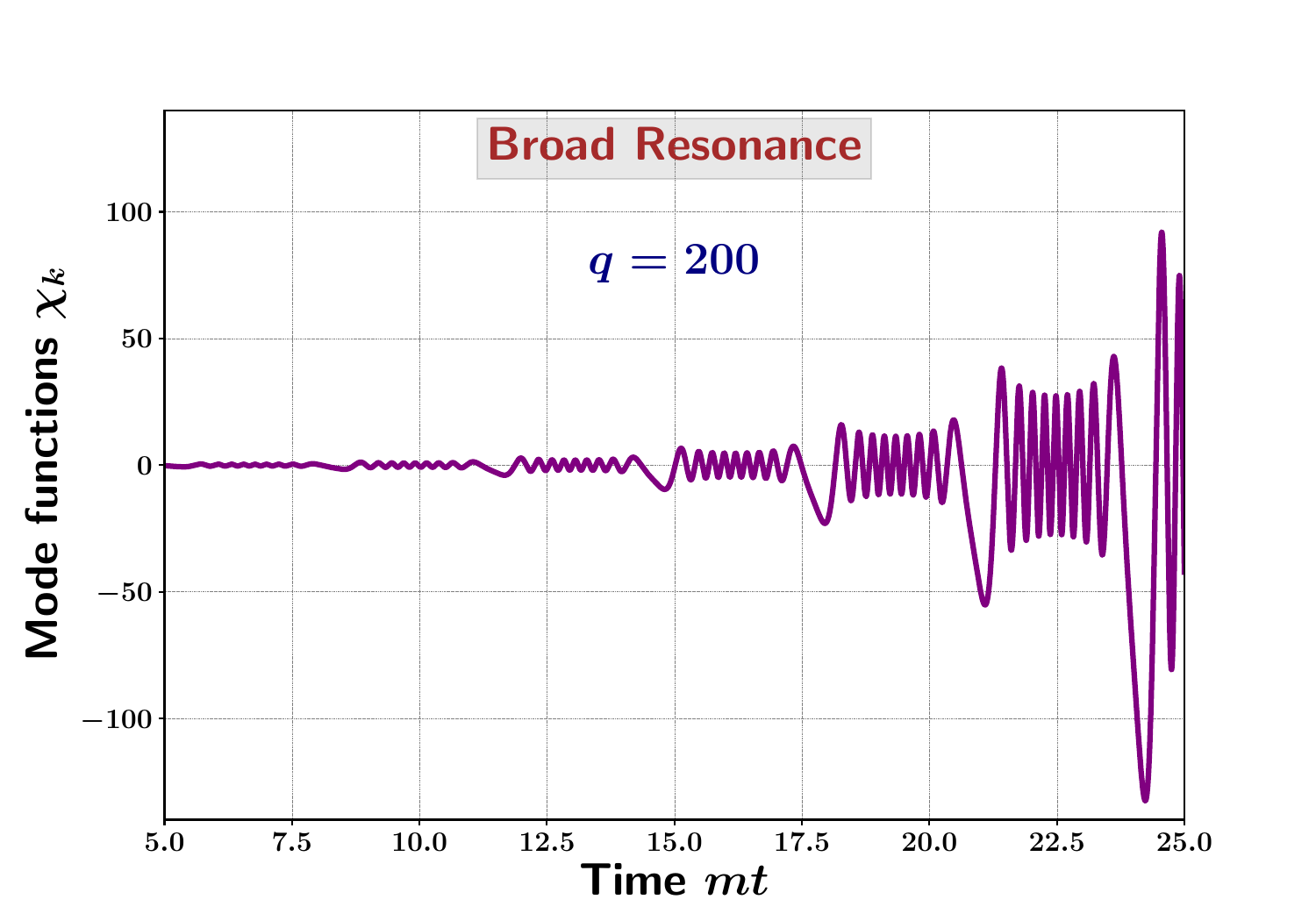}}
\subfigure[][]{
\includegraphics[width=0.46\textwidth]{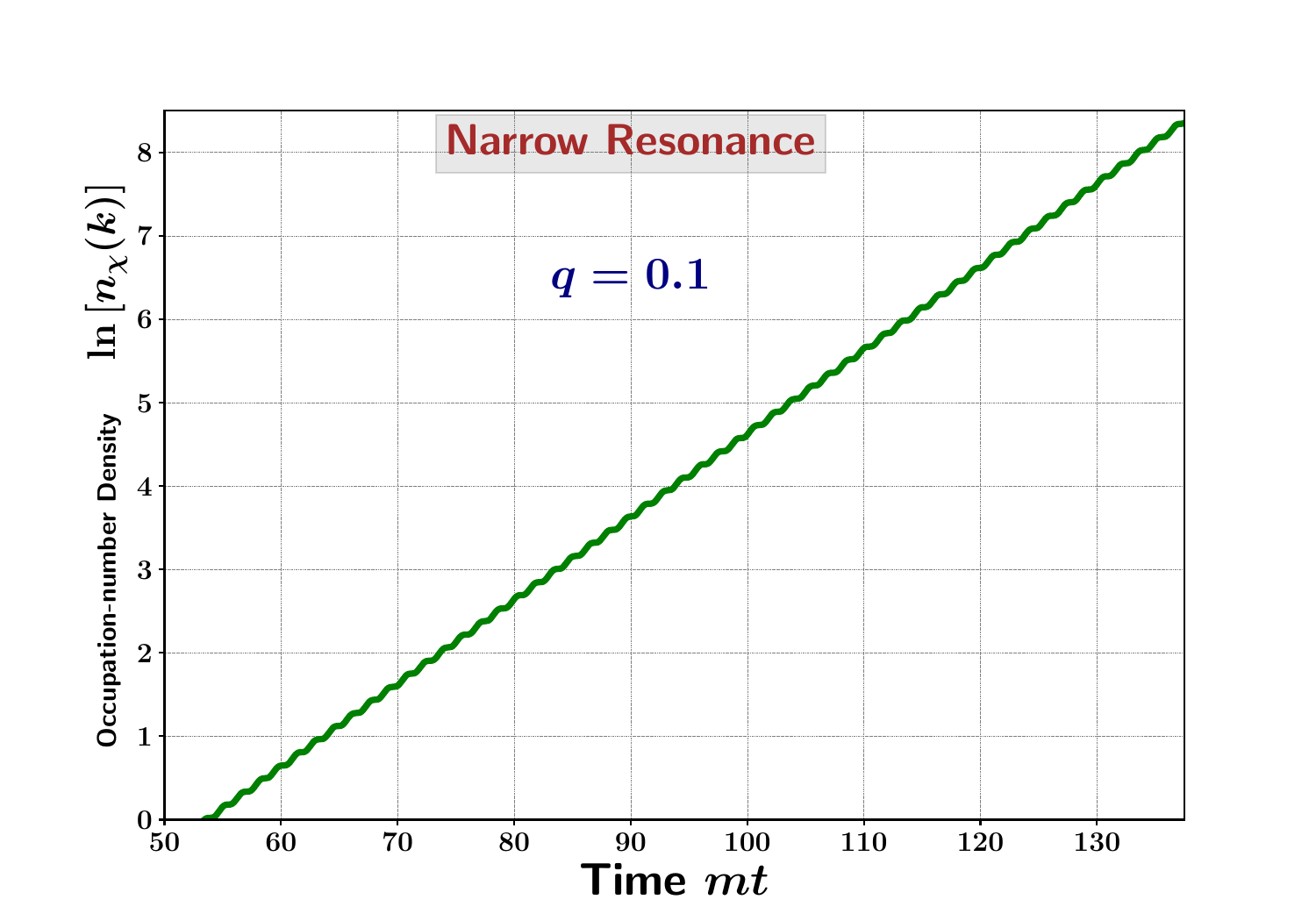}}
\subfigure[][]{
\includegraphics[width=0.46\textwidth]{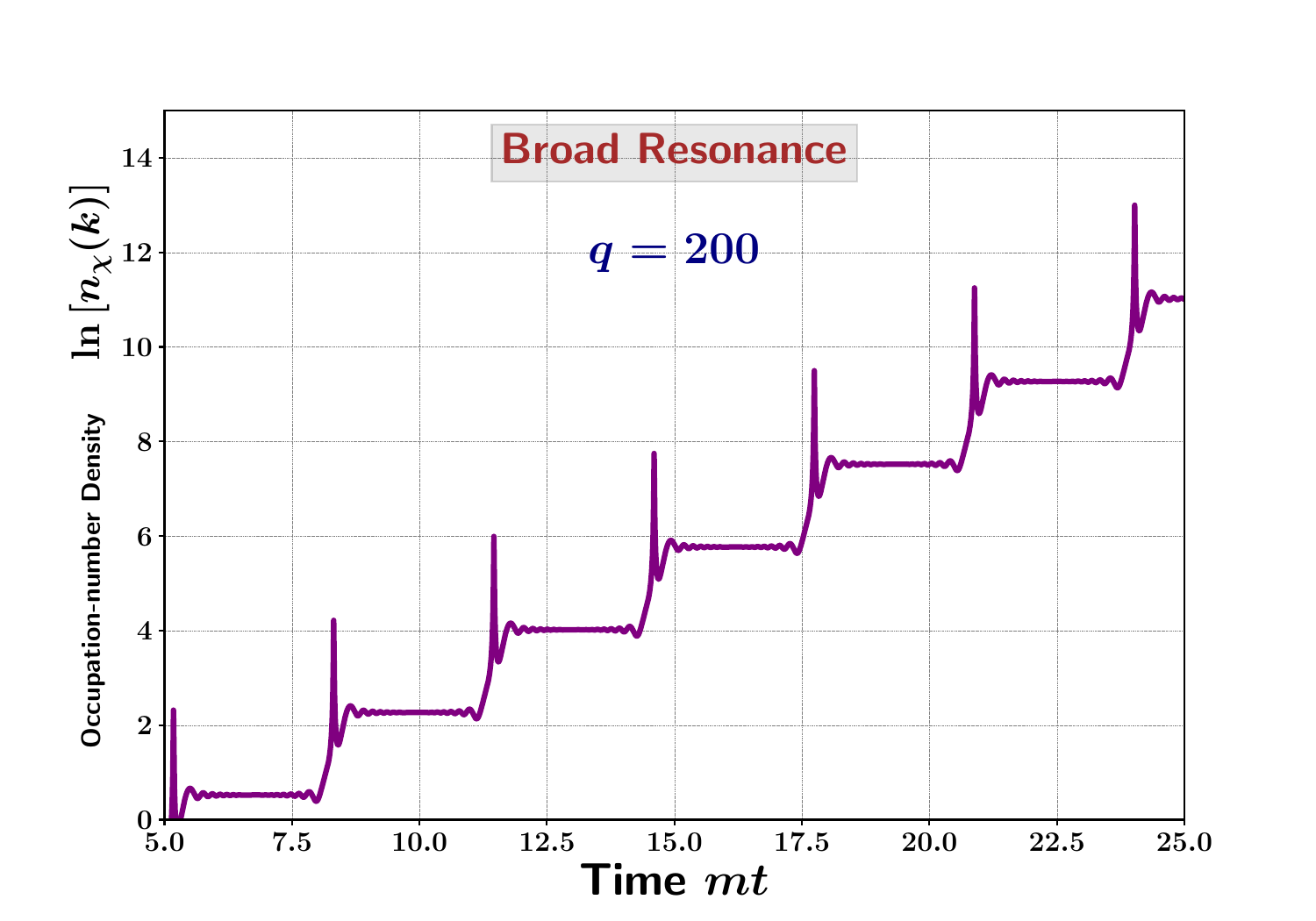}}
    \caption{Narrow parametric resonance of the $\chi$-field for $q \sim 0.1$ is shown in green  (\textbf{left}) and broad parametric resonance for $q \sim 200$ (\textbf{right}) is shown in purple  for modes lying inside the narrow and broad resonance bands, respectively, corresponding to the Mathieu Eq.~(\ref{eq:Mathieu}). The upper figures show the growth in the amplitude of  the mode functions, while the lower ones show the growth of the (logarithm of the)  occupation-number density, $n_\chi(k)$. The particle number densities grow exponentially for such  resonant modes, with broad resonance being more efficient than the narrow one. Note that  in  the broad band resonance,  peaks in the particle number density occur whenever the inflaton field value vanishes, i.e., $\phi(t) = 0$.}
    \label{fig:Resonance_broad_narrow}
\end{center}    
\end{figure}
\begin{enumerate}
\item {\em Narrow Resonance}:

Narrow band resonance occurs when the instability band width is quite narrow. From  Fig.~\ref{fig:Resonance_bands} for Eq.~(\ref{eq:Mathieu}), the narrow resonance is observed for $q \rightarrow 0$ and $A_k \rightarrow n^2$ ($n \in \mathbb{Z}^+ $). The width of a narrow band is given by $\Delta k/m \sim q^n$. Such resonance occurs only due to the dense occupation of $\chi$ modes leading to a smooth exponential increase of $n_k$ as shown in Fig.~\ref{fig:Resonance_broad_narrow}. The first narrow  instability  band corresponds to $n = 1$, which is described by $A_k \sim 1\pm q$. It is the widest and most important band in this narrow regime, which    can be interpreted as the production of  two $\chi$-particles with momentum $k \approx m$ from the decay of two inflaton particles, which is an excellent match with the perturbative analysis. Indeed, narrow resonance bands with higher $n$ can be interpreted as the simultaneous non-perturbative decay of $2n$-inflaton particles into a pair of $\chi$-particles with momentum  $k \approx nm \geq m $. In the narrow regime, since the interactions are weak, $\Omega_\chi^2(k) \approx k^2$ stays almost a  constant. Hence the inflaton condensate continues to oscillate coherently. 
\item {\em Broad Resonance}:

Broad band parametric resonance corresponds to  the broader instability regions in  Fig.~\ref{fig:Resonance_bands}. For Eq.~\eqref{eq:Mathieu}, such resonance occurs in the non-perturbative limit $q \gtrsim 1$. Inflaton decay in the broad resonance regime is much more efficient than its narrow counterpart, as a continuous region of $k$ modes is available for particle production. As shown in  Fig.~\ref{fig:Resonance_broad_narrow}, particle production occurs in bursts (rather than smoothly as in narrow resonance) due to the non-adiabatic change in the effective frequency, $\Omega^2_\chi(k,T) = A_k - 2q \, \cos(2T)$, of the parametric oscillator, namely 
\beq
\f{1}{\Omega^2_\chi(k,T)} \, \Big\vert \f{{\rm d} \Omega_\chi }{{\rm d}T} \Big\vert \, \geq \, 1 \, .
\label{eq:Omega_non_adiabatic}
\eeq
Using Eq.~(\ref{eq:mathieu_q_Ak}), the above condition for non-adiabatic particle production becomes 
\beq
2q \, |\sin{(2T)}| \, \geq \, \l[ \l(\f{k}{m}\r)^2 + 4q \, \sin^2(T)\r]^{3/2} \, ,
\label{eq:Omega_non_adiabatic_derived}
\eeq
which is easy to satisfy in the neighbourhood of $T = n\pi$ for $ k/m \ll q$, even for $k\geq m$, as shown in Fig.~\ref{fig:Mathieu_adiabaticity_violation}, also see Ref~\cite{Kofman:1997yn}. 
\end{enumerate}
\begin{figure}[htb]
\begin{center}
\includegraphics[width=0.7\textwidth]{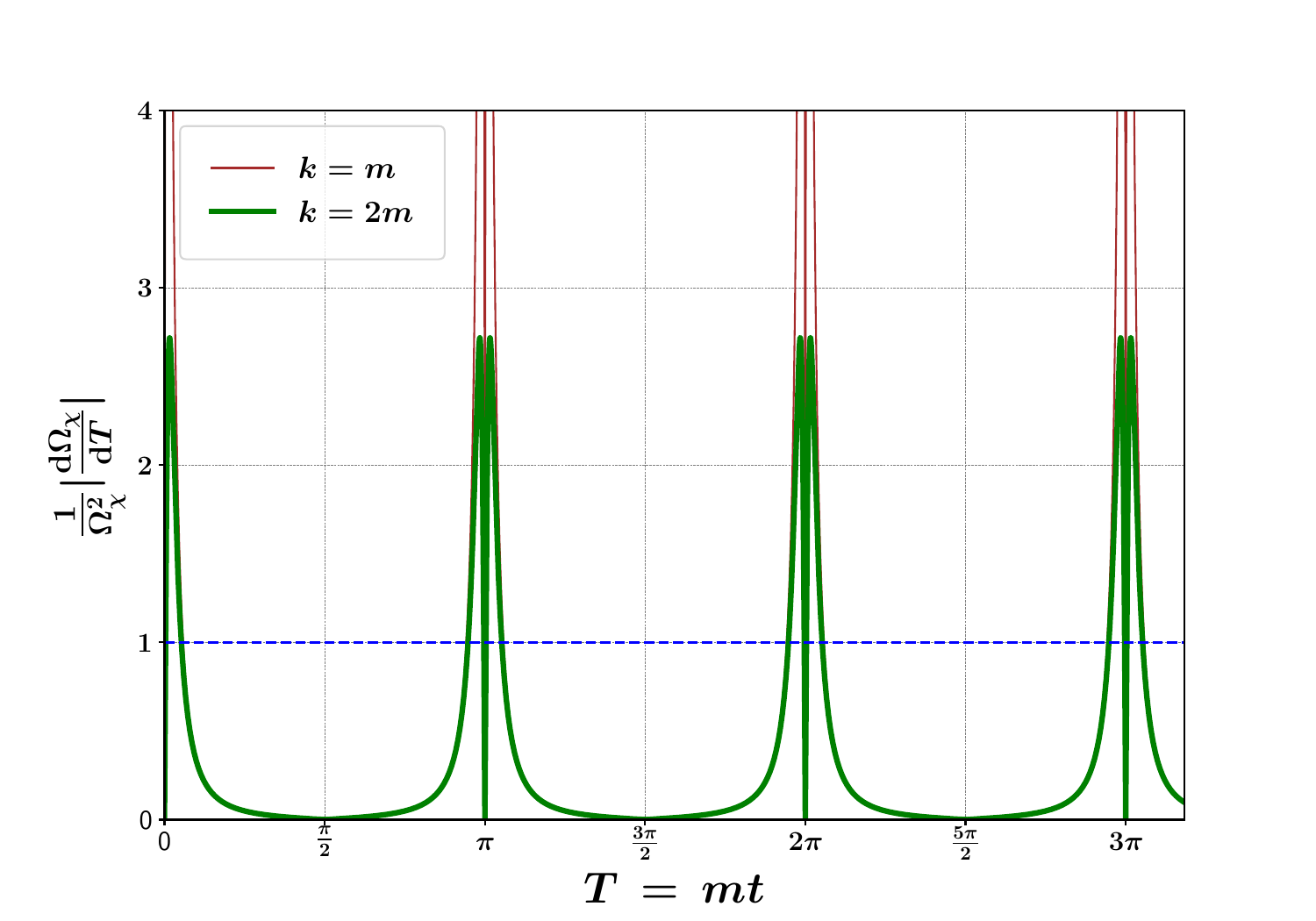}
\caption{Violation of the adiabaticity condition, as stated in Eq.~(\ref{eq:Omega_non_adiabatic_derived}), is demonstrated in the broad parametric resonance regime of the Mathieu Eq.~(\ref{eq:Mathieu}) with a resonance parameter $q=200$. The brown and green  curves correspond to $k=m$ and $k=2m$ respectively. Particle production in the broad resonance regime takes place in the neighbourhood of $T=n\pi$ where the adiabaticity condition is violated, \textit{i.e} $\f{1}{\Omega^2_\chi} \, \Big\vert \f{{\rm d} \Omega_\chi }{{\rm d}T} \Big\vert \, \geq \, 1$ (for the segments of the green and brown curves above the blue dashed line).  }
\label{fig:Mathieu_adiabaticity_violation}
\end{center}
\end{figure}
Hence in the   broad resonance regime, it is possible to produce  quanta of the offspring field with momenta greater than the rest mass of the inflaton, as can be seen from Fig.~\ref{fig:Resonance_bands}. This non-perturbative effect is a distinctive feature of the broad parametric resonance~\cite{Kofman:1997yn}. 
\subsubsection*{(2) \underline{Backreaction:}}
The second stage witnesses the backreaction of  explosively growing $\chi$ particles on the homogeneous inflaton condensate $\phi(t)$.  Backreaction becomes significant  when  the energy density of the fluctuations becomes comparable to that of the condensate \textit{i.e.}, $\rho_{\chi},\rho_{\delta\varphi} \sim \rho_\phi$, leading to the fragmentation of the inflaton condensate.  Since the coherent oscillations are hindered, parametric resonance is quenched \cite{Kofman:1997yn} and the production of $\chi$ particles slows down appreciably. 
\subsubsection*{(3) \underline{Scattering and thermalization:}}
In the third and final phase, the remaining inflaton condensate $\phi(t)$ decays perturbatively, leading
to a large number of decay products. This in turn leads to the  re-scattering
off these new decay products, which further decay into the quanta of other fields that they are (very
weakly) coupled to. Eventually the decay products thermalize resulting in the commencement of the
familiar radiation-dominated hot Big Bang phase, acquiring a reheating temperature $T_{\rm re}$ (which is different from the result obtained in the perturbative approach given  in Eq.~(\ref{eq:T_reheating_perturbative})).
Thus the end of the third stage leads to the commencement of
 the radiative hot Big Bang  phase of expansion during which the EoS in the universe is
$p \simeq \rho/3$. Note that the exact mechanism by which the universe thermalized after the inflaton decay is not currently known, however, see  Refs.~\cite{Micha:2002ey,Micha:2003ws,Micha:2004bv}.

\medskip

In the above discussion of preheating {\em via} parametric resonance, we ignored the effects of the background space which continues to expand as reheating proceeds. In practice, since  $H\neq 0$, the physical momenta of the produced $\chi$ field quanta, given by $k_p = k/a$, get redshifted away from the broad resonance band, as shown by the white  flow lines in Fig.~\ref{fig:Resonance_bands}. Consequently, resonant production of $\chi$ particles are quenched and the inflaton eventually proceeds to decay perturbatively. In certain cases, the effects of expansion of space leads to {\em stochastic resonance} during preheating, as discussed in Ref.~\cite{Kofman:1997yn}. However, we will not  elaborate further on such complex phenomenology associated with reheating. 

\bigskip

The dynamics of the  aforementioned three stages of reheating is quite complex and usually requires a numerical
treatment using lattice simulations of the full non-linear field equations \cite{Antusch:2021aiw,Figueroa:2021yhd,Figueroa:2023xmq}.
Reheating dynamics is one of the important open problems in theoretical and observational cosmology. For a detailed account of reheating, we direct the interested readers to Ref.~\cite{Lozanov:2019jxc}.

\section{Discussions}
\label{sec:Discussion}
The physics of inflation is quite  broad and rich, consisting of topics ranging from fundamental physics, String Theory and UV completeness to the observational aspects of CMB and LSS. Furthermore, 
 the inflationary cosmology involves 
 signals in the form of primordial features such as the  relic  gravitational waves, primordial non-Gaussianity, primordial black holes (PBHs) \textit{etc}. Additionally,  a number of important topics, such as the initial conditions for inflation, eternal inflation and multiverse, loop corrections to inflationary correlators, effective field theory of inflation and  alternatives to inflation, are of great interest to  researchers in the field.

\medskip
 
 In these lecture notes we  covered only some aspects of inflationary cosmology, although we  made an attempt to be comprehensive and pedagogical in our approach to these topics. Consequently, we have left out some of the aforementioned  key aspects of inflation in the present version of these lecture notes. Therefore, before concluding, we provide  brief discussions on some of the important topics in inflationary cosmology that were not covered in the preceding sections.

\begin{itemize}

\item {\em Relativistic cosmological perturbation theory}: While computing the power spectra of inflationary scalar and tensor perturbations, we worked in the framework of  gauge-invariant  relativistic perturbation theory~\cite{Bardeen:1980kt}, and made use of a number of standard results from  perturbation theory, especially in Secs.~\ref{sec:inf_dyn_scalar_QF}, \ref{sec:inf_dyn_tensor_QF} and \ref{sec:Non_Gaussianity}. However, we did not provide a detailed introduction to the subject, for which we point the readers to Refs.~\cite{Mukhanov:1990me,Dodelson:2003ft,Brandenberger:2003vk,Malik:2008im,Baumann_TASI,Lyth:2009zz,Gorbunov:2011zzc,Baumann:2022mni}.

\item {\em Beyond two-point correlators -- primordial non-Gaussianity}: The imprints of high energy primordial interactions during inflation are encoded in the higher-order correlation functions of inflationary scalar and tensor fluctuations. This makes the study of  primordial non-Gaussianity one of the most important frontiers in  early universe cosmology~\cite{Komatsu:2009kd,Meerburg:2019qqi}. However, our treatment of the subject was  minimal in Sec.~\ref{sec:Non_Gaussianity}. In fact, apart from the standard In-In formalism (interaction picture in QFT) \cite{Maldacena:2002vr,Weinberg:2005vy,Adshead:2009cb,Chen:2010xka,Giddings:2010ui,Wang:2013zva,Arkani-Hamed:2015bza}, there are a number of alternate techniques to compute the higher order primordial correlators, such as the $\delta$N formalism \cite{Sasaki:1995aw,Sugiyama:2012tj,Seery:2005wm,Seery:2005gb,Seery:2006vu,Seery:2006js,Seery:2008qj,Abolhasani:2018gyz,Wands:2010af,Langlois:2008vk,Tzavara:2010ge},   wave functional $\Psi[\zeta]$ approach~\cite{Arkani-Hamed:2017fdk,Hillman:2021bnk} and  cosmological bootstrap~\cite{Benincasa:2022gtd,Baumann:2022jpr,Maldacena:2011nz,Arkani-Hamed:2018kmz,Baumann:2019oyu,Salcedo:2022aal,Goodhew:2020hob,Goodhew:2021oqg,Melville:2021lst,Stefanyszyn:2023qov,Jazayeri:2021fvk}. A thorough account of the subject can be found in the \href{https://github.com/ddbaumann/cosmo-correlators/blob/main/LectureNotes-July2023.pdf}{\purple {\bf lecture notes}}.  

\item {\em Inflationary model building}:  In Sec.~\ref{sec:Observation_implications}, we stressed that  in the single field slow-roll inflationary 
 framework, the latest CMB observations favour asymptotically flat potentials. Furthermore, in Sec.~\ref{sec:inf_models}, we discussed the dynamics of  a number of asymptotically flat potentials, in light of the recent observations. However, our presentation in Sec.~\ref{sec:inf_dyn_obs} was rather phenomenological, without reference to the microphysical origin of the scalar field and its potential. Inflationary model building from the perspective  of (potentially UV-complete) fundamental physics, both in the context of quantum field theory and String Theory is an important and active field of research, see Refs.~\cite{Linde:1990flp,Lyth:1998xn,Baumann:2014nda}.

\item {\em Effective field theory (EFT) of inflation}: A convenient way to study inflationary dynamics for a large class of models in a model-independent way is the formalism of EFT applied to inflation \cite{Cheung:2007st,Weinberg:2008hq,Senatore:2016aui}.  Since the time-translational symmetry of pure de Sitter spacetime is spontaneously broken during inflation (in order to ensure the end of inflation), working  in the unitary gauge, a general EFT can be constructed  keeping the lowest dimensional operators invariant under  spatial diffeomorphisms~\cite{Cheung:2007st}. The EFT approach enables us to systematically incorporate high energy corrections to slow-roll inflationary dynamics.

\item {\em Observational probes of inflation}: Since the inflationary correlators constitute one of the  primary  probes of ultra-high energy physics of the early universe, a plethora of precision observational missions have been dedicated to  scrutinize various predictions of the inflationary scenario. This is an important topic in  fundamental  observational cosmology, especially  in the context of CMB, galaxy clustering, and 21 cm intensity fluctuations.  Since our treatment of the observational aspects of inflation in  Sec.~\ref{sec:evolution_fluct_post} was  minimal,  we direct the readers to Refs.~\cite{Dodelson:2003ft,Tegmark:2004qd,Dodelson:2003ip,Mukhanov:2005sc,Bond:1984fp,Hu:2008hd,Challinor_2009,Bernardeau:2001qr,Blumenthal:1984bp,Pritchard:2008da,Furlanetto:2019jso,Baumann:2022mni,Baumann:2018muz} 
 for a more elaborate discussion on the subject.

\item {\em Beyond slow-roll inflation}: The latest CMB observations provide support for the single field slow-roll paradigm of inflation and favour asymptotically flat potentials, as discussed in Sec.~\ref{sec:Observation_implications}. However, it is important to stress that the CMB and LSS observations are  sensitive to the comoving  wavenumbers in the range $k_{\rm CMB} \in [0.0005, 0.5]~{\rm Mpc}^{-1}$, which correspond to only $7$-$8$ e-folds of accelerated expansion during inflation, and hence a relatively small region of the inflaton potential, see Fig.~\ref{fig:inf_pot_toy_vanilla}.
 
On smaller scales, possible deviations from the slow-roll dynamics may lead to interesting changes in the spectra of primordial perturbations~\cite{Kinney:2005vj,Motohashi:2014ppa,Dimopoulos:2017ged,Tasinato:2020vdk}. In  particular, if the scalar perturbations are sufficiently large on small scales, then Primordial Black Holes (PBHs) may form when these modes enter the Hubble radius during 
the post-inflationary epoch~\cite{Hawking:1971ei,Carr:1974nx,Carr:1975qj,Sasaki:2018dmp,Escriva:2022duf,Byrnes:2021jka,Ozsoy:2023ryl}.  PBHs are therefore a powerful probe of the inflaton potential over the full range of field values. Large, potentially PBH forming, fluctuations can be generated by a feature in the inflationary potential~\cite{Starobinsky:1992ts,Ivanov:1994pa,Mishra:2019pzq,Ragavendra:2020sop,Karam:2022nym,Cole:2023wyx}, such as a flat inflection point shown in Fig.~\ref{fig:inf_pot_toy_USR_feature}. Such a feature can substantially slow down the already slowly rolling inflaton field, causing the inflaton to enter into an ultra slow-roll (USR) phase, which leads to an enhancement of the power spectrum, ${\cal P}_{\zeta}$, of the primordial curvature perturbations, see Refs.~\cite{Kinney:2005vj,Motohashi:2014ppa,Dimopoulos:2017ged,Cook:2015hma,Pattison:2018bct,Byrnes:2018txb,Mishra:2019pzq,Cole:2022xqc,Ozsoy:2019lyy,Ozsoy:2021pws,Tasinato:2023ukp,Tasinato:2023ioq}.  Although scalar and tensor fluctuations are decoupled at linear order in perturbation theory, second-order tensor fluctuations can be sourced by the enhanced first-order scalar fluctuations and hence they are called the {\em Scalar-induced Gravitational Waves} (SIGWs).  The study of enhancement of 
 the small-scale primordial fluctuations is an active topic of research at present, especially in the context of PBH formation, since PBHs are a natural candidate for dark matter, see Refs.~\cite{Chapline:1975ojl,Meszaros:1975ef,1979A&A....80..104N,Ivanov:1994pa,Carr:2016drx,Green:2020jor,Carr:2020xqk,Green:2024bam}.
\begin{figure}[htb]
\begin{center}
\includegraphics[width=0.75\textwidth]{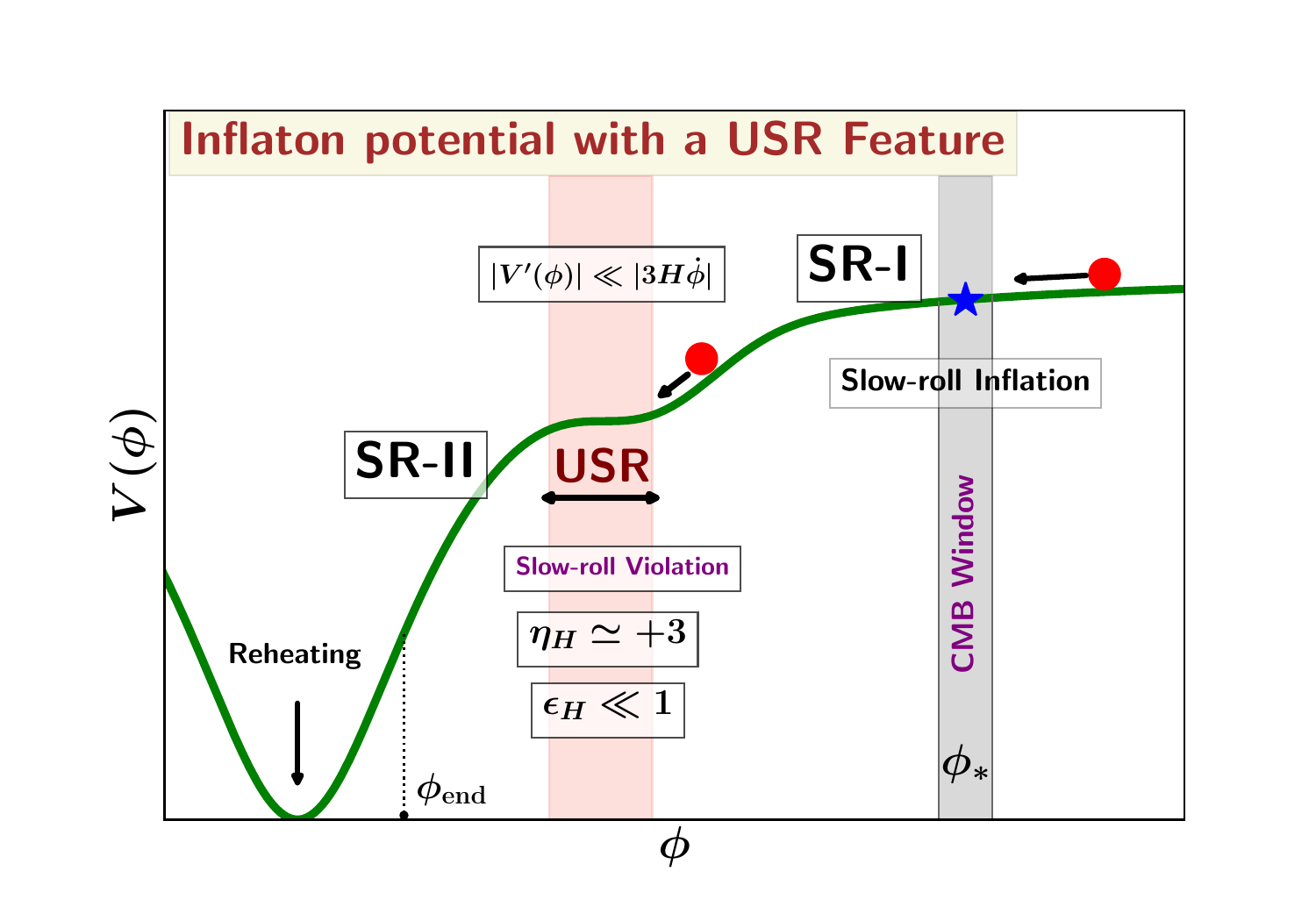}
\caption{A schematic illustration of a plateau potential (solid green line). The `CMB Window' represents field values corresponding to cosmological scales  $k_{\rm CMB}\in \l[0.0005,0.5\r]~{\rm Mpc}^{-1}$ that are probed by CMB observations. The blue star represents the CMB pivot scale $k_* = 0.05 \, {\rm Mpc}^{-1}$. The potential has a flat inflection-point like segment (highlighted with pink shading) which results in ultra slow-roll (USR) inflation. After the first slow-roll phase (SR-I) near the CMB Window, the inflaton enters into an USR phase. During this transient phase of USR, the second slow-roll condition is violated, specifically $\eta_H\simeq +3$. This leads to an enhancement in the primordial perturbations on small scales. Later, the inflaton emerges from the USR into another slow-roll phase (SR-II) before inflation ends at $\phi_{\rm end}$.}
\label{fig:inf_pot_toy_USR_feature}
\end{center}
\end{figure}
\item {\em Large quantum fluctuations and stochastic inflation}: Under certain conditions, the comoving curvature fluctuations during inflation can be quite large, \textit{e.g.} in the scenarios of eternal inflation~\cite{Linde:1990flp,Linde:1986fd,Linde:1986fc,Guth:2007ng,Creminelli:2008es,Rudelius:2019cfh,Linde:2015edk,Celoria:2021vjw,Cohen:2022clv,Cohen:2021jbo} (on length scales larger than $k_{\rm CMB}^{-1}$)  and  PBH formation (on length scales smaller than $k_{\rm CMB}^{-1}$) as discussed above. In such situations,  one needs a convenient formalism to resum the non-linearities associated with the fluctuations in order to carry out the computation of inflationary correlators accurately. The {\em stochastic inflation} formalism~\cite{Starobinsky:1982ee,Starobinsky:1986fx,Salopek:1990jq,Salopek:1990re,Starobinsky:1994bd,Grain:2017dqa} is  an useful approach in this direction.  It  is an effective treatment of  the dynamics of the  long-wavelength (IR)  part of the inflaton field  coarse-grained on scales much greater than the  Hubble radius \textit{i.e.}~$k \leq \sigma \, aH$, with  the constant $\sigma \ll 1$.  In this framework, the evolution of the  coarse-grained inflaton field is governed by two first-order  non-linear  stochastic {\em Langevin-type} differential equations  which receive constant quantum kicks from  the small scale UV modes that are exiting the Hubble radius due to the accelerated expansion during inflation. Hence the small-scale fluctuations constitute  classical stochastic noise terms in the Langevin equations. The formalism is usually combined  with  the classical $\delta N$ formalism \cite{Starobinsky:1982ee,Sasaki:1995aw,Lyth:2004gb,Wands:2000dp,Lyth:2005fi}  in order to compute cosmological correlators in this framework.  This leads to the emergence of  the  stochastic $\delta {\cal N}$ formalism, see Refs.~ \cite{Starobinsky:1986fx,Fujita:2013cna,Fujita:2014tja,Vennin:2015hra,Pattison:2017mbe,Pattison:2021oen,Mishra:2023lhe}.

 Typical fluctuations, which induce temperature anisotropies  in the CMB and seed the formation of large scale structure of the universe, belong to the head of the primordial PDF, which is predominantly Gaussian \cite{Maldacena:2002vr,Creminelli:2003iq,Meerburg:2019qqi,Planck:2019kim}. However, rare fluctuations, such as those leading to eternal inflation and primordial black hole formation, belong to the tail of the PDF which is expected to be highly non-Gaussian~\cite{Pattison:2017mbe,Ezquiaga:2019ftu,Figueroa:2020jkf,Celoria:2021vjw,Cohen:2022clv,Ferrante:2022mui,Gow:2022jfb}, as illustrated schematically in Fig.~\ref{fig:inf_PDF_gen}.  The tail of the primordial PDF can be determined in the framework of stochastic $\delta {\cal N}$ formalism  by using the techniques of first-passage time analysis~\cite{Vennin:2015hra,Pattison:2017mbe,Ezquiaga:2019ftu,Pattison:2021oen,De:2020hdo,Figueroa:2020jkf,Pattison:2021oen,Mishra:2023lhe,Vennin:2024yzl}. 
 \begin{figure}[htb]
\begin{center}
\includegraphics[width=0.7\textwidth]{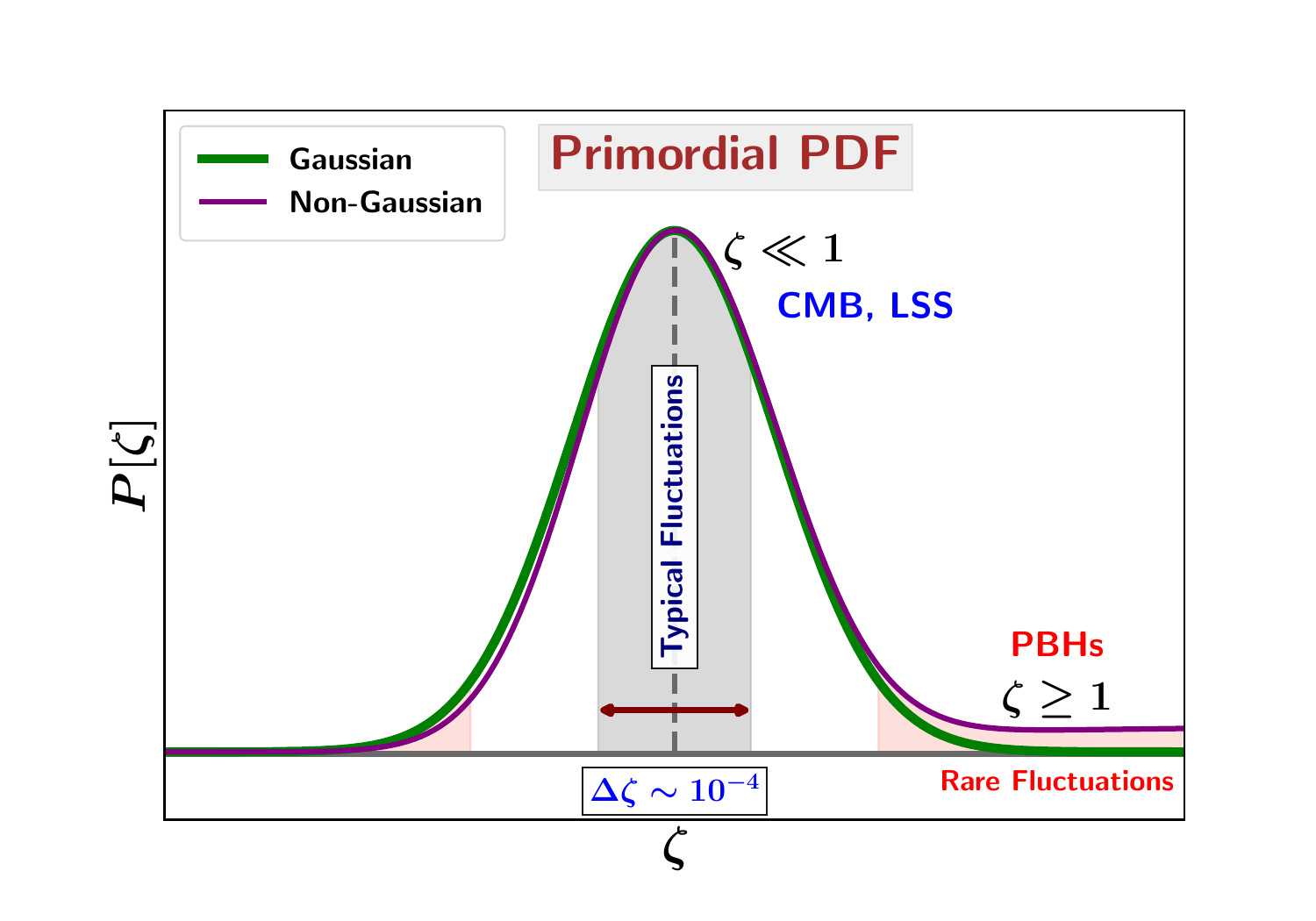}
\caption{A schematic illustration of the probability distribution (PDF), $P[\zeta]$, of the primordial curvature fluctuations $\zeta$. The typical fluctuations, which induce temperature fluctuations in the CMB and facilitate the formation of large scale structure of the universe, belong to the head of the primordial PDF (represented by grey colour shaded region) which is predominantly Gaussian (green colour curve). However, rare fluctuations, such as those leading to eternal inflation and primordial black hole formation, belong to the tail of the PDF (represented by pink colour shaded region) which can be highly non-Gaussian (purple colour curve).}
\label{fig:inf_PDF_gen}
\end{center}
\end{figure}
\item {\em Multi-field inflation}: In these lecture notes we primarily focused on the simplest scenario of single field inflationary dynamics. Since  inflation is supposed  to have taken place in the very early universe at energy scales much higher than that of the Standard Model (SM) of particle physics, it is expected to be described in the
framework of beyond SM physics, such as supersymmetry, supergravity and string theory. Most  beyond SM theories feature multiple scalar degrees of freedom which might have been relevant during inflation.
Hence the multi-field inflationary framework have been investigated extensively  over the past three decades~\cite{Linde:1993cn,Hwang:2000jh,GrootNibbelink:2000vx,Gordon:2000hv,Hwang:2001fb,GrootNibbelink:2001qt,Hwang:2007ni,Wands:2007bd,Senatore:2010wk,Hwang:2015prq,Gong:2016qmq,}. They have distinctive signatures in the form of primordial  non-Gaussianity and isocurvature  perturbations, which make testing  the multi-field inflationary scenario  an important observational target for the upcoming decades~\cite{Seery:2005gb,Peterson:2010np,Achucarro:2010da,Peterson:2011yt,Elliston:2012ab,Burgess:2012dz,Kaiser:2013sna,Schutz:2013fua}.

\item {\em Dissipative effects during inflation}: In the standard inflationary scenario, the inflaton couplings to external fields are assumed to be very small. Particle production during inflation can be ignored because the  accelerated expansion of space rapidly dilutes the number densities of the produced particles. Hence, the inflaton decay becomes significant only after the end of inflation when the expansion of space begins to  decelerate, ultimately leading to reheating, as discussed in Sec.~\ref{sec:Reheating}. However, if the inflaton couplings to other degrees of freedom are  strong enough, then particle production during inflation leads to a non-negligible backreaction on the inflaton dynamics, modifying the inflationary predictions of the  primordial perturbations. In general, such scenarios are collectively referred to as {\em dissipative inflation}~\cite{Brandenberger:1984cz,Lyth:1996yj,Berera:1998gx,LopezNacir:2011kk,Anber:2009ua,Adshead:2015kza,LopezNacir:2012rm,Cook:2011hg,Cook:2013oaa,Creminelli:2023aly}. In certain regimes, the produced light degrees of freedom 
  behave as a (sub-dominant) thermal radiation bath during inflation, leading to the emergence of  {\em warm  inflation}~\cite{Berera:1995ie,Berera:1998px,Berera:2008ar,Berghaus:2019whh,Mirbabayi:2022cbt,}. The warm inflationary paradigm exhibits certain advantages over the standard dissipationless  {\em cold inflationary paradigm} from the perspective of fundamental model building as well as reheating,  see Refs.~\cite{Ooguri:2018wrx,Das:2018rpg,Berera:2023liv}.  

\item {\em Initial conditions for Inflation}: Our discussions in Sec.~\ref{sec:inf_resolution} stressed that once inflation begins, and lasts for a sufficiently long number of  e-folds of expansion, it successfully addresses the initial conditions for the hot Big Bang phase. However, we did not make any comment on whether inflation can begin starting from generic initial conditions in the pre-inflationary universe, closer to the Planck scale. In fact, the investigation of initial conditions for inflation is an important conceptual exercise, and there has been a great deal  of research work on the topic, both analytically and computationally, see Refs.~\cite{Belinsky:1985zd,Mukhanov:2005sc,Baumann_TASI,Remmen:2013eja,Brandenberger:2016uzh, Mishra:2018dtg,Starobinsky:1982mr,Jensen:1986nf,Kitada:1991ih,Maleknejad:2012as,Goldwirth:1991rj,Hartle:1983ai,Linde:1983mx,Vilenkin:1982de,Albrecht:1986pi,Easther:2014zga,Goldwirth:1989pr,East:2015ggf,Clough:2016ymm}.

\item {\em Alternatives to inflation}: While cosmic inflation is certainly  the leading scenario in explaining  the initial conditions for the hot Big Bang, there exit a number of alternatives to inflation~\cite{Brandenberger:2009jq,Brandenberger:2018wbg,Creminelli:2007aq,Ijjas:2019pyf,Ijjas:2018qbo,Graham:2017hfr} which can  also address the initial conditions, and are consistent with the observations of primordial perturbations at present. Amongst these are the matter bounce scenario~\cite{Finelli:2001sr,Brandenberger:2016vhg,Wands:1998yp}, pre-Big Bang and the ekpyrotic cosmology~\cite{Gasperini:1992em,Lidsey:1999mc,Gasperini:2002bn,Khoury:2001wf,Raveendran:2018yyh,Raveendran:2017vfx,Levy:2015awa,Ijjas:2014fja}, and emergent string gas  cosmology~\cite{Brandenberger:1988aj,Brandenberger:2011et,Brandenberger:2008nx}, to mention a few. However, many of the  future observational probes will be dedicated  to provide a decisive direction, either for, or against the inflationary paradigm, see Refs.~\cite{Tegmark:2004qd,Dodelson:2003ip,Brandenberger:2011eq,Vagnozzi:2022qmc,Achucarro:2022qrl,Green:2022bre,CMB-S4:2022ght}. 

\end{itemize}

\section{Supplementary sources}
\label{sec:Supp}

There exist a number of excellent sources (lecture notes/reviews) in the literature  on  cosmic inflation. In the following, I refer to some of the sources that I have personally found useful  over the years. 
\begin{enumerate}
\item I originally learned inflation from three sources: (i) Handwritten lecture notes by Paolo Creminelli given at ICTP in 2013/14, (ii) a set of lectures on inflation given by Nima Arkani-Hamed (\href{https://indico.ictp.it/event/a13193/session/2/contribution/9/material/video/}{\purple {\bf link to the videos}}) and Juan Maldacena (\href{https://indico.ictp.it/event/a13193/session/3/contribution/17/material/video/}{\purple {\bf link to the videos}})  at the ICTP Spring School in 2014, and finally (iii)  `\textit{TASI Lectures on Inflation}'  by Daniel Baumann  in Ref.~\cite{Baumann_TASI}. 
\item Lecture notes on `\textit{Inflation and the theory of cosmological perturbations}' by Antonio Riotto  in Ref.~\cite{Riotto:2002yw}.
\item `\textit{TASI lectures on Inflation}' by William Kinney in Ref.~\cite{kinney2009tasi}.
\item Lecture notes titled `\textit{An introduction to inflation and cosmological perturbation theory}' by L. Sriramkumar in Ref.~\cite{Sriramkumar:2009kg}.
\item TASI `\textit{Lectures on Inflation}' by Leonardo Senatore in Ref.~\cite{Senatore:2016aui}.
\item Standard cosmology textbooks such as the `\textit{Physical Foundations of Cosmology}' by Mukhanov \cite{Mukhanov:2005sc}, `\textit{Introduction to the theory of the early universe: Cosmological perturbations and inflationary theory}'  by Gorbunov and Rubakov \cite{Gorbunov:2011zzc} and `\textit{Cosmology}' by Daniel Baumann \cite{Baumann:2022mni}. 
\item `\textit{TASI lectures on Primordial Cosmology}' by Daniel Baumann in Ref.~\cite{Baumann:2018muz} which have been quite useful to me in the past few years.  
\item A set of lectures on inflation by Matthew Kleban at the ICTP Summer School 2018 (link to the \href{https://www.youtube.com/playlist?list=PL11nn_gZTAYrI7HOc3RS6E4n4YQyRR7Ry}{\purple {\bf  YouTube playlist}}). 
\item `\textit{Lectures  on Cosmological Correlations}'  by Daniel Baumann and Austin Joyce, given in the \href{https://github.com/ddbaumann/cosmo-correlators/blob/main/LectureNotes-July2023.pdf}{\purple {\bf github link}},  covering more recent developments in the field. 
\end{enumerate}

\section*{Appendices}
\appendix
\section{Hubble parameter in the early universe}
\label{app:Hubble_z}
The early universe is assumed to be thermal and radiation dominated, at least up to a temperature of $T_{\rm EW} \sim 200~$GeV.  The radiation density of the thermal plasma is given by
\beq \rho_r(T) = \frac{\pi^2}{30} \, g_{*}(T) \, T^4 ,
\label{rho_r}
\eeq
where, $g_*(T)$ is  the effective number of relativistic degrees of freedom in the energy density at temperature $T$.
In the flat ${\rm \Lambda}$CDM model, matter, radiation and dark energy evolve without interacting with each other at lower redshifts (late times). However,  that was not  the case in the very early universe. Due to the presence of high density and temperature conditions, the universe was in an extremely hot and dense plasma state. Hence the rate of interactions were higher, and  matter and radiation could convert into each other more easily. 
Since entropy is conserved for a thermal distribution, keeping track of the entropy is quite  useful in determining the evolution of radiation density and temperature  in the hot Big Bang phase. 
Using the conservation of entropy \cite{Kolb:1990vq},
\beq
S_0 = S_T,
\eeq
where, $S_0$ is the entropy today and $S_T$ is the entropy measured at some temperature T in the  past. This leads to
\beq
s_0 \, a_0^3 = s_T \, a^3 ,
\label{eq:entropy_density}
\eeq
where $s$ is the entropy density given by
$$ s = \frac{2 \pi^2}{45}\, g^s_*(T)\, T^3 \, , $$ 
and $g^s_*$ is the effective number of relativistic degrees of freedom in the entropy density.\footnote{$g_*$ and $g^s_*$ were exactly same until the electron-positron annihilation when our universe was about 6 seconds old.} Hence, Eq.~(\ref{eq:entropy_density}) can be written as
$$
\frac{2 \pi^2}{45}\, T_0^3 \, g^s_{0,*} \, a_0^3 = \frac{2 \pi^2}{45} \, {T^3}\, g^s_{*} \, {a}^3 \, ,
$$
yielding
\beq
\boxed{ T = \l(\frac{g^s_{0,*}}{g^s_*}\r)^{1/3}\,\l(\frac{a_0}{a}\r)\, T_0 = \l(\frac{g^s_{0,*}}{g^s_*}\r)^{1/3}\,\l(1+z\r)\, T_0 } \, .
\label{eq:T_app_1}
\eeq
Inserting Eq.~(\ref{eq:T_app_1}) into Eq.~(\ref{rho_r}), we obtain the expression for the radiation energy density at any given redshift $z$ to be 
\beq
\rho_r(T) = \frac{\pi^2}{30} \, g_*(T)\,\l(\frac{g^s_{0,*}}{g^s_*}\r)^{4/3} \, (1+z)^4 \, T_0^4 \, ,
\label{rh_r}
\eeq
while the radiation energy density at the present epoch is given by
\beq
\rho_{0r} = \frac{\pi^2}{30}\,g_{0,*}\,T_0^4 \, .
\label{rh_0}
\eeq
Accordingly, using the Friedmann Eq.~(\ref{eq:Friedmann_1_gen}) with $K=0$,  we obtain the expression for the Hubble parameter as a function of redshift in the radiation dominated epoch to be
\beq
\boxed{ H(z) \simeq H_0  \, \l(\Omega_{0r} \, \frac{g_*}{g_{0,*}} \r)^{1/2} \, \l(\frac{g^s_{0,*}}{g^s_{*}}\r)^{2/3} \, (1+z)^2 } \, .
\label{H_RD}
\eeq

\section{Kinematics of reheating and the duration of inflation}
\label{app:N*_duration}
The epoch of reheating is usually characterized by a set of  three parameters $\lbrace \wre,\Nre,\Tre \rbrace$, namely, the effective equation of state (EoS) during reheating $\wre$, the duration of reheating  $\Nre$ and the  temperature at the end of reheating $\Tre$, when the universe transits to a thermalized 
radiation dominated  hot Big Bang phase. The duration of reheating can be defined by the number of $e$-folds of expansion  between the end of inflation $a_e$ and the end of reheating 
(or equivalently, the commencement of radiation domination) $a_{re}$, given by 
\beq
\Nre=\ln \l(\f{a_{\rm re}}{a_e}\r) \, .
\label{eq:N_re_definition}
\eeq
While $\Nre$ and $\Tre$ are interesting physical quantities, in their own right, describing the epoch of reheating, they are also potentially important for correctly interpreting the bounds on the CMB observables such as the scalar spectral index $n_{_S}$  and the tensor-to-scalar ratio $r$.

\begin{figure}[t]
\begin{center}
\includegraphics[width=0.7\textwidth]{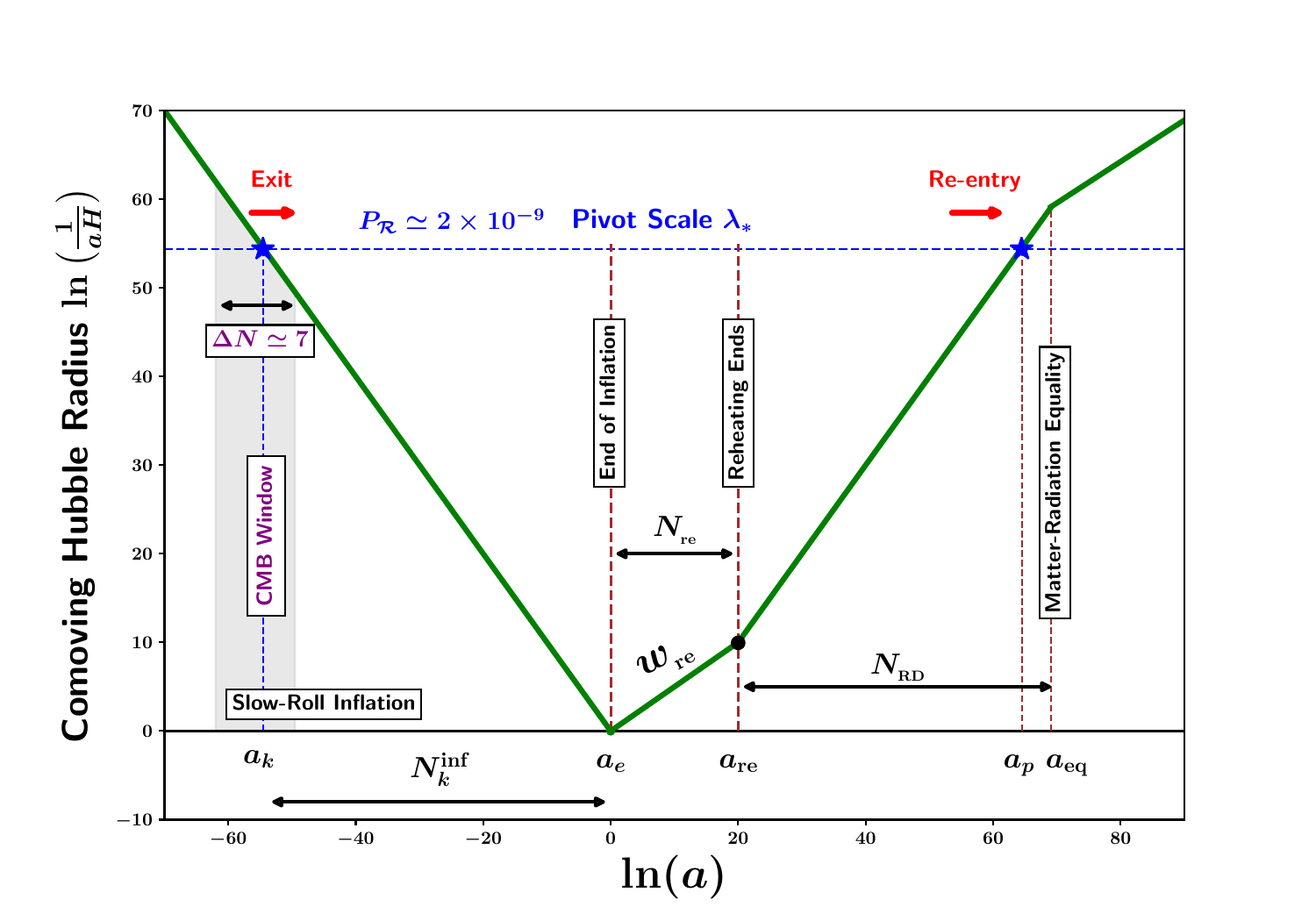}
\caption{Schematic illustration of  the evolution of the comoving Hubble radius
$(aH)^{-1}$ with scale factor. During inflation $(aH)^{-1}$ decreases which causes physical scales to
exit the Hubble radius. After inflation ends $(aH)^{-1}$ increases, and physical scales begin to enter
the Hubble radius.
The CMB pivot scale, as used by the Planck mission, is set at $k_*=0.05~{\rm Mpc}^{-1}$.
It enters the 
Hubble radius during the radiation dominated epoch when $a_p \sim 4\times 10^{-5}\, a_0$. Note that the duration of reheating $\Nre$, and hence the duration of the radiation dominated epoch $\Nrd$, changes for different values of the reheating equation of state $\wre$. Recall 
 that $(aH)^{-1} \propto a$ during the radiative regime and
$(aH)^{-1} \propto a^{-1}$ during
inflation.
}
\label{fig:RH_causal}
\end{center}
\end{figure}

Following the evolution of the comoving Hubble radius from the epoch of Hubble-exit, at $a=a_k$, of a comoving scale $k$,  through its post-inflationary Hubble-entry at $a=a_p$, until the present epoch $a=a_0$, we obtain 
\beq
\ln {\l( \f{k}{a_0 H_0} \r)} = -N_k^{\rm inf} - \Nre - \Nrd - \ln {\l(1+z_{\rm eq}\r)} + \ln {\f{H_k^{\rm inf}}{H_0}}~,
\label{eq:CMB_reheat_k}
\eeq
where $H_k^{\rm inf}$ is the Hubble parameter at the time of the Hubble-exit of the scale $k$,  $N_k^{\rm inf}=\ln {\l( a_e/a_k \r)}$ is 
the number of $e$-folds between the Hubble-exit (of scale $k$) and the end of inflation, $\Nrd$ is the duration of the radiation dominated epoch and $z_{\rm eq}$ is the redshift at the epoch of  matter-radiation equality. In general, $k$ may correspond to any observable CMB scale in the range $k\in \l[0.0005,0.5\r]~{\rm Mpc}^{-1}$. The CMB pivot scale, namely $k\equiv k_*=0.05~{\rm Mpc}^{-1}$, makes its Hubble-entry 
during the radiative epoch at $a_p\sim 4\times 10^{-5}\, a_0$, see Fig.~\ref{fig:RH_causal}. 

The primary goal here is to characterize the epoch of reheating between the end of inflation $a_e$ and the commencement of the radiative epoch $a_{\rm re}$. Assuming the  effective EoS $\wre$ during reheating to be a constant 
allows us to match the density at the beginning of the radiative epoch to the density at the end of inflation by 
\beq
\Rre = \rho_e \l(\f{a_e}{a_{\rm re}} \r)^{3(1+\wre)}~,
\label{eq:CMB_reheat_rho}
\eeq
which yields the following expression for the duration of reheating \cite{Mishra:2021wkm}:
\beq
\Nre \equiv \ln {\l(\f{a_{\rm re}}{a_e}\r)} = \f{1}{3(1+\wre)}\ln {\l(\f{\rho_e}{\Rre}\r)}~.
\label{eq:CMB_reheat_Nre}
\eeq
Expressing $\Rre$ in terms of the reheating temperature $\Tre$, we obtain
\beq
\Nre = \f{1}{3(1+\wre)}\ln {\l(\f{\rho_e}{\f{\pi^2}{30} \, g_{\rm re}\Tre^4}\r)}~,
\label{eq:CMB_reheat_Nre_Tre}
\eeq
 where $g_{\rm re} \equiv g(\Tre)$ is the effective number of relativistic degrees of freedom at the end of reheating. Applying entropy conservation, as discussed in App.~\ref{app:Hubble_z}, to express $\Tre$ in terms of $a_{\rm re}$, we obtain  
\beq
\Tre = \l(\f{g_{\rm eq}^s}{g_{\rm re}^s}\r)^{\f{1}{3}} \l(\f{a_{\rm eq}}{a_{\rm re}}\r)T_{\rm eq}~,
\label{eq:CMB_reheat_Tre}
\eeq
where $g_{\rm eq}^s$ and $g_{\rm re}^s$ are the effective number of relativistic degrees of freedom in
the entropy at the epoch of matter-radiation equality and at the end of reheating respectively,
 while $T_{\rm eq}$  is the temperature at the matter-radiation equality. Incorporating (\ref{eq:CMB_reheat_Tre}) into (\ref{eq:CMB_reheat_Nre_Tre}), we obtain (see Ref.~\cite{Mishra:2021wkm})
\beq
\Nre = \f{4}{3(1+\wre)}\l[ \f{1}{4} \ln {\l(\f{30}{\pi^2 g_{\rm re}}\r)} + \f{1}{3}\ln {\l(\f{g_{\rm re}^s}{g_{\rm eq}^s}\r)} + \ln {\l(\f{\rho_e^{\f{1}{4}}}{T_{\rm eq}}\r)}-\Nrd\r]~.
\label{eq:CMB_reheat_Nre_1}
\eeq
Substituting $\Nrd$ from (\ref{eq:CMB_reheat_k}) into (\ref{eq:CMB_reheat_Nre_1}), we arrive at an
 important expression for the duration of reheating, namely (see Ref.~\cite{Mishra:2021wkm}) 
\beq
\Nre = \f{4}{3(1+\wre)}\l[ \f{1}{4} \ln {\l(\f{30}{\pi^2 g_{\rm re}}\r)} + \f{1}{3}\ln {\l(\f{g_{\rm re}^s}{g_0^s}\r)} + \ln {\l(\f{\rho_e^{\f{1}{4}}}{H_k^{\rm inf}}\r)} + \ln {\l(\f{k}{a_0 T_0}\r)}+N_k^{\rm inf}+\Nre \r]~.
\label{eq:CMB_reheat_Nre_2}
\eeq
Note that if $\wre = 1/3$, then the term $\Nre$ cancels from both sides of (\ref{eq:CMB_reheat_Nre_2}), yielding the 
following expression for $N_k^{\rm inf}$
\beq
N_k^{\rm inf} = - \l[ \ln {\l(\f{k}{a_0 T_0}\r)} + \ln {\l(\f{\rho_e^{\f{1}{4}}}{H_k^{\rm inf}}\r)} + \f{1}{4} \ln {\l(\f{30}{\pi^2 g_{\rm re}}\r)} + \f{1}{3}\ln {\l(\f{g_{\rm re}^s}{g_0^s}\r)} \r] ~.
\label{eq:CMB_reheat_Nk}
\eeq
This arises because the end of reheating, and hence the beginning of the radiative epoch, cannot be strictly defined 
within this framework if $\wre = 1/3$. However for
  $\wre\neq 1/3$ one obtains the following final expression 
for $\Nre$ from Eq.~(\ref{eq:CMB_reheat_Nre_2})
\beq
\Nre = -\f{4}{1-3\wre}\l[N_k^{\rm inf} + \ln {\l(\f{\rho_e^{\f{1}{4}}}{H_k^{\rm inf}}\r)}  + \ln {\l(\f{k}{a_0 T_0}\r) + \f{1}{4} \ln {\l(\f{30}{\pi^2 g_{\rm re}}\r)} + \f{1}{3}\ln {\l(\f{g_{\rm re}^s}{g_0^s}\r)} } \r]~.
\label{eq:CMB_reheat_Nre_3}
\eeq
Accordingly the expression for the reheating temperature $\Tre$ in terms of the duration of reheating $\Nre$ and effective reheating EoS, $\wre$,
 follows from (\ref{eq:CMB_reheat_Nre_Tre}) to be 
\beq
\Tre = \l(\f{30 \rho_e}{\pi^2 g_{\rm re}}\r)^\f{1}{4} e^{-\f{3}{4}\l(1+\wre\r)\Nre}~.
\label{eq:CMB_reheat_Tre2}
\eeq
  Note that  the expressions (\ref{eq:CMB_reheat_Nre_3}) and ({\ref{eq:CMB_reheat_Tre2}}) are valid only for $\wre\neq 1/3$. We return to the case $\wre = 1/3$ at the end of this subsection, for which the relevant final expression for $N_k^{\rm inf}$, following Eq.~(\ref{eq:CMB_reheat_Nk}), is given in Eq.~(\ref{eq:CMB_reheat_Nk_SR}).

\begin{figure}[hbt]
\begin{center}
\subfigure[][]{
\includegraphics[width=0.475\textwidth]{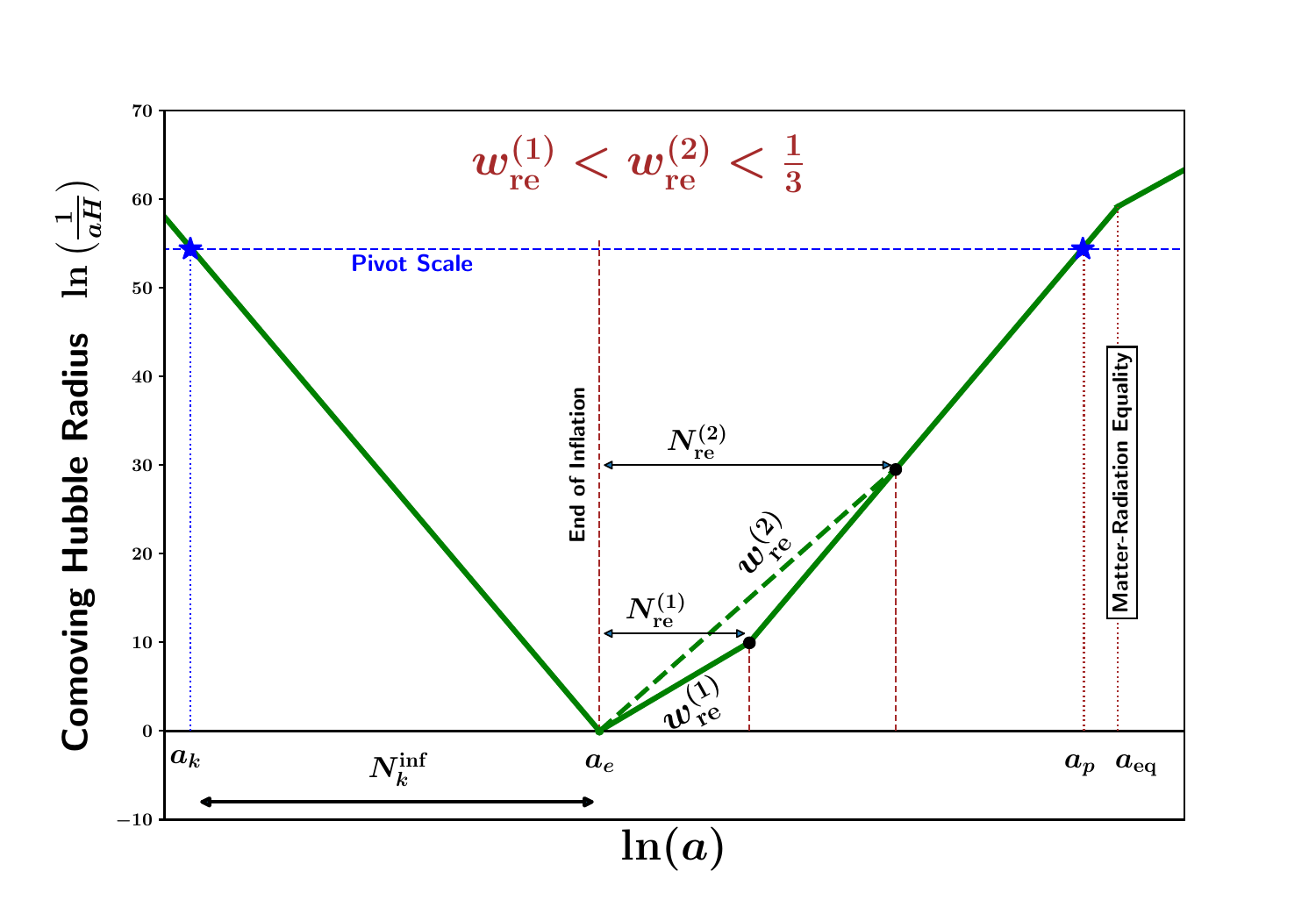}}
\subfigure[][]{
\includegraphics[width=0.475\textwidth]{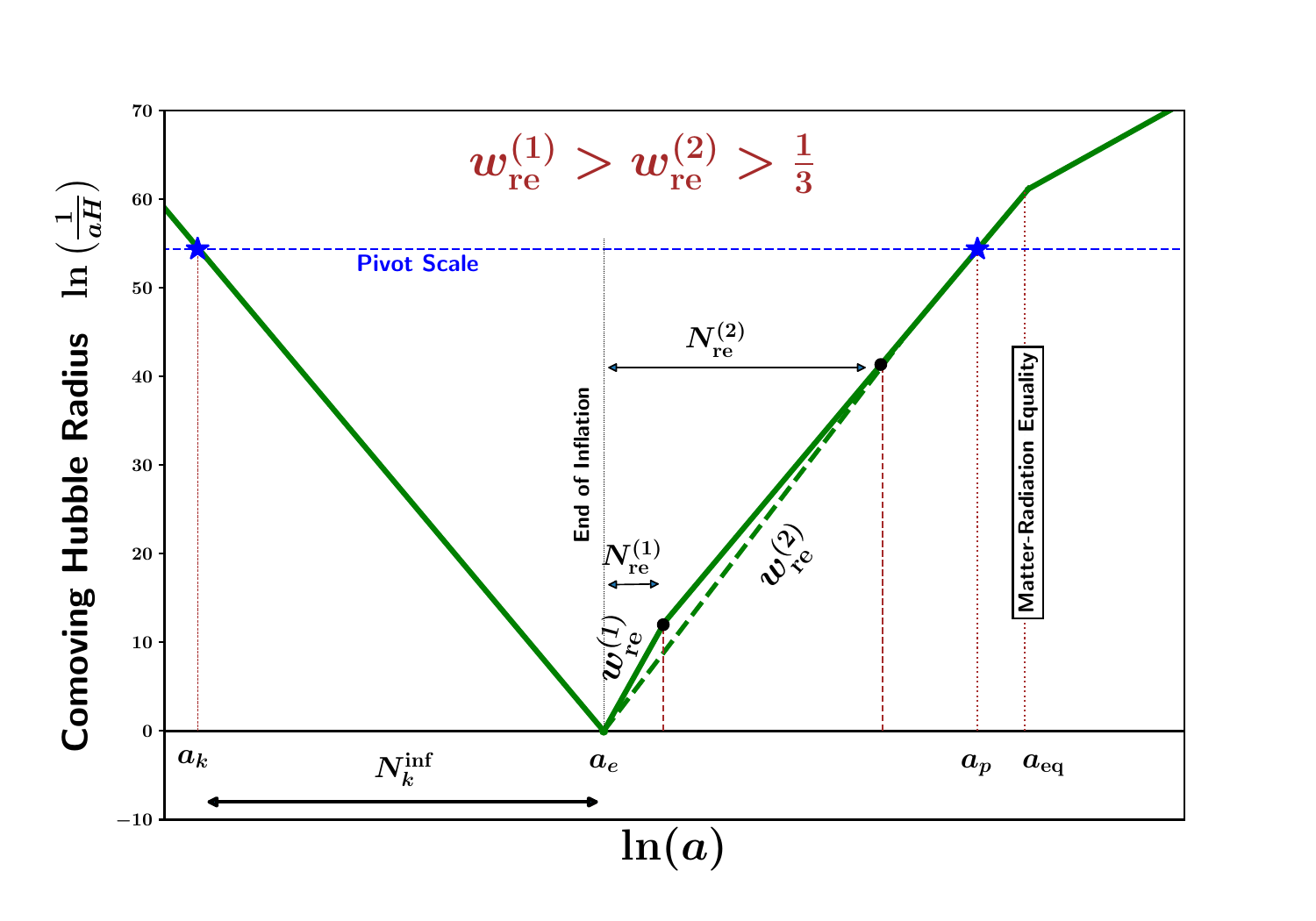}}
\caption{This figure schematically illustrates the evolution of the comoving Hubble radius
$(aH)^{-1}$ with scale factor of the universe and explicitly depicts the dependence of the duration of reheating  on the reheating equation of state for a particular inflationary model with a given $N_k^{\rm inf}$. 
The {\bf left panel} shows  that for shallow reheating EoS $\wre<1/3$, the duration of reheating $\Nre$ is longer for a higher value of $\wre$, namely $\Nre^{(1)} < \Nre^{(2)}$ for
$\wre^{(1)}<\wre^{(2)}$. The {\bf right panel} demonstrates  that for a
 stiffer reheating EoS $\wre>1/3$, the duration of reheating $\Nre$ is shorter for a higher 
value of $\wre$, namely $\Nre^{(1)} < \Nre^{(2)}$ for
$\wre^{(1)}>\wre^{(2)}$, in accordance with equation (\ref{eq:CMB_reheat_Nre_SR}). 
Note that $(aH)^{-1} \propto a$ during the radiative regime and
$(aH)^{-1} \propto a^{-1}$ during 
inflation.
For comparison see Fig.~\ref{fig:RH_causal}.}
\label{fig:inf_causal_w_Nre}
\end{center}
\end{figure}

In the context of single field slow-roll inflation, the value of the inflaton field at the end of inflation $\phi_e$ can be determined   from the condition
\beq
\epsilon_{_V}(\phi_e) = \f{\mpl^2}{2}\l(\f{V_{,\phi}}{V}\r)^2\bigg\vert_{\phi_e} \simeq 1~,
 \label{eq:inf_end}
\eeq
and the corresponding inflaton density at the end of inflation is given by 
$$\rho_e \equiv \rho_\phi\bigg\vert_{\phi_e} = \f{1}{2}\dot{\phi}^2+V(\phi)\bigg\vert_{\phi_e} \simeq \f{3}{2}V(\phi_e) 
\equiv \f{3}{2}V_e~.$$
Substituting $\rho_e = \f{3}{2}V_e$ in
(\ref{eq:CMB_reheat_Tre2}) results in the following expression for the reheating temperature 
\beq
\Tre =  \l(\f{45}{\pi^2 g_{\rm re}}\r)^\f{1}{4} V_e^{\f{1}{4}}\, e^{-\f{3}{4}\l(1+\wre\r)\Nre}~.
\label{eq:CMB_reheat_Tre_SR}
\eeq
Assuming $k$ to be the CMB pivot scale, $k\equiv k_* = a_k H_k^{\rm inf} = a_p H_p = 0.05~{\rm Mpc}^{-1}$ in 
(\ref{eq:CMB_reheat_Nre_3}) and inserting the values of $T_0$, $g_0^s$, $g_{\rm re}$ and $g_{\rm re}^s$, one arrives at the following formula 
which expresses the duration of reheating $\Nre$  as a function of the reheating EoS, $\wre$,
on the  one hand, and parameters of the inflationary potential $V_e$, $H_k^{\rm inf}$, on the other: 
\beq
\Nre = \f{4}{1-3\wre}\l[61.55 -N_k^{\rm inf} - \ln {\l(\f{V_e^{\f{1}{4}}}{H_k^{\rm inf}}\r)}  \r]~, ~~~~\wre \neq 1/3 .
\label{eq:CMB_reheat_Nre_SR}
\eeq
 Eqs.~(\ref{eq:CMB_reheat_Nre_SR})  and (\ref{eq:CMB_reheat_Tre_SR})  capture some of the 
essential implications of reheating kinematics on CMB observables  and possess  important physical significance. For example, it is easy to see, from Eq.~(\ref{eq:CMB_reheat_Nre_SR}), that  
for a  softer reheating EoS with $\wre< 1/3$, a higher value of $N_k^{\rm inf}$ corresponds to a shorter reheating period $\Nre$, for a given model of inflation. 
Exactly the opposite is true for
a stiffer EoS with  $\wre > 1/3$. In this case the RHS of Eq.~(\ref{eq:CMB_reheat_Nre_SR}) flips sign
so that a larger
 value of $N_k^{\rm inf}$ implies a larger $\Nre$ and hence a longer duration of reheating. Similarly 
 Eq.~(\ref{eq:CMB_reheat_Tre_SR}) implies that the longer is
 the duration of reheating $\Nre$, the lower will be the reheating temperature $\Tre$.
Moreover this result is independent of the value of $\wre$ simply because
 $1+\wre>0$ (since $\wre > -1/3$ by definition).  Another interesting aspect of Eq.~(\ref{eq:CMB_reheat_Nre_SR}) is that,  given an inflationary potential with a fixed value of $N_k^{\rm inf}$ (which satisfies the CMB bound on $n_{_S}\in \l[0.957,0.976\r]$), the duration of reheating $\Nre$ increases with an increase in the effective EoS $\wre$ as long as $\wre<1/3$. This is demonstrated in the left 
panel of Fig. \ref{fig:inf_causal_w_Nre} in which $\Nre^{(1)} < \Nre^{(2)}$ for 
$\wre^{(1)}<\wre^{(2)}<1/3$.  Similarly  $\Nre$ increases with a {\em decrease}
 in the effective EoS, $\wre$, if $\wre>1/3$. This is shown
 in the right panel of fig. \ref{fig:inf_causal_w_Nre} where $\Nre^{(1)} < \Nre^{(2)}$
 for $\wre^{(1)}>\wre^{(2)}>1/3$.
These arguments also indicate that $w=1/3$ is a critical value of the EoS during reheating.
 
Turning attention to $\Tre$, one notes that conservative  upper and lower bounds on this
quantity can be placed from the following considerations.
 In Sec.~\ref{sec:Observation_implications}, we learned  that the CMB upper bound on the tensor-to-scalar ratio, namely $r\leq 0.036$, 
translates into an upper bound on the inflationary Hubble scale $H_k^{\rm inf} \leq  4.7\times 10^{13}~{\rm GeV}$, which in turn sets an upper bound on the energy scale of inflation $T_{\inf} \leq 1.4\times 10^{16}~{\rm GeV}$. Since reheating happens after the end of 
inflation, one gets  $\Tre \leq 1.4\times 10^{16}~{\rm GeV}$ as an absolute upper bound on the reheating temperature. Similarly, in order to preserve the success of the hot Big Bang phase, reheating must terminate before the beginning of Big Bang Nucleosynthesis (BBN) yielding the absolute lower bound $\Tre \gg  1~{\rm MeV}$. 
Hence the most conservative bounds on the reheating temperature are
\beq
1 \, {\rm MeV} \ll \Tre \leq 10^{16}\, {\rm GeV} \, .
\label{eq:bound_Tre}
\eeq  
Note that Eq.~(\ref{eq:CMB_reheat_Nre_SR}) is  only applicable for reheating EoS
with $\wre \neq 1/3$.
For $\wre = 1/3$, using Eq.~(\ref{eq:CMB_reheat_Nk}), assuming $k$ to be the CMB pivot scale, i.e $k\equiv k_* = 0.05~{\rm Mpc}^{-1}$ and inserting the values of $T_0$, $g_0^s$, $g_{\rm re}$ and $g_{\rm re}^s$, as was previously done for the case $\wre\neq 1/3$,  one obtains the following strong prediction for $N_k^{\rm inf}$.
\beq
N_k^{\rm inf} =  61.55 - \ln {\l(\f{V_e^{\f{1}{4}}}{H_k^{\rm inf}}\r)} \, .
\label{eq:CMB_reheat_Nk_SR}
\eeq
Since $H_e^{\rm inf} = \sqrt{3m_p^2 \, V_e}$, Eq.~(\ref{eq:CMB_reheat_Nk_SR}) becomes
\beq
N_k^{\rm inf} =  61.3 - \f{1}{2} \, \ln {\l[\f{H_e \, m_p}{\l(H_k^{\rm inf}\r)^2}\r]} \, .
\label{eq:CMB_reheat_Nk_H}
\eeq
 For the standard quasi-dS inflation, since $H_e \simeq H_k^{\rm inf}$, we obtain the final expression
\beq
\boxed{ N_k^{\rm inf} =  61.3 - \f{1}{2} \,  \ln {\l(\f{m_p}{H_k^{\rm inf}}\r)} } \, ,
\label{eq:CMB_reheat_Nk_Hk}
\eeq
which results in
 \beq
N_k^{\rm inf} = \begin{cases}  55.5 \, ; &  H_k^{\rm inf} = 2.4\times 10^{13} \, {\rm GeV} \, , \\ 52.1 \, ; &  H_k^{\rm inf} = 2.4 \times 10^{10} \, {\rm GeV} \, , \\ 48.6 \, ; &  H_k^{\rm inf} = 2.4 \times 10^{7} \, {\rm GeV} \, . \end{cases} 
\label{eq:CMB_reheat_Nk_Hk_max_min}
 \eeq
If reheating lasts  longer with an EoS $\wre \neq 1/3$, then $N_k^{\rm inf}$  would span a larger range.

\section{Inflaton-clock dynamics:~Hamilton-Jacobi formalism}
\label{app:Hamilton_Jacobi}
In the standard scenario of single field slow-roll inflation, the inflaton field $\phi$ evolves monotonically during inflation\footnote{The same is not true after inflation, because the inflaton oscillates around the minimum of its potential}. Hence, one can use $\phi$ as a clock to describe the dynamics of the background spacetime. This $\phi$-clock evolution is particularly useful in obtaining the functional form of the potential $V(\phi)$, given an inflationary evolution $H(t)$, or vice versa. 

In order to achieve this, let us convert the time variable in the Friedmann equations from cosmic time $t$ to the inflaton clock time $\phi$. From Eq.~(\ref{eq:friedmann2}), we get 
$$ \dot{H} \equiv \f{{\rm d} H}{{\rm d}\phi} \, \dot{\phi} = -\frac{1}{2m_p^2}\, \dot{\phi}^2 \, ,$$
leading to 
\beq
\l( \f{{\rm d} H}{{\rm d} \phi} \r)^2 = \f{\dot{\phi}^2}{4\, m_p^4} \, .
\label{eq:H_phi_sqr}
\eeq
Replacing the value of $\dot{\phi}^2$ from Eq.~(\ref{eq:friedmann1}), we obtain 
\beq
\boxed{ ~\l( \f{{\rm d} H}{{\rm d} \phi} \r)^2 ~ = ~ \f{3}{2} \, \l[ \f{H(\phi)}{m_p}\r]^2 -  \, \f{1}{2} \, \f{V(\phi)}{m_p^4}~ } \, ,
\label{eq:H_phi_HJ}
\eeq
 known as the {\em Hamilton-Jacobi} equation which is a first-order non-linear ordinary differential equation that  describes the evolution of $H(\phi)$ during inflation for a given potential $V(\phi)$. The corresponding expressions for the slow-roll parameters $\epsilon_H, \, \eta_H$ becomes
\ber
\epsilon_H &=& 2\, m_p^2 \, \l(\f{H_{,\phi}}{H}\r)^2  \, ,
\label{eq:epsilon_H_HJ}\\
\eta_H &=&  2\, m_p^2 \, \l(\f{H_{,\phi\phi}}{H}\r) \, .
\label{eq:eta_H_HJ}
\eer
The Hamilton-Jacobi formalism has a number of important applications in inflationary cosmology. In particular, it is a convenient way to demonstrate that the  slow-roll trajectory given in Eq.~(\ref{eq:friedmann_SR2}) is a local  attractor, see Refs.~\cite{Liddle:1994dx,Baumann_TASI}. Another important application of Eq.~(\ref{eq:H_phi_HJ}) is to determine the functional form of $V(\phi)$ which leads to a particular expansion history $H(t)$ during inflation~\cite{Lidsey:1995np}, as mentioned before. Let us proceed to demonstrate this for {\em power-law inflation}.

\medskip

In Sec.~\ref{inf_dyn_background} we discussed  inflation in the quasi-de Sitter paradigm where the expansion of space was nearly exponential, $a(t) \propto a^{Ht}$. However, it is also possible to realise inflationary expansion which is not exponential. A key example of such a non-exponential inflation is the so called power-law inflation where the scale factor of the universe evolves as a monomial or power-law function of time, namely
\beq
a(t) \propto t^p \, ; \quad {\rm with} \quad p>1 ~~ \Rightarrow ~~ H(t) = \f{p}{t} \, .
\label{eq:expansion_power-law}
\eeq
Since $p>1$, we have $\ddot{a}>0$, leading to accelerated expansion or inflation. We are then interested in determining the functional form of $V(\phi)$ which facilitates such a power-law expansion of space.  To determine $V(\phi)$, we first replace the time dependence of $H$ with the $\phi$ dependence {\em via} the following steps. Using Eq.~(\ref{eq:friedmann2}) and (\ref{eq:expansion_power-law}), we get
$$\dot{\phi}^2 \equiv - 2 \, m_p^2 \, \dot{H} = \f{2p \, m_p^2 }{t^2} \, ,$$
which leads to
$$ \f{\phi(t)}{m_p} = \sqrt{2p} \, \ln\l(\f{t}{t_1}\r) \, ,$$
 where $t_1$ is an integration constant. By inverting the above expression, we obtain cosmic time in terms of $\phi$ to be
 \beq
t = t_1 \, e^{\f{1}{\sqrt{2p}} \, \f{\phi}{m_p}} \, .
\label{eq:t_phi_power-law}
 \eeq
Incorporating this into the expression for $H(t)$ in Eq.~(\ref{eq:expansion_power-law}), we obtain the final expression for the Hubble parameter and its derivative in terms of $\phi$-clock to be
 \ber
H(\phi) &=&   H_1 \, e^{-\f{1}{\sqrt{2p}} \, \f{\phi}{m_p}} \, , \label{eq:H_phi_power-law} \\
H_{,\phi} (\phi) &=& -\f{1}{ \sqrt{2p}} \, \l( \f{H_1}{m_p} \r) \,  e^{-\f{1}{\sqrt{2p}} \, \f{\phi}{m_p}} \, , \label{eq:H_der_phi_power-law}
 \eer
where $H_1 = p/t_1$. The expression for the scalar field potential $V(\phi)$ from the Hamilton-Jacobi Eq.~(\ref{eq:H_phi_HJ}) is given by 
\beq
\boxed{~ \f{V(\phi)}{m_p^4}    ~ = ~ 3 \, \l[ \f{H(\phi)}{m_p}\r]^2 - 2 \, \l( \f{{\rm d} H}{{\rm d} \phi} \r)^2 ~ } \, ,
\label{eq:V_phi_HJ}
\eeq
which, using Eq.~(\ref{eq:H_der_phi_power-law}), becomes 
\beq
\boxed{ V(\phi) =  V_0 \, e^{-\lambda \, \f{\phi}{m_p}} }\, ,
\label{eq:V_phi_HJ_power-law}
\eeq
with
\beq
V_0 = \l( 3 - \f{1}{p}\r) \l(\f{H_1}{m_p} \r)^2 \, m_p^4 \, , \quad  \boxed{ \lambda \, = \, \sqrt{\f{2}{p}} } \, .
\label{eq:power-law_inf_parameters}
\eeq
 Power-law inflation was amongst the earliest studied inflationary models, see Ref.~\cite{Lucchin:1984yf}. Unfortunately, their slow-roll predictions for $\lbrace n_{_S}, \, r \rbrace$ do not satisfy the CMB constraints, as can be seen from the purple colour curve in Fig.~\ref{fig:inf_ns_r_latest}.
 
 \bigskip

 From the above analysis,  we conclude that an exponential potential leads to the power-law expansion of space. Since we did not make use of the fact that $p>1$, the conclusion can be extended to decelerated expansion of space for $p<1$. In fact exponential potentials play an important roll in the post-inflationary scalar field dynamics. They appear in String Theory \cite{Apers:2024ffe} and have interesting  tracker properties \cite{Copeland:1997et} in the presence of a dominant background matter-energy component with EoS $w_B$, provided  they are steep enough, \textit{i.e.} $\lambda^2 > 3(1+w_B)$. Consequently, they have been used extensively  both in the context of dark energy \cite{Barreiro:1999zs,Bag:2017vjp} and dark matter \cite{Mishra:2017ehw} with  appropriate modifications.
 
\section{Mukhanov-Sasaki equation for scalar fluctuations}
\label{app:MS}

 In this Appendix we outline the derivation of the Mukhanov-Sasaki equation, first  in the comoving gauge and then in the  spatially flat gauge. Starting from the action of a canonical scalar field minimally coupled to gravity 
$$ S[g_{\mu\nu},\varphi] = \int \,  {\rm d}^4x \,  \sqrt{-g} \,  \l( \, \f{m_p^2}{2} \, R - \f{1}{2}  \, \partial_{\mu}\varphi \,  \partial_{\nu}\varphi  \, g^{\mu\nu}-V(\varphi)\r)~, $$
we consider linear field fluctuations $\varphi(t,\vec{x}) = \phi(t) + \delta\varphi(t,\vec{x})$ and the linearly perturbed  ADM metric of the form \cite{Arnowitt:1962hi,Maldacena:2002vr}
$$ {\rm d}s^2 = -  \alpha^2 \, {\rm d}t^2 + \gamma_{ij} \l( {\rm d}x^i + \beta^i \, {\rm d}t \r) \l( {\rm d}x^j + \beta^j \, {\rm d}t \r) \, ,$$
where $\alpha = 1 + \delta\alpha$ and $\beta^i$ are the lapse and shift functions, which appear as Lagrange multipliers in the perturbed action.  However, $\gamma_{ij}$ are the dynamical metric perturbations in the ADM formalism. 

\subsection{Comoving gauge}
\label{app:MS_comov_gauge}
In the comoving gauge in which  inflaton fluctuations vanish, we have
$$
\delta\varphi(t,\vec{x}) = 0 \, ; \quad \gamma_{ij}(t,\vec{x}) = a^2 \bigl[ \bigl(  1 + 2 \, \zeta(t,\vec{x}) \bigr) \, \delta_{ij} \, + \, h_{ij}(t,\vec{x}) \bigr] \, .
$$ 
By varying the perturbed action \textit{w.r.t} the lapse and shift, we obtain the GR momentum and Hamiltonian constraints (see Refs.~\cite{Maldacena:2002vr,Baumann:2018muz}) to be
\beq
\delta\alpha = \f{\dot{\zeta}}{H} \, ; ~~~~ \partial_i \beta^i = - \f{\partial^2\zeta}{H} + \f{1}{2} \, a^2 \l( \f{\dot{\phi}^2}{H^2} \r) \, \dot{\zeta} \, .
\label{eq:GR_constraints_comoving}
\eeq
Incorporating the above expressions into the action, expanding around the background, the quadratic action for $\zeta$  becomes \cite{Maldacena:2002vr,Baumann:2018muz}
\beq
S^{(2)}[\zeta] = \int {\rm d}t \, {\rm d}^3\vec{x} ~~  a^3 \, \epsilon_H \, \l[  \,   \dot{\zeta}^2 - \l(\f{\partial_i  \zeta}{a}\r)^2 \,  \r] \, ,
\label{eq:Quad_action_comov}
\eeq
which, in terms of conformal time $\tau$,  takes the form
$$
 S^{(2)}[\zeta] = \f{1}{2} \int {\rm d}\tau \, {\rm d}^3\vec{x} ~  2 \, a^2 \, m_p^2 \, \epsilon_H  \,  \l[ (\zeta')^2 - (\partial_i \zeta)^2 \r]  \, .
$$
With the change of variable $v = am_p\sqrt{2\epsilon_H} \, \zeta$, the Fourier modes $v_k$ satisfy the Mukhanov-Sasaki equation 
$$v_k'' + \l[ k^2 + {\cal M}_{\rm eff}^2(\tau) \r] \, v_k = 0 ~,$$
with 
\begin{equation}
{\cal M}_{\rm eff}^2(\tau) = - \f{\l( a \, m_p \, \sqrt{2\epsilon_H}\r)''}{ a \, m_p \, \sqrt{2\epsilon_H}}  = -(aH)^2 \l[ 2 + 2 \epsilon_H - 3 \eta_H + 2 \epsilon_H^2 + \eta_H^2 - 3 \epsilon_H \eta_H  - \f{1}{aH} \, \eta'_H \r] \, .
\end{equation}

\subsection{Spatially flat gauge}
\label{app:MS_flat_gauge}
 In the spatially flat gauge 
 $$
 \gamma_{ij}(t,\vec{x}) = a^2(t) \bigl[ \delta_{ij} \, +  \, h_{ij} (t,\vec{x}) \bigr] \, .
 $$
 Imposing the GR momentum and Hamiltonian constraints, one obtains (see Ref.~\cite{Maldacena:2002vr})
$$ \delta\alpha = \sqrt{\f{\epsilon_H}{2}} \, \f{\delta\varphi}{m_p} \, ; ~~~~ \partial_i \beta^i = - \epsilon_H \, \f{{\rm d}}{{\rm d}t} \l( \f{1}{\sqrt{2\epsilon_H}} \, \f{\delta\varphi}{m_p}\r)~.$$
Incorporating the above expressions into the action, expanding around the background, the quadratic action for $\delta\varphi$ fluctuations becomes \cite{Maldacena:2002vr,Baumann:2018muz}
\beq
S^{(2)}[\delta\varphi] = \f{1}{2} \, \int \,  {\rm d}t {\rm d}^3{\vec x} ~ a^3  \l[ \delta \dot \varphi^2  - \f{(\partial_i\delta\varphi)^2}{a^2} - \l( \f{{\rm d}^2 V(\phi)}{{\rm d} \phi^2} + 2\, \epsilon_H \, H^2 \, \l( 2 \, \eta_H - \epsilon_H - 3 \r) \r)  \delta\varphi^2 \r] \, .
\label{eq:Quad_action_spat_flat}
\eeq
The Euler-Lagrange equation for $\delta\varphi$ is given by
\beq
\delta \ddot \varphi + 3 \, H \, \delta \dot \varphi - \f{\nabla^2}{a^2} \, \delta\varphi - a^2H^2 \l( 2 - \epsilon_H - \f{1}{H^2} \, \f{{\rm d}^2 V(\phi)}{{\rm d} \phi^2}  - \f{2 \dot{\phi}}{m_p^2 H^3} \, \f{{\rm d} V(\phi)}{{\rm d} \phi}  - \f{\dot{\phi}^2}{m_p^4 H^4}  \, V \r)  \, \delta\varphi = 0 ~.
\label{eq:El_deltaphi_spat_flat}
\eeq
With the change of variable $v = a \, \delta\varphi$, the Fourier modes $v_k$ satisfy the Mukhanov-Sasaki equation 
$$v_k'' + \l[ k^2 + {\cal M}_{\rm eff}^2(\tau) \r] \, v_k = 0 ~,$$
with
\begin{align}
{\cal M}_{\rm eff}^2(\tau) &= -a^2H^2 \l( 2 - \epsilon_H - \f{1}{H^2} \, \f{{\rm d}^2 V(\phi)}{{\rm d} \phi^2}  - \f{2 \dot{\phi}}{m_p^2 H^3} \, \f{{\rm d} V(\phi)}{{\rm d} \phi}  - \f{\dot{\phi}^2}{m_p^4 H^4}  \, V \r) \, , \nonumber \\
&= -(aH)^2 \l[ 2 + 2 \epsilon_H - 3 \eta_H + 2 \epsilon_H^2 + \eta_H^2 - 3 \epsilon_H \eta_H  - \f{1}{aH} \, \eta'_H \r] \, .
\end{align}
which is the same as that  in the comoving gauge.

\section{Analytical solution of the Mukhanov-Sasaki equation}
\label{app:MS_analyt_sol}
 For  slow-roll potentials,  the Mukhanov-Sasaki (MS) Eq.~(\ref{eq:MS_modes}) can be written as a  Bessel equation assuming  
$$- (aH)^{-2} \, {\cal M}_{\rm eff}^2= \nu^2 - 1/4$$
to be a constant, which can be solved analytically (see Refs.~\cite{Birrell:1982ix}). In this Appendix we present this solution in terms of both Hankel functions (in App.~\ref{app:MS_analyt_sol_hankel}) and Bessel functions (in App.~\ref{app:MS_analyt_sol_bessel}).

For the analytical treatment, we write the Mukhanov-Sasaki (MS) Eq.~(\ref{eq:MS_modes})  as 
\beq
 \f{{\rm d}^2 v_k}{{\rm d} \tau^2} + \l[ \, k^2 - \f{\nu^2 - \f{1}{4}}{\tau^2}  \, \r] v_k = 0 ~.
\label{eq:MS_modes_nu_tau}
\eeq
It is convenient to define  a new time variable
\beq
T = -k\tau = \f{k}{a \, H} \, ,
\label{eq:time_Hankel}
\eeq
in terms of which  the MS Eq.~(\ref{eq:MS_modes_nu_tau}) takes the form
\beq
\boxed{ \, \f{{\rm d}^2 v_k}{{\rm d} T^2} + \l[ \, 1 - \f{\nu^2 - \f{1}{4}}{T^2}  \, \r] v_k = 0 \, } \, .
\label{eq:MS_modes_nu_T}
\eeq
All modes undergo Hubble-exit at $T  = 1$, with sub (super)-Hubble scales corresponding to $T \gg (\ll) 1$. 
Using the variable redefinition $F =   v_k/\sqrt{T}$, this equation can be transformed into the more familiar  Bessel equation:
\beq
\f{{\rm d}^2 F}{{\rm d} T^2} +   \f{1}{T} \, \f{{\rm d} F}{{\rm d} T}  + \l[ \, 1 - \f{\nu^2}{T^2}  \, \r] F = 0 ~.
\label{eq:MS_modes_nu_Bessel}
\eeq
The general solution to Eq.~(\ref{eq:MS_modes_nu_Bessel}) (when $\nu$ is not an integer) can be written either as a linear combination of Hankel functions of the first and second kind $\lbrace  H_\nu^{(1)}(T), \,  H_\nu^{(2)}(T)   \rbrace$ or as a linear combination of positive and negative order ($\pm \nu$) Bessel functions of the first kind  $ \lbrace J_{-\nu}(T), \, J_{\nu}(T) \rbrace$. The functions are related by~\cite{NIST:DLMF}
\beq
H_\nu^{(1,2)}(T)  = \f{ \pm  J_{-\nu}(T) \mp e^{\mp i\pi\nu} \, J_{\nu}(T) }{ i\sin{(\pi\nu)}  }    \,. \label{eq:Hank_Bess12} 
\eeq
\subsection{In terms of Hankel functions}
\label{app:MS_analyt_sol_hankel}
The general solution to the Bessel Eq.~(\ref{eq:MS_modes_nu_Bessel}) in terms of the  Hankel functions is given by
\beq
F(T) = C_1 \, H_\nu^{(1)}(T) \, +   \, C_2 \, H_\nu^{(2)}(T)~,
\label{eq:Bessel_sol_Hankel}
\eeq
where the coefficients $C_1$ and $C_2$ are fixed by initial/boundary conditions. Hence the solution to the MS  equation  can be written as 
\beq
 v_k (T) = \sqrt{T} \, \l[  \, C_1 \, H_\nu^{(1)}(T) \, +   \, C_2 \, H_\nu^{(2)}(T) \, \r] \,.
\label{eq:MS_sol_Hankel}
\eeq
In the sub-Hubble limit, $T \gg 1$, the Hankel functions take the form~\cite{NIST:DLMF}
\begin{align}
 H_\nu^{(1)}(T) \bigg \vert_{T \to \infty} &\simeq \sqrt{\f{2}{\pi}} \, \f{1}{\sqrt{T}}  \, e^{ i T}  \, e^{-i \l(  \nu + \f{1}{2}  \r) \f{\pi}{2}} \, , \label{eq:hankel1_sub} \\
H_\nu^{(2)}(T) \bigg \vert_{T \to \infty} &\simeq \sqrt{\f{2}{\pi}} \, \f{1}{\sqrt{T}}  \, e^{ - i T} \, e^{i \l(  \nu + \f{1}{2}  \r) \f{\pi}{2}} \, , \label{eq:hankel2_sub}
\end{align}
while in the super-Hubble limit, $T \ll 1$, the Hankel functions take the form~\cite{NIST:DLMF} 
\begin{align}
 H_\nu^{(1)}(T) \bigg \vert_{T \to 0} &\simeq \sqrt{\f{2}{\pi}} \, e^{- i  \f{\pi}{2}} \,  2^{\nu - \f{3}{2}}   \, \f{\Gamma(\nu)}{\Gamma(\f{3}{2})} \,  T^{-\nu} \, , \label{eq:hankel1_super} \\
H_\nu^{(2)}(T) \bigg \vert_{T \to 0} &\simeq ~ - \sqrt{\f{2}{\pi}} \, e^{- i  \f{\pi}{2}} \,  2^{\nu - \f{3}{2}}   \, \f{\Gamma(\nu)}{\Gamma(\f{3}{2})} \,  T^{-\nu} \, . \label{eq:hankel2_super}
\end{align}
The Bunch-Davies conditions, for  the mode functions take the form
$$v_k(T) \bigg \vert_{T \to \infty}  \to  \f{1}{\sqrt{2k}} \, e^{ i T} = \sqrt{T}  \, C_1  H_\nu^{(1)}(T) \bigg \vert_{T \to \infty} ~,$$
which yields 
$$C_1 = \f{1}{\sqrt{2k}}  \, \sqrt{\f{\pi}{2}} \, e^{i \l(  \nu + \f{1}{2}  \r) \f{\pi}{2}}  ~,~~~{\rm and}~~~  C_2 =0~,$$
and hence the final expression for the mode functions becomes 
\beq
\boxed{ \, v_k (T) = e^{i \l(  \nu + \f{1}{2}  \r) \f{\pi}{2}} \, \sqrt{\f{\pi}{2}}  \, \f{1}{\sqrt{2k}} \, \sqrt{T} \,  H_\nu^{(1)}(T) \, }  \, .
\label{eq_MS_mode_fun_Hankel}
\eeq
Using Eq.~(\ref{eq:MS_Ps}), the power spectrum for comoving curvature perturbations can be found to be 
\beq
{\cal P}_\zeta(k) \equiv \f{k^3}{2\pi^2} \, \f{|v_k|^2}{a^2m_p^2 \, 2 \epsilon_H} = \f{1}{8\pi^2 \, \epsilon_H} \l(\f{H}{m_p}\r)^2 \times \f{\pi}{2} \, T^3 \, | H_\nu^{(1)}(T) |^2 \, . 
\label{eq:Ps_Hankel_1}
\eeq
Using the super-Hubble limit of the Hankel function of the first kind from Eq.~(\ref{eq:hankel1_super}) and replacing $T=k/(aH)$ from Eq.~(\ref{eq:time_Hankel}), we obtain the power spectrum on super-Hubble scales to be 
\beq
\boxed{~ {\cal P}_\zeta(k) \big\vert_{k \ll aH}  =2^{2\l(\nu-3/2\r)} \, \l[\f{\Gamma(\nu)}{\Gamma \l( 3/2 \r)}\r]^2 \times  \f{1}{8\pi^2 \, \epsilon_H} \l(\f{H}{m_p}\r)^2   \l( \f{k}{aH} \r)^{-2\l(\nu-3/2\r)}~ } \, , 
\label{eq:Ps_Hankel_2}
\eeq
which gets reduced to Eq.~(\ref{eq:P_S_SR}) in the quasi-dS limit $\nu \to 3/2$.

\subsection{In terms of Bessel functions}
\label{app:MS_analyt_sol_bessel}

The general solution to the Bessel equation, Eq.~(\ref{eq:MS_modes_nu_Bessel}), in terms  of the  Bessel functions  of the first kind of order $\pm \nu$ is given by

\beq
F(T) = \sqrt{\f{\pi}{2}} \, \f{1}{\sin{(\pi\nu)}}  \, \l[ \, C_{+} \, J_{-\nu}(T) \, +   \, C_{-} \, J_{\nu}(T) \, \r] ~,
\label{eq:Bessel_sol_Bessel_1st}
\eeq
where  the coefficients $C_{+}$ and $C_{-}$ are again  to be fixed by initial/boundary conditions. Hence the solution to MS Eq.~(\ref{eq:MS_modes}) can be written as

\beq
 v_k (T) =  \sqrt{\f{\pi}{2}} \, \f{1}{\sin{(\pi\nu)}}  \, \sqrt{T}  \, \l[  \,  C_{+} \, J_{-\nu}(T) \, +   \, C_{-} \, J_{\nu}(T)  \, \r] \, .
\label{eq:MS_sol_Bessel_1st}
\eeq

Imposing Bunch-Davies initial conditions we get
$$C_{+} = -i \, \f{1}{\sqrt{2k}} \, e^{i \l(  \nu + \f{1}{2}  \r) \f{\pi}{2}}  ~,~~~{\rm and}~~~  C_{-} = i \, \f{1}{\sqrt{2k}} \,e^{i \l(  \nu + \f{1}{2}  \r) \f{\pi}{2}} \, e^{-i \pi \nu }  ~,$$
and hence the final expression for the mode functions becomes 
\beq
 v_k (T) = - i  \, \sqrt{\f{\pi}{2}} \,   \f{e^{i \l(  \nu + \f{1}{2}  \r) \f{\pi}{2}} }{ \sin{(\pi\nu)}} \, \f{1}{\sqrt{2k}} \, \sqrt{T} \l[ \, J_{-\nu}(T) \, -  e^{-i \pi \nu } \, J_{\nu}(T)  \,   \r]  ~.
\label{eq_MS_mode_fun_Bessel}
\eeq
 With the help of Eq.~(\ref{eq:Hank_Bess12}), we see by equating Eqs.~(\ref{eq:MS_sol_Hankel})~and~(\ref{eq:MS_sol_Bessel_1st}), that the relation between the Hankel coefficients $ \lbrace C_1, \, C_2 \rbrace $ and Bessel coefficients $ \lbrace  C_{+}, \,  C_{-}  \rbrace $ is given by
\beq
C_1 = i \sqrt{\f{\pi}{2}}  \, \l[ \,   \f{C_{+} + e^{-i2\pi\nu} C_{-}}{1-e^{-i2\pi\nu} }    \, \r] \, , ~~~
C_2 = i \sqrt{\f{\pi}{2}} \,  \l[ \,   \f{C_{+} + e^{i\pi\nu} C_{-}}{1-e^{-i2\pi\nu} }    \, \r] e^{-i2\pi\nu}  \, . 
\label{eq:ceff_Hank_Bess12}
\eeq
In Eq.~(\ref{eq:Ps_Hankel_2}), we expressed the power spectrum  in terms of Hankel functions.
However, the same can alternatively be  expressed in terms of the Bessel functions by using Eqs.~(\ref{eq:Hank_Bess12}) and (\ref{eq:ceff_Hank_Bess12}).

\section{Inflationary fluctuations as a massive scalar  in pure dS}
\label{app:massive_dS}
In this appendix, we will show that the inflationary fluctuations can be thought of as the fluctuations of a massive field in pure de Sitter spacetime. The action of a massive scalar field ({\em Klein-Gordon}  field) $\chi$ in dS is given by
\beq
 S_{\rm dS}[\chi] = \int {\rm d}^4x \, \sqrt{-g}  \,  \l[ -\f{1}{2} \, \partial_{\mu}\chi \; \partial_{\nu}\chi \, g^{\mu\nu} - \f{1}{2}  \, M^2 \, \chi^2 \r]  \, .
\label{eq:Action_massive_dS}
\eeq
Using the expression for the FLRW metric in terms of the conformal time, as given in Eq.~(\ref{eq:flat_FLRW_tau}),  for which $\sqrt{-g} = a^4$, the above action can be written as 
\beq
 S_{\rm dS}[\chi] = \f{1}{2} \int {\rm d}\tau \, {\rm d}^3\vec{x} ~  a^2   \l[(\chi')^2 - (\partial_i \chi)^2  -  a^2  M^2  \, \chi^2\r] \, .
\label{eq:Action_massive_dS_tau}
\eeq
In order to convert the above expression into the action of a canonical scalar field in Minkowski spacetime, we define the new field variable
\beq
u(\tau, \, \vec{x}) = a(\tau)\, \chi(\tau, \, \vec{x})  \quad \Rightarrow \quad \chi'(\tau,\, \vec{x}) =  \f{u'}{a} - \f{a'}{a}\, \f{u}{a} \, ,
\label{eq:MS_massive _dS_variable}
\eeq
which reduces the action~(\ref{eq:Action_massive_dS_tau}) to
\beq
 S_{\rm dS}[u] = \f{1}{2} \int {\rm d}\tau \, {\rm d}^3\vec{x} ~  \l[(u')^2 - (\partial_i u)^2  -  a^2 M^2 \,  u^2 - 2 \, \f{a'}{a} \, uu' + \l(\f{a'}{a}\r)^2 \, u^2\r] \, .
\label{eq:Action_massive_dS_MS1}
\eeq
Incorporating 
$$2 \, \f{a'}{a} \, uu' = \l[\l(\f{a'}{a}\r) \, u^2 \r]' - \f{a''}{a}\, u^2 $$
into Eq.~(\ref{eq:Action_massive_dS_MS1}),  and dropping the boundary (total time derivative) term, we get the final expression for the scalar field action to be
\beq
\boxed{ ~  S_{\rm dS}[u] = \f{1}{2} \int {\rm d}\tau \, {\rm d}^3\vec{x} ~  \l[(u')^2 - (\partial_i u)^2  -  \l( a^2 M^2 \,   - \,  \f{a''}{a} \r) u^2 \r] ~ } \, ,
\label{eq:Action_massive_dS_MS2}
\eeq
which represents the action of a scalar field $u(\tau,\vec{x})$ in the Minkowski spacetime 
with a time-dependent effective mass given by
\beq
\boxed{ \, {\cal M}_{\rm mdS}^2(\tau)  \equiv M^2 \, a^2(\tau) \,   -  \, \f{a''}{a} = \f{\l(M/H\r)^2-2}{\tau^2} \, } \, ,
\label{eq:dS_massive _eff}
\eeq
where we used 
$$a = -\f{1}{H\tau} \, , \quad \text{and} \quad \f{a''}{a}=\f{2}{\tau^2} \, ,$$
for the dS spacetime. If $\chi$ was a massless field with $M=0$, then the effective mass term~(\ref{eq:dS_massive _eff}) gets reduced to
\beq
\boxed{ \, {\cal M}_{\rm mdS}^2(\tau) = - \f{a''}{a} = -\f{2}{\tau^2} \, } \, ,
\label{eq:dS_massive _eff_massless}
\eeq
 which demonstrates that {\em a massless field in de Sitter spacetime is analogous to a  massive field, with time-dependent tachyonic mass, in the Minkowski spacetime}. The first equality in Eq.~(\ref{eq:dS_massive _eff_massless}) is true in general for any FLRW spacetime, see Ref.~\cite{Birrell:1982ix,Baumann:2018muz}. In fact, the effect of expansion of space has appeared as a time dependent mass term~(\ref{eq:dS_massive _eff}) or (\ref{eq:dS_massive _eff_massless}) in the effective Minkowski action~(\ref{eq:Action_massive_dS_MS2}). 
 
 Nevertheless, going back to the massive field case with action~(\ref{eq:Action_massive_dS_MS2}), the corresponding equation of dynamics of the Fourier mode functions of the canonical  scalar $u(\tau,\vec{x})$ becomes
 \beq
 \f{{\rm d}^2 u_k}{{\rm d} \tau^2} + \l[ \, k^2 - \f{\nu_{\rm mdS}^2 - \f{1}{4}}{\tau^2}  \, \r] u_k = 0 \, ,
\label{eq:MS_modes_massive_dS}
 \eeq
where 
\beq
\boxed{ \, \nu_{\rm mdS}^2 = \f{9}{4} - \f{M^2}{H^2} \, }
\label{eq:mdS_nu}
\eeq
is a constant. Following the discussion in App.~(\ref{app:MS_analyt_sol_hankel}), the Bunch-Davies initial condition imposed solution to Eq.~(\ref{eq:MS_modes_massive_dS}) is given by
\beq
  v_k (\tau) = e^{i \l(  \nu_{\rm mdS} + \f{1}{2}  \r) \f{\pi}{2}} \, \sqrt{\f{\pi}{2}}  \, \f{1}{\sqrt{2k}} \, \sqrt{-k\tau} \,  H_{\nu \, {\rm mdS}}^{(1)}(-k\tau)  \, .
\label{eq_MS_mdS_fun_Hankel} 
\eeq
The power spectrum of the massive $\chi$ field becomes 
\beq
{\cal P}_\chi(k) \equiv \f{k^3}{2\pi^2} \, \f{|u_k|^2}{a^2} = \l(\f{H}{2\pi}\r)^2 \times \f{\pi}{2} \, (-k\tau)^3 \, | H_{\nu \, {\rm mdS}}^{(1)}(-k\tau) |^2 \, . 
\label{eq:Ps_mdS_Hankel_1}
\eeq
Using the super-Hubble limit of the Hankel function of the first kind from Eq.~(\ref{eq:hankel1_super}), we obtain the power spectrum of $\chi$ on super-Hubble scales to be 
\beq
\boxed{~ {\cal P}_\chi(k) \big\vert_{-k\tau \ll 1}  =2^{2\l(\nu_{\rm mdS}-3/2\r)} \, \l[\f{\Gamma(\nu_{\rm mdS})}{\Gamma \l( 3/2 \r)}\r]^2 \times   \l(\f{H}{2\pi}\r)^2   \l( \f{k}{aH} \r)^{\l(3- 2\nu_{\rm mdS}\r)}~ } \, , 
\label{eq:Ps_mdS_Hankel_2}
\eeq
which, in the massless limit $\nu_{\rm mdS} \to 3/2$, gets reduced to the famous result
\beq
\boxed{~ {\cal P}_\chi(k) \big\vert_{-k\tau \ll 1}  \, =  \, \l(\f{H}{2\pi}\r)^2 ~ } \, . 
\label{eq:Ps_massless_dS_Hankel_2}
\eeq
The above expression shows that the power spectrum of a massless scalar field in the dS spacetime is {\em scale-invariant}, which is a reflection of the conformal symmetry  of dS spacetime.  

\medskip

Let us turn our attention back to the massive case. Assuming $M < \f{3}{2} \, H$ (known as the {\em principal series}),  for which $\nu_{\rm mdS}^2 > 0$ and hence $\nu_{\rm mdS}$ is real,  Eq.~(\ref{eq:Ps_mdS_Hankel_2}) demonstrates that the power spectrum of a massive field in the dS spacetime has a  blue tilt, given by
\beq
n_{_\chi} =  3 - 2\, \nu_{\rm mdS} \,  .
\label{eq:n_S_mdS}
\eeq
If the mass of the scalar field is much smaller compared to the Hubble scale, then the scalar spectral index is given by
$$ n_{_\chi} \big\vert_{m \ll H} = \f{2}{3} \,  \f{M^2}{H^2} \, .$$
However, if $M > \f{3}{2} \, H$ (known as the {\em complementary series}), then $\nu_{\rm mdS}^2 < 0$ and hence $\nu_{\rm mdS}$ becomes imaginary. An imaginary $\nu_{\rm mdS}$ results in an exponentially suppressed power spectrum due to the presence of the term $e^{i \l(  \nu_{\rm mdS} + \f{1}{2}  \r) \f{\pi}{2}}$ in Eq.~(\ref{eq_MS_mdS_fun_Hankel}). Therefore, the dominant contributions to the primordial power spectra are expected from light degrees of freedom. This concludes our discussion on  massive scalar field in pure dS spacetime. 

 \bigskip

Coming back to inflationary scalar fluctuations, the Mukhanov-Sasaki Eq.~(\ref{eq:MS_modes}) for the inflationary scalar fluctuations  can be written as 
\beq
 \f{{\rm d}^2 v_k}{{\rm d} \tau^2} + \l[ \, k^2 - \f{\nu_{\rm inf}^2 - \f{1}{4}}{\tau^2}  \, \r] v_k = 0 ~,
\label{eq:MS_modes_nu_tau_inf}
\eeq
where (the Hankel exponent) $\nu_{\rm inf}$  is given by
\beq
\nu_{\rm inf}^2 \equiv - {\cal M}_{\rm eff}^2 \, \tau^2 + \f{1}{4} =  \f{9}{4} + 2 \,  \epsilon_H - 3  \, \eta_H + 2 \,  \epsilon_H^2 + \eta_H^2 - 3 \,  \epsilon_H  \, \eta_H  - \f{1}{aH} \, \eta'_H \, .
\label{eq:inf_nu_full}
\eeq
Since the recent CMB observations favour asymptotically flat potentials with $\epsilon_H \ll |\eta_H|$, we can  impose the  qdS approximation and ignore $\epsilon_H$. Furthermore, assuming $\eta_H$ to be roughly a constant, Eq.~(\ref{eq:inf_nu_full}) takes the form
\beq
\nu_{\rm inf}^2 \simeq \f{9}{4} - \eta_H \l( 3- \eta_H \r) = \l( \f{3}{2} - \eta_H \r)^2 \, .
\label{eq:inf_nu_dS}
\eeq
Comparing Eqs.~(\ref{eq:inf_nu_dS}) and  (\ref{eq:mdS_nu}), we can define the mass of inflationary scalar fluctuations to be 
\beq
{\cal M}_{\rm inf}^2 = \eta_H \l( 3 - \eta_H \r) H^2  \, .
\label{eq:mdS_minf}
\eeq
Note that ${\cal M}_{\rm inf}^2 > 0$ for $\eta_H \in \l( 0,3 \r)$ while the fluctuations are tachyonic for $\eta_H < 0$ and $\eta_H >3 $.

\newpage

\printbibliography

@article{Planck:2018vyg,
    author = "Aghanim, N. and others",
    collaboration = "Planck",
    title = "{Planck 2018 results. VI. Cosmological parameters}",
    eprint = "1807.06209",
    archivePrefix = "arXiv",
    primaryClass = "astro-ph.CO",
    doi = "10.1051/0004-6361/201833910",
    journal = "Astron. Astrophys.",
    volume = "641",
    pages = "A6",
    year = "2020",
    note = "[Erratum: Astron.Astrophys. 652, C4 (2021)]"
}

@article{Starobinsky:1980te,
    author = "Starobinsky, Alexei A.",
    editor = "Khalatnikov, I. M. and Mineev, V. P.",
    title = "{A New Type of Isotropic Cosmological Models Without Singularity}",
    doi = "10.1016/0370-2693(80)90670-X",
    journal = "Phys. Lett. B",
    volume = "91",
    pages = "99--102",
    year = "1980"
}

@article{Guth:1980zm,
    author = "Guth, Alan H.",
    editor = "Fang, Li-Zhi and Ruffini, R.",
    title = "{The Inflationary Universe: A Possible Solution to the Horizon and Flatness Problems}",
    reportNumber = "SLAC-PUB-2576",
    doi = "10.1103/PhysRevD.23.347",
    journal = "Phys. Rev. D",
    volume = "23",
    pages = "347--356",
    year = "1981"
}

@article{Linde:1981mu,
    author = "Linde, Andrei D.",
    editor = "Fang, Li-Zhi and Ruffini, R.",
    title = "{A New Inflationary Universe Scenario: A Possible Solution of the Horizon, Flatness, Homogeneity, Isotropy and Primordial Monopole Problems}",
    reportNumber = "LEBEDEV-81-229",
    doi = "10.1016/0370-2693(82)91219-9",
    journal = "Phys. Lett. B",
    volume = "108",
    pages = "389--393",
    year = "1982"
}

@article{Albrecht:1982wi,
    author = "Albrecht, Andreas and Steinhardt, Paul J.",
    editor = "Fang, Li-Zhi and Ruffini, R.",
    title = "{Cosmology for Grand Unified Theories with Radiatively Induced Symmetry Breaking}",
    reportNumber = "UPR-0185T",
    doi = "10.1103/PhysRevLett.48.1220",
    journal = "Phys. Rev. Lett.",
    volume = "48",
    pages = "1220--1223",
    year = "1982"
}

@article{Linde:1983gd,
    author = "Linde, Andrei D.",
    title = "{Chaotic Inflation}",
    doi = "10.1016/0370-2693(83)90837-7",
    journal = "Phys. Lett. B",
    volume = "129",
    pages = "177--181",
    year = "1983"
}

@article{Mukhanov:1981xt,
    author = "Mukhanov, Viatcheslav F. and Chibisov, G. V.",
    title = "{Quantum Fluctuations and a Nonsingular Universe}",
    journal = "JETP Lett.",
    volume = "33",
    pages = "532--535",
    year = "1981"
}

@article{Kinney:2005vj,
    author = "Kinney, William H.",
    title = "{Horizon crossing and inflation with large eta}",
    eprint = "gr-qc/0503017",
    archivePrefix = "arXiv",
    doi = "10.1103/PhysRevD.72.023515",
    journal = "Phys. Rev. D",
    volume = "72",
    pages = "023515",
    year = "2005"
}

@misc{kinney2009tasi,
      title={TASI Lectures on Inflation}, 
      author={William H. Kinney},
      year={2009},
      eprint={0902.1529},
      archivePrefix={arXiv},
      primaryClass={astro-ph.CO}
}

@article{Hawking:1982cz,
    author = "Hawking, S. W.",
    title = "{The Development of Irregularities in a Single Bubble Inflationary Universe}",
    reportNumber = "Print-83-0015 (CAMBRIDGE)",
    doi = "10.1016/0370-2693(82)90373-2",
    journal = "Phys. Lett. B",
    volume = "115",
    pages = "295",
    year = "1982"
}

@article{Starobinsky:1982ee,
    author = "Starobinsky, Alexei A.",
    title = "{Dynamics of Phase Transition in the New Inflationary Universe Scenario and Generation of Perturbations}",
    doi = "10.1016/0370-2693(82)90541-X",
    journal = "Phys. Lett. B",
    volume = "117",
    pages = "175--178",
    year = "1982"
}

@article{Guth:1982ec,
    author = "Guth, Alan H. and Pi, S. Y.",
    title = "{Fluctuations in the New Inflationary Universe}",
    doi = "10.1103/PhysRevLett.49.1110",
    journal = "Phys. Rev. Lett.",
    volume = "49",
    pages = "1110--1113",
    year = "1982"
}

@article{Starobinsky:1979ty,
    author = "Starobinsky, Alexei A.",
    editor = "Khalatnikov, I. M. and Mineev, V. P.",
    title = "{Spectrum of relict gravitational radiation and the early state of the universe}",
    journal = "JETP Lett.",
    volume = "30",
    pages = "682--685",
    year = "1979"
}

@inproceedings{Baumann_TASI,
    author = "Baumann, Daniel",
    title = "{Inflation}",
    booktitle = "{Theoretical Advanced Study Institute in Elementary Particle Physics}: {Physics of the Large and the Small}",
    eprint = "0907.5424",
    archivePrefix = "arXiv",
    primaryClass = "hep-th",
    reportNumber = "TASI-2009",
    doi = "10.1142/9789814327183_0010",
    pages = "523--686",
    year = "2011"
}

@article{Baumann:2018muz,
    author = "Baumann, Daniel",
    title = "{Primordial Cosmology}",
    eprint = "1807.03098",
    archivePrefix = "arXiv",
    primaryClass = "hep-th",
    doi = "10.22323/1.305.0009",
    journal = "PoS",
    volume = "TASI2017",
    pages = "009",
    year = "2018"
}

@article{Riotto:2002yw,
    author = "Riotto, Antonio",
    editor = "Dvali, G. and Perez-Lorenzana, Abdel and Senjanovic, G. and Thompson, G. and Vissani, F.",
    title = "{Inflation and the theory of cosmological perturbations}",
    eprint = "hep-ph/0210162",
    archivePrefix = "arXiv",
    reportNumber = "DFPD-TH-02-22",
    journal = "ICTP Lect. Notes Ser.",
    volume = "14",
    pages = "317--413",
    year = "2003"
}

@book{Baumann:2022mni,
    author = "Baumann, Daniel",
    title = "{Cosmology}",
    doi = "10.1017/9781108937092",
    isbn = "978-1-108-93709-2",
    publisher = "Cambridge University Press",
    month = "7",
    year = "2022"
}

@book{Birrell:1982ix,
    author = "Birrell, N. D. and Davies, P. C. W.",
    title = "{Quantum Fields in Curved Space}",
    doi = "10.1017/CBO9780511622632",
    isbn = "978-0-521-27858-4, 978-0-521-27858-4",
    publisher = "Cambridge Univ. Press",
    address = "Cambridge, UK",
    series = "Cambridge Monographs on Mathematical Physics",
    month = "2",
    year = "1984"
}

@misc{NIST:DLMF,
         key = "{\relax DLMF}",
       title = "{\it NIST Digital Library of Mathematical Functions}",
howpublished = "\url{https://dlmf.nist.gov/}, Release 1.1.12 of 2023-12-15",
         url = "https://dlmf.nist.gov/",
        note = "F.~W.~J. Olver, A.~B. {Olde Daalhuis}, D.~W. Lozier, B.~I. Schneider,
                R.~F. Boisvert, C.~W. Clark, B.~R. Miller, B.~V. Saunders,
                H.~S. Cohl, and M.~A. McClain, eds."}

@book{Mukhanov:2005sc,
    author = "Mukhanov, V.",
    title = "{Physical Foundations of Cosmology}",
    doi = "10.1017/CBO9780511790553",
    isbn = "978-0-521-56398-7",
    publisher = "Cambridge University Press",
    address = "Oxford",
    year = "2005"
}

@article{Sotiriou:2008rp,
    author = "Sotiriou, Thomas P. and Faraoni, Valerio",
    title = "{f(R) Theories Of Gravity}",
    eprint = "0805.1726",
    archivePrefix = "arXiv",
    primaryClass = "gr-qc",
    doi = "10.1103/RevModPhys.82.451",
    journal = "Rev. Mod. Phys.",
    volume = "82",
    pages = "451--497",
    year = "2010"
}

@article{Maldacena:2002vr,
    author = "Maldacena, Juan Martin",
    title = "{Non-Gaussian features of primordial fluctuations in single field inflationary models}",
    eprint = "astro-ph/0210603",
    archivePrefix = "arXiv",
    doi = "10.1088/1126-6708/2003/05/013",
    journal = "JHEP",
    volume = "05",
    pages = "013",
    year = "2003"
}

@article{Whitt:1984pd,
    author = "Whitt, Brian",
    title = "{Fourth Order Gravity as General Relativity Plus Matter}",
    reportNumber = "Print-84-0715 (CAMBRIDGE)",
    doi = "10.1016/0370-2693(84)90332-0",
    journal = "Phys. Lett. B",
    volume = "145",
    pages = "176--178",
    year = "1984"
}

@article{Bunch:1978yq,
    author = "Bunch, T. S. and Davies, P. C. W.",
    title = "{Quantum Field Theory in de Sitter Space: Renormalization by Point Splitting}",
    doi = "10.1098/rspa.1978.0060",
    journal = "Proc. Roy. Soc. Lond. A",
    volume = "360",
    pages = "117--134",
    year = "1978"
}

@article{Sasaki:1986hm,
    author = "Sasaki, Misao",
    title = "{Large Scale Quantum Fluctuations in the Inflationary Universe}",
    reportNumber = "RRK-86-29",
    doi = "10.1143/PTP.76.1036",
    journal = "Prog. Theor. Phys.",
    volume = "76",
    pages = "1036",
    year = "1986"
}

@article{Mukhanov:1988jd,
    author = "Mukhanov, Viatcheslav F.",
    title = "{Quantum Theory of Gauge Invariant Cosmological Perturbations}",
    journal = "Sov. Phys. JETP",
    volume = "67",
    pages = "1297--1302",
    year = "1988"
}

@article{Mukhanov:1990me,
    author = "Mukhanov, Viatcheslav F. and Feldman, H. A. and Brandenberger, Robert H.",
    title = "{Theory of cosmological perturbations. Part 1. Classical perturbations. Part 2. Quantum theory of perturbations. Part 3. Extensions}",
    reportNumber = "BROWN-HET-796, BROWN-HET-800, BROWN-HET-780",
    doi = "10.1016/0370-1573(92)90044-Z",
    journal = "Phys. Rept.",
    volume = "215",
    pages = "203--333",
    year = "1992"
}

@proceedings{Gibbons:1984hx,
    editor = "Gibbons, G. W. and Hawking, S. W. and Siklos, S. T. C.",
    title = "{THE VERY EARLY UNIVERSE. PROCEEDINGS, NUFFIELD WORKSHOP, CAMBRIDGE, UK, JUNE 21 - JULY 9, 1982}",
    year = "1984"
}

@book{Guth:1997wk,
    author = "Guth, Alan H.",
    title = "{The inflationary universe: The quest for a new theory of cosmic origins}",
    year = "1997"
}

@article{Wands:2000dp,
    author = "Wands, David and Malik, Karim A. and Lyth, David H. and Liddle, Andrew R.",
    title = "{A New approach to the evolution of cosmological perturbations on large scales}",
    eprint = "astro-ph/0003278",
    archivePrefix = "arXiv",
    doi = "10.1103/PhysRevD.62.043527",
    journal = "Phys. Rev. D",
    volume = "62",
    pages = "043527",
    year = "2000"
}

@article{Liddle:2003as,
    author = "Liddle, Andrew R and Leach, Samuel M",
    title = "{How long before the end of inflation were observable perturbations produced?}",
    eprint = "astro-ph/0305263",
    archivePrefix = "arXiv",
    doi = "10.1103/PhysRevD.68.103503",
    journal = "Phys. Rev. D",
    volume = "68",
    pages = "103503",
    year = "2003"
}

@article{Lyth:2004gb,
    author = "Lyth, David H. and Malik, Karim A. and Sasaki, Misao",
    title = "{A General proof of the conservation of the curvature perturbation}",
    eprint = "astro-ph/0411220",
    archivePrefix = "arXiv",
    reportNumber = "YITP-04-67",
    doi = "10.1088/1475-7516/2005/05/004",
    journal = "JCAP",
    volume = "05",
    pages = "004",
    year = "2005"
}

@article{Planck_overview,
    author = "Aghanim, N. and others",
    collaboration = "Planck",
    title = "{Planck 2018 results. I. Overview and the cosmological legacy of Planck}", 
    eprint = "1807.06205",
    archivePrefix = "arXiv",
    primaryClass = "astro-ph.CO",
    doi = "10.1051/0004-6361/201833880",
    journal = "Astron. Astrophys.",
    volume = "641",
    pages = "A1",
    year = "2020"
}

@article{Planck_inflation,
    author = "Akrami, Y. and others",
    collaboration = "Planck",
    title = "{Planck 2018 results. X. Constraints on inflation}",
    eprint = "1807.06211",
    archivePrefix = "arXiv",
    primaryClass = "astro-ph.CO",
    doi = "10.1051/0004-6361/201833887",
    journal = "Astron. Astrophys.",
    volume = "641",
    pages = "A10",
    year = "2020"
}

@article{Brandenberger:2016uzh,
    author = "Brandenberger, Robert",
    title = "{Initial conditions for inflation \textemdash{} A short review}",
    eprint = "1601.01918",
    archivePrefix = "arXiv",
    primaryClass = "hep-th",
    doi = "10.1142/S0218271817400028",
    journal = "Int. J. Mod. Phys. D",
    volume = "26",
    number = "01",
    pages = "1740002",
    year = "2016"
}

@article{Martin:2013tda,
    author = "Martin, Jerome and Ringeval, Christophe and Vennin, Vincent",
    title = "{Encyclop\ae{}dia Inflationaris}",
    eprint = "1303.3787",
    archivePrefix = "arXiv",
    primaryClass = "astro-ph.CO",
    doi = "10.1016/j.dark.2014.01.003",
    journal = "Phys. Dark Univ.",
    volume = "5-6",
    pages = "75--235",
    year = "2014"
}

@article{Shtanov:2022pdx,
    author = "Shtanov, Yuri and Sahni, Varun and Mishra, Swagat S.",
    title = "{Tabletop potentials for inflation from f(R) gravity}",
    eprint = "2210.01828",
    archivePrefix = "arXiv",
    primaryClass = "gr-qc",
    doi = "10.1088/1475-7516/2023/03/023",
    journal = "JCAP",
    volume = "03",
    pages = "023",
    year = "2023"
}

@article{Mishra:2018dtg,
    author = "Mishra, Swagat S. and Sahni, Varun and Toporensky, Alexey V.",
    title = "{Initial conditions for Inflation in an FRW Universe}",
    eprint = "1801.04948",
    archivePrefix = "arXiv",
    primaryClass = "gr-qc",
    doi = "10.1103/PhysRevD.98.083538",
    journal = "Phys. Rev. D",
    volume = "98",
    number = "8",
    pages = "083538",
    year = "2018"
}

@article{Mishra2023,
    author = "Mishra, Swagat S. and Copeland, Edmund J. and Green, Anne M.",
    title = "{Primordial black holes and stochastic inflation beyond slow roll: I -- noise matrix elements}",
    eprint = "2303.17375",
    archivePrefix = "arXiv",
    primaryClass = "astro-ph.CO",
    month = "3",
    year = "2023"
}

@article{Mishra:2021wkm,
    author = "Mishra, Swagat S. and Sahni, Varun and Starobinsky, Alexei A.",
    title = "{Curing inflationary degeneracies using reheating predictions and relic gravitational waves}",
    eprint = "2101.00271",
    archivePrefix = "arXiv",
    primaryClass = "gr-qc",
    doi = "10.1088/1475-7516/2021/05/075",
    journal = "JCAP",
    volume = "05",
    pages = "075",
    year = "2021"
}

@article{Mishra:2022ijb,
    author = "Mishra, Swagat S. and Sahni, Varun",
    title = "{Canonical and Non-canonical Inflation in the light of the recent BICEP/Keck results}",
    eprint = "2202.03467",
    archivePrefix = "arXiv",
    primaryClass = "astro-ph.CO",
    month = "2",
    year = "2022"
}

@article{BICEP:2021xfz,
    author = "Ade, P. A. R. and others",
    collaboration = "BICEP, Keck",
    title = "{Improved Constraints on Primordial Gravitational Waves using Planck, WMAP, and BICEP/Keck Observations through the 2018 Observing Season}",
    eprint = "2110.00483",
    archivePrefix = "arXiv",
    primaryClass = "astro-ph.CO",
    doi = "10.1103/PhysRevLett.127.151301",
    journal = "Phys. Rev. Lett.",
    volume = "127",
    number = "15",
    pages = "151301",
    year = "2021"
}

@article{Kachru:2003aw,
    author = "Kachru, Shamit and Kallosh, Renata and Linde, Andrei D. and Trivedi, Sandip P.",
    title = "{De Sitter vacua in string theory}",
    eprint = "hep-th/0301240",
    archivePrefix = "arXiv",
    reportNumber = "SLAC-PUB-9630, SU-ITP-03-01, TIFR-TH-03-03",
    doi = "10.1103/PhysRevD.68.046005",
    journal = "Phys. Rev. D",
    volume = "68",
    pages = "046005",
    year = "2003"
}

@article{Kachru:2003sx,
    author = "Kachru, Shamit and Kallosh, Renata and Linde, Andrei D. and Maldacena, Juan Martin and McAllister, Liam P. and Trivedi, Sandip P.",
    title = "{Towards inflation in string theory}",
    eprint = "hep-th/0308055",
    archivePrefix = "arXiv",
    reportNumber = "SLAC-PUB-9669, SU-ITP-03-18, TIFR-TH-03-06",
    doi = "10.1088/1475-7516/2003/10/013",
    journal = "JCAP",
    volume = "10",
    pages = "013",
    year = "2003"
}

@article{Kallosh:2019jnl,
    author = "Kallosh, Renata and Linde, Andrei",
    title = "{On hilltop and brane inflation after Planck}",
    eprint = "1906.02156",
    archivePrefix = "arXiv",
    primaryClass = "hep-th",
    doi = "10.1088/1475-7516/2019/09/030",
    journal = "JCAP",
    volume = "09",
    pages = "030",
    year = "2019"
}

@article{Turner:1983he,
    author = "Turner, Michael S.",
    title = "{Coherent Scalar Field Oscillations in an Expanding Universe}",
    reportNumber = "EFI-83-29-CHICAGO",
    doi = "10.1103/PhysRevD.28.1243",
    journal = "Phys. Rev. D",
    volume = "28",
    pages = "1243",
    year = "1983"
}

@article{Kallosh:2013yoa,
    author = "Kallosh, Renata and Linde, Andrei and Roest, Diederik",
    title = "{Superconformal Inflationary $\alpha$-Attractors}",
    eprint = "1311.0472",
    archivePrefix = "arXiv",
    primaryClass = "hep-th",
    doi = "10.1007/JHEP11(2013)198",
    journal = "JHEP",
    volume = "11",
    pages = "198",
    year = "2013"
}

@article{Kofman:1994rk,
    author = "Kofman, Lev and Linde, Andrei D. and Starobinsky, Alexei A.",
    title = "{Reheating after inflation}",
    eprint = "hep-th/9405187",
    archivePrefix = "arXiv",
    reportNumber = "UH-IFA-94-35, SU-ITP-94-13, YITP-U-94-15",
    doi = "10.1103/PhysRevLett.73.3195",
    journal = "Phys. Rev. Lett.",
    volume = "73",
    pages = "3195--3198",
    year = "1994"
}

@article{Shtanov:1994ce,
    author = "Shtanov, Y. and Traschen, Jennie H. and Brandenberger, Robert H.",
    title = "{Universe reheating after inflation}",
    eprint = "hep-ph/9407247",
    archivePrefix = "arXiv",
    reportNumber = "BROWN-HET-957",
    doi = "10.1103/PhysRevD.51.5438",
    journal = "Phys. Rev. D",
    volume = "51",
    pages = "5438--5455",
    year = "1995"
}

@inproceedings{Kofman:1996mv,
    author = "Kofman, Lev A.",
    title = "{The Origin of matter in the universe: Reheating after inflation}",
    eprint = "astro-ph/9605155",
    archivePrefix = "arXiv",
    reportNumber = "UH-IFA-96-28",
    month = "5",
    year = "1996"
}

@article{Kofman:1997yn,
    author = "Kofman, Lev and Linde, Andrei D. and Starobinsky, Alexei A.",
    title = "{Towards the theory of reheating after inflation}",
    eprint = "hep-ph/9704452",
    archivePrefix = "arXiv",
    reportNumber = "IFA-97-28, SU-ITP-97-18",
    doi = "10.1103/PhysRevD.56.3258",
    journal = "Phys. Rev. D",
    volume = "56",
    pages = "3258--3295",
    year = "1997"
}

@article{Greene:1997fu,
    author = "Greene, Patrick B. and Kofman, Lev and Linde, Andrei D. and Starobinsky, Alexei A.",
    title = "{Structure of resonance in preheating after inflation}",
    eprint = "hep-ph/9705347",
    archivePrefix = "arXiv",
    reportNumber = "SU-ITP-97-19, IFA-97-29",
    doi = "10.1103/PhysRevD.56.6175",
    journal = "Phys. Rev. D",
    volume = "56",
    pages = "6175--6192",
    year = "1997"
}

@article{Allahverdi:2010xz,
    author = "Allahverdi, Rouzbeh and Brandenberger, Robert and Cyr-Racine, Francis-Yan and Mazumdar, Anupam",
    title = "{Reheating in Inflationary Cosmology: Theory and Applications}",
    eprint = "1001.2600",
    archivePrefix = "arXiv",
    primaryClass = "hep-th",
    doi = "10.1146/annurev.nucl.012809.104511",
    journal = "Ann. Rev. Nucl. Part. Sci.",
    volume = "60",
    pages = "27--51",
    year = "2010"
}

@article{Amin:2014eta,
    author = "Amin, Mustafa A. and Hertzberg, Mark P. and Kaiser, David I. and Karouby, Johanna",
    title = "{Nonperturbative Dynamics Of Reheating After Inflation: A Review}",
    eprint = "1410.3808",
    archivePrefix = "arXiv",
    primaryClass = "hep-ph",
    doi = "10.1142/S0218271815300037",
    journal = "Int. J. Mod. Phys. D",
    volume = "24",
    pages = "1530003",
    year = "2014"
}

@article{Lozanov:2019jxc,
    author = "Lozanov, Kaloian D.",
    title = "{Lectures on Reheating after Inflation}",
    eprint = "1907.04402",
    archivePrefix = "arXiv",
    primaryClass = "astro-ph.CO",
    month = "7",
    year = "2019"
}

@article{Mishra:2017ehw,
    author = "Mishra, Swagat S. and Sahni, Varun and Shtanov, Yuri",
    title = "{Sourcing Dark Matter and Dark Energy from $\alpha$-attractors}",
    eprint = "1703.03295",
    archivePrefix = "arXiv",
    primaryClass = "gr-qc",
    doi = "10.1088/1475-7516/2017/06/045",
    journal = "JCAP",
    volume = "06",
    pages = "045",
    year = "2017"
}

@article{Amin:2011hj,
    author = "Amin, Mustafa A. and Easther, Richard and Finkel, Hal and Flauger, Raphael and Hertzberg, Mark P.",
    title = "{Oscillons After Inflation}",
    eprint = "1106.3335",
    archivePrefix = "arXiv",
    primaryClass = "astro-ph.CO",
    doi = "10.1103/PhysRevLett.108.241302",
    journal = "Phys. Rev. Lett.",
    volume = "108",
    pages = "241302",
    year = "2012"
}

@article{Zhou:2013tsa,
    author = "Zhou, Shuang-Yong and Copeland, Edmund J. and Easther, Richard and Finkel, Hal and Mou, Zong-Gang and Saffin, Paul M.",
    title = "{Gravitational Waves from Oscillon Preheating}",
    eprint = "1304.6094",
    archivePrefix = "arXiv",
    primaryClass = "astro-ph.CO",
    doi = "10.1007/JHEP10(2013)026",
    journal = "JHEP",
    volume = "10",
    pages = "026",
    year = "2013"
}

@article{Lozanov:2014zfa,
    author = "Lozanov, Kaloian D. and Amin, Mustafa A.",
    title = "{End of inflation, oscillons, and matter-antimatter asymmetry}",
    eprint = "1408.1811",
    archivePrefix = "arXiv",
    primaryClass = "hep-ph",
    doi = "10.1103/PhysRevD.90.083528",
    journal = "Phys. Rev. D",
    volume = "90",
    number = "8",
    pages = "083528",
    year = "2014"
}

@article{Lozanov:2017hjm,
    author = "Lozanov, Kaloian D. and Amin, Mustafa A.",
    title = "{Self-resonance after inflation: oscillons, transients and radiation domination}",
    eprint = "1710.06851",
    archivePrefix = "arXiv",
    primaryClass = "astro-ph.CO",
    doi = "10.1103/PhysRevD.97.023533",
    journal = "Phys. Rev. D",
    volume = "97",
    number = "2",
    pages = "023533",
    year = "2018"
}

@article{Lozanov:2019ylm,
    author = "Lozanov, Kaloian D. and Amin, Mustafa A.",
    title = "{Gravitational perturbations from oscillons and transients after inflation}",
    eprint = "1902.06736",
    archivePrefix = "arXiv",
    primaryClass = "astro-ph.CO",
    doi = "10.1103/PhysRevD.99.123504",
    journal = "Phys. Rev. D",
    volume = "99",
    number = "12",
    pages = "123504",
    year = "2019"
}

@article{Mahbub:2023faw,
    author = "Mahbub, Rafid and Mishra, Swagat S.",
    title = "{Oscillon formation from preheating in asymmetric inflationary potentials}",
    eprint = "2303.07503",
    archivePrefix = "arXiv",
    primaryClass = "astro-ph.CO",
    doi = "10.1103/PhysRevD.108.063524",
    journal = "Phys. Rev. D",
    volume = "108",
    number = "6",
    pages = "063524",
    year = "2023"
}

@article{Kim:2017duj,
    author = "Kim, Jinsu and McDonald, John",
    title = "{Inflaton Condensate Fragmentation: Analytical Conditions and Application to $\alpha$-Attractor Models}",
    eprint = "1702.08777",
    archivePrefix = "arXiv",
    primaryClass = "astro-ph.CO",
    doi = "10.1103/PhysRevD.95.123537",
    journal = "Phys. Rev. D",
    volume = "95",
    number = "12",
    pages = "123537",
    year = "2017"
}

@article{Kim:2021ipz,
    author = "Kim, Jinsu and McDonald, John",
    title = "{General analytical conditions for inflaton fragmentation: Quick and easy tests for its occurrence}",
    eprint = "2111.12474",
    archivePrefix = "arXiv",
    primaryClass = "astro-ph.CO",
    reportNumber = "CERN-TH-2021-202",
    doi = "10.1103/PhysRevD.105.063508",
    journal = "Phys. Rev. D",
    volume = "105",
    number = "6",
    pages = "063508",
    year = "2022"
}

@article{Zhang:2020bec,
    author = "Zhang, Hong-Yi and Amin, Mustafa A. and Copeland, Edmund J. and Saffin, Paul M. and Lozanov, Kaloian D.",
    title = "{Classical Decay Rates of Oscillons}",
    eprint = "2004.01202",
    archivePrefix = "arXiv",
    primaryClass = "hep-th",
    doi = "10.1088/1475-7516/2020/07/055",
    journal = "JCAP",
    volume = "07",
    pages = "055",
    year = "2020"
}

@article{Amin:2018xfe,
    author = "Amin, Mustafa A. and Braden, Jonathan and Copeland, Edmund J. and Giblin, John T. and Solorio, Christian and Weiner, Zachary J. and Zhou, Shuang-Yong",
    title = "{Gravitational waves from asymmetric oscillon dynamics?}",
    eprint = "1803.08047",
    archivePrefix = "arXiv",
    primaryClass = "astro-ph.CO",
    doi = "10.1103/PhysRevD.98.024040",
    journal = "Phys. Rev. D",
    volume = "98",
    pages = "024040",
    year = "2018"
}

@article{Albrecht:1982mp,
    author = "Albrecht, Andreas and Steinhardt, Paul J. and Turner, Michael S. and Wilczek, Frank",
    title = "{Reheating an Inflationary Universe}",
    reportNumber = "UPR-0189T, EFI-82-09-CHICAGO",
    doi = "10.1103/PhysRevLett.48.1437",
    journal = "Phys. Rev. Lett.",
    volume = "48",
    pages = "1437",
    year = "1982"
}

@book{Kolb:1990vq,
    author = "Kolb, Edward W. and Turner, Michael S.",
    title = "{The Early Universe}",
    reportNumber = "FERMILAB-BOOK-1990-01",
    doi = "10.1201/9780429492860",
    isbn = "978-0-201-62674-2",
    volume = "69",
    year = "1990"
}

@article{Antusch:2021aiw,
    author = "Antusch, Stefan and Figueroa, Daniel G. and Marschall, Kenneth and Torrenti, Francisco",
    title = "{Characterizing the postinflationary reheating history: Single daughter field with quadratic-quadratic interaction}",
    eprint = "2112.11280",
    archivePrefix = "arXiv",
    primaryClass = "astro-ph.CO",
    doi = "10.1103/PhysRevD.105.043532",
    journal = "Phys. Rev. D",
    volume = "105",
    number = "4",
    pages = "043532",
    year = "2022"
}

@article{Figueroa:2021yhd,
    author = "Figueroa, Daniel G. and Florio, Adrien and Torrenti, Francisco and Valkenburg, Wessel",
    title = "{CosmoLattice: A modern code for lattice simulations of scalar and gauge field dynamics in an expanding universe}",
    eprint = "2102.01031",
    archivePrefix = "arXiv",
    primaryClass = "astro-ph.CO",
    doi = "10.1016/j.cpc.2022.108586",
    journal = "Comput. Phys. Commun.",
    volume = "283",
    pages = "108586",
    year = "2023"
}

@article{Figueroa:2023xmq,
    author = "Figueroa, Daniel G. and Florio, Adrien and Torrenti, Francisco",
    title = "{Present and future of CosmoLattice}",
    eprint = "2312.15056",
    archivePrefix = "arXiv",
    primaryClass = "astro-ph.CO",
    month = "12",
    year = "2023"
}

@book{mclachlan1947theory,
  title={Theory and Application of Mathieu Functions},
  author={McLachlan, N.W.},
  lccn={47005683},
  url={https://books.google.co.in/books?id=pCXEvwEACAAJ},
  year={1947},
  publisher={Clarendon Press}
}

@book{magnus2004hill,
  title={Hill's Equation},
  author={Magnus, W. and Winkler, S.},
  isbn={9780486495651},
  lccn={2003062011},
  series={Dover Books on Mathematics Series},
  url={https://books.google.co.uk/books?id=ML5wm-T4RVQC},
  year={2004},
  publisher={Dover Publications}
}

@article{Berera:2023liv,
    author = "Berera, Arjun",
    title = "{The Warm Inflation Story}",
    eprint = "2305.10879",
    archivePrefix = "arXiv",
    primaryClass = "hep-ph",
    doi = "10.3390/universe9060272",
    journal = "Universe",
    volume = "9",
    number = "6",
    pages = "272",
    year = "2023"
}

@article{Dodelson:2003ip,
    author = "Dodelson, Scott",
    editor = "Nieves, J. F. and Volkas, R. R.",
    title = "{Coherent phase argument for inflation}",
    eprint = "hep-ph/0309057",
    archivePrefix = "arXiv",
    reportNumber = "FERMILAB-CONF-03-450-A",
    doi = "10.1063/1.1627736",
    journal = "AIP Conf. Proc.",
    volume = "689",
    number = "1",
    pages = "184--196",
    year = "2003"
}

@article{Malik:2008im,
    author = "Malik, Karim A. and Wands, David",
    title = "{Cosmological perturbations}",
    eprint = "0809.4944",
    archivePrefix = "arXiv",
    primaryClass = "astro-ph",
    doi = "10.1016/j.physrep.2009.03.001",
    journal = "Phys. Rept.",
    volume = "475",
    pages = "1--51",
    year = "2009"
}

@article{Arnowitt:1962hi,
    author = "Arnowitt, Richard L. and Deser, Stanley and Misner, Charles W.",
    title = "{The Dynamics of general relativity}",
    eprint = "gr-qc/0405109",
    archivePrefix = "arXiv",
    doi = "10.1007/s10714-008-0661-1",
    journal = "Gen. Rel. Grav.",
    volume = "40",
    pages = "1997--2027",
    year = "2008"
}

@ARTICLE{1986ApJ...304...15B,
       author = {{Bardeen}, J.~M. and {Bond}, J.~R. and {Kaiser}, N. and {Szalay}, A.~S.},
        title = "{The Statistics of Peaks of Gaussian Random Fields}",
      journal = {\apj},
         year = 1986,
        month = may,
       volume = {304},
        pages = {15},
          doi = {10.1086/164143},
       adsurl = {https://ui.adsabs.harvard.edu/abs/1986ApJ...304...15B}
}

@article{Arkani-Hamed:2015bza,
    author = "Arkani-Hamed, Nima and Maldacena, Juan",
    title = "{Cosmological Collider Physics}",
    eprint = "1503.08043",
    archivePrefix = "arXiv",
    primaryClass = "hep-th",
    month = "3",
    year = "2015"
}

@article{Creminelli:2003iq,
    author = "Creminelli, Paolo",
    title = "{On non-Gaussianities in single-field inflation}",
    eprint = "astro-ph/0306122",
    archivePrefix = "arXiv",
    reportNumber = "HUTP-03-A036",
    doi = "10.1088/1475-7516/2003/10/003",
    journal = "JCAP",
    volume = "10",
    pages = "003",
    year = "2003"
}

@article{Chen:2010xka,
    author = "Chen, Xingang",
    title = "{Primordial Non-Gaussianities from Inflation Models}",
    eprint = "1002.1416",
    archivePrefix = "arXiv",
    primaryClass = "astro-ph.CO",
    doi = "10.1155/2010/638979",
    journal = "Adv. Astron.",
    volume = "2010",
    pages = "638979",
    year = "2010"
}

@article{Komatsu:2010hc,
    author = "Komatsu, Eiichiro",
    title = "{Hunting for Primordial Non-Gaussianity in the Cosmic Microwave Background}",
    eprint = "1003.6097",
    archivePrefix = "arXiv",
    primaryClass = "astro-ph.CO",
    reportNumber = "TCC-011-10",
    doi = "10.1088/0264-9381/27/12/124010",
    journal = "Class. Quant. Grav.",
    volume = "27",
    pages = "124010",
    year = "2010"
}

@article{Wang:2013zva,
    author = "Wang, Yi",
    title = "{Inflation, Cosmic Perturbations and Non-Gaussianities}",
    eprint = "1303.1523",
    archivePrefix = "arXiv",
    primaryClass = "hep-th",
    doi = "10.1088/0253-6102/62/1/19",
    journal = "Commun. Theor. Phys.",
    volume = "62",
    pages = "109--166",
    year = "2014"
}

@article{Lee:2016vti,
    author = "Lee, Hayden and Baumann, Daniel and Pimentel, Guilherme L.",
    title = "{Non-Gaussianity as a Particle Detector}",
    eprint = "1607.03735",
    archivePrefix = "arXiv",
    primaryClass = "hep-th",
    doi = "10.1007/JHEP12(2016)040",
    journal = "JHEP",
    volume = "12",
    pages = "040",
    year = "2016"
}

@article{Meerburg:2019qqi,
    author = "Meerburg, P. Daniel and others",
    title = "{Primordial Non-Gaussianity}",
    eprint = "1903.04409",
    archivePrefix = "arXiv",
    primaryClass = "astro-ph.CO",
    reportNumber = "FERMILAB-PUB-19-140-A",
    journal = "Bull. Am. Astron. Soc.",
    volume = "51",
    number = "3",
    pages = "107",
    year = "2019"
}

@article{Allen:1985ux,
    author = "Allen, Bruce",
    title = "{Vacuum States in de Sitter Space}",
    reportNumber = "UCSB-TH-3-1985",
    doi = "10.1103/PhysRevD.32.3136",
    journal = "Phys. Rev. D",
    volume = "32",
    pages = "3136",
    year = "1985"
}

@article{Starobinsky:2005ab,
    author = "Starobinsky, Alexei A.",
    title = "{Inflaton field potential producing the exactly flat spectrum of adiabatic perturbations}",
    eprint = "astro-ph/0507193",
    archivePrefix = "arXiv",
    doi = "10.1134/1.2121807",
    journal = "JETP Lett.",
    volume = "82",
    pages = "169--173",
    year = "2005"
}

@book{Linde:1990flp,
    author = "Linde, Andrei D.",
    title = "{Particle physics and inflationary cosmology}",
    eprint = "hep-th/0503203",
    archivePrefix = "arXiv",
    volume = "5",
    year = "1990"
}

@article{Sahni:1990tx,
    author = "Sahni, Varun",
    title = "{The Energy Density of Relic Gravity Waves From Inflation}",
    reportNumber = "PRINT-90-0282 (CANADA)",
    doi = "10.1103/PhysRevD.42.453",
    journal = "Phys. Rev. D",
    volume = "42",
    pages = "453--463",
    year = "1990"
}

@article{Allen:1987bk,
    author = "Allen, Bruce",
    title = "{The Stochastic Gravity Wave Background in Inflationary Universe Models}",
    reportNumber = "PRINT-88-0063 (TUFTS)",
    doi = "10.1103/PhysRevD.37.2078",
    journal = "Phys. Rev. D",
    volume = "37",
    pages = "2078",
    year = "1988"
}

@article{Dodelson:1997hr,
    author = "Dodelson, Scott and Kinney, William H. and Kolb, Edward W.",
    title = "{Cosmic microwave background measurements can discriminate among inflation models}",
    eprint = "astro-ph/9702166",
    archivePrefix = "arXiv",
    reportNumber = "FERMILAB-PUB-97-037-A",
    doi = "10.1103/PhysRevD.56.3207",
    journal = "Phys. Rev. D",
    volume = "56",
    pages = "3207--3215",
    year = "1997"
}

@article{Caprini:2018mtu,
    author = "Caprini, Chiara and Figueroa, Daniel G.",
    title = "{Cosmological Backgrounds of Gravitational Waves}",
    eprint = "1801.04268",
    archivePrefix = "arXiv",
    primaryClass = "astro-ph.CO",
    doi = "10.1088/1361-6382/aac608",
    journal = "Class. Quant. Grav.",
    volume = "35",
    number = "16",
    pages = "163001",
    year = "2018"
}

@article{Micha:2002ey,
    author = "Micha, Raphael and Tkachev, Igor I.",
    title = "{Relativistic turbulence: A Long way from preheating to equilibrium}",
    eprint = "hep-ph/0210202",
    archivePrefix = "arXiv",
    doi = "10.1103/PhysRevLett.90.121301",
    journal = "Phys. Rev. Lett.",
    volume = "90",
    pages = "121301",
    year = "2003"
}

@inproceedings{Micha:2003ws,
    author = "Micha, Raphael and Tkachev, Igor I.",
    title = "{Preheating and thermalization after inflation}",
    booktitle = "{5th Internationa Conference on Strong and Electroweak Matter}",
    eprint = "hep-ph/0301249",
    archivePrefix = "arXiv",
    doi = "10.1142/9789812704498_0020",
    pages = "210--219",
    year = "2003"
}

@article{Micha:2004bv,
    author = "Micha, Raphael and Tkachev, Igor I.",
    title = "{Turbulent thermalization}",
    eprint = "hep-ph/0403101",
    archivePrefix = "arXiv",
    doi = "10.1103/PhysRevD.70.043538",
    journal = "Phys. Rev. D",
    volume = "70",
    pages = "043538",
    year = "2004"
}

@article{Liddle:1994dx,
    author = "Liddle, Andrew R. and Parsons, Paul and Barrow, John D.",
    title = "{Formalizing the slow roll approximation in inflation}",
    eprint = "astro-ph/9408015",
    archivePrefix = "arXiv",
    reportNumber = "SUSSEX-AST-94-8-1",
    doi = "10.1103/PhysRevD.50.7222",
    journal = "Phys. Rev. D",
    volume = "50",
    pages = "7222--7232",
    year = "1994"
}

@book{Rubakov:2017xzr,
    author = "Rubakov, Valery A. and Gorbunov, Dmitry S.",
    title = "{Introduction to the Theory of the Early Universe}: {Hot big bang theory}",
    doi = "10.1142/10447",
    isbn = "978-981-320-987-9, 978-981-320-988-6, 978-981-322-005-8",
    publisher = "World Scientific",
    address = "Singapore",
    year = "2017"
}

@book{Gorbunov:2011zzc,
    author = "Gorbunov, Dmitry S. and Rubakov, Valery A.",
    title = "{Introduction to the theory of the early universe: Cosmological perturbations and inflationary theory}",
    doi = "10.1142/7873",
    year = "2011"
}

@book{Lyth:2009zz,
    author = "Lyth, David H. and Liddle, Andrew R.",
    title = "{The primordial density perturbation: Cosmology, inflation and the origin of structure}",
    year = "2009"
}

@ARTICLE{2021ApJ...919...16F,
       author = {{Freedman}, Wendy L.},
        title = "{Measurements of the Hubble Constant: Tensions in Perspective}",
      journal = {\apj},
         year = 2021,
        month = sep,
       volume = {919},
       number = {1},
          eid = {16},
        pages = {16},
          doi = {10.3847/1538-4357/ac0e95},
archivePrefix = {arXiv},
       eprint = {2106.15656},
 primaryClass = {astro-ph.CO},
       adsurl = {https://ui.adsabs.harvard.edu/abs/2021ApJ...919...16F}
}

@ARTICLE{2022MNRAS.511..662H,
       author = {{Hagstotz}, Steffen and {Reischke}, Robert and {Lilow}, Robert},
        title = "{A new measurement of the Hubble constant using fast radio bursts}",
      journal = {\mnras},
         year = 2022,
        month = mar,
       volume = {511},
       number = {1},
        pages = {662-667},
          doi = {10.1093/mnras/stac077},
archivePrefix = {arXiv},
       eprint = {2104.04538},
 primaryClass = {astro-ph.CO},
       adsurl = {https://ui.adsabs.harvard.edu/abs/2022MNRAS.511..662H}
}

@ARTICLE{2022MNRAS.515L...1W,
       author = {{Wu}, Qin and {Zhang}, Guo-Qiang and {Wang}, Fa-Yin},
        title = "{An 8 per cent determination of the Hubble constant from localized fast radio bursts}",
      journal = {\mnras},
         year = 2022,
        month = sep,
       volume = {515},
       number = {1},
        pages = {L1-L5},
          doi = {10.1093/mnrasl/slac022},
archivePrefix = {arXiv},
       eprint = {2108.00581},
 primaryClass = {astro-ph.CO},
       adsurl = {https://ui.adsabs.harvard.edu/abs/2022MNRAS.515L...1W}
}

@ARTICLE{2016ApJ...826...56R,
       author = {{Riess}, Adam G. and {Macri}, Lucas M. and {Hoffmann}, Samantha L. and {Scolnic}, Dan and {Casertano}, Stefano and {Filippenko}, Alexei V. and {Tucker}, Brad E. and {Reid}, Mark J. and {Jones}, David O. and {Silverman}, Jeffrey M. and {Chornock}, Ryan and {Challis}, Peter and {Yuan}, Wenlong and {Brown}, Peter J. and {Foley}, Ryan J.},
        title = "{A 2.4\% Determination of the Local Value of the Hubble Constant}",
      journal = {\apj},
         year = 2016,
        month = jul,
       volume = {826},
       number = {1},
          eid = {56},
        pages = {56},
          doi = {10.3847/0004-637X/826/1/56},
archivePrefix = {arXiv},
       eprint = {1604.01424},
 primaryClass = {astro-ph.CO},
       adsurl = {https://ui.adsabs.harvard.edu/abs/2016ApJ...826...56R}
}

@ARTICLE{2018ApJ...855..136R,
       author = {{Riess}, Adam G. and {Casertano}, Stefano and {Yuan}, Wenlong and {Macri}, Lucas and {Anderson}, Jay and {MacKenty}, John W. and {Bowers}, J. Bradley and {Clubb}, Kelsey I. and {Filippenko}, Alexei V. and {Jones}, David O. and {Tucker}, Brad E.},
        title = "{New Parallaxes of Galactic Cepheids from Spatially Scanning the Hubble Space Telescope: Implications for the Hubble Constant}",
      journal = {\apj},
         year = 2018,
        month = mar,
       volume = {855},
       number = {2},
          eid = {136},
        pages = {136},
          doi = {10.3847/1538-4357/aaadb7},
archivePrefix = {arXiv},
       eprint = {1801.01120},
 primaryClass = {astro-ph.SR},
       adsurl = {https://ui.adsabs.harvard.edu/abs/2018ApJ...855..136R}
}

@ARTICLE{2019ApJ...876...85R,
       author = {{Riess}, Adam G. and {Casertano}, Stefano and {Yuan}, Wenlong and {Macri}, Lucas M. and {Scolnic}, Dan},
        title = "{Large Magellanic Cloud Cepheid Standards Provide a 1\% Foundation for the Determination of the Hubble Constant and Stronger Evidence for Physics beyond {\ensuremath{\Lambda}}CDM}",
      journal = {\apj},
         year = 2019,
        month = may,
       volume = {876},
       number = {1},
          eid = {85},
        pages = {85},
          doi = {10.3847/1538-4357/ab1422},
archivePrefix = {arXiv},
       eprint = {1903.07603},
 primaryClass = {astro-ph.CO},
       adsurl = {https://ui.adsabs.harvard.edu/abs/2019ApJ...876...85R}
}

@ARTICLE{2021ApJ...908L...6R,
       author = {{Riess}, Adam G. and {Casertano}, Stefano and {Yuan}, Wenlong and {Bowers}, J. Bradley and {Macri}, Lucas and {Zinn}, Joel C. and {Scolnic}, Dan},
        title = "{Cosmic Distances Calibrated to 1\% Precision with Gaia EDR3 Parallaxes and Hubble Space Telescope Photometry of 75 Milky Way Cepheids Confirm Tension with {\ensuremath{\Lambda}}CDM}",
      journal = {\apl},
         year = 2021,
        month = feb,
       volume = {908},
       number = {1},
          eid = {L6},
        pages = {L6},
          doi = {10.3847/2041-8213/abdbaf},
archivePrefix = {arXiv},
       eprint = {2012.08534},
 primaryClass = {astro-ph.CO},
       adsurl = {https://ui.adsabs.harvard.edu/abs/2021ApJ...908L...6R}
}

@article{Bertone:2016nfn,
    author = "Bertone, Gianfranco and Hooper, Dan",
    title = "{History of dark matter}",
    eprint = "1605.04909",
    archivePrefix = "arXiv",
    primaryClass = "astro-ph.CO",
    reportNumber = "FERMILAB-PUB-16-157-A",
    doi = "10.1103/RevModPhys.90.045002",
    journal = "Rev. Mod. Phys.",
    volume = "90",
    number = "4",
    pages = "045002",
    year = "2018"
}

@article{Peebles:2017bzw,
    author = "Peebles, P. J. E.",
    title = "{Growth of the nonbaryonic dark matter theory}",
    eprint = "1701.05837",
    archivePrefix = "arXiv",
    primaryClass = "astro-ph.CO",
    doi = "10.1038/s41550-017-0057",
    journal = "Nature Astron.",
    volume = "1",
    number = "3",
    pages = "0057",
    year = "2017"
}

@article{Green:2021jrr,
    author = "Green, Anne M.",
    title = "{Dark matter in astrophysics/cosmology}",
    eprint = "2109.05854",
    archivePrefix = "arXiv",
    primaryClass = "hep-ph",
    doi = "10.21468/SciPostPhysLectNotes.37",
    journal = "SciPost Phys. Lect. Notes",
    volume = "37",
    pages = "1",
    year = "2022"
}

@article{Sahni:2004ai,
    author = "Sahni, Varun",
    editor = "Papantonopoulos, E.",
    title = "{Dark matter and dark energy}",
    eprint = "astro-ph/0403324",
    archivePrefix = "arXiv",
    doi = "10.1007/b99562",
    journal = "Lect. Notes Phys.",
    volume = "653",
    pages = "141--180",
    year = "2004"
}

@article{Sahni:1999gb,
    author = "Sahni, Varun and Starobinsky, Alexei A.",
    title = "{The Case for a positive cosmological Lambda term}",
    eprint = "astro-ph/9904398",
    archivePrefix = "arXiv",
    reportNumber = "IUCAA-25-2000",
    doi = "10.1142/S0218271800000542",
    journal = "Int. J. Mod. Phys. D",
    volume = "9",
    pages = "373--444",
    year = "2000"
}

@article{Peebles:2002gy,
    author = "Peebles, P. J. E. and Ratra, Bharat",
    editor = "Hsu, Jong-Ping and Fine, D.",
    title = "{The Cosmological Constant and Dark Energy}",
    eprint = "astro-ph/0207347",
    archivePrefix = "arXiv",
    reportNumber = "KSUPT-02-3",
    doi = "10.1103/RevModPhys.75.559",
    journal = "Rev. Mod. Phys.",
    volume = "75",
    pages = "559--606",
    year = "2003"
}

@article{Copeland:2006wr,
    author = "Copeland, Edmund J. and Sami, M. and Tsujikawa, Shinji",
    title = "{Dynamics of dark energy}",
    eprint = "hep-th/0603057",
    archivePrefix = "arXiv",
    doi = "10.1142/S021827180600942X",
    journal = "Int. J. Mod. Phys. D",
    volume = "15",
    pages = "1753--1936",
    year = "2006"
}

@article{Padmanabhan:2002ji,
    author = "Padmanabhan, T.",
    title = "{Cosmological constant: The Weight of the vacuum}",
    eprint = "hep-th/0212290",
    archivePrefix = "arXiv",
    doi = "10.1016/S0370-1573(03)00120-0",
    journal = "Phys. Rept.",
    volume = "380",
    pages = "235--320",
    year = "2003"
}

@article{Bousso:2007gp,
    author = "Bousso, Raphael",
    title = "{TASI Lectures on the Cosmological Constant}",
    eprint = "0708.4231",
    archivePrefix = "arXiv",
    primaryClass = "hep-th",
    doi = "10.1007/s10714-007-0557-5",
    journal = "Gen. Rel. Grav.",
    volume = "40",
    pages = "607--637",
    year = "2008"
}

@article{Sahni:2006pa,
    author = "Sahni, Varun and Starobinsky, Alexei",
    title = "{Reconstructing Dark Energy}",
    eprint = "astro-ph/0610026",
    archivePrefix = "arXiv",
    doi = "10.1142/S0218271806009704",
    journal = "Int. J. Mod. Phys. D",
    volume = "15",
    pages = "2105--2132",
    year = "2006"
}

@article{Carroll:2000fy,
    author = "Carroll, Sean M.",
    title = "{The Cosmological constant}",
    eprint = "astro-ph/0004075",
    archivePrefix = "arXiv",
    reportNumber = "EFI-2000-13",
    doi = "10.12942/lrr-2001-1",
    journal = "Living Rev. Rel.",
    volume = "4",
    pages = "1",
    year = "2001"
}

@article{Weinberg:1988cp,
    author = "Weinberg, Steven",
    editor = "Hsu, Jong-Ping and Fine, D.",
    title = "{The Cosmological Constant Problem}",
    reportNumber = "UTTG-12-88",
    doi = "10.1103/RevModPhys.61.1",
    journal = "Rev. Mod. Phys.",
    volume = "61",
    pages = "1--23",
    year = "1989"
}

@article{Weinberg:1987dv,
    author = "Weinberg, Steven",
    title = "{Anthropic Bound on the Cosmological Constant}",
    reportNumber = "UTTG-06-87",
    doi = "10.1103/PhysRevLett.59.2607",
    journal = "Phys. Rev. Lett.",
    volume = "59",
    pages = "2607",
    year = "1987"
}

@article{SupernovaCosmologyProject:1998vns,
    author = "Perlmutter, S. and others",
    collaboration = "Supernova Cosmology Project",
    title = "{Measurements of $\Omega$ and $\Lambda$ from 42 High Redshift Supernovae}",
    eprint = "astro-ph/9812133",
    archivePrefix = "arXiv",
    reportNumber = "LBNL-41801, LBL-41801",
    doi = "10.1086/307221",
    journal = "Astrophys. J.",
    volume = "517",
    pages = "565--586",
    year = "1999"
}

@article{SupernovaCosmologyProject:1997zqe,
    author = "Perlmutter, S. and others",
    collaboration = "Supernova Cosmology Project",
    title = "{Discovery of a supernova explosion at half the age of the Universe and its cosmological implications}",
    eprint = "astro-ph/9712212",
    archivePrefix = "arXiv",
    reportNumber = "LBL-41172, LBNL-41172",
    doi = "10.1038/34124",
    journal = "Nature",
    volume = "391",
    pages = "51--54",
    year = "1998"
}

@article{SupernovaSearchTeam:1998fmf,
    author = "Riess, Adam G. and others",
    collaboration = "Supernova Search Team",
    title = "{Observational evidence from supernovae for an accelerating universe and a cosmological constant}",
    eprint = "astro-ph/9805201",
    archivePrefix = "arXiv",
    doi = "10.1086/300499",
    journal = "Astron. J.",
    volume = "116",
    pages = "1009--1038",
    year = "1998"
}

@article{Krauss:1995yb,
    author = "Krauss, Lawrence M. and Turner, Michael S.",
    title = "{The Cosmological constant is back}",
    eprint = "astro-ph/9504003",
    archivePrefix = "arXiv",
    reportNumber = "CWRU-P6-95, FERMILAB-PUB-95-063-A",
    doi = "10.1007/BF02108229",
    journal = "Gen. Rel. Grav.",
    volume = "27",
    pages = "1137--1144",
    year = "1995"
}

@article{Einstein:1917ce,
    author = "Einstein, Albert",
    title = "{Cosmological Considerations in the General Theory of Relativity}",
    journal = "Sitzungsber. Preuss. Akad. Wiss. Berlin (Math. Phys. )",
    volume = "1917",
    pages = "142--152",
    year = "1917"
}

@article{Zeldovich:1968ehl,
    author = "Zel'dovich, Ya. B. and Krasinski, Andrzej and Zeldovich, Ya. B.",
    title = "{The Cosmological constant and the theory of elementary particles}",
    doi = "10.1007/s10714-008-0624-6",
    journal = "Sov. Phys. Usp.",
    volume = "11",
    pages = "381--393",
    year = "1968"
}

@article{Hawking:1970zqf,
    author = "Hawking, S. W. and Penrose, R.",
    title = "{The Singularities of gravitational collapse and cosmology}",
    doi = "10.1098/rspa.1970.0021",
    journal = "Proc. Roy. Soc. Lond. A",
    volume = "314",
    pages = "529--548",
    year = "1970"
}

@article{Guth:1985ya,
    author = "Guth, Alan H. and Pi, So-Young",
    title = "{The Quantum Mechanics of the Scalar Field in the New Inflationary Universe}",
    reportNumber = "MIT-CTP-1246",
    doi = "10.1103/PhysRevD.32.1899",
    journal = "Phys. Rev. D",
    volume = "32",
    pages = "1899--1920",
    year = "1985"
}

@article{Hartle:1983ai,
    author = "Hartle, J. B. and Hawking, S. W.",
    editor = "Fang, Li-Zhi and Ruffini, R.",
    title = "{Wave Function of the Universe}",
    reportNumber = "PRINT-83-0937 (CAMBRIDGE)",
    doi = "10.1103/PhysRevD.28.2960",
    journal = "Phys. Rev. D",
    volume = "28",
    pages = "2960--2975",
    year = "1983"
}

@article{Hawking:1983hj,
    author = "Hawking, S. W.",
    editor = "Fang, Li-Zhi and Ruffini, R.",
    title = "{The Quantum State of the Universe}",
    reportNumber = "PRINT-84-0117 (CAMBRIDGE)",
    doi = "10.1016/0550-3213(84)90093-2",
    journal = "Nucl. Phys. B",
    volume = "239",
    pages = "257",
    year = "1984"
}

@article{Vilenkin:1982de,
    author = "Vilenkin, Alexander",
    title = "{Creation of Universes from Nothing}",
    reportNumber = "TUTP-82-8",
    doi = "10.1016/0370-2693(82)90866-8",
    journal = "Phys. Lett. B",
    volume = "117",
    pages = "25--28",
    year = "1982"
}

@inproceedings{Sasaki:1993kt,
    author = "Sasaki, M. and Tanaka, T. and Yamamoto, K. and Yokoyama, J.",
    title = "{Quantum state inside a vacuum bubble and creation of an open thermal universe}",
    booktitle = "{37th Yamada Conference: Evolution of the Universe and its Observational Quest}",
    pages = "93--98",
    month = "6",
    year = "1993"
}

@article{Sasaki:1993ha,
    author = "Sasaki, Misao and Tanaka, Takahiro and Yamamoto, Kazuhiro and Yokoyama, Jun'ichi",
    title = "{Quantum state inside a vacuum bubble and creation of an open universe}",
    reportNumber = "YITP-U-93-19, KUNS-1205",
    doi = "10.1016/0370-2693(93)91364-S",
    journal = "Phys. Lett. B",
    volume = "317",
    pages = "510--516",
    year = "1993"
}

@article{Remmen:2013eja,
    author = "Remmen, Grant N. and Carroll, Sean M.",
    title = "{Attractor Solutions in Scalar-Field Cosmology}",
    eprint = "1309.2611",
    archivePrefix = "arXiv",
    primaryClass = "gr-qc",
    reportNumber = "CALT-68-2853",
    doi = "10.1103/PhysRevD.88.083518",
    journal = "Phys. Rev. D",
    volume = "88",
    pages = "083518",
    year = "2013"
}

@article{Belinsky:1985zd,
    author = "Belinsky, V. A. and Khalatnikov, I. M. and Grishchuk, L. P. and Zeldovich, Ya. B.",
    title = "{INFLATIONARY STAGES IN COSMOLOGICAL MODELS WITH A SCALAR FIELD}",
    doi = "10.1016/0370-2693(85)90644-6",
    journal = "Phys. Lett. B",
    volume = "155",
    pages = "232--236",
    year = "1985"
}

@article{Pattison:2018bct,
    author = "Pattison, Chris and Vennin, Vincent and Assadullahi, Hooshyar and Wands, David",
    title = "{The attractive behaviour of ultra-slow-roll inflation}",
    eprint = "1806.09553",
    archivePrefix = "arXiv",
    primaryClass = "astro-ph.CO",
    doi = "10.1088/1475-7516/2018/08/048",
    journal = "JCAP",
    volume = "08",
    pages = "048",
    year = "2018"
}

@article{Motohashi:2014ppa,
    author = "Motohashi, Hayato and Starobinsky, Alexei A. and Yokoyama, Jun'ichi",
    title = "{Inflation with a constant rate of roll}",
    eprint = "1411.5021",
    archivePrefix = "arXiv",
    primaryClass = "astro-ph.CO",
    reportNumber = "RESCEU-51-14",
    doi = "10.1088/1475-7516/2015/09/018",
    journal = "JCAP",
    volume = "09",
    pages = "018",
    year = "2015"
}

@article{Dimopoulos:2017ged,
    author = "Dimopoulos, Konstantinos",
    title = "{Ultra slow-roll inflation demystified}",
    eprint = "1707.05644",
    archivePrefix = "arXiv",
    primaryClass = "hep-ph",
    doi = "10.1016/j.physletb.2017.10.066",
    journal = "Phys. Lett. B",
    volume = "775",
    pages = "262--265",
    year = "2017"
}

@article{Mishra:2023lhe,
    author = "Mishra, Swagat S. and Copeland, Edmund J. and Green, Anne M.",
    title = "{Primordial black holes and stochastic inflation beyond slow roll. Part I. Noise matrix elements}",
    eprint = "2303.17375",
    archivePrefix = "arXiv",
    primaryClass = "astro-ph.CO",
    doi = "10.1088/1475-7516/2023/09/005",
    journal = "JCAP",
    volume = "09",
    pages = "005",
    year = "2023"
}

@inproceedings{Spradlin:2001pw,
    author = "Spradlin, Marcus and Strominger, Andrew and Volovich, Anastasia",
    title = "{Les Houches lectures on de Sitter space}",
    booktitle = "{Les Houches Summer School: Session 76: Euro Summer School on Unity of Fundamental Physics: Gravity, Gauge Theory and Strings}",
    eprint = "hep-th/0110007",
    archivePrefix = "arXiv",
    pages = "423--453",
    month = "10",
    year = "2001"
}

@inproceedings{Bousso:2002fq,
    author = "Bousso, Raphael",
    title = "{Adventures in de Sitter space}",
    booktitle = "{Workshop on Conference on the Future of Theoretical Physics and Cosmology in Honor of Steven Hawking's 60th Birthday}",
    eprint = "hep-th/0205177",
    archivePrefix = "arXiv",
    pages = "539--569",
    month = "5",
    year = "2002"
}

@article{Anninos:2012qw,
    author = "Anninos, Dionysios",
    title = "{De Sitter Musings}",
    eprint = "1205.3855",
    archivePrefix = "arXiv",
    primaryClass = "hep-th",
    doi = "10.1142/S0217751X1230013X",
    journal = "Int. J. Mod. Phys. A",
    volume = "27",
    pages = "1230013",
    year = "2012"
}

@book{Baumann:2014nda,
    author = "Baumann, Daniel and McAllister, Liam",
    title = "{Inflation and String Theory}",
    eprint = "1404.2601",
    archivePrefix = "arXiv",
    primaryClass = "hep-th",
    doi = "10.1017/CBO9781316105733",
    isbn = "978-1-107-08969-3, 978-1-316-23718-2",
    publisher = "Cambridge University Press",
    series = "Cambridge Monographs on Mathematical Physics",
    month = "5",
    year = "2015"
}

@article{Linde:2017pwt,
    author = "Linde, Andrei",
    title = "{On the problem of initial conditions for inflation}",
    eprint = "1710.04278",
    archivePrefix = "arXiv",
    primaryClass = "hep-th",
    doi = "10.1007/s10701-018-0177-9",
    journal = "Found. Phys.",
    volume = "48",
    number = "10",
    pages = "1246--1260",
    year = "2018"
}

@article{Bhatt:2022mmn,
    author = "Bhatt, Siddharth S. and Mishra, Swagat S. and Basak, Soumen and Sahoo, Surya N.",
    title = "{Numerical simulations of inflationary dynamics: slow roll and beyond}",
    eprint = "2212.00529",
    archivePrefix = "arXiv",
    primaryClass = "gr-qc",
    month = "12",
    year = "2022"
}

@article{Kallosh:2022feu,
    author = "Kallosh, Renata and Linde, Andrei",
    title = "{Polynomial \ensuremath{\alpha}-attractors}",
    eprint = "2202.06492",
    archivePrefix = "arXiv",
    primaryClass = "astro-ph.CO",
    doi = "10.1088/1475-7516/2022/04/017",
    journal = "JCAP",
    volume = "04",
    number = "04",
    pages = "017",
    year = "2022"
}

@article{Iacconi:2023mnw,
    author = {Iacconi, Laura and Fasiello, Matteo and V\"aliviita, Jussi and Wands, David},
    title = "{Novel CMB constraints on the \ensuremath{\alpha} parameter in alpha-attractor models}",
    eprint = "2306.00918",
    archivePrefix = "arXiv",
    primaryClass = "astro-ph.CO",
    doi = "10.1088/1475-7516/2023/10/015",
    journal = "JCAP",
    volume = "10",
    pages = "015",
    year = "2023"
}

@article{Roest:2013fha,
    author = "Roest, Diederik",
    title = "{Universality classes of inflation}",
    eprint = "1309.1285",
    archivePrefix = "arXiv",
    primaryClass = "hep-th",
    doi = "10.1088/1475-7516/2014/01/007",
    journal = "JCAP",
    volume = "01",
    pages = "007",
    year = "2014"
}

@article{Mukhanov:2013tua,
    author = "Mukhanov, Viatcheslav",
    title = "{Quantum Cosmological Perturbations: Predictions and Observations}",
    eprint = "1303.3925",
    archivePrefix = "arXiv",
    primaryClass = "astro-ph.CO",
    doi = "10.1140/epjc/s10052-013-2486-7",
    journal = "Eur. Phys. J. C",
    volume = "73",
    pages = "2486",
    year = "2013"
}

@article{Mukhanov:2014uwa,
    author = "Mukhanov, Viatcheslav",
    title = "{Inflation without Selfreproduction}",
    eprint = "1409.2335",
    archivePrefix = "arXiv",
    primaryClass = "astro-ph.CO",
    doi = "10.1002/prop.201400074",
    journal = "Fortsch. Phys.",
    volume = "63",
    pages = "36--41",
    year = "2015"
}

@article{Gobbetti:2015cya,
    author = "Gobbetti, Roberto and Pajer, Enrico and Roest, Diederik",
    title = "{On the Three Primordial Numbers}",
    eprint = "1505.00968",
    archivePrefix = "arXiv",
    primaryClass = "astro-ph.CO",
    doi = "10.1088/1475-7516/2015/09/058",
    journal = "JCAP",
    volume = "09",
    pages = "058",
    year = "2015"
}

@article{Creminelli:2014nqa,
    author = "Creminelli, Paolo and Dubovsky, Sergei and L\'opez Nacir, Diana and Simonovi\'c, Marko and Trevisan, Gabriele and Villadoro, Giovanni and Zaldarriaga, Matias",
    title = "{Implications of the scalar tilt for the tensor-to-scalar ratio}",
    eprint = "1412.0678",
    archivePrefix = "arXiv",
    primaryClass = "astro-ph.CO",
    doi = "10.1103/PhysRevD.92.123528",
    journal = "Phys. Rev. D",
    volume = "92",
    number = "12",
    pages = "123528",
    year = "2015"
}

@article{Martin:2016iqo,
    author = "Martin, Jerome and Ringeval, Christophe and Vennin, Vincent",
    title = "{Shortcomings of New Parametrizations of Inflation}",
    eprint = "1609.04739",
    archivePrefix = "arXiv",
    primaryClass = "astro-ph.CO",
    doi = "10.1103/PhysRevD.94.123521",
    journal = "Phys. Rev. D",
    volume = "94",
    number = "12",
    pages = "123521",
    year = "2016"
}

@article{Bhattacharyya:2024duw,
    author = "Bhattacharyya, Arpan and Brahma, Suddhasattwa and Haque, S. Shajidul and Lund, Jacob S. and Paul, Arpon",
    title = "{The Early Universe as an Open Quantum System: Complexity and Decoherence}",
    eprint = "2401.12134",
    archivePrefix = "arXiv",
    primaryClass = "hep-th",
    month = "1",
    year = "2024"
}

@phdthesis{Colas:2023wxa,
    author = "Colas, Thomas",
    title = "{Open Effective Field Theories for primordial cosmology : dissipation, decoherence and late-time resummation of cosmological inhomogeneities}",
    reportNumber = "tel-04195628, 2023UPASP075",
    school = "Institut d'astrophysique spatiale, France, AstroParticule et Cosmologie, France, APC, Paris",
    year = "2023"
}

@article{Martin:2018zbe,
    author = "Martin, Jerome and Vennin, Vincent",
    title = "{Observational constraints on quantum decoherence during inflation}",
    eprint = "1801.09949",
    archivePrefix = "arXiv",
    primaryClass = "astro-ph.CO",
    doi = "10.1088/1475-7516/2018/05/063",
    journal = "JCAP",
    volume = "05",
    pages = "063",
    year = "2018"
}

@article{Burgess:2017ytm,
    author = "Burgess, C. P.",
    title = "{Intro to Effective Field Theories and Inflation}",
    eprint = "1711.10592",
    archivePrefix = "arXiv",
    primaryClass = "hep-th",
    month = "11",
    year = "2017"
}

@article{Martin:2015qta,
    author = "Martin, Jerome and Vennin, Vincent",
    title = "{Quantum Discord of Cosmic Inflation: Can we Show that CMB Anisotropies are of Quantum-Mechanical Origin?}",
    eprint = "1510.04038",
    archivePrefix = "arXiv",
    primaryClass = "astro-ph.CO",
    doi = "10.1103/PhysRevD.93.023505",
    journal = "Phys. Rev. D",
    volume = "93",
    number = "2",
    pages = "023505",
    year = "2016"
}

@article{Burgess:2014eoa,
    author = "Burgess, C. P. and Holman, R. and Tasinato, G. and Williams, M.",
    title = "{EFT Beyond the Horizon: Stochastic Inflation and How Primordial Quantum Fluctuations Go Classical}",
    eprint = "1408.5002",
    archivePrefix = "arXiv",
    primaryClass = "hep-th",
    reportNumber = "CERN-PH-TH-2014-142",
    doi = "10.1007/JHEP03(2015)090",
    journal = "JHEP",
    volume = "03",
    pages = "090",
    year = "2015"
}

@article{Lochan:2014dca,
    author = "Lochan, Kinjalk and Parattu, Krishnamohan and Padmanabhan, T.",
    title = "{Quantum Evolution Leading to Classicality: A Concrete Example}",
    eprint = "1404.2605",
    archivePrefix = "arXiv",
    primaryClass = "gr-qc",
    doi = "10.1007/s10714-014-1841-9",
    journal = "Gen. Rel. Grav.",
    volume = "47",
    number = "1",
    pages = "1841",
    year = "2015"
}

@article{Martin:2012ua,
    author = "Martin, Jerome",
    editor = "Pal, Supratik and Basu, Banasri",
    title = "{The Quantum State of Inflationary Perturbations}",
    eprint = "1209.3092",
    archivePrefix = "arXiv",
    primaryClass = "hep-th",
    doi = "10.1088/1742-6596/405/1/012004",
    journal = "J. Phys. Conf. Ser.",
    volume = "405",
    pages = "012004",
    year = "2012"
}

@article{Kiefer:2008ku,
    author = "Kiefer, Claus and Polarski, David",
    title = "{Why do cosmological perturbations look classical to us?}",
    eprint = "0810.0087",
    archivePrefix = "arXiv",
    primaryClass = "astro-ph",
    doi = "10.1166/asl.2009.1023",
    journal = "Adv. Sci. Lett.",
    volume = "2",
    pages = "164--173",
    year = "2009"
}

@article{Kiefer:2007zza,
    author = "Kiefer, Claus and Lohmar, Ingo and Polarski, David and Starobinsky, Alexei A.",
    editor = "Diosi, Lajos and Elze, Hans-Thomas and Vitiello, Giuseppe",
    title = "{Origin of classical structure in the Universe}",
    doi = "10.1088/1742-6596/67/1/012023",
    journal = "J. Phys. Conf. Ser.",
    volume = "67",
    pages = "012023",
    year = "2007"
}

@article{Burgess:2006jn,
    author = "Burgess, Cliff P. and Holman, R. and Hoover, D.",
    title = "{Decoherence of inflationary primordial fluctuations}",
    eprint = "astro-ph/0601646",
    archivePrefix = "arXiv",
    reportNumber = "MCMASTER-06-01",
    doi = "10.1103/PhysRevD.77.063534",
    journal = "Phys. Rev. D",
    volume = "77",
    pages = "063534",
    year = "2008"
}

@article{Kiefer:1998pb,
    author = "Kiefer, Claus and Lesgourgues, Julien and Polarski, David and Starobinsky, Alexei A.",
    title = "{The Coherence of primordial fluctuations produced during inflation}",
    eprint = "gr-qc/9806066",
    archivePrefix = "arXiv",
    reportNumber = "LMPT-06-98",
    doi = "10.1088/0264-9381/15/10/002",
    journal = "Class. Quant. Grav.",
    volume = "15",
    pages = "L67--L72",
    year = "1998"
}

@article{Kiefer:1998jk,
    author = "Kiefer, Claus and Polarski, David",
    title = "{Emergence of classicality for primordial fluctuations: Concepts and analogies}",
    eprint = "gr-qc/9805014",
    archivePrefix = "arXiv",
    reportNumber = "FREIBURG-THEP-98-7, FREIBURG-THP-98-7",
    doi = "10.1002/andp.2090070302",
    journal = "Annalen Phys.",
    volume = "7",
    pages = "137--158",
    year = "1998"
}

@article{Kiefer:1998qe,
    author = "Kiefer, Claus and Polarski, David and Starobinsky, Alexei A.",
    title = "{Quantum to classical transition for fluctuations in the early universe}",
    eprint = "gr-qc/9802003",
    archivePrefix = "arXiv",
    reportNumber = "THEP-97-33, FREIBURG-THEP-97-33",
    doi = "10.1142/S0218271898000292",
    journal = "Int. J. Mod. Phys. D",
    volume = "7",
    pages = "455--462",
    year = "1998"
}

@article{Lucchin:1984yf,
    author = "Lucchin, F. and Matarrese, S.",
    title = "{Power Law Inflation}",
    reportNumber = "SISSA-82/84/EP",
    doi = "10.1103/PhysRevD.32.1316",
    journal = "Phys. Rev. D",
    volume = "32",
    pages = "1316",
    year = "1985"
}

@article{Bag:2017vjp,
    author = "Bag, Satadru and Mishra, Swagat S. and Sahni, Varun",
    title = "{New tracker models of dark energy}",
    eprint = "1709.09193",
    archivePrefix = "arXiv",
    primaryClass = "gr-qc",
    doi = "10.1088/1475-7516/2018/08/009",
    journal = "JCAP",
    volume = "08",
    pages = "009",
    year = "2018"
}

@article{Copeland:1997et,
    author = "Copeland, Edmund J. and Liddle, Andrew R and Wands, David",
    title = "{Exponential potentials and cosmological scaling solutions}",
    eprint = "gr-qc/9711068",
    archivePrefix = "arXiv",
    reportNumber = "SUSX-TH-97-022, SUSSEX-AST-97-11-1, PU-RCG-97-20",
    doi = "10.1103/PhysRevD.57.4686",
    journal = "Phys. Rev. D",
    volume = "57",
    pages = "4686--4690",
    year = "1998"
}

@article{Barreiro:1999zs,
    author = "Barreiro, T. and Copeland, Edmund J. and Nunes, N. J.",
    title = "{Quintessence arising from exponential potentials}",
    eprint = "astro-ph/9910214",
    archivePrefix = "arXiv",
    reportNumber = "SUSX-TH-016",
    doi = "10.1103/PhysRevD.61.127301",
    journal = "Phys. Rev. D",
    volume = "61",
    pages = "127301",
    year = "2000"
}

@article{Apers:2024ffe,
    author = "Apers, Fien and Conlon, Joseph P. and Copeland, Edmund J. and Mosny, Martin and Revello, Filippo",
    title = "{String Theory and the First Half of the Universe}",
    eprint = "2401.04064",
    archivePrefix = "arXiv",
    primaryClass = "hep-th",
    month = "1",
    year = "2024"
}

@article{Tegmark:2004qd,
    author = "Tegmark, Max",
    title = "{What does inflation really predict?}",
    eprint = "astro-ph/0410281",
    archivePrefix = "arXiv",
    doi = "10.1088/1475-7516/2005/04/001",
    journal = "JCAP",
    volume = "04",
    pages = "001",
    year = "2005"
}

@article{Kamionkowski:1996zd,
    author = "Kamionkowski, Marc and Kosowsky, Arthur and Stebbins, Albert",
    title = "{A Probe of primordial gravity waves and vorticity}",
    eprint = "astro-ph/9609132",
    archivePrefix = "arXiv",
    reportNumber = "CU-TP-767, CAL-615, FERMILAB-PUB-96-327-A",
    doi = "10.1103/PhysRevLett.78.2058",
    journal = "Phys. Rev. Lett.",
    volume = "78",
    pages = "2058--2061",
    year = "1997"
}

@article{Seljak:1996gy,
    author = "Seljak, Uros and Zaldarriaga, Matias",
    title = "{Signature of gravity waves in polarization of the microwave background}",
    eprint = "astro-ph/9609169",
    archivePrefix = "arXiv",
    doi = "10.1103/PhysRevLett.78.2054",
    journal = "Phys. Rev. Lett.",
    volume = "78",
    pages = "2054--2057",
    year = "1997"
}

@article{Kamionkowski:1996ks,
    author = "Kamionkowski, Marc and Kosowsky, Arthur and Stebbins, Albert",
    title = "{Statistics of cosmic microwave background polarization}",
    eprint = "astro-ph/9611125",
    archivePrefix = "arXiv",
    reportNumber = "FERMILAB-PUB-96-426-A, CU-TP-787, CAL-617",
    doi = "10.1103/PhysRevD.55.7368",
    journal = "Phys. Rev. D",
    volume = "55",
    pages = "7368--7388",
    year = "1997"
}

@article{Planck:2019kim,
    author = "Akrami, Y. and others",
    collaboration = "Planck",
    title = "{Planck 2018 results. IX. Constraints on primordial non-Gaussianity}",
    eprint = "1905.05697",
    archivePrefix = "arXiv",
    primaryClass = "astro-ph.CO",
    doi = "10.1051/0004-6361/201935891",
    journal = "Astron. Astrophys.",
    volume = "641",
    pages = "A9",
    year = "2020"
}

@article{Celoria:2021vjw,
    author = "Celoria, Marco and Creminelli, Paolo and Tambalo, Giovanni and Yingcharoenrat, Vicharit",
    title = "{Beyond perturbation theory in inflation}",
    eprint = "2103.09244",
    archivePrefix = "arXiv",
    primaryClass = "hep-th",
    doi = "10.1088/1475-7516/2021/06/051",
    journal = "JCAP",
    volume = "06",
    pages = "051",
    year = "2021"
}

@article{Borde:2001nh,
    author = "Borde, Arvind and Guth, Alan H. and Vilenkin, Alexander",
    title = "{Inflationary space-times are incompletein past directions}",
    eprint = "gr-qc/0110012",
    archivePrefix = "arXiv",
    reportNumber = "MIT-CTP-3183",
    doi = "10.1103/PhysRevLett.90.151301",
    journal = "Phys. Rev. Lett.",
    volume = "90",
    pages = "151301",
    year = "2003"
}

@article{Cook:2015vqa,
    author = "Cook, Jessica L. and Dimastrogiovanni, Emanuela and Easson, Damien A. and Krauss, Lawrence M.",
    title = "{Reheating predictions in single field inflation}",
    eprint = "1502.04673",
    archivePrefix = "arXiv",
    primaryClass = "astro-ph.CO",
    doi = "10.1088/1475-7516/2015/04/047",
    journal = "JCAP",
    volume = "04",
    pages = "047",
    year = "2015"
}

@article{Dai:2014jja,
    author = "Dai, Liang and Kamionkowski, Marc and Wang, Junpu",
    title = "{Reheating constraints to inflationary models}",
    eprint = "1404.6704",
    archivePrefix = "arXiv",
    primaryClass = "astro-ph.CO",
    doi = "10.1103/PhysRevLett.113.041302",
    journal = "Phys. Rev. Lett.",
    volume = "113",
    pages = "041302",
    year = "2014"
}

@article{Munoz:2014eqa,
    author = "Munoz, Julian B. and Kamionkowski, Marc",
    title = "{Equation-of-State Parameter for Reheating}",
    eprint = "1412.0656",
    archivePrefix = "arXiv",
    primaryClass = "astro-ph.CO",
    doi = "10.1103/PhysRevD.91.043521",
    journal = "Phys. Rev. D",
    volume = "91",
    number = "4",
    pages = "043521",
    year = "2015"
}

@article{Figueroa:2019paj,
    author = "Figueroa, Daniel G. and Tanin, Erwin H.",
    title = "{Ability of LIGO and LISA to probe the equation of state of the early Universe}",
    eprint = "1905.11960",
    archivePrefix = "arXiv",
    primaryClass = "astro-ph.CO",
    doi = "10.1088/1475-7516/2019/08/011",
    journal = "JCAP",
    volume = "08",
    pages = "011",
    year = "2019"
}

@article{Sahni:2001qp,
    author = "Sahni, Varun and Sami, M. and Souradeep, Tarun",
    title = "{Relic gravity waves from brane world inflation}",
    eprint = "gr-qc/0105121",
    archivePrefix = "arXiv",
    doi = "10.1103/PhysRevD.65.023518",
    journal = "Phys. Rev. D",
    volume = "65",
    pages = "023518",
    year = "2002"
}

@article{Ema:2020ggo,
    author = "Ema, Yohei and Jinno, Ryusuke and Nakayama, Kazunori",
    title = "{High-frequency Graviton from Inflaton Oscillation}",
    eprint = "2006.09972",
    archivePrefix = "arXiv",
    primaryClass = "astro-ph.CO",
    reportNumber = "DESY-20-104",
    doi = "10.1088/1475-7516/2020/09/015",
    journal = "JCAP",
    volume = "09",
    pages = "015",
    year = "2020"
}

@article{Creminelli:2014fca,
    author = "Creminelli, Paolo and L\'opez Nacir, Diana and Simonovi\'c, Marko and Trevisan, Gabriele and Zaldarriaga, Matias",
    title = "{$\phi^2$ Inflation at its Endpoint}",
    eprint = "1405.6264",
    archivePrefix = "arXiv",
    primaryClass = "astro-ph.CO",
    doi = "10.1103/PhysRevD.90.083513",
    journal = "Phys. Rev. D",
    volume = "90",
    number = "8",
    pages = "083513",
    year = "2014"
}

@article{Boyle_2008,
   title={Probing the early universe with inflationary gravitational waves},
   volume={77},
   ISSN={1550-2368},
   url={http://dx.doi.org/10.1103/PhysRevD.77.063504},
   DOI={10.1103/physrevd.77.063504},
   number={6},
   journal={Physical Review D},
   publisher={American Physical Society (APS)},
   author={Boyle, Latham A. and Steinhardt, Paul J.},
   year={2008},
   month=mar }

@article{LIGOScientific:2003jxj,
    author = "Abbott, B. and others",
    collaboration = "LIGO Scientific",
    title = "{Analysis of first LIGO science data for stochastic gravitational waves}",
    eprint = "gr-qc/0312088",
    archivePrefix = "arXiv",
    reportNumber = "FERMILAB-PUB-03-416",
    doi = "10.1103/PhysRevD.69.122004",
    journal = "Phys. Rev. D",
    volume = "69",
    pages = "122004",
    year = "2004"
}

@article{LISACosmologyWorkingGroup:2022jok,
    author = "Auclair, Pierre and others",
    collaboration = "LISA Cosmology Working Group",
    title = "{Cosmology with the Laser Interferometer Space Antenna}",
    eprint = "2204.05434",
    archivePrefix = "arXiv",
    primaryClass = "astro-ph.CO",
    reportNumber = "LISA CosWG-22-03, FERMILAB-PUB-22-349-SCD",
    doi = "10.1007/s41114-023-00045-2",
    journal = "Living Rev. Rel.",
    volume = "26",
    number = "1",
    pages = "5",
    year = "2023"
}

@article{Harry:2006fi,
    author = "Harry, G. M. and Fritschel, P. and Shaddock, D. A. and Folkner, W. and Phinney, E. S.",
    title = "{Laser interferometry for the big bang observer}",
    doi = "10.1088/0264-9381/23/15/008",
    journal = "Class. Quant. Grav.",
    volume = "23",
    pages = "4887--4894",
    year = "2006",
    note = "[Erratum: Class.Quant.Grav. 23, 7361 (2006)]"
}

@article{NANOGrav:2023hvm,
    author = "Afzal, Adeela and others",
    collaboration = "NANOGrav",
    title = "{The NANOGrav 15 yr Data Set: Search for Signals from New Physics}",
    eprint = "2306.16219",
    archivePrefix = "arXiv",
    primaryClass = "astro-ph.HE",
    reportNumber = "FERMILAB-PUB-23-589-T",
    doi = "10.3847/2041-8213/acdc91",
    journal = "Astrophys. J. Lett.",
    volume = "951",
    number = "1",
    pages = "L11",
    year = "2023"
}

@book{Dodelson:2003ft,
    author = "Dodelson, Scott",
    title = "{Modern Cosmology}",
    isbn = "978-0-12-219141-1",
    publisher = "Academic Press",
    address = "Amsterdam",
    year = "2003"
}

@article{Weinberg:2005vy,
    author = "Weinberg, Steven",
    title = "{Quantum contributions to cosmological correlations}",
    eprint = "hep-th/0506236",
    archivePrefix = "arXiv",
    reportNumber = "UTTG-01-05",
    doi = "10.1103/PhysRevD.72.043514",
    journal = "Phys. Rev. D",
    volume = "72",
    pages = "043514",
    year = "2005"
}

@article{Giddings:2010ui,
    author = "Giddings, Steven B. and Sloth, Martin S.",
    title = "{Cosmological diagrammatic rules}",
    eprint = "1005.3287",
    archivePrefix = "arXiv",
    primaryClass = "hep-th",
    reportNumber = "CERN-PH-TH-2010-108",
    doi = "10.1088/1475-7516/2010/07/015",
    journal = "JCAP",
    volume = "07",
    pages = "015",
    year = "2010"
}

@article{Arkani-Hamed:2017fdk,
    author = "Arkani-Hamed, Nima and Benincasa, Paolo and Postnikov, Alexander",
    title = "{Cosmological Polytopes and the Wavefunction of the Universe}",
    eprint = "1709.02813",
    archivePrefix = "arXiv",
    primaryClass = "hep-th",
    month = "9",
    year = "2017"
}

@article{Benincasa:2022gtd,
    author = "Benincasa, Paolo",
    title = "{Amplitudes meet Cosmology: A (Scalar) Primer}",
    eprint = "2203.15330",
    archivePrefix = "arXiv",
    primaryClass = "hep-th",
    doi = "10.1142/S0217751X22300101",
    month = "3",
    year = "2022"
}

@inproceedings{Baumann:2022jpr,
    author = "Baumann, Daniel and Green, Daniel and Joyce, Austin and Pajer, Enrico and Pimentel, Guilherme L. and Sleight, Charlotte and Taronna, Massimo",
    title = "{Snowmass White Paper: The Cosmological Bootstrap}",
    booktitle = "{Snowmass 2021}",
    eprint = "2203.08121",
    archivePrefix = "arXiv",
    primaryClass = "hep-th",
    month = "3",
    year = "2022"
}

@article{Lyth:1998xn,
    author = "Lyth, David H. and Riotto, Antonio",
    title = "{Particle physics models of inflation and the cosmological density perturbation}",
    eprint = "hep-ph/9807278",
    archivePrefix = "arXiv",
    reportNumber = "LANCS-TH-9720, FERMILAB-PUB-97-292-A, CERN-TH-97-383, OUTP-98-39-P",
    doi = "10.1016/S0370-1573(98)00128-8",
    journal = "Phys. Rept.",
    volume = "314",
    pages = "1--146",
    year = "1999"
}

@article{Adshead:2009cb,
    author = "Adshead, Peter and Easther, Richard and Lim, Eugene A.",
    title = "{The 'in-in' Formalism and Cosmological Perturbations}",
    eprint = "0904.4207",
    archivePrefix = "arXiv",
    primaryClass = "hep-th",
    doi = "10.1103/PhysRevD.80.083521",
    journal = "Phys. Rev. D",
    volume = "80",
    pages = "083521",
    year = "2009"
}

@article{Maldacena:2011nz,
    author = "Maldacena, Juan M. and Pimentel, Guilherme L.",
    title = "{On graviton non-Gaussianities during inflation}",
    eprint = "1104.2846",
    archivePrefix = "arXiv",
    primaryClass = "hep-th",
    reportNumber = "PUPT-2371",
    doi = "10.1007/JHEP09(2011)045",
    journal = "JHEP",
    volume = "09",
    pages = "045",
    year = "2011"
}

@article{Arkani-Hamed:2018kmz,
    author = "Arkani-Hamed, Nima and Baumann, Daniel and Lee, Hayden and Pimentel, Guilherme L.",
    title = "{The Cosmological Bootstrap: Inflationary Correlators from Symmetries and Singularities}",
    eprint = "1811.00024",
    archivePrefix = "arXiv",
    primaryClass = "hep-th",
    doi = "10.1007/JHEP04(2020)105",
    journal = "JHEP",
    volume = "04",
    pages = "105",
    year = "2020"
}

@article{Baumann:2019oyu,
    author = "Baumann, Daniel and Duaso Pueyo, Carlos and Joyce, Austin and Lee, Hayden and Pimentel, Guilherme L.",
    title = "{The cosmological bootstrap: weight-shifting operators and scalar seeds}",
    eprint = "1910.14051",
    archivePrefix = "arXiv",
    primaryClass = "hep-th",
    doi = "10.1007/JHEP12(2020)204",
    journal = "JHEP",
    volume = "12",
    pages = "204",
    year = "2020"
}

@article{Salcedo:2022aal,
    author = "Salcedo, Santiago Agui and Lee, Mang Hei Gordon and Melville, Scott and Pajer, Enrico",
    title = "{The Analytic Wavefunction}",
    eprint = "2212.08009",
    archivePrefix = "arXiv",
    primaryClass = "hep-th",
    doi = "10.1007/JHEP06(2023)020",
    journal = "JHEP",
    volume = "06",
    pages = "020",
    year = "2023"
}

@article{Goodhew:2020hob,
    author = "Goodhew, Harry and Jazayeri, Sadra and Pajer, Enrico",
    title = "{The Cosmological Optical Theorem}",
    eprint = "2009.02898",
    archivePrefix = "arXiv",
    primaryClass = "hep-th",
    doi = "10.1088/1475-7516/2021/04/021",
    journal = "JCAP",
    volume = "04",
    pages = "021",
    year = "2021"
}

@article{Goodhew:2021oqg,
    author = "Goodhew, Harry and Jazayeri, Sadra and Lee, Mang Hei Gordon and Pajer, Enrico",
    title = "{Cutting cosmological correlators}",
    eprint = "2104.06587",
    archivePrefix = "arXiv",
    primaryClass = "hep-th",
    doi = "10.1088/1475-7516/2021/08/003",
    journal = "JCAP",
    volume = "08",
    pages = "003",
    year = "2021"
}

@article{Melville:2021lst,
    author = "Melville, Scott and Pajer, Enrico",
    title = "{Cosmological Cutting Rules}",
    eprint = "2103.09832",
    archivePrefix = "arXiv",
    primaryClass = "hep-th",
    doi = "10.1007/JHEP05(2021)249",
    journal = "JHEP",
    volume = "05",
    pages = "249",
    year = "2021"
}

@article{Stefanyszyn:2023qov,
    author = "Stefanyszyn, David and Tong, Xi and Zhu, Yuhang",
    title = "{Cosmological Correlators Through the Looking Glass: Reality, Parity, and Factorisation}",
    eprint = "2309.07769",
    archivePrefix = "arXiv",
    primaryClass = "hep-th",
    month = "9",
    year = "2023"
}

@article{Jazayeri:2021fvk,
    author = "Jazayeri, Sadra and Pajer, Enrico and Stefanyszyn, David",
    title = "{From locality and unitarity to cosmological correlators}",
    eprint = "2103.08649",
    archivePrefix = "arXiv",
    primaryClass = "hep-th",
    doi = "10.1007/JHEP10(2021)065",
    journal = "JHEP",
    volume = "10",
    pages = "065",
    year = "2021"
}

@article{Hillman:2021bnk,
    author = "Hillman, Aaron and Pajer, Enrico",
    title = "{A differential representation of cosmological wavefunctions}",
    eprint = "2112.01619",
    archivePrefix = "arXiv",
    primaryClass = "hep-th",
    doi = "10.1007/JHEP04(2022)012",
    journal = "JHEP",
    volume = "04",
    pages = "012",
    year = "2022"
}

@article{Seery:2005wm,
    author = "Seery, David and Lidsey, James E.",
    title = "{Primordial non-Gaussianities in single field inflation}",
    eprint = "astro-ph/0503692",
    archivePrefix = "arXiv",
    doi = "10.1088/1475-7516/2005/06/003",
    journal = "JCAP",
    volume = "06",
    pages = "003",
    year = "2005"
}

@article{Seery:2005gb,
    author = "Seery, David and Lidsey, James E.",
    title = "{Primordial non-Gaussianities from multiple-field inflation}",
    eprint = "astro-ph/0506056",
    archivePrefix = "arXiv",
    doi = "10.1088/1475-7516/2005/09/011",
    journal = "JCAP",
    volume = "09",
    pages = "011",
    year = "2005"
}

@article{Komatsu:2009kd,
    author = "Komatsu, E. and others",
    title = "{Non-Gaussianity as a Probe of the Physics of the Primordial Universe and the Astrophysics of the Low Redshift Universe}",
    eprint = "0902.4759",
    archivePrefix = "arXiv",
    primaryClass = "astro-ph.CO",
    reportNumber = "SLAC-PUB-14737",
    month = "2",
    year = "2009"
}

@article{Seery:2006vu,
    author = "Seery, David and Lidsey, James E. and Sloth, Martin S.",
    title = "{The inflationary trispectrum}",
    eprint = "astro-ph/0610210",
    archivePrefix = "arXiv",
    doi = "10.1088/1475-7516/2007/01/027",
    journal = "JCAP",
    volume = "01",
    pages = "027",
    year = "2007"
}

@article{Seery:2006js,
    author = "Seery, David and Lidsey, James E.",
    title = "{Non-Gaussianity from the inflationary trispectrum}",
    eprint = "astro-ph/0611034",
    archivePrefix = "arXiv",
    doi = "10.1088/1475-7516/2007/01/008",
    journal = "JCAP",
    volume = "01",
    pages = "008",
    year = "2007"
}

@article{Seery:2008qj,
    author = "Seery, David and Malik, Karim A. and Lyth, David H.",
    title = "{Non-gaussianity of inflationary field perturbations from the field equation}",
    eprint = "0802.0588",
    archivePrefix = "arXiv",
    primaryClass = "astro-ph",
    doi = "10.1088/1475-7516/2008/03/014",
    journal = "JCAP",
    volume = "03",
    pages = "014",
    year = "2008"
}

@article{Lyth:2006qz,
    author = "Lyth, David H. and Seery, David",
    title = "{Classicality of the primordial perturbations}",
    eprint = "astro-ph/0607647",
    archivePrefix = "arXiv",
    doi = "10.1016/j.physletb.2008.03.010",
    journal = "Phys. Lett. B",
    volume = "662",
    pages = "309--313",
    year = "2008"
}

@article{Sugiyama:2012tj,
    author = "Sugiyama, Naonori S. and Komatsu, Eiichiro and Futamase, Toshifumi",
    title = "{$\delta$N formalism}",
    eprint = "1208.1073",
    archivePrefix = "arXiv",
    primaryClass = "gr-qc",
    doi = "10.1103/PhysRevD.87.023530",
    journal = "Phys. Rev. D",
    volume = "87",
    number = "2",
    pages = "023530",
    year = "2013"
}

@article{Abolhasani:2018gyz,
    author = "Abolhasani, Ali Akbar and Sasaki, Misao",
    title = "{Single-field consistency relation and $\delta N$-formalism}",
    eprint = "1805.11298",
    archivePrefix = "arXiv",
    primaryClass = "astro-ph.CO",
    reportNumber = "IPMU18-0098, YITP-18-57",
    doi = "10.1088/1475-7516/2018/08/025",
    journal = "JCAP",
    volume = "08",
    pages = "025",
    year = "2018"
}

@article{Wands:2010af,
    author = "Wands, David",
    title = "{Local non-Gaussianity from inflation}",
    eprint = "1004.0818",
    archivePrefix = "arXiv",
    primaryClass = "astro-ph.CO",
    reportNumber = "YITP-10-25",
    doi = "10.1088/0264-9381/27/12/124002",
    journal = "Class. Quant. Grav.",
    volume = "27",
    pages = "124002",
    year = "2010"
}

@article{Bardeen:1980kt,
    author = "Bardeen, James M.",
    title = "{Gauge Invariant Cosmological Perturbations}",
    doi = "10.1103/PhysRevD.22.1882",
    journal = "Phys. Rev. D",
    volume = "22",
    pages = "1882--1905",
    year = "1980"
}

@article{Langlois:2008vk,
    author = "Langlois, David and Vernizzi, Filippo and Wands, David",
    title = "{Non-linear isocurvature perturbations and non-Gaussianities}",
    eprint = "0809.4646",
    archivePrefix = "arXiv",
    primaryClass = "astro-ph",
    doi = "10.1088/1475-7516/2008/12/004",
    journal = "JCAP",
    volume = "12",
    pages = "004",
    year = "2008"
}

@article{Tzavara:2010ge,
    author = "Tzavara, Eleftheria and van Tent, Bartjan",
    title = "{Bispectra from two-field inflation using the long-wavelength formalism}",
    eprint = "1012.6027",
    archivePrefix = "arXiv",
    primaryClass = "astro-ph.CO",
    reportNumber = "LPT-10-108",
    doi = "10.1088/1475-7516/2011/06/026",
    journal = "JCAP",
    volume = "06",
    pages = "026",
    year = "2011"
}

@article{Cheung:2007st,
    author = "Cheung, Clifford and Creminelli, Paolo and Fitzpatrick, A. Liam and Kaplan, Jared and Senatore, Leonardo",
    title = "{The Effective Field Theory of Inflation}",
    eprint = "0709.0293",
    archivePrefix = "arXiv",
    primaryClass = "hep-th",
    reportNumber = "IC-2007-032",
    doi = "10.1088/1126-6708/2008/03/014",
    journal = "JHEP",
    volume = "03",
    pages = "014",
    year = "2008"
}

@article{Weinberg:2008hq,
    author = "Weinberg, Steven",
    title = "{Effective Field Theory for Inflation}",
    eprint = "0804.4291",
    archivePrefix = "arXiv",
    primaryClass = "hep-th",
    reportNumber = "UTTG-01-08",
    doi = "10.1103/PhysRevD.77.123541",
    journal = "Phys. Rev. D",
    volume = "77",
    pages = "123541",
    year = "2008"
}

@inproceedings{Senatore:2016aui,
    author = "Senatore, Leonardo",
    title = "{Lectures on Inflation}",
    booktitle = "{Theoretical Advanced Study Institute in Elementary Particle Physics}: {New Frontiers in Fields and Strings}",
    eprint = "1609.00716",
    archivePrefix = "arXiv",
    primaryClass = "hep-th",
    doi = "10.1142/9789813149441_0008",
    pages = "447--543",
    year = "2017"
}

@article{Hu:2008hd,
    author = "Hu, Wayne",
    title = "{Lecture Notes on CMB Theory: From Nucleosynthesis to Recombination}",
    eprint = "0802.3688",
    archivePrefix = "arXiv",
    primaryClass = "astro-ph",
    month = "2",
    year = "2008"
}

@inproceedings{Challinor_2009,
   title={Lecture notes on the physics of cosmic microwave background anisotropies},
   url={http://dx.doi.org/10.1063/1.3151849},
   DOI={10.1063/1.3151849},
   booktitle={AIP Conference Proceedings},
   publisher={AIP},
   author={Challinor, Anthony and Peiris, Hiranya and Novello, Mario and Perez, Santiago},
   year={2009} }

@article{Bernardeau:2001qr,
    author = "Bernardeau, F. and Colombi, S. and Gaztanaga, E. and Scoccimarro, R.",
    title = "{Large scale structure of the universe and cosmological perturbation theory}",
    eprint = "astro-ph/0112551",
    archivePrefix = "arXiv",
    reportNumber = "SACLAY-T01-142",
    doi = "10.1016/S0370-1573(02)00135-7",
    journal = "Phys. Rept.",
    volume = "367",
    pages = "1--248",
    year = "2002"
}

@article{Blumenthal:1984bp,
    author = "Blumenthal, George R. and Faber, S. M. and Primack, Joel R. and Rees, Martin J.",
    editor = "Srednicki, M. A.",
    title = "{Formation of Galaxies and Large Scale Structure with Cold Dark Matter}",
    reportNumber = "SLAC-PUB-3307",
    doi = "10.1038/311517a0",
    journal = "Nature",
    volume = "311",
    pages = "517--525",
    year = "1984"
}

@article{Bond:1984fp,
    author = "Bond, J. R. and Efstathiou, G.",
    title = "{Cosmic background radiation anisotropies in universes dominated by nonbaryonic dark matter}",
    doi = "10.1086/184362",
    journal = "Astrophys. J. Lett.",
    volume = "285",
    pages = "L45--L48",
    year = "1984"
}

@article{Pritchard:2008da,
    author = "Pritchard, Jonathan R. and Loeb, Abraham",
    title = "{Evolution of the 21 cm signal throughout cosmic history}",
    eprint = "0802.2102",
    archivePrefix = "arXiv",
    primaryClass = "astro-ph",
    doi = "10.1103/PhysRevD.78.103511",
    journal = "Phys. Rev. D",
    volume = "78",
    pages = "103511",
    year = "2008"
}

@article{Furlanetto:2019jso,
    author = "Furlanetto, Steven and others",
    title = "{Astro 2020 Science White Paper: Fundamental Cosmology in the Dark Ages with 21-cm Line Fluctuations}",
    eprint = "1903.06212",
    archivePrefix = "arXiv",
    primaryClass = "astro-ph.CO",
    month = "3",
    year = "2019"
}

@article{Starobinsky:1982mr,
    author = "Starobinsky, Alexei A.",
    title = "{Isotropization of arbitrary cosmological expansion given an effective cosmological constant}",
    journal = "JETP Lett.",
    volume = "37",
    pages = "66--69",
    year = "1983"
}

@article{Jensen:1986nf,
    author = "Jensen, Lars Gerhard and Stein-Schabes, Jaime A.",
    title = "{Is Inflation Natural?}",
    reportNumber = "FERMILAB-PUB-86-116-A",
    doi = "10.1103/PhysRevD.35.1146",
    journal = "Phys. Rev. D",
    volume = "35",
    pages = "1146",
    year = "1987"
}

@article{Kitada:1991ih,
    author = "Kitada, Yuichi and Maeda, Kei-ichi",
    title = "{Cosmic no hair theorem in power law inflation}",
    reportNumber = "WU-AP-12-91, UTAP-130-91",
    doi = "10.1103/PhysRevD.45.1416",
    journal = "Phys. Rev. D",
    volume = "45",
    pages = "1416--1419",
    year = "1992"
}

@article{Maleknejad:2012as,
    author = "Maleknejad, A. and Sheikh-Jabbari, M. M.",
    title = "{Revisiting Cosmic No-Hair Theorem for Inflationary Settings}",
    eprint = "1203.0219",
    archivePrefix = "arXiv",
    primaryClass = "hep-th",
    reportNumber = "IPM-P-2012-006",
    doi = "10.1103/PhysRevD.85.123508",
    journal = "Phys. Rev. D",
    volume = "85",
    pages = "123508",
    year = "2012"
}

@article{Goldwirth:1991rj,
    author = "Goldwirth, Dalia S. and Piran, Tsvi",
    title = "{Initial conditions for inflation}",
    reportNumber = "CFA-3336",
    doi = "10.1016/0370-1573(92)90073-9",
    journal = "Phys. Rept.",
    volume = "214",
    pages = "223--291",
    year = "1992"
}

@article{Linde:1983mx,
    author = "Linde, Andrei D.",
    title = "{Quantum Creation of the Inflationary Universe}",
    reportNumber = "LEBEDEV-84-8",
    doi = "10.1007/BF02790571",
    journal = "Lett. Nuovo Cim.",
    volume = "39",
    pages = "401--405",
    year = "1984"
}

@article{Albrecht:1986pi,
    author = "Albrecht, Andreas and Brandenberger, Robert H. and Matzner, Richard",
    title = "{Inflation With Generalized Initial Conditions}",
    reportNumber = "PRINT-86-1223 (CAMBRIDGE)",
    doi = "10.1103/PhysRevD.35.429",
    journal = "Phys. Rev. D",
    volume = "35",
    pages = "429",
    year = "1987"
}

@article{Easther:2014zga,
    author = "Easther, Richard and Price, Layne C. and Rasero, Javier",
    title = "{Inflating an Inhomogeneous Universe}",
    eprint = "1406.2869",
    archivePrefix = "arXiv",
    primaryClass = "astro-ph.CO",
    doi = "10.1088/1475-7516/2014/08/041",
    journal = "JCAP",
    volume = "08",
    pages = "041",
    year = "2014"
}

@article{Goldwirth:1989pr,
    author = "Goldwirth, Dalia S. and Piran, Tsvi",
    title = "{Inhomogeneity and the Onset of Inflation}",
    reportNumber = "HEBREW-2",
    doi = "10.1103/PhysRevLett.64.2852",
    journal = "Phys. Rev. Lett.",
    volume = "64",
    pages = "2852--2855",
    year = "1990"
}

@article{East:2015ggf,
    author = "East, William E. and Kleban, Matthew and Linde, Andrei and Senatore, Leonardo",
    title = "{Beginning inflation in an inhomogeneous universe}",
    eprint = "1511.05143",
    archivePrefix = "arXiv",
    primaryClass = "hep-th",
    doi = "10.1088/1475-7516/2016/09/010",
    journal = "JCAP",
    volume = "09",
    pages = "010",
    year = "2016"
}

@article{Clough:2016ymm,
    author = "Clough, Katy and Lim, Eugene A. and DiNunno, Brandon S. and Fischler, Willy and Flauger, Raphael and Paban, Sonia",
    title = "{Robustness of Inflation to Inhomogeneous Initial Conditions}",
    eprint = "1608.04408",
    archivePrefix = "arXiv",
    primaryClass = "hep-th",
    reportNumber = "KCL-PH-TH-2016-53",
    doi = "10.1088/1475-7516/2017/09/025",
    journal = "JCAP",
    volume = "09",
    pages = "025",
    year = "2017"
}

@article{Hawking:1971ei,
    author = "Hawking, Stephen",
    title = "{Gravitationally collapsed objects of very low mass}",
    doi = "10.1093/mnras/152.1.75",
    journal = "Mon. Not. Roy. Astron. Soc.",
    volume = "152",
    pages = "75",
    year = "1971"
}

@article{Carr:1974nx,
    author = "Carr, Bernard J. and Hawking, S. W.",
    title = "{Black holes in the early Universe}",
    doi = "10.1093/mnras/168.2.399",
    journal = "Mon. Not. Roy. Astron. Soc.",
    volume = "168",
    pages = "399--415",
    year = "1974"
}

@article{Carr:1975qj,
    author = "Carr, Bernard J.",
    title = "{The Primordial black hole mass spectrum}",
    doi = "10.1086/153853",
    journal = "Astrophys. J.",
    volume = "201",
    pages = "1--19",
    year = "1975"
}

@article{Sasaki:2018dmp,
    author = "Sasaki, Misao and Suyama, Teruaki and Tanaka, Takahiro and Yokoyama, Shuichiro",
    title = "{Primordial black holes\textemdash{}perspectives in gravitational wave astronomy}",
    eprint = "1801.05235",
    archivePrefix = "arXiv",
    primaryClass = "astro-ph.CO",
    doi = "10.1088/1361-6382/aaa7b4",
    journal = "Class. Quant. Grav.",
    volume = "35",
    number = "6",
    pages = "063001",
    year = "2018"
}

@article{Chapline:1975ojl,
    author = "Chapline, George F.",
    title = "{Cosmological effects of primordial black holes}",
    doi = "10.1038/253251a0",
    journal = "Nature",
    volume = "253",
    number = "5489",
    pages = "251--252",
    year = "1975"
}

@article{Meszaros:1975ef,
    author = "Meszaros, P.",
    title = "{Primeval black holes and galaxy formation}",
    journal = "Astron. Astrophys.",
    volume = "38",
    pages = "5--13",
    year = "1975"
}

@article{Ivanov:1994pa,
    author = "Ivanov, P. and Naselsky, P. and Novikov, I.",
    title = "{Inflation and primordial black holes as dark matter}",
    reportNumber = "NORDITA-94-12-A",
    doi = "10.1103/PhysRevD.50.7173",
    journal = "Phys. Rev. D",
    volume = "50",
    pages = "7173--7178",
    year = "1994"
}

@article{Carr:2016drx,
    author = "Carr, Bernard and Kuhnel, Florian and Sandstad, Marit",
    title = "{Primordial Black Holes as Dark Matter}",
    eprint = "1607.06077",
    archivePrefix = "arXiv",
    primaryClass = "astro-ph.CO",
    reportNumber = "NORDITA-2016-83",
    doi = "10.1103/PhysRevD.94.083504",
    journal = "Phys. Rev. D",
    volume = "94",
    number = "8",
    pages = "083504",
    year = "2016"
}

@article{Green:2020jor,
    author = "Green, Anne M. and Kavanagh, Bradley J.",
    title = "{Primordial Black Holes as a dark matter candidate}",
    eprint = "2007.10722",
    archivePrefix = "arXiv",
    primaryClass = "astro-ph.CO",
    doi = "10.1088/1361-6471/abc534",
    journal = "J. Phys. G",
    volume = "48",
    number = "4",
    pages = "043001",
    year = "2021"
}

@article{Carr:2020xqk,
    author = "Carr, Bernard and Kuhnel, Florian",
    title = "{Primordial Black Holes as Dark Matter: Recent Developments}",
    eprint = "2006.02838",
    archivePrefix = "arXiv",
    primaryClass = "astro-ph.CO",
    doi = "10.1146/annurev-nucl-050520-125911",
    journal = "Ann. Rev. Nucl. Part. Sci.",
    volume = "70",
    pages = "355--394",
    year = "2020"
}

@inproceedings{Green:2024bam,
    author = "Green, Anne M.",
    title = "{Primordial Black Holes as a dark matter candidate -- a brief overview}",
    eprint = "2402.15211",
    archivePrefix = "arXiv",
    primaryClass = "astro-ph.CO",
    month = "2",
    year = "2024"
}

@article{Escriva:2022duf,
    author = "Escriv\`a, Albert and Kuhnel, Florian and Tada, Yuichiro",
    title = "{Primordial Black Holes}",
    eprint = "2211.05767",
    archivePrefix = "arXiv",
    primaryClass = "astro-ph.CO",
    month = "11",
    year = "2022"
}

@article{Starobinsky:1986fx,
    author = "Starobinsky, Alexei A.",
    title = "{STOCHASTIC DE SITTER (INFLATIONARY) STAGE IN THE EARLY UNIVERSE}",
    doi = "10.1007/3-540-16452-9_6",
    journal = "Lect. Notes Phys.",
    volume = "246",
    pages = "107--126",
    year = "1986"
}

@article{Sasaki:1995aw,
    author = "Sasaki, Misao and Stewart, Ewan D.",
    title = "{A General analytic formula for the spectral index of the density perturbations produced during inflation}",
    eprint = "astro-ph/9507001",
    archivePrefix = "arXiv",
    reportNumber = "LANCS-TH-9504, OU-TAP-22",
    doi = "10.1143/PTP.95.71",
    journal = "Prog. Theor. Phys.",
    volume = "95",
    pages = "71--78",
    year = "1996"
}

@article{Lyth:2005fi,
    author = "Lyth, David H. and Rodriguez, Yeinzon",
    title = "{The Inflationary prediction for primordial non-Gaussianity}",
    eprint = "astro-ph/0504045",
    archivePrefix = "arXiv",
    doi = "10.1103/PhysRevLett.95.121302",
    journal = "Phys. Rev. Lett.",
    volume = "95",
    pages = "121302",
    year = "2005"
}

@article{Grain:2017dqa,
    author = "Grain, Julien and Vennin, Vincent",
    title = "{Stochastic inflation in phase space: Is slow roll a stochastic attractor?}",
    eprint = "1703.00447",
    archivePrefix = "arXiv",
    primaryClass = "gr-qc",
    doi = "10.1088/1475-7516/2017/05/045",
    journal = "JCAP",
    volume = "05",
    pages = "045",
    year = "2017"
}

@article{Salopek:1990jq,
    author = "Salopek, D. S. and Bond, J. R.",
    title = "{Nonlinear evolution of long wavelength metric fluctuations in inflationary models}",
    reportNumber = "FERMILAB-PUB-90-131-A",
    doi = "10.1103/PhysRevD.42.3936",
    journal = "Phys. Rev. D",
    volume = "42",
    pages = "3936--3962",
    year = "1990"
}

@article{Salopek:1990re,
    author = "Salopek, D. S. and Bond, J. R.",
    title = "{Stochastic inflation and nonlinear gravity}",
    reportNumber = "FERMILAB-PUB-90-167-A",
    doi = "10.1103/PhysRevD.43.1005",
    journal = "Phys. Rev. D",
    volume = "43",
    pages = "1005--1031",
    year = "1991"
}

@article{Starobinsky:1994bd,
    author = "Starobinsky, Alexei A. and Yokoyama, Junichi",
    title = "{Equilibrium state of a selfinteracting scalar field in the De Sitter background}",
    eprint = "astro-ph/9407016",
    archivePrefix = "arXiv",
    reportNumber = "YITP-U-94-12",
    doi = "10.1103/PhysRevD.50.6357",
    journal = "Phys. Rev. D",
    volume = "50",
    pages = "6357--6368",
    year = "1994"
}

@article{Fujita:2013cna,
    author = "Fujita, Tomohiro and Kawasaki, Masahiro and Tada, Yuichiro and Takesako, Tomohiro",
    title = "{A new algorithm for calculating the curvature perturbations in stochastic inflation}",
    eprint = "1308.4754",
    archivePrefix = "arXiv",
    primaryClass = "astro-ph.CO",
    reportNumber = "IPMU-13-0154, ICRR-REPORT-658-2013-7",
    doi = "10.1088/1475-7516/2013/12/036",
    journal = "JCAP",
    volume = "12",
    pages = "036",
    year = "2013"
}

@article{Fujita:2014tja,
    author = "Fujita, Tomohiro and Kawasaki, Masahiro and Tada, Yuichiro",
    title = "{Non-perturbative approach for curvature perturbations in stochastic $\delta N$ formalism}",
    eprint = "1405.2187",
    archivePrefix = "arXiv",
    primaryClass = "astro-ph.CO",
    reportNumber = "ICRR-REPORT-680-2014-6, IPMU-14-0113",
    doi = "10.1088/1475-7516/2014/10/030",
    journal = "JCAP",
    volume = "10",
    pages = "030",
    year = "2014"
}

@article{Vennin:2015hra,
    author = "Vennin, Vincent and Starobinsky, Alexei A.",
    title = "{Correlation Functions in Stochastic Inflation}",
    eprint = "1506.04732",
    archivePrefix = "arXiv",
    primaryClass = "hep-th",
    doi = "10.1140/epjc/s10052-015-3643-y",
    journal = "Eur. Phys. J. C",
    volume = "75",
    pages = "413",
    year = "2015"
}

@article{Cohen:2022clv,
    author = "Cohen, Timothy and Green, Daniel and Premkumar, Akhil",
    title = "{Large deviations in the early Universe}",
    eprint = "2212.02535",
    archivePrefix = "arXiv",
    primaryClass = "hep-th",
    reportNumber = "CERN-TH-2022-206",
    doi = "10.1103/PhysRevD.107.083501",
    journal = "Phys. Rev. D",
    volume = "107",
    number = "8",
    pages = "083501",
    year = "2023"
}

@article{Ferrante:2022mui,
    author = "Ferrante, Giacomo and Franciolini, Gabriele and Iovino, Junior., Antonio and Urbano, Alfredo",
    title = "{Primordial non-Gaussianity up to all orders: Theoretical aspects and implications for primordial black hole models}",
    eprint = "2211.01728",
    archivePrefix = "arXiv",
    primaryClass = "astro-ph.CO",
    doi = "10.1103/PhysRevD.107.043520",
    journal = "Phys. Rev. D",
    volume = "107",
    number = "4",
    pages = "043520",
    year = "2023"
}

@article{Gow:2022jfb,
    author = "Gow, Andrew D. and Assadullahi, Hooshyar and Jackson, Joseph H. P. and Koyama, Kazuya and Vennin, Vincent and Wands, David",
    title = "{Non-perturbative non-Gaussianity and primordial black holes}",
    eprint = "2211.08348",
    archivePrefix = "arXiv",
    primaryClass = "astro-ph.CO",
    doi = "10.1209/0295-5075/acd417",
    journal = "EPL",
    volume = "142",
    number = "4",
    pages = "49001",
    year = "2023"
}

@article{Vennin:2024yzl,
    author = "Vennin, Vincent and Wands, David",
    title = "{Quantum diffusion and large primordial perturbations from inflation}",
    eprint = "2402.12672",
    archivePrefix = "arXiv",
    primaryClass = "astro-ph.CO",
    month = "2",
    year = "2024"
}

@article{Pattison:2017mbe,
    author = "Pattison, Chris and Vennin, Vincent and Assadullahi, Hooshyar and Wands, David",
    title = "{Quantum diffusion during inflation and primordial black holes}",
    eprint = "1707.00537",
    archivePrefix = "arXiv",
    primaryClass = "hep-th",
    doi = "10.1088/1475-7516/2017/10/046",
    journal = "JCAP",
    volume = "10",
    pages = "046",
    year = "2017"
}

@article{Ezquiaga:2019ftu,
    author = "Ezquiaga, Jose Mar\'\i{}a and Garc\'\i{}a-Bellido, Juan and Vennin, Vincent",
    title = "{The exponential tail of inflationary fluctuations: consequences for primordial black holes}",
    eprint = "1912.05399",
    archivePrefix = "arXiv",
    primaryClass = "astro-ph.CO",
    doi = "10.1088/1475-7516/2020/03/029",
    journal = "JCAP",
    volume = "03",
    pages = "029",
    year = "2020"
}

@article{De:2020hdo,
    author = "De, Aritra and Mahbub, Rafid",
    title = "{Numerically modeling stochastic inflation in slow-roll and beyond}",
    eprint = "2010.12685",
    archivePrefix = "arXiv",
    primaryClass = "astro-ph.CO",
    doi = "10.1103/PhysRevD.102.123509",
    journal = "Phys. Rev. D",
    volume = "102",
    number = "12",
    pages = "123509",
    year = "2020"
}

@article{Figueroa:2020jkf,
    author = "Figueroa, Daniel G. and Raatikainen, Sami and Rasanen, Syksy and Tomberg, Eemeli",
    title = "{Non-Gaussian Tail of the Curvature Perturbation in Stochastic Ultraslow-Roll Inflation: Implications for Primordial Black Hole Production}",
    eprint = "2012.06551",
    archivePrefix = "arXiv",
    primaryClass = "astro-ph.CO",
    reportNumber = "HIP-2020-32/TH",
    doi = "10.1103/PhysRevLett.127.101302",
    journal = "Phys. Rev. Lett.",
    volume = "127",
    number = "10",
    pages = "101302",
    year = "2021"
}

@article{Pattison:2021oen,
    author = "Pattison, Chris and Vennin, Vincent and Wands, David and Assadullahi, Hooshyar",
    title = "{Ultra-slow-roll inflation with quantum diffusion}",
    eprint = "2101.05741",
    archivePrefix = "arXiv",
    primaryClass = "astro-ph.CO",
    doi = "10.1088/1475-7516/2021/04/080",
    journal = "JCAP",
    volume = "04",
    pages = "080",
    year = "2021"
}

@article{Linde:1986fd,
    author = "Linde, Andrei D.",
    title = "{Eternally Existing Selfreproducing Chaotic Inflationary Universe}",
    reportNumber = "Print-86-0418 (LEBEDEV INST), LEBEDEV-86-106",
    doi = "10.1016/0370-2693(86)90611-8",
    journal = "Phys. Lett. B",
    volume = "175",
    pages = "395--400",
    year = "1986"
}

@article{Linde:1986fc,
    author = "Linde, Andrei D.",
    title = "{ETERNAL CHAOTIC INFLATION}",
    reportNumber = "Print-86-0417 (LEBEDEV INST), IC/86/74",
    doi = "10.1142/S0217732386000129",
    journal = "Mod. Phys. Lett. A",
    volume = "1",
    pages = "81",
    year = "1986"
}

@article{Linde:2015edk,
    author = "Linde, Andrei",
    title = "{A brief history of the multiverse}",
    eprint = "1512.01203",
    archivePrefix = "arXiv",
    primaryClass = "hep-th",
    doi = "10.1088/1361-6633/aa50e4",
    journal = "Rept. Prog. Phys.",
    volume = "80",
    number = "2",
    pages = "022001",
    year = "2017"
}

@article{Creminelli:2008es,
    author = "Creminelli, Paolo and Dubovsky, Sergei and Nicolis, Alberto and Senatore, Leonardo and Zaldarriaga, Matias",
    title = "{The Phase Transition to Slow-roll Eternal Inflation}",
    eprint = "0802.1067",
    archivePrefix = "arXiv",
    primaryClass = "hep-th",
    reportNumber = "IC-2008-002",
    doi = "10.1088/1126-6708/2008/09/036",
    journal = "JHEP",
    volume = "09",
    pages = "036",
    year = "2008"
}

@article{Rudelius:2019cfh,
    author = "Rudelius, Tom",
    title = "{Conditions for (No) Eternal Inflation}",
    eprint = "1905.05198",
    archivePrefix = "arXiv",
    primaryClass = "hep-th",
    doi = "10.1088/1475-7516/2019/08/009",
    journal = "JCAP",
    volume = "08",
    pages = "009",
    year = "2019"
}

@article{Guth:2007ng,
    author = "Guth, Alan H.",
    editor = "Sola, Joan",
    title = "{Eternal inflation and its implications}",
    eprint = "hep-th/0702178",
    archivePrefix = "arXiv",
    reportNumber = "MIT-CTP-3811",
    doi = "10.1088/1751-8113/40/25/S25",
    journal = "J. Phys. A",
    volume = "40",
    pages = "6811--6826",
    year = "2007"
}

@article{Gong:2016qmq,
    author = "Gong, Jinn-Ouk",
    title = "{Multi-field inflation and cosmological perturbations}",
    eprint = "1606.06971",
    archivePrefix = "arXiv",
    primaryClass = "gr-qc",
    reportNumber = "APCTP-PRE2016-012",
    doi = "10.1142/S021827181740003X",
    journal = "Int. J. Mod. Phys. D",
    volume = "26",
    number = "01",
    pages = "1740003",
    year = "2016"
}

@article{Wands:2007bd,
    author = "Wands, David",
    title = "{Multiple field inflation}",
    eprint = "astro-ph/0702187",
    archivePrefix = "arXiv",
    doi = "10.1007/978-3-540-74353-8_8",
    journal = "Lect. Notes Phys.",
    volume = "738",
    pages = "275--304",
    year = "2008"
}

@article{Hwang:2000jh,
    author = "Hwang, Jai-chan and Noh, Hyerim",
    title = "{Cosmological perturbations with multiple scalar fields}",
    eprint = "astro-ph/0009268",
    archivePrefix = "arXiv",
    reportNumber = "KNU-2000-9",
    doi = "10.1016/S0370-2693(00)01253-3",
    journal = "Phys. Lett. B",
    volume = "495",
    pages = "277--283",
    year = "2000"
}

@article{Hwang:2001fb,
    author = "Hwang, Jai-chan and Noh, Hyerim",
    title = "{Cosmological perturbations with multiple fluids and fields}",
    eprint = "astro-ph/0103244",
    archivePrefix = "arXiv",
    doi = "10.1088/0264-9381/19/3/308",
    journal = "Class. Quant. Grav.",
    volume = "19",
    pages = "527--550",
    year = "2002"
}

@article{Hwang:2007ni,
    author = "Hwang, Jai-chan and Noh, Hyerim",
    title = "{Second-order perturbations of cosmological fluids: Relativistic effects of pressure, multi-component, curvature, and rotation}",
    eprint = "0704.1927",
    archivePrefix = "arXiv",
    primaryClass = "astro-ph",
    doi = "10.1103/PhysRevD.76.103527",
    journal = "Phys. Rev. D",
    volume = "76",
    pages = "103527",
    year = "2007"
}

@article{Hwang:2015prq,
    author = "Hwang, Jai-chan and Noh, Hyerim and Park, Chan-Gyung",
    title = "{Fully non-linear cosmological perturbations of multicomponent fluid and field systems}",
    eprint = "1511.01360",
    archivePrefix = "arXiv",
    primaryClass = "gr-qc",
    doi = "10.1093/mnras/stw1505",
    journal = "Mon. Not. Roy. Astron. Soc.",
    volume = "461",
    number = "3",
    pages = "3239--3258",
    year = "2016"
}

@article{GrootNibbelink:2000vx,
    author = "Groot Nibbelink, S. and van Tent, B. J. W.",
    title = "{Density perturbations arising from multiple field slow roll inflation}",
    eprint = "hep-ph/0011325",
    archivePrefix = "arXiv",
    reportNumber = "SPIN-2000-31, ITP-UU-00-35, NIKHEF-00-038",
    month = "11",
    year = "2000"
}

@article{GrootNibbelink:2001qt,
    author = "Groot Nibbelink, S. and van Tent, B. J. W.",
    title = "{Scalar perturbations during multiple field slow-roll inflation}",
    eprint = "hep-ph/0107272",
    archivePrefix = "arXiv",
    reportNumber = "SPIN-2001-19, ITP-UU-01-27",
    doi = "10.1088/0264-9381/19/4/302",
    journal = "Class. Quant. Grav.",
    volume = "19",
    pages = "613--640",
    year = "2002"
}

@article{Peterson:2010np,
    author = "Peterson, Courtney M. and Tegmark, Max",
    title = "{Testing Two-Field Inflation}",
    eprint = "1005.4056",
    archivePrefix = "arXiv",
    primaryClass = "astro-ph.CO",
    doi = "10.1103/PhysRevD.83.023522",
    journal = "Phys. Rev. D",
    volume = "83",
    pages = "023522",
    year = "2011"
}

@article{Peterson:2011yt,
    author = "Peterson, Courtney M. and Tegmark, Max",
    title = "{Testing multifield inflation: A geometric approach}",
    eprint = "1111.0927",
    archivePrefix = "arXiv",
    primaryClass = "astro-ph.CO",
    doi = "10.1103/PhysRevD.87.103507",
    journal = "Phys. Rev. D",
    volume = "87",
    number = "10",
    pages = "103507",
    year = "2013"
}

@article{Gordon:2000hv,
    author = "Gordon, Christopher and Wands, David and Bassett, Bruce A. and Maartens, Roy",
    title = "{Adiabatic and entropy perturbations from inflation}",
    eprint = "astro-ph/0009131",
    archivePrefix = "arXiv",
    doi = "10.1103/PhysRevD.63.023506",
    journal = "Phys. Rev. D",
    volume = "63",
    pages = "023506",
    year = "2000"
}

@article{Achucarro:2010da,
    author = "Achucarro, Ana and Gong, Jinn-Ouk and Hardeman, Sjoerd and Palma, Gonzalo A. and Patil, Subodh P.",
    title = "{Features of heavy physics in the CMB power spectrum}",
    eprint = "1010.3693",
    archivePrefix = "arXiv",
    primaryClass = "hep-ph",
    reportNumber = "LPTENS-10-36, CPHT-RR-080.0910",
    doi = "10.1088/1475-7516/2011/01/030",
    journal = "JCAP",
    volume = "01",
    pages = "030",
    year = "2011"
}

@article{Elliston:2012ab,
    author = "Elliston, Joseph and Seery, David and Tavakol, Reza",
    title = "{The inflationary bispectrum with curved field-space}",
    eprint = "1208.6011",
    archivePrefix = "arXiv",
    primaryClass = "astro-ph.CO",
    doi = "10.1088/1475-7516/2012/11/060",
    journal = "JCAP",
    volume = "11",
    pages = "060",
    year = "2012"
}

@article{Burgess:2012dz,
    author = "Burgess, C. P. and Horbatsch, M. W. and Patil, Subodh. P.",
    title = "{Inflating in a Trough: Single-Field Effective Theory from Multiple-Field Curved Valleys}",
    eprint = "1209.5701",
    archivePrefix = "arXiv",
    primaryClass = "hep-th",
    reportNumber = "CERN-PH-TH-2012-246, CPHT-RR-021.0512",
    doi = "10.1007/JHEP01(2013)133",
    journal = "JHEP",
    volume = "01",
    pages = "133",
    year = "2013"
}

@article{Kaiser:2013sna,
    author = "Kaiser, David I. and Sfakianakis, Evangelos I.",
    title = "{Multifield Inflation after Planck: The Case for Nonminimal Couplings}",
    eprint = "1304.0363",
    archivePrefix = "arXiv",
    primaryClass = "astro-ph.CO",
    reportNumber = "PREPRINT-MIT-CTP-4451",
    doi = "10.1103/PhysRevLett.112.011302",
    journal = "Phys. Rev. Lett.",
    volume = "112",
    number = "1",
    pages = "011302",
    year = "2014"
}

@article{Schutz:2013fua,
    author = "Schutz, Katelin and Sfakianakis, Evangelos I. and Kaiser, David I.",
    title = "{Multifield Inflation after Planck: Isocurvature Modes from Nonminimal Couplings}",
    eprint = "1310.8285",
    archivePrefix = "arXiv",
    primaryClass = "astro-ph.CO",
    reportNumber = "PREPRINT-MIT-CTP-4509",
    doi = "10.1103/PhysRevD.89.064044",
    journal = "Phys. Rev. D",
    volume = "89",
    number = "6",
    pages = "064044",
    year = "2014"
}

@article{Linde:1993cn,
    author = "Linde, Andrei D.",
    title = "{Hybrid inflation}",
    eprint = "astro-ph/9307002",
    archivePrefix = "arXiv",
    reportNumber = "SU-ITP-93-17",
    doi = "10.1103/PhysRevD.49.748",
    journal = "Phys. Rev. D",
    volume = "49",
    pages = "748--754",
    year = "1994"
}

@article{Berera:1998gx,
    author = "Berera, Arjun and Gleiser, Marcelo and Ramos, Rudnei O.",
    title = "{Strong dissipative behavior in quantum field theory}",
    eprint = "hep-ph/9803394",
    archivePrefix = "arXiv",
    reportNumber = "VAND-TH-98-02, DART-HEP-98-02, IF-UERJ-98-10",
    doi = "10.1103/PhysRevD.58.123508",
    journal = "Phys. Rev. D",
    volume = "58",
    pages = "123508",
    year = "1998"
}

@article{Berera:1998px,
    author = "Berera, Arjun and Gleiser, Marcelo and Ramos, Rudnei O.",
    title = "{A First principles warm inflation model that solves the cosmological horizon / flatness problems}",
    eprint = "hep-ph/9809583",
    archivePrefix = "arXiv",
    reportNumber = "VAND-TH-98-16, DART-HEP-98-04, IF-UERJ-DFT-13-98",
    doi = "10.1103/PhysRevLett.83.264",
    journal = "Phys. Rev. Lett.",
    volume = "83",
    pages = "264--267",
    year = "1999"
}

@article{Berera:2008ar,
    author = "Berera, Arjun and Moss, Ian G. and Ramos, Rudnei O.",
    title = "{Warm Inflation and its Microphysical Basis}",
    eprint = "0808.1855",
    archivePrefix = "arXiv",
    primaryClass = "hep-ph",
    doi = "10.1088/0034-4885/72/2/026901",
    journal = "Rept. Prog. Phys.",
    volume = "72",
    pages = "026901",
    year = "2009"
}

@article{Berera:1995ie,
    author = "Berera, Arjun",
    title = "{Warm inflation}",
    eprint = "astro-ph/9509049",
    archivePrefix = "arXiv",
    reportNumber = "PSU-TH-159",
    doi = "10.1103/PhysRevLett.75.3218",
    journal = "Phys. Rev. Lett.",
    volume = "75",
    pages = "3218--3221",
    year = "1995"
}

\end{document}